\documentclass[preprint,12pt,authoryear]{elsarticle}

\usepackage{amssymb}

\makeatletter
\def\ps@pprintTitle{%
   \let\@oddhead\@empty
   \let\@evenhead\@empty
   \def\@oddfoot{\reset@font\hfil\thepage\hfil}
   \let\@evenfoot\@oddfoot
}
\makeatother

\usepackage{latexsym,bm}
\usepackage{amsmath,amsfonts}
\usepackage[usenames]{color}
\usepackage{graphicx}
\usepackage{url}
\usepackage[markup=default]{changes}
\usepackage{xr}
\newcommand{\beginsupplement}{%
  \setcounter{figure}{0}
  \renewcommand{\thesection}{S\arabic{section}}  
  \renewcommand{\thetable}{S\arabic{table}}  
  \renewcommand{\thefigure}{S\arabic{figure}}
  \renewcommand{\theequation}{S\arabic{equation}}
}

\begin{document}

\begin{frontmatter}



\title{False Discovery Rates to Detect Signals from Incomplete Spatially Aggregated Data}


\author[as]{Hsin-Cheng Huang\corref{cor1}}
  \ead{hchuang@stat.sinica.edu.tw}
\author[uow]{Noel Cressie}
  \ead{ncressie@uow.edu.au}
\author[uow]{Andrew Zammit-Mangion}
  \ead{azm@uow.edu.au}
\author[ut]{Guowen Huang}
  \ead{hgw0610209@gmail.com}
\address[as]{Institute of Statistical Science, Academia Sinica,
  Taiwan, R.O.C.}
\address[uow]{National Institute for Applied Statistics Research Australia,
  University of Wollongong, Australia}
\address[ut]{Department of Statistical Science, University of Toronto, Canada}

\cortext[cor1]{Corresponding author}

\date{}

\begin{abstract}
\baselineskip=20pt
There are a number of ways to test for the absence/presence of a spatial signal in a completely observed fine-resolution image.
One of these is a powerful nonparametric procedure called Enhanced False Discovery Rate (EFDR).
A drawback of EFDR is that it requires the data to be defined on regular pixels in a rectangular spatial domain.
Here, we develop an EFDR procedure for possibly incomplete data defined on irregular small areas.
Motivated by statistical learning, we
use conditional simulation (CS) to condition on the available data and simulate the full rectangular image at its finest resolution many times ($M$, say).
EFDR is then applied to each of these simulations resulting in $M$ estimates of the signal and $M$ statistically dependent $p$-values.
Averaging over these estimates yields a single, combined estimate of a possible signal,
but inference is needed to determine whether there really is a signal present.
We test the original null hypothesis of no signal by combining the $M$ $p$-values into a single $p$-value using copulas and a composite likelihood.
If the null hypothesis of no signal is rejected, we use the combined estimate.
We call this new procedure EFDR-CS and, to demonstrate its effectiveness, we show results from a simulation study;
an experiment where we introduce aggregation and incompleteness into temperature-change data in the Asia-Pacific;
and an application to total-column carbon dioxide from satellite remote sensing data over a region of the Middle East,
Afghanistan, and the western part of Pakistan.
\bigskip
\end{abstract}


\end{frontmatter}

\baselineskip=24pt

\section{Introduction}

Spatial statistical data have been classified as geostatistical, lattice, or point pattern (Cressie, 1993).
Here, our interest is in detecting a spatial signal from irregular lattice data, sometimes called small area data,
which we consider to be the result of aggregation of pixel values of a fine-resolution image
where it is possible that not all these values are included in the aggregation.
We emphasize that the fine-resolution pixel values are not observed, only the irregular lattice data.
We formalize this below.

\subsection{Statistical learning and inference for a spatial signal}

We consider first a signal-detection problem in a standard rectangular image.
That is, $D$ is a rectangular lattice of locations defined on $n=n_1\times n_2$ nonoverlapping, fine-resolution
areas (or pixels) $\{A_j:j=1,\dots,n\}$ with area $h_x h_y$,
so $D=\{(i_1 h_x, i_2 h_y)': i_1=1,\dots,n_1;\,i_2=1,\dots,n_2\} \equiv\{\bm{s}_j:j=1,\dots,n\}$.
Then the $j$-th pixel value is $Z(\bm{s}_j)$,
and the fine-resolution image values are denoted as $\bm{Z}\equiv(Z(\bm{s}_1),\dots,Z(\bm{s}_n))'$.
If $\bm{Z}$ were observed, this would result in regular-lattice data.
In a geostatistical context, where there is an underlying continuously indexed spatial process,
the $\{A_j:j=1,\dots,n\}$ have been called Basic Areal Units (BAUs); see Nguyen \textit{et al.}~(2012).

In what follows, we write
\begin{equation}
\bm{Z}=\bm{\mu}+\bm{\delta},
\label{eq:data2}
\end{equation}

\noindent where $\bm{\delta}\sim\mathrm{Gau}(\bm{0},\bm{\Sigma})$ with
``Gau" denoting an $n$-variate Gaussian distribution,
and we are interested in detecting if there is a spatial signal in the image's mean vector $\bm{\mu}$.
We model spatial statistical dependence in the noise term $\bm{\delta}$ through a spatial covariance function $C(\cdot)$.
For example, suppose we are comparing two noisy images and want a way to declare
whether the images have the same underlying signal or not.
This problem can be formulated in terms of a hypothesis test, where
\begin{equation}
H_0:\bm{\mu}=\bm{0}~~\mbox{versus}~~H_1:\bm{\mu}\neq\bm{0}
\label{eq:hypothesis}
\end{equation}

\noindent is tested, and the regular-lattice data $\bm{Z}$ is defined to be the pixel-wise difference between the two images.

If the data vector $\bm{Z}$ at the pixel resolution is completely observed,
then a powerful nonparametric hypothesis-testing method based on the false discovery rate (FDR),
called the enhanced FDR (EFDR) procedure (Shen \textit{et al}., 2002),
can be applied to test $H_0$: $\bm{\mu}=\bm{\mu}_0$ and, if it is rejected, to estimate the spatial signal, $\bm{\mu}$.
The EFDR procedure is described in more detail in Section \ref{sec:method}.

Benjamini and Heller (2007) introduced another procedure for estimating the spatial signal using the FDR when
repeated measurements of $\bm{Z}$ are available and, in more recent literature,
Martinez \textit{et al.}~(2013) considered a similar testing problem
with $\bm{Z}$ corresponding to a completely observed two-dimensional image of a moving-window spectrogram.
They took a Bayesian approach and generated posterior samples to control a Bayesian FDR.
Sun \text{et al}.~(2015)
and Risser \textit{et al.}~(2019)
developed procedures to test $H_0$: $\mu_j=\mu_0$ (or $\mu_j\leq \mu_0$), for $j=1,\dots,n$,
that do not require $\bm{Z}$ to be completely observed.
The former paper relies on a posterior sample of $\bm{\mu}$ from Bayesian modeling, and it is sensitive to model misspecification.
The latter paper applies a hierarchical Bayesian model and controls the FDR in a Bayesian decision-theoretical framework
when repeated measurements of $\bm{Z}$ are available.

The hypothesis-testing method developed by Hering and Genton (2011) tests an average (over a spatial domain $D$) effect
of $\bm{\mu}-\bm{\mu}_0$, and it does not rely directly on a Gaussian model for $\bm{\delta}$.
The trade-off taken by the authors to achieve a valid procedure is to integrate out ``space." When $H_0$ is rejected
their procedure provides no local information about where the spatial signal might be.
Gilleland~(2013) used it to test competing weather forecasts, and he provides software for it in his R package ``SpatialVx".
Yun \textit{et al.}~(2018) considered testing the equality of the spatial means (or spatial covariances) 
between two spatio-temporal random fields.
While nonparametric in nature, their approach requires $p$-values to be available at individual locations based on data observed at multiple time points.
Lei \textit{et al.}~(2017) developed a sequential testing procedure by gradually pruning a candidate rejection set,
which can be applied to identify some spatial signal.
None of the papers reviewed above is able to address the change-of-support problem that is central to our research,
and which we describe in Section \ref{sec:model}.

We consider the general problem where possibly coarser-resolution irregular-lattice data are observed:
\begin{equation}
Z(B_k)=\frac{\displaystyle\sum_{\bm{s}}Z(\bm{s})I(\bm{s}\in D\cap B_k)}{\displaystyle\sum_{\bm{s}}I(\bm{s}\in D\cap B_k)};\quad k=1,\dots,K,
\label{eq:block data}
\end{equation}

\noindent where $B_k$ is made up of one or more pixels $\{A_j\}$,
and we wish to make inference on the spatial signal $\bm{\mu}=(\mu_1,\cdots,\mu_n)'$ at the finest resolution.
In what follows, we allow for a general type of coarsening where there might be some overlap of the $\{B_k\}$,
or where there is no coarsening but not every pixel in $\{A_j:j=1,\dots,n\}$ is included.
That is, our approach can handle situations where
$B_k\cap B_\ell\neq\emptyset$ for some $k\neq\ell$ and $\displaystyle\cup_{k=1}^K B_k\subsetneqq D$.

\subsection{The spatial statistical model}
\label{sec:model}

Consider a spatial Gaussian process, $\{Z(\bm{s}):\bm{s}\in D\subset\mathbb{R}^d\}$,
defined on a finite regular lattice of locations $D$, which are in $d$-dimensional Euclidean space $\mathbb{R}^d$. From \eqref{eq:data2},
\begin{equation}
Z(\bm{s})=\mu(\bm{s})+\delta(\bm{s});\quad \bm{s}\in D,
\label{eq:signal+noise}
\end{equation}

\noindent where $\mu(\cdot)$ is a deterministic mean function,
$\delta(\cdot)$ is a zero-mean stationary Gaussian process with a covariance function,
$C(\bm{u})\equiv \mathrm{cov}(\delta(\bm{s}),\delta(\bm{s}+\bm{u}))=\mathrm{cov}(Z(\bm{s}),Z(\bm{s}+\bm{u}))$.
In what follows, we consider the two-dimensional Euclidean space where $d=2$,
although our approach is general and applies to any $d\in\{1,2,3,\dots\}$.

The spatial statistical model \eqref{eq:signal+noise} can be written as,
\begin{equation}
\bm{Z}\sim\mathrm{Gau}(\bm{\mu},\bm{\Sigma}),
\label{eq:data}
\end{equation}

\noindent where $\bm{\mu}\equiv(\mu(\bm{s}_1),\dots,\mu(\bm{s}_n))' \equiv(\mu_1,\dots,\mu_n)'$,
$\bm{\Sigma}\equiv\mathrm{cov}(\bm{Z})=\mathrm{cov}(\bm{\delta})$, and
the $(i,j)$-th element of $\bm{\Sigma}$ is $C(\bm{s}_j-\bm{s}_i)$, which is generally non-zero.
By applying the linear operator \eqref{eq:block data} to the model \eqref{eq:data}, we obtain
the aggregated data vector $\bm{\tilde{Z}}\,\equiv\,(Z(B_1),\dots,$ $Z(B_K))'$. Hence, $\tilde{\bm{Z}}$ can be written as:
\begin{equation}
\tilde{\bm{Z}}=\bm{HZ} \sim\mathrm{Gau}(\bm{H}\bm{\mu},\bm{H\Sigma H}'),
\label{eq:aggregation}
\end{equation}

\noindent for some known $K\times n$ matrix $\bm{H}$ that represents the spatial averaging in \eqref{eq:block data}.
Note that $K$, the dimension of $\tilde{\bm{Z}}$, is usually smaller than $n$ but, in the case of overlapping $\{B_k\}$, it could be larger than $n$.
Importantly, the mean in \eqref{eq:aggregation} is $\bm{H}\bm{\mu}$,
although we still wish to make inference on $\bm{\mu}$ at the finest resolution.
Specifically, we wish to estimate $\bm{\mu}$ from $\tilde{\bm{Z}}$ and,
to see whether it is estimating something non-zero, we carry out inference by testing $H_0$ versus $H_1$ given by \eqref{eq:hypothesis}.
Examples of $\tilde{\bm{Z}}$ in \eqref{eq:aggregation} are many:
the $\{B_k\}$ that define $\tilde{\bm{Z}}$ might correspond to provinces/states in a country, or counties in a state,
or blocks of an image at a coarser resolution than the image's native resolution,
or the observed areas/pixels in an image with missing data
(e.g., an image of Earth's surface partially obscured by cloud).

The aggregation matrix $\bm{H}$ can also be written as $\bm{H}=\bm{\Lambda}^{-1}\bm{H}^*$,
where $\bm{H}^*$ is a $K \times n$ incidence matrix that describes the aggregation relationship between $\bm{Z}$ and $\bm{\tilde{Z}}$,
and $\bm{\Lambda}$ is a $K\times K$ diagonal matrix with its $k$-th diagonal element equal to the number of pixels in $B_k$, for $k=1,\dots,K$.
If the original image $\{A_j\}$ has some pixels not observed (i.e., missing),
then the original pixels that are observed can be represented as $\{B_1,\dots,B_K\}$ made up of $K<n$ distinct pixels from $\{A_j\}$.
In that case, $\tilde{\bm{Z}}$ is a sub-vector of $\bm{Z}$,
$\bm{H}$ is a sub-matrix of the $n$-dimensional identity matrix, and $\bm{\Lambda}$ is the $K\times K$ identity matrix.

We now summarize the organization of our paper.
In Section~\ref{sec:method}, we introduce our proposed signal-detection procedure EFDR-CS, which includes conditional simulation (CS), EFDR,
and the combining of dependent $p$-values to test for spatial signal.
Section~\ref{sec:simulation} gives a simulation study with results that demonstrate the validity and relative efficiency of EFDR-CS
to the problem of signal detection from incomplete spatially aggregated data.
In the first part of Section~\ref{sec:application}, we apply the EFDR-CS procedure to a temperature dataset over the Asia-Pacific region
generated by a climate model from the National Center for Atmospheric Research (NCAR),
in an experiment where we purposely introduce both aggregation and incompleteness into the data.
In the second part of Section \ref{sec:application},
remote sensing data from NASA's Orbiting Carbon Observatory-2 satellite is used to illustrate EFDR-CS
inference for total-column carbon dioxide over a region of the Middle East,
Afghanistan, and the western part of Pakistan.
Finally, discussion and conclusions are given in Section~\ref{sec:discussion}.
Additional material is given in the on-line Supplementary Material.


\section{Inferring spatial signal from data on an irregular lattice}
\label{sec:method}
\bigskip

\subsection{The EFDR procedure}

When all the data, $\bm{Z}$, in \eqref{eq:data} are available,
we can apply the EFDR procedure (Shen \textit{et al.}, 2002) to test
$H_0$: $\bm{\mu}=\bm{\mu}_0$ versus $H_1$: $\bm{\mu}\neq\bm{\mu}_0$.
Since $\bm{\mu}_0$ is specified,
without loss of generality we can assume that the null hypothesis is $H_0:\bm{\mu}=\bm{0}$.
Then the EFDR procedure is performed in four steps.

First, $\bm{Z}$ is transformed into a vector of wavelet coefficients by applying an orthogonal discrete wavelet transform,
\[
\bm{\nu}=\mathcal{W}\bm{Z}=\mathcal{W}\bm{\mu}+\mathcal{W}\bm{\delta},
\]
where $\mathcal{W}$ is a known $n\times n$ orthogonal discrete-wavelet-transform (DWT) matrix (see Daubechies, 1992),
and $\bm{\delta}$ is given in \eqref{eq:data2}.
The wavelet coefficients of the noise, $\mathcal{W}\bm{\delta}$, can be written as
\begin{equation}
\mathcal{W}\bm{\delta}=(\bm{d}'_{-1},\dots,\bm{d}'_{-J}, \bm{c}'_{-J})'
\label{eq:dwt delta}
\end{equation}

\noindent where, for $j=-1,\dots,-J$, the wavelet coefficients at the $j$-th scale are
\[
\bm{d}_{j}\,\equiv\,\big((\bm{d}^{(1)}_j)',(\bm{d}^{(2)}_j)',(\bm{d}^{(3)}_j)'\big)',\quad\mbox{and}\quad
\bm{d}^{(m)}_j\,\equiv\,\big(d^{(m)}_{j,k_1,k_2}:k_1=1,\dots,n_1 2^{j},\,k_2=1,\dots,n_2 2^{j}\big)',
\]
$n_1$ and $n_2$ are powers of two, and $\log_2(\min(n_1,n_2))\geq J$.
Each component, $m=1,2,3$, corresponds to the horizontal, vertical, and diagonal spatial orientations, respectively;
and $\bm{c}_{-J} \,\equiv\,\{c_{-J,k_1,k_2}:k_1=1, \dots,n_1 2^{-J},\,k_2=1,\dots,n_2 2^{-J}\}$
comprise the scaling-function coefficients.
In ``wavelet space," the signal can be identified more easily,
since typically it has a sparse wavelet representation
(i.e., only a few components of $\mathcal{W}\bm{\mu}$ are non-zero), and the error has been decorrelated.

Second, by utilizing this property that
the elements of $\mathcal{W}\bm{\delta}$ tend to be uncorrelated and have a homogeneous variance within each wavelet scale/orientation
(Shen \textit{et al.}, 2002),
$\bm{\nu}=\mathcal{W}\bm{Z}$ is standardized by scale/orientation.
Under $H_0:\bm{\mu}=\bm{0}$,
the resulting standardized coefficients are assumed to be independent and identically distributed $\mathrm{Gau}(0,1)$ random variables.

Third, both statistical and computational efficiencies are increased by reducing the number of tests on the wavelet coefficients.
This is achieved by ordering all the individual wavelet-coefficient hypotheses using the network structure of wavelets
and then selecting the number of hypotheses based on the generalized-degrees-of-freedom criterion of Ye (1998).

Fourth, the false-discovery-rate (FDR) procedure of Benjamini and Hochberg (1995) is applied to the selected wavelet coefficients
to obtain a $p$-value for the hypothesis test of $H_0$ and an estimate of the spatial signal $\bm{\mu}$ through the inverse DWT.
If the $p$-value is larger than a pre-specified level $\alpha$, it is concluded that $\bm{\mu}=\bm{0}$.
Here, the $p$-value is interpreted as the strictest level of FDR control with at least one null hypothesis rejected
and, in Yekutieli and Benjamini (1999), it is called the smallest FDR-adjusted $p$-value.
Notice that the EFDR procedure controls the FDR (at level $\alpha$) with tests on the multiple coefficients in the wavelet space;
EFDR also controls the Type-I error of testing a global null hypothesis of $H_0:\bm{\mu} =\bm{0}$ at the same level $\alpha$.
Importantly, we do not claim to control the FDR of the individual hypothesis tests that constitute \eqref{eq:hypothesis}.
\vspace{0.5cm}

\subsection{Testing for signal in conditionally simulated images}
\label{sec:testing}

Our goal is to test $H_0$: $\bm{\mu}=\bm{0}$, based on the model \eqref{eq:data},
however we only observe $\tilde{\bm{Z}}$ in \eqref{eq:aggregation}, not $\bm{Z}$.
Our methodology is based on conditionally simulating $\bm{Z}$, conditional on $\tilde{\bm{Z}}$,
which takes into account the spatial dependence given by $\bm{\Sigma}=\mathrm{var}(\bm{Z})$ in \eqref{eq:data}.
This approach is very similar to multiple imputation that has been developed in a non-spatial context (Little and Rubin, 2002).
Henceforth, we write $\bm{\Sigma}$ as $\bm{\Sigma}(\bm{\theta})$,
where 
\begin{equation}
\bm{\theta}\,\equiv\,(\theta_1,\dots,\theta_{3J+1})'
\label{eq:theta}
\end{equation}

\noindent parameterizes the individual variances of $\mathcal{W}\bm{\delta}=(\bm{d}'_{-1},\dots,\bm{d}'_{-J}, \bm{c}'_{-J})'$ with
\[
\bm{d}_{-j}^{(m)}\sim \mathrm{Gau}(\bm{0},\theta_{3(j-1)+m}\bm{I});\,m=1,2,3;\,j=1,\dots,J,\quad\mbox{and}\quad
\bm{c}_{-J}\sim \mathrm{Gau}(\bm{0},\theta_{3J+1}\bm{I}).
\]
Details of the estimation procedure in our methodology are given in Section \ref{sec:estimation} of the Supplementary Material.

Once the spatial covariance parameters have been estimated, we are ready to apply
our methodology to the very general problem of detecting spatial signal.
Our new procedure, EFDR-CS, consists of the following steps.
First, we simulate $M$ times the $n$-dimensional vector $\bm{Z}$ \textit{conditional on the data} $\bm{\tilde{Z}}$
(with $\hat{\bm{\theta}}$ substituted in for $\bm{\theta}$), via
\begin{equation}
\bm{Z}\big|\tilde{\bm{Z}}\sim \mathrm{Gau}\big(\bm{\Sigma}(\hat{\bm{\theta}})\bm{H}'\big(\bm{H}\bm{\Sigma}(\hat{\bm{\theta}})\bm{H}'\big)^{-1}\bm{\tilde{Z}},\,
\bm{\Sigma}(\hat{\bm{\theta}})-\bm{\Sigma}(\hat{\bm{\theta}})\bm{H}'\big(\bm{H}\bm{\Sigma}(\hat{\bm{\theta}})\bm{H}'\big)^{-1}
\bm{H}\bm{\Sigma}(\hat{\bm{\theta}})\big),
\label{eq:conditional simulation}
\end{equation} 

\noindent resulting in the $M$ simulated outcomes, $\bm{Z}_1,\dots,\bm{Z}_M$. Then we
apply the EFDR procedure to each $\bm{Z}_1,\dots,\bm{Z}_M$ separately,
from which we obtain corresponding $p$-values, $p_1,\dots,p_M$, and estimates, $\hat{\bm{\mu}}_1,\dots,\hat{\bm{\mu}}_M$, of $\bm{\mu}$.

In our implementation of EFDR-CS, we estimate $\bm{\mu}$ with
\begin{equation}
\hat{\bm{\mu}}\,\equiv\,\sum_{i=1}^M\hat{\bm{\mu}}_i\big/M,
\label{eq:muhat}
\end{equation}

\noindent where $\{\hat{\bm{\mu}}_i:i=1,\dots,M\}$ are given by the EFDR procedure of Shen \textit{et al.}~(2002) applied to
each of the $M$ conditional simulations.
Now $\hat{\bm{\mu}}$ is an estimate of a spatial signal $\bm{\mu}$ that may in fact be zero, so inference on $\bm{\mu}$ is needed.

We now show how $\hat{\bm{\mu}}$ can be accompanied by a $p$-value that allows one to infer whether $\bm{\mu}=\bm{0}$ or not.
We combine $\{p_i:i=1,\dots,M\}$ into a single $p$-value,
although doing so is not straightforward since the $M$ $p$-values are statistically dependent,
each being a function of $\tilde{\bm{Z}}$.
Even if they were independent,
the na\"{i}ve approach of taking the sample mean of $\{p_i:i=1,\dots,M\}$ tends to produce a $p$-value for testing \eqref{eq:hypothesis} that is too large (Brown, 1975).
For independent $\{p_i\}$, Fisher (1925) proposed using the test statistic,
\begin{equation}
T\,\equiv\,-2\displaystyle\sum_{i=1}^M\log p_i,
\label{eq:Brown}
\end{equation}

\noindent to test $H_0$. Brown (1975) used the same test statistic for dependent $\{p_i\}$
obtained from multiple one-sided location tests in a multivariate Gaussian setting with known covariance matrix.
In the next subsection, we develop new distribution theory for $T$ to account for the special dependence between the $\{p_i\}$
that is a consequence of the CS.

Finally, a single $p$-value is obtained with regard to $T$ and its distribution, from which $H_0$ is tested.
A succinct summary of these steps is given in Section \ref{sec:our method}.

\subsection{Distribution theory for combining dependent $p$-values}

From \eqref{eq:Brown}, we write $T=\displaystyle\sum_{i=1}^M t_i$, where $t_i\equiv -2\log p_i$, and
we use the flexible Gamma family of distributions to approximate the distribution of $T$.
That is, we fit $T$ to a $\Gamma(a,b)$ distribution whose probability density function is
$f(x)=\displaystyle\frac{b^a}{\Gamma(a)}x^{a-1}\exp(-bx)$, for $x\geq 0$, and $0$ for $x<0$,
where our proposed methodology determines $a$ and $b$.
Under $H_0$, the marginal distribution of $t_i$ is $\Gamma(1,1/2)$,
which is a chi-squared distribution on $2$ degrees of freedom (e.g., Littell and Folks, 1971).
If $\{t_i\}$ were independent, then $T\sim\Gamma(M,1/2)$, so that $a=M$ and $b=1/2$
(resulting in Fisher's combined probability test).
In our case, $\{t_i\}$ are not independent, which leads to the need for estimates of
the Gamma parameters $a$ and $b$.

The dependence in $\{p_i\}$ is caused by dependence between the replicates from the conditional simulation:
Each $t_i$ depends on the original $K$-dimensional data vector $\bm{\tilde{Z}}$,
and hence they are not independent.
However, they are exchangeable (e.g., Section 3.17 of Spiegelhalter \textit{et al}., 2004),
and hence $\mathrm{cov}(t_i,t_j)=\sigma^2\rho$, for $i\neq j$.
We call $\rho$ the level of exchangeability, and
\begin{equation}
\bm{U}\,\equiv\,\mathrm{cov}((t_1,\dots,t_M)')\,=\,\sigma^2
\left(
\begin{matrix}
1 & \rho & \cdots & \rho\\
\rho & \ddots & \ddots & \vdots\\
\vdots & \ddots & \ddots & \rho\\
\rho & \cdots & \rho & 1
\end{matrix}
\right),
\label{eq:IC}
\end{equation}

\noindent which is a matrix of constant intra-class correlations.
Note that $\bm{U}\bm{1}=\sigma^2(1+(M-1)\rho)\bm{1}$,
which implies that an eigenvalue of $\bm{U}$ is $\sigma^2(1+(M-1)\rho)$.
The covariance matrix $\bm{U}$ is known to have only two eigenvalues with the second one being $\sigma^2(1-\rho)$
(e.g., see Example 3.9 of Schott, 2017).
Since $\bm{U}$ is nonnegative-definite, the eigenvalues must be nonnegative,
and hence $-1/(M-1)\leq\rho\leq 1$.
Since $\rho$ does not depend on $M$, we do not wish the parameter space to depend on $M$, and hence the parameter space for $\rho$ is
\begin{equation}
0\leq\rho<1,
\label{eq:rho}
\end{equation}

\noindent in which \eqref{eq:IC} is positive-definite.

Recall that $T=\displaystyle\sum_{i=1}^M t_i$.
Then under $H_0$, the first moment is $\mathrm{E}(T)=\displaystyle\sum_{i=1}^M\mathrm{E}(t_i)=2M$,
and the second central moment is,
\begin{align*}
  \mathrm{var}(T)
\,=\, \sum_{i=1}^M\mathrm{var}(t_i)+\sum_{1\leq i\neq j\leq M}\mathrm{cov}(t_i,t_j)
\,=\, 4M\{1+(M-1)\rho\},
\end{align*}
    
\noindent since $\sigma^2\equiv\mathrm{var}(t_i)=4$.
These relations for $\mathrm{E}(T)$ and $\mathrm{var}(T)$,
together with $\mathrm{E}(V)=a/b$ and $\mathrm{var}(V)=a/b^2$, where $V\sim\Gamma(a,b)$,
give the following estimating equations for the Gamma parameters $a$ and $b$:
\begin{align}
  a
=&~ (\mathrm{E}(T))^2/\mathrm{var}(T)=M/(1+(M-1)\rho),
\label{eq:a}\\
  b
=&~ \mathrm{E}(T)/\mathrm{var}(T)=1/\{2(1+(M-1)\rho)\}.
\label{eq:b}
\end{align}

\noindent In practice, $\rho$ is not known. We discuss its estimation in the next subsection.

\subsection{Estimation of the level of exchangeability, $\rho$}
\label{sec:exchangeability}

From \eqref{eq:a}, \eqref{eq:b}, and for $\rho$ given, the distribution of $T$ can be fitted to a $\Gamma(a,b)$ distribution.
In this subsection, we present a method for estimating $\rho\in[0,1)$ based on bivariate copulas.
We use copulas because we know that $t_i\geq 0$ has a marginal distribution that is exactly $\chi_2^2$
(an exponential distribution with rate parameter $2$)
for all $i=1,\dots,M$.
Therefore, the cumulative distribution function of $t_i$ is $P(t_i\leq x)=F(x)=1-\exp(-x/2)$, for $x\geq 0$, and $=\,0$ for $x<0$.
For $i\neq j$, we use Gaussian copulas to model the bivariate distribution of $(t_i,t_j)$.
That is, $(t_i,t_j)$ is modeled as a bivariate exponential distribution with cumulative distribution function,
\begin{equation}
P(t_i\leq x_1,t_j\leq x_2)\,=\,G(F(x_1),F(x_2);r)\,\equiv\,F_2(x_1,x_2;r)
\label{eq:copula}
\end{equation}

\noindent where, for $u_1=F(x_1)$ and $u_2=F(x_2)$,
$G(u_1,u_2;r)$ is the bivariate Gaussian copula generated by a bivariate standard Gaussian distribution
with correlation $r\in[0,1)$ (Song, 2000).

The probability density function of $G(u_1,u_2;r)$ is given by
\[
g(u_1,u_2;r)=\frac{1}{\sqrt{1-r^2}}\exp\bigg\{-\frac{1}{2}(\Phi^{-1}(u_1),\Phi^{-1}(u_2))
\bigg(\bigg(
\begin{matrix}
1 & r\\
r & 1
\end{matrix}
\bigg)^{-1}-\,\bm{I}\bigg)\bigg(
\begin{matrix}
\Phi^{-1}(u_1)\\
\Phi^{-1}(u_2)
\end{matrix}
\bigg)
\bigg\};~u_1,u_2\in[0,1],
\]
where $\Phi$ is the cumulative distribution function of the standard $\mathrm{Gau}(0,1)$ distribution.
It follows that $\rho=\mathrm{corr}(t_i,t_j)=\rho(r)$ is a function of the Gaussian-based correlation $r$, where recall that $\rho\in[0,1)$.

Let $f_2(x_1,x_2;r)$ be the resulting bivariate exponential probability density function obtained from \eqref{eq:copula}.
Since $\mathrm{E}(t_i)=2$ and $\mathrm{var}(t_i)=4$, for $i=1,\dots,M$, we have
\[
\mathrm{corr}(t_i,t_j)=\,\rho(r)=\frac{1}{4}\int\!\!\int x_1 x_2 f_2(x_1,x_2;r)\mathrm{d}x_1\mathrm{d}x_2-1.
\]
In our procedure, we estimate $r$ by maximizing the composite likelihood function,
\[
L(r)\,\equiv\,\prod_{1\leq i<j\leq M}f_2(t_i,t_j;r),
\]
where recall that $t_i=-2\log p_i$; $i=1,\dots,M$.
Denote this estimator by $\hat{r}$, and hence denote the maximum composite likelihood (MCL) estimator of $\rho$ as $\rho(\hat{r})$.
We use simulation to obtain $\rho(\hat{r})$; that is, we sample from \eqref{eq:copula}
with $\hat{r}$ in place of $r$ and, from the simulations, we compute the sample correlation, which we denote by $\hat{\rho}$.
If $\hat{\rho}$ lies outside the parameter space $[0,1)$ given by \eqref{eq:rho}, we put it equal to the nearest value in the parameter space.
Then we obtain the following estimates of the Gamma parameters, $a$ and $b$:
\begin{equation}
\hat{a}=M/(1+(M-1)\hat{\rho})\quad\mbox{and}\quad\hat{b}=1/\{2(1+(M-1)\hat{\rho})\}.
\label{eq:hat_ab}
\end{equation}

Other estimates of $a$ and $b$ are possible. For example,
there is a simple method-of-moments estimate that can be used, which we now present. Under $H_0$,
$\sum_{i=1}^M(t_i-2)^2\big/M$ is an unbiased estimator of $\sigma^2$, and
$\sum_{1\leq i<j\leq M}(t_i-t_j)^2\big/\{M(M-1)/2\}$ is an unbiased estimator of $2\sigma^2(1-\rho)$.
Then a method-of-moments estimator is:
\begin{equation}
\tilde{\rho}=1-\displaystyle\frac{\sum_{1\leq i<j\leq M}(t_i-t_j)^2/(M-1)}{\sum_{i=1}^M(t_i-2)^2}.
\label{eq:rho tilde}
\end{equation}

\noindent Again, if $\tilde{\rho}$ lies outside the parameter space $[0,1)$, we put it equal to the nearest value in the parameter space.
Upon substituting $\tilde{\rho}$ for $\rho$ in \eqref{eq:a} and \eqref{eq:b}, we obtain
\begin{equation}
\tilde{a}\,=\,M\big/\big(1+(M-1)\tilde{\rho}\big))\quad\mbox{and}\quad\tilde{b}\,=\,1\big/\big\{2(1+(M-1)\tilde{\rho})\big\}.
\label{eq:tilde_ab}
\end{equation}

\subsection{A final $p$-value to infer the presence of the spatial signal}
\label{sec:pvalue}

Recall that our goal is to detect signals from incomplete spatially aggregated data $\tilde{\bm{Z}}$.
The final $p$-value of our EFDR-CS procedure for testing \eqref{eq:hypothesis} depends on $a$ and $b$ through
\begin{equation}
p=1-F_{\Gamma(a,b)}(T),
\label{eq:pvalue}
\end{equation}
    
\noindent where $T$ is given by \eqref{eq:Brown} and
$F_{\Gamma(a,b)}$ is the cumulative distribution function of a $\Gamma(a,b)$ random variable.
Hence, once $a$ and $b$ are specified or estimated, $p$ in \eqref{eq:pvalue}
can be easily obtained from Gamma-distribution tables.
For $\hat{a}$ and $\hat{b}$ given by \eqref{eq:hat_ab}, we obtain the final $p$-value,
\begin{equation}
\hat{p}~\equiv~1-F_{\Gamma(\hat{a},\hat{b})}(T).
\label{eq:hat_pvalue}
\end{equation}

\noindent In Section \ref{sec:simulation}, we call the EFDR-CS procedure based on \eqref{eq:hat_pvalue}, ``CPL," which is an abbreviation of ``copula."

For $\tilde{a}$ and $\tilde{b}$ given by \eqref{eq:tilde_ab}, we obtain the final $p$-value,
\begin{equation}
\tilde{p}~\equiv~1-F_{\Gamma(\tilde{a},\tilde{b})}(T).
\label{eq:tilde_pvalue}
\end{equation}

\noindent In Section \ref{sec:simulation}, we call the EFDR-CS procedure based on \eqref{eq:tilde_pvalue}, ``MOM," which is
an abbreviation of ``method-of-moments." MOM is easier to implement than CPL, although it is typically not as statistically efficient.

For $a$ and $b$ specified or estimated, the significance test of \eqref{eq:hypothesis} at level $\alpha$ is:
\begin{equation}
\mbox{reject $H_0$ if}~T>F^{-1}_{\Gamma(a,b)}(1-\alpha),
\label{eq:test}
\end{equation}

\noindent where $T$ is defined by \eqref{eq:Brown}, and $\alpha$ is a pre-specified significance level between $0$ and $1$ (e.g., $\alpha=0.05$).

In Section~\ref{sec:simulation},
we present simulation experiments for inference on the spatial signal $\bm{\mu}$ based on the significance test \eqref{eq:test}.
The applications we give in Section \ref{sec:application} use CPL, where $(a,b)=(\hat{a},\hat{b})$ given by \eqref{eq:hat_ab},
since in Section \ref{sec:simulation} we found it more statistically efficient (although not substantially so) than MOM given by \eqref{eq:tilde_ab}.
In some circumstances, the computational simplicity of the estimates given by \eqref{eq:tilde_ab}
may be preferred to the more involved estimates given by \eqref{eq:hat_ab}.
Results from a simulation experiment that demonstrate the validity of this hypothesis-testing procedure (in a simple non-spatial setting) are given in Section
\ref{sec:Type-I} of the Supplementary Material.

\subsection{A summary of our proposed procedure for inferring spatial signal}
\label{sec:our method}

For detecting and estimating pixel-scale signal from incomplete spatially aggregated data,
we propose the following six steps:
\begin{enumerate}
\item Estimate $\bm{\theta}$ in $\bm{\Sigma}(\bm{\theta})$ (given by \eqref{eq:theta}) by $\hat{\bm{\theta}}$ through \eqref{eq:theta.hat} in the Supplementary Material,
  based on the data $\bm{\tilde{Z}}$ and under $H_0$: $\bm{\mu}=\bm{0}$.

\item Using $\bm{\Sigma}(\hat{\bm{\theta}})$,
  simulate $M$ times the $n$-dimensional vector $\bm{Z}$ conditional on $\bm{\tilde{Z}}$ via \eqref{eq:conditional simulation},
  and obtain the conditional simulations $\{\bm{Z}_1,\dots,\bm{Z}_M\}$.

\item Apply EFDR to each of $\bm{Z}_1,\dots,\bm{Z}_M$, from which the $p$-values $p_1,\dots,p_M$,
  and the corresponding estimates $\hat{\bm{\mu}}_1,\dots,\hat{\bm{\mu}}_M$ of $\bm{\mu}$, are obtained.

\item Estimate $\rho$, the level of exchangeability in $\{p_1,\dots,p_M\}$, from which the
 Gamma parameters $a$ and $b$ are estimated, for example by \eqref{eq:hat_ab} or \eqref{eq:tilde_ab}.

\item Obtain an estimate $\hat{\bm{\mu}}$ of the spatial signal from \eqref{eq:muhat}.

\item Combine $\{p_1,\dots,p_M\}$ into the final $p$-value using \eqref{eq:pvalue}, and use it in \eqref{eq:test}
  to test the hypothesis $H_0:\bm{\mu}=\bm{0}$ at the 100\,$\alpha\%$ level of significance.
\end{enumerate}


\section{Simulation studies}
\label{sec:simulation}

To evaluate the performance of our proposed procedure summarized in Section~\ref{sec:our method},
we performed three experiments in a factorial design.
Our proposed procedure was evaluated
under three scenarios involving complete data at different scales of aggregation and two types of incomplete data
(missing in a contiguous block and missing at random) at different scales of aggregation.
%
%
The ``responses" used in the study were the Type-I error rate (i.e., the probability of incorrectly rejecting a true $H_0$),
the power (i.e., the probability of correctly rejecting a false $H_0$),
and the receiver operating characteristic (ROC) curve (i.e., a plot of the power as a function of the Type-I error rate).

In our experimental set-up, we let the finest pixel resolution be of size $64\times 64$.
That is, $\bm{Z}$ in \eqref{eq:data} is a vector of length $n=n_1\times n_2=64\times 64=\mbox{4,096}$.
To check the power of our proposed procedure, we generated data with a signal given by
\begin{equation}
\mu(\bm{s})=h\times I(\bm{s}\in\Delta_r);\quad\bm{s}\in D,
\label{eq:signal}
\end{equation}

\noindent where $D=\{(i_1,i_2):i_1,i_2=1,\dots,64\}$, and we considered four different $r\times r$ square regions $\Delta_r\subset D$ of width $r\in\{4,6,8,10\}$.
Here, all squares were centered at the juncture of the middle four pixels in the $64\times 64$ region.
For each $\Delta_r$, we considered six different signal magnitudes $h\in\{0,1,\dots,5\}$,
where $h=0$ corresponds to no signal and is used to compute the Type-I error rate.
We generated spatially correlated errors using an exponential covariance function
(i.e., $C(\bm{u})=\exp(\|\bm{u}\|/\phi)$),
and we checked how the power depends on the degree of spatial correlation by considering $\phi\in\{0,5,10\}$.
For each setting of $r$, $h$, and $\phi$, we simulated 400 datasets $\tilde{\bm{Z}}$,
from which we obtained the empirical power curve and the empirical ROC curve.

Throughout the simulation, we chose two wavelet-decomposition levels (i.e., $J=2$),
resulting in seven (i.e., $3J+1=7$) wavelet classes corresponding to different scales and orientations.
For each of the 400 simulated datasets, we estimated $\hat{\bm{\theta}}(\hat{\tau}^2,\hat{\phi})=\hat{\bm{\theta}}=(\hat{\theta}_1,\dots,\hat{\theta}_7)'$ through \eqref{eq:theta.hat}
in the Supplementary Material,
where $\hat{\tau}^2$ and $\hat{\phi}$ are the ML estimators based on the exponential covariance model, $C(\bm{u})=\tau^2\exp(\|\bm{u}\|/\phi)$.
Using the estimate $\hat{\bm{\theta}}$ for a given dataset, we generated $M=100$ conditional simulations through \eqref{eq:conditional simulation}.
We then used the R package ``EFDR" (Zammit-Mangion and Huang, 2015) on each conditionally simulated
$64\times 64$ image,
implemented with the Daubechies least asymmetric wavelet filter of length $8$ (Daubechies, 1992),
and we let the number of hypotheses to be tested in the wavelet space be $100$.
For each dataset, these conditional simulations produced $100$ $p$-values,
which were combined using the statistic $T$ in \eqref{eq:Brown} and the final $p$-value given by \eqref{eq:pvalue}.
Then the hypothesis test \eqref{eq:hypothesis} was performed using \eqref{eq:test}.

We compared the performance of the EFDR-CS procedure, CPL (and its variant MOM), with the na\"{i}ve approach
where the $p$-values are combined na\"{i}vely through their simple average (NVE).
In addition, we considered an ideal setting (IDL) where
it is assumed that all fine-resolution pixels were observed and $\tilde{\bm{Z}}_{64\times 64}\equiv\bm{Z}$,
so that the EFDR procedure can be directly applied without CS.
Three experiments and an analysis of their responses are now presented.

\subsection{Experiment 1: Complete data at different scales of aggregation}
\label{sec:Exp1}

\begin{figure}[tb]\centering
\begin{tabular}{ccccc}
~~(a) & (b) & (c) & (d) & (e)\\
\rotatebox{90}{$\quad \quad \phi=0$}~\includegraphics[scale=0.16,trim={1cm 2cm 0.5cm 1.5cm},clip]{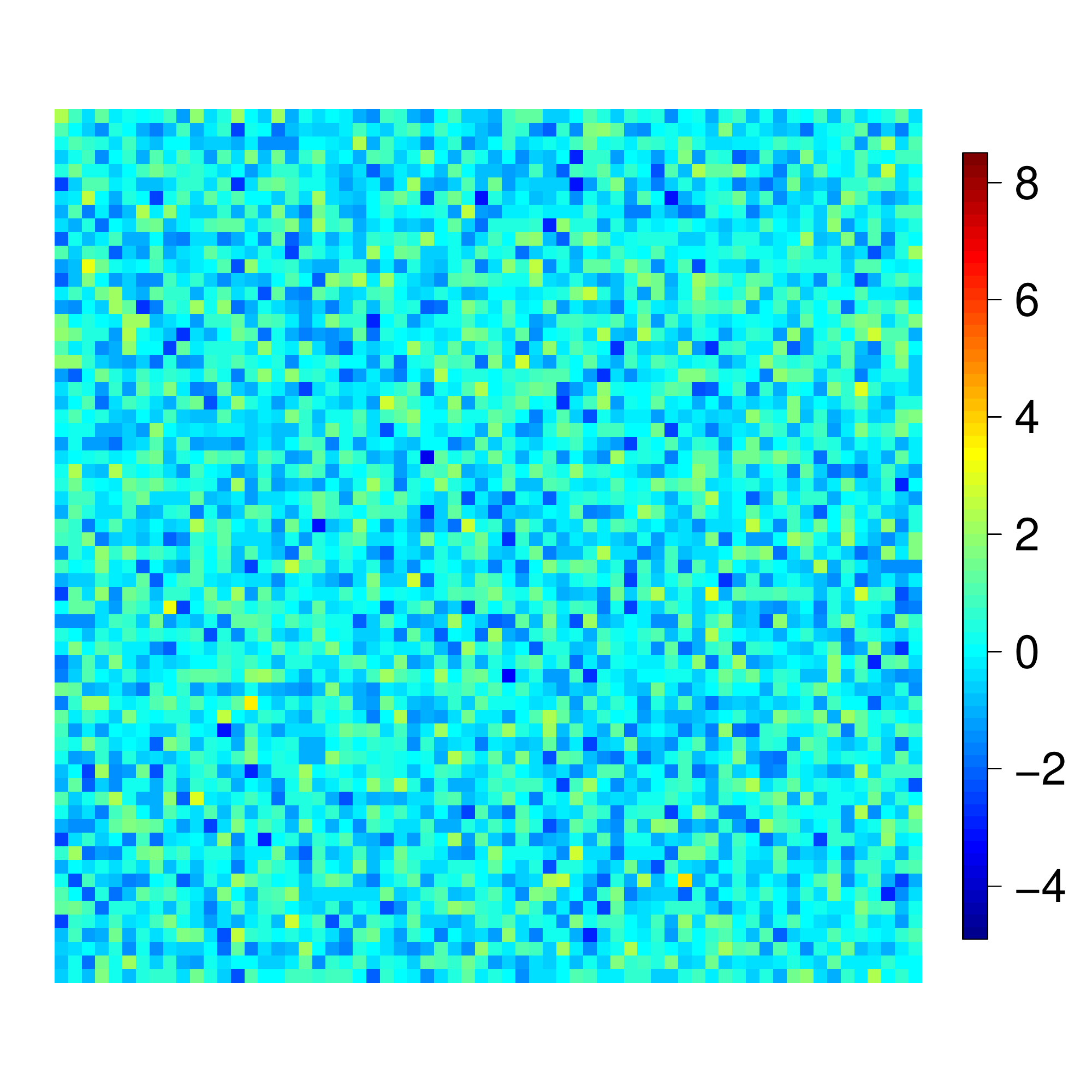} &
\!\!\!\includegraphics[scale=0.16,trim={1cm 2cm 0.5cm 1.5cm},clip]{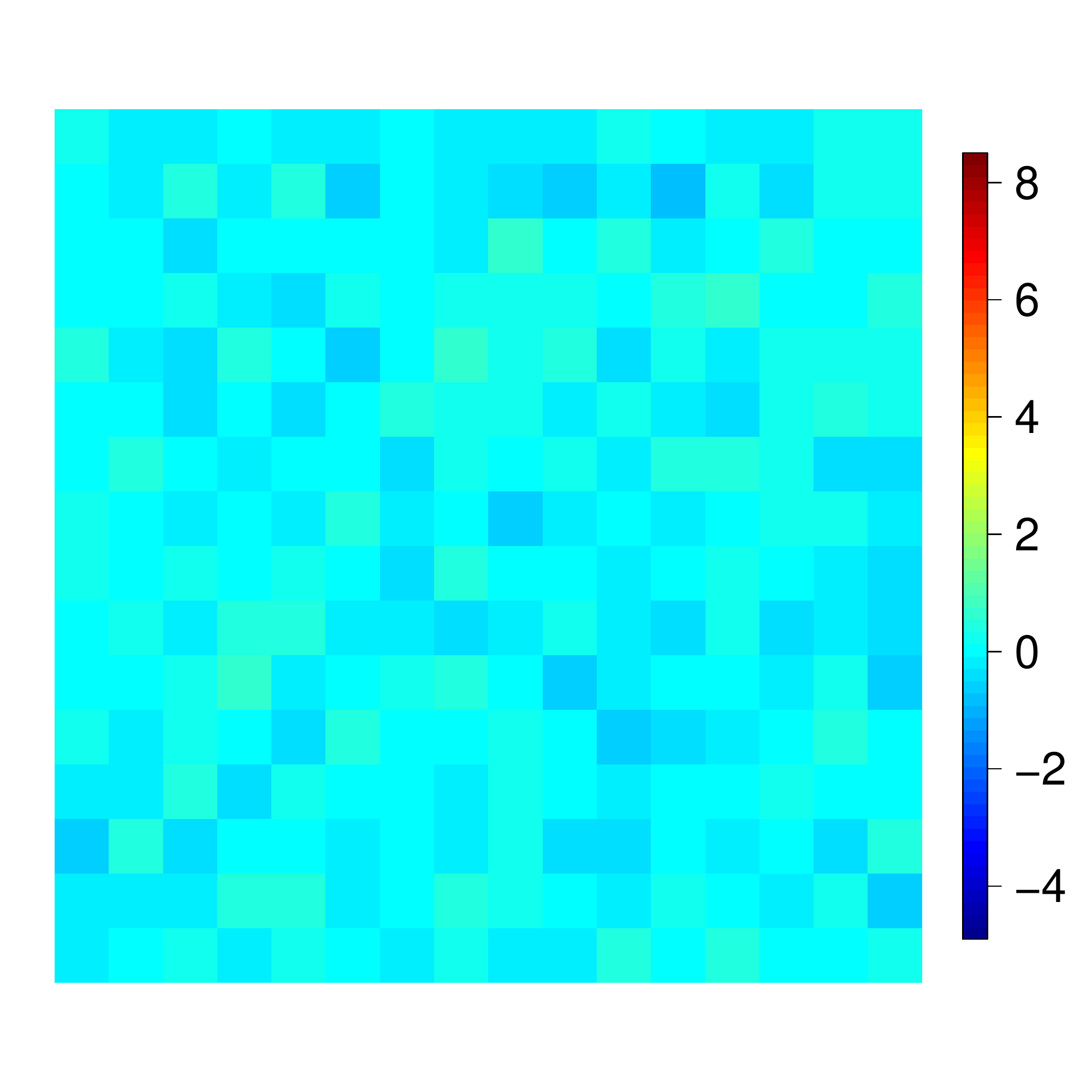}&
\!\!\!\includegraphics[scale=0.16,trim={1cm 2cm 0.5cm 1.5cm},clip]{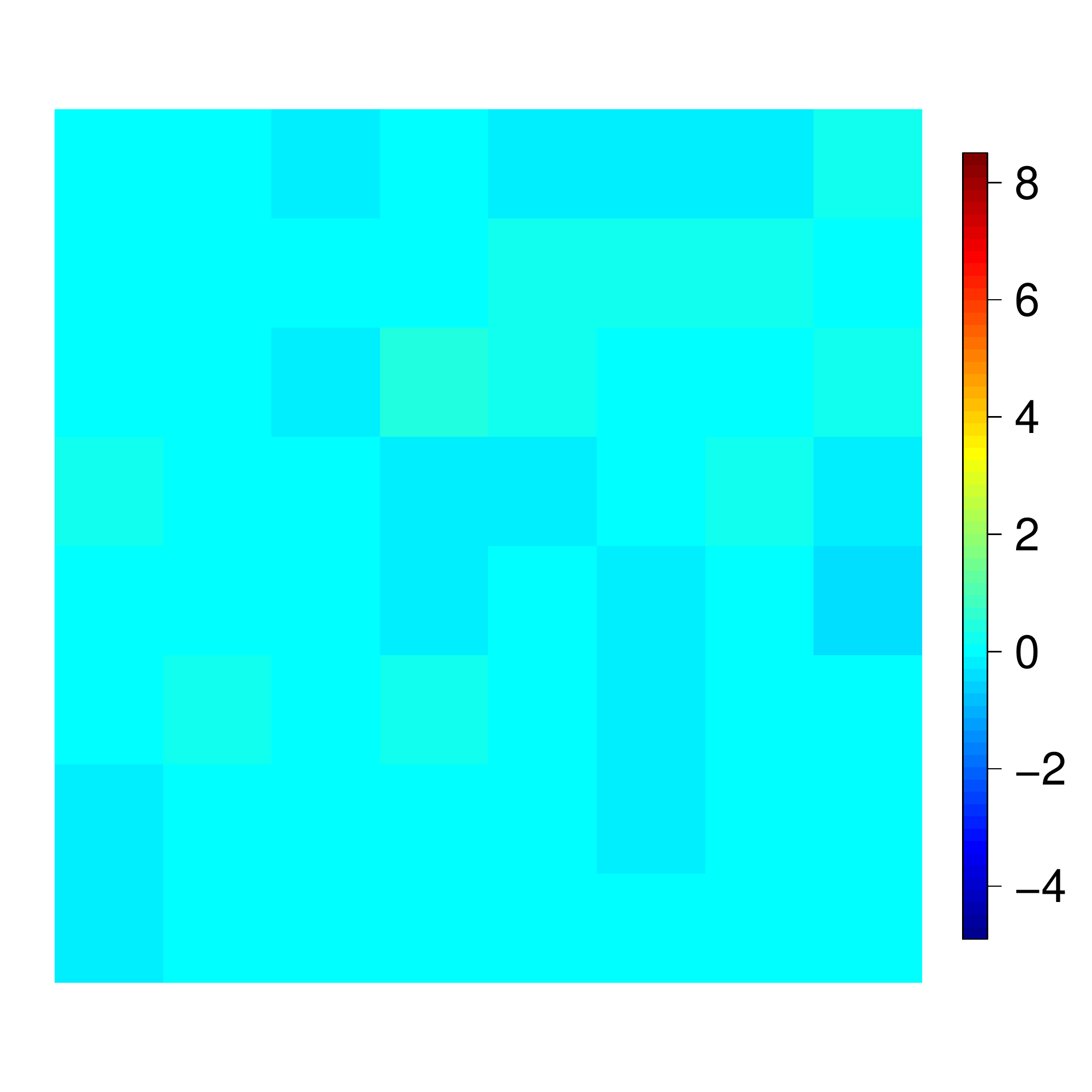}&
\!\!\!\includegraphics[scale=0.16,trim={1cm 2cm 0.5cm 1.5cm},clip]{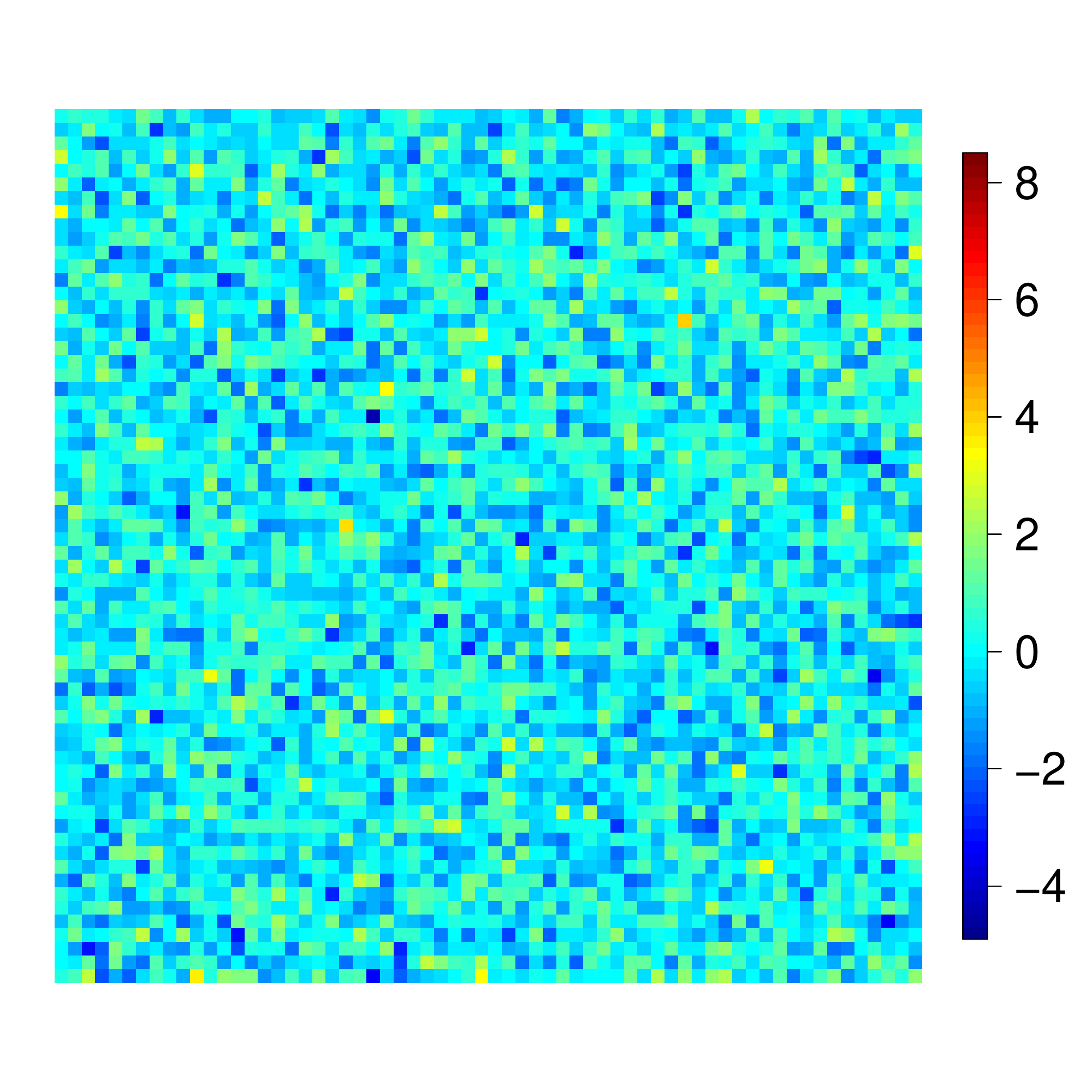}&
\!\!\!\includegraphics[scale=0.16,trim={1cm 2cm 0.5cm 1.5cm},clip]{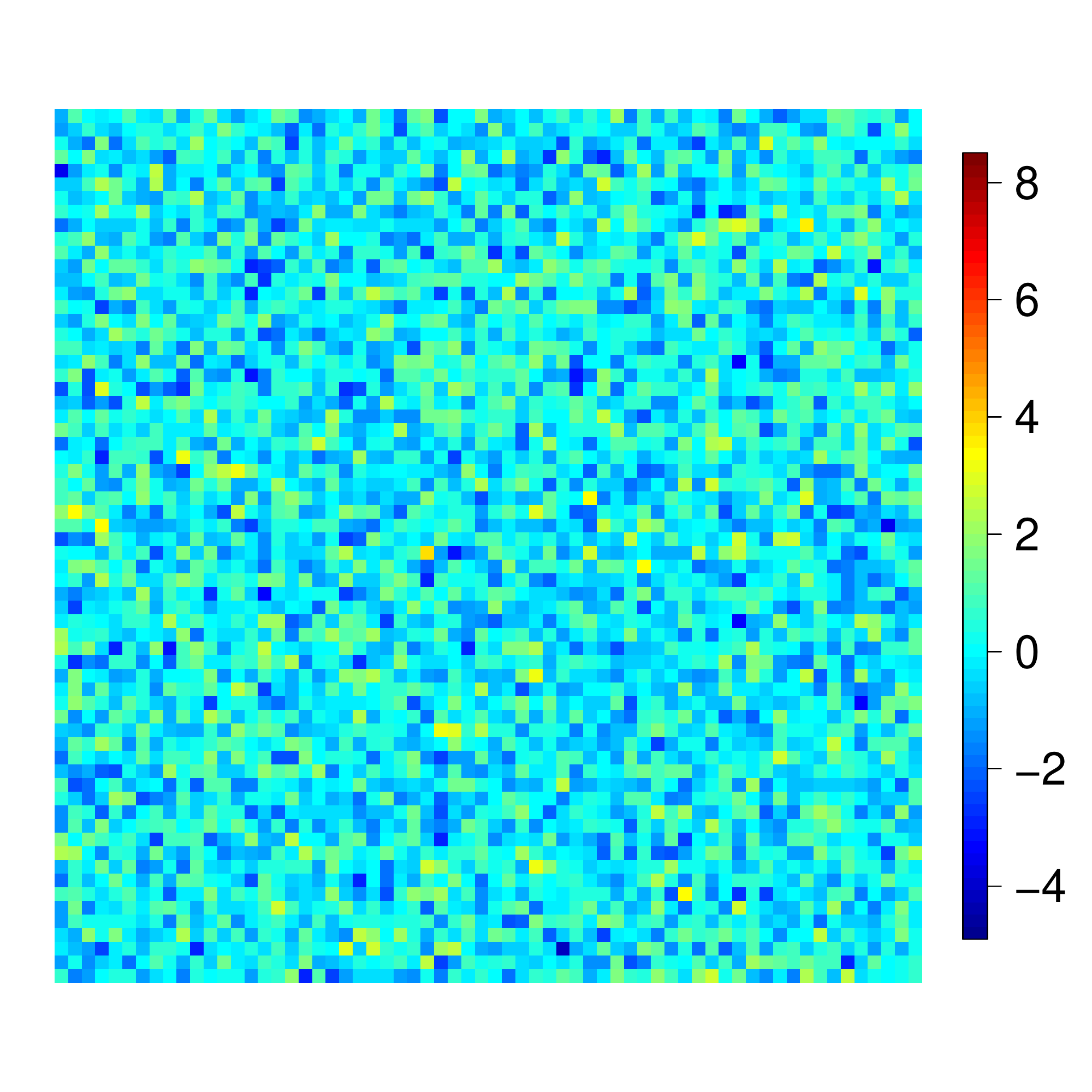}\\
\rotatebox{90}{$\quad \quad \phi=5$}~\includegraphics[scale=0.16,trim={1cm 2cm 0.5cm 1.5cm},clip]{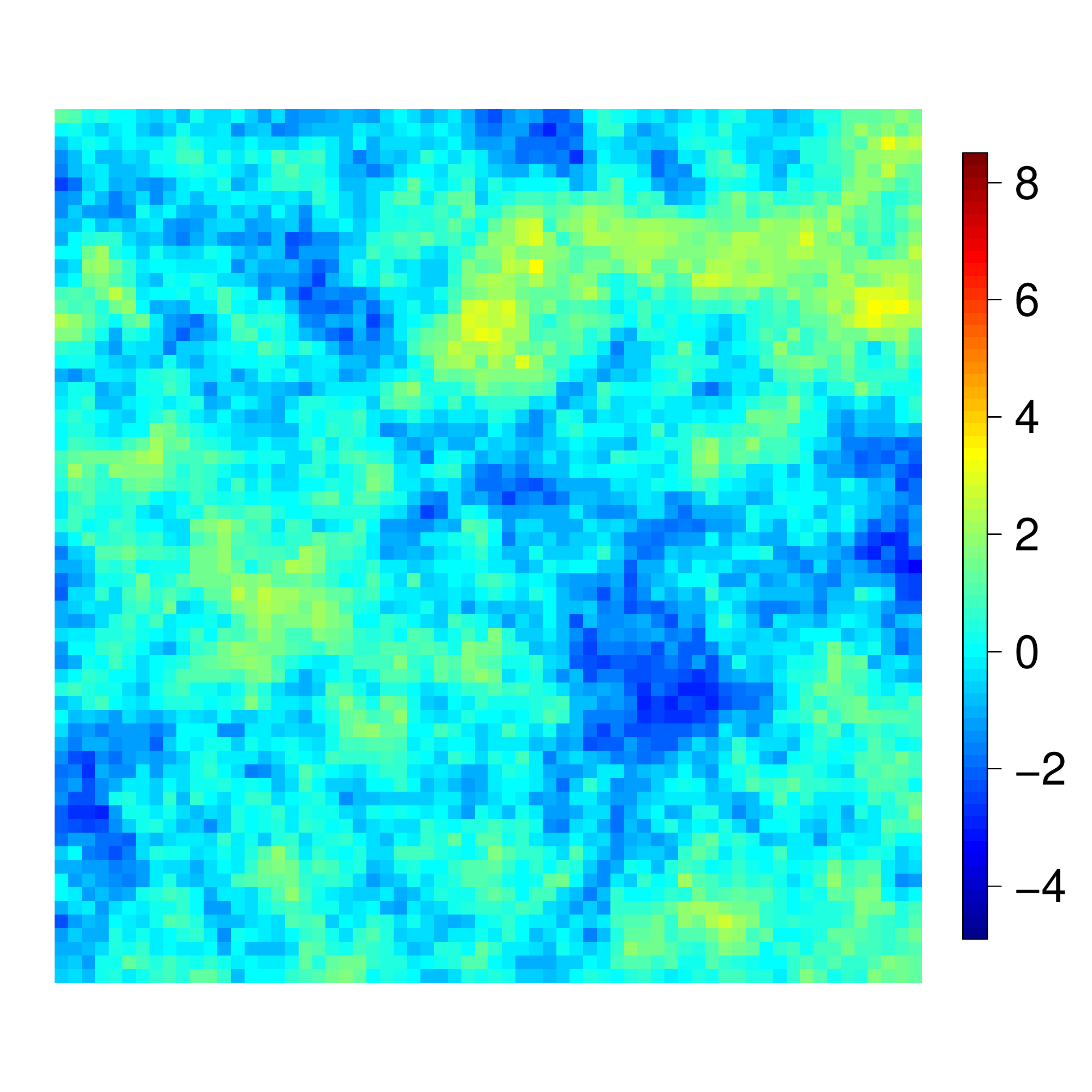} &
\!\!\!\includegraphics[scale=0.16,trim={1cm 2cm 0.5cm 1.5cm},clip]{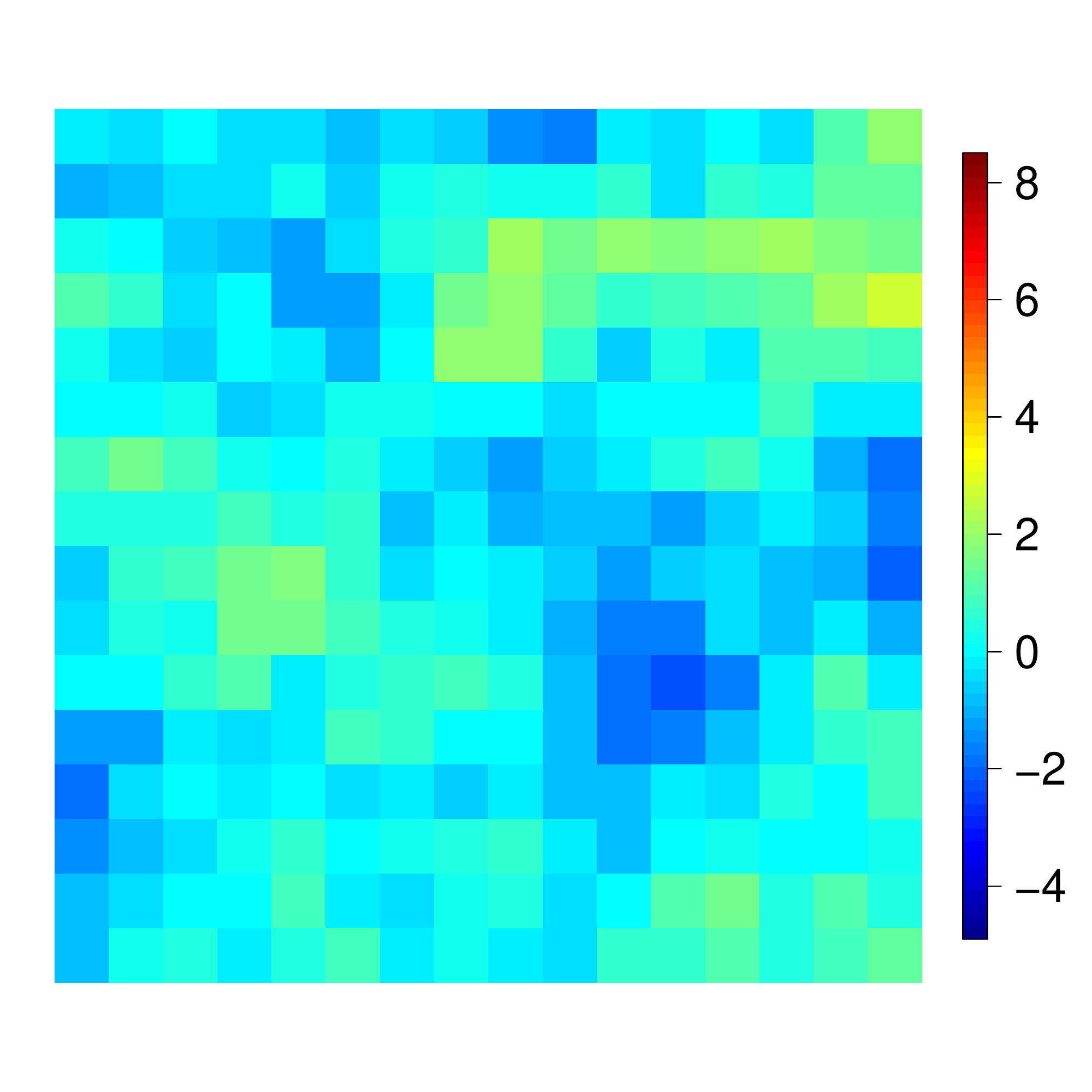}&
\!\!\!\includegraphics[scale=0.16,trim={1cm 2cm 0.5cm 1.5cm},clip]{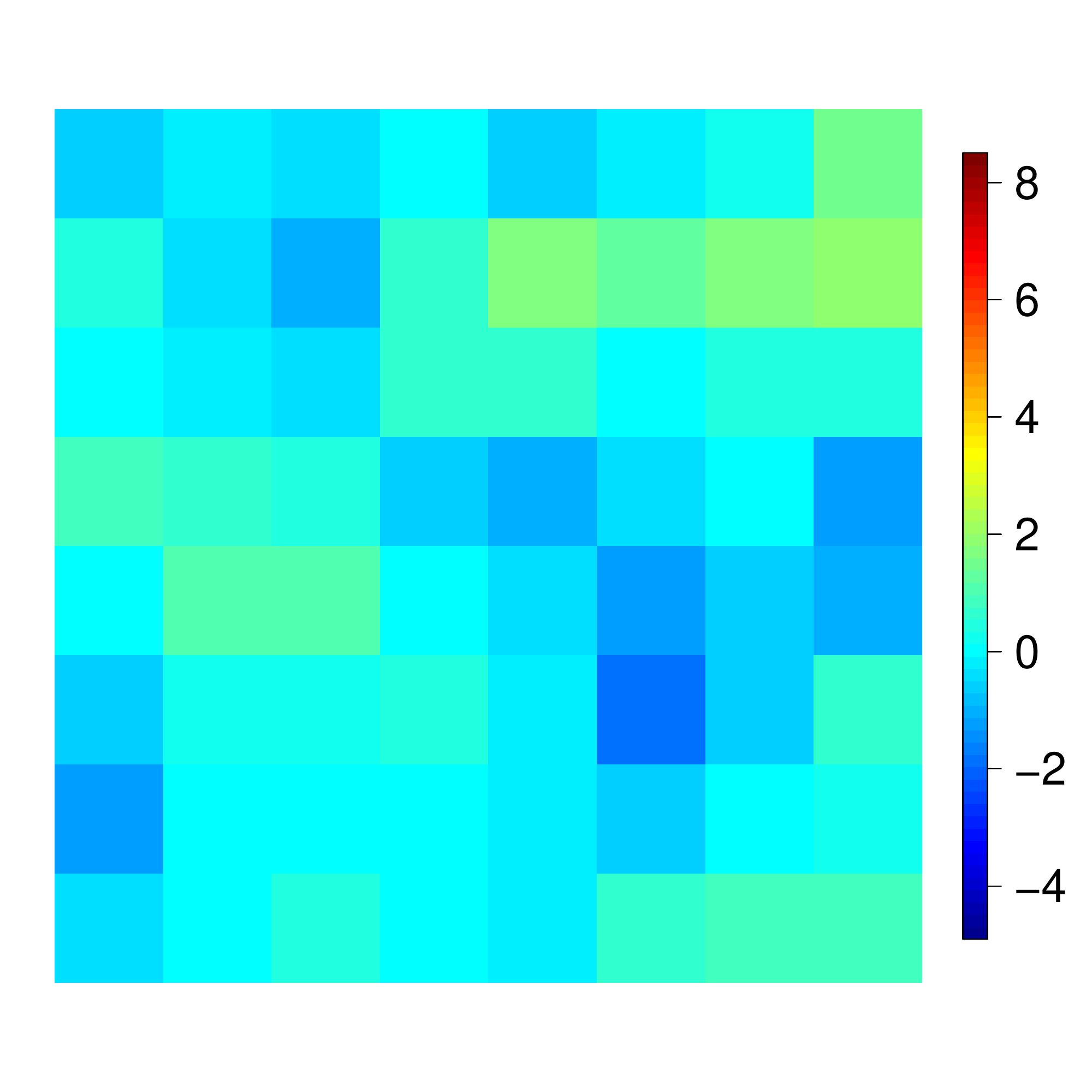}&
\!\!\!\includegraphics[scale=0.16,trim={1cm 2cm 0.5cm 1.5cm},clip]{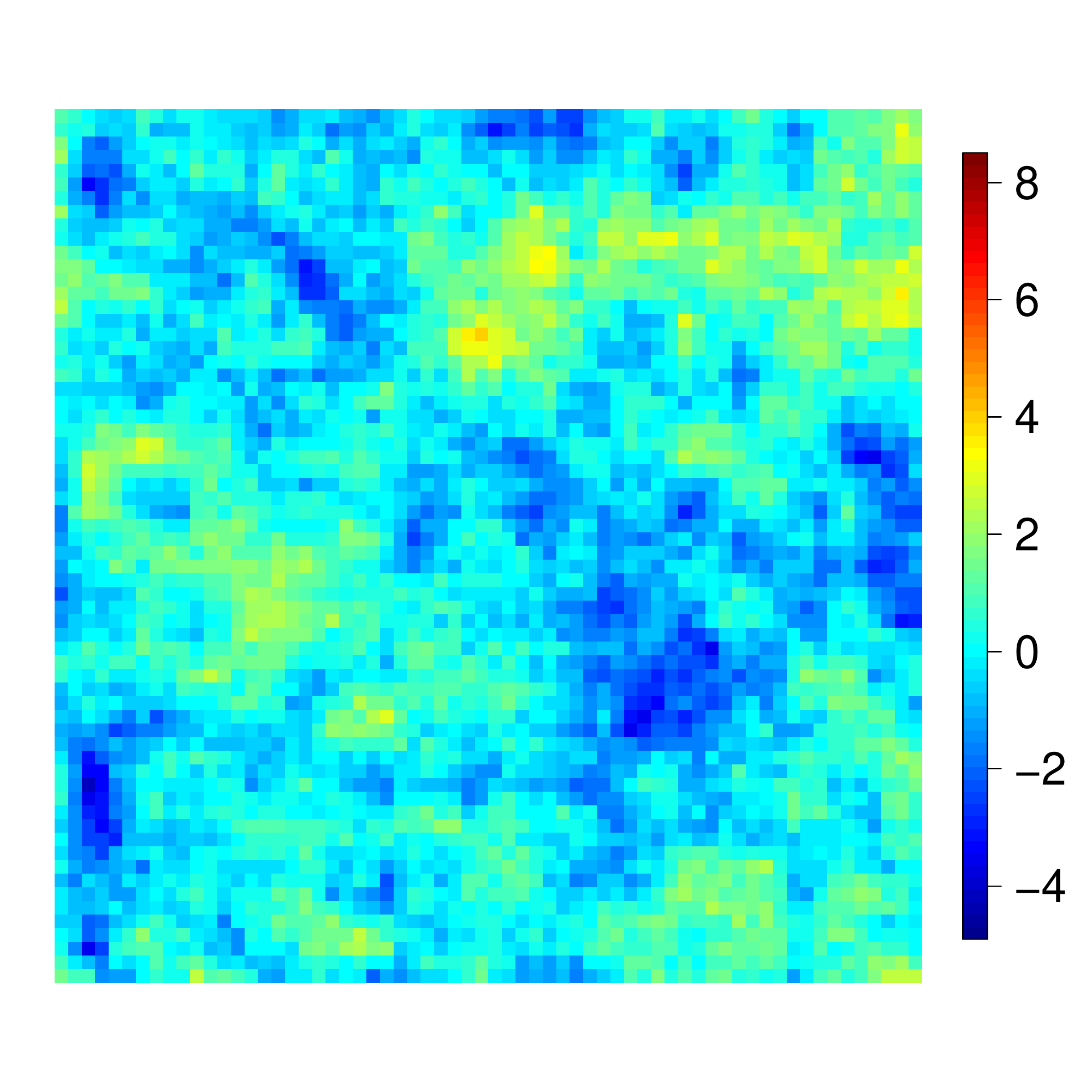}&
\!\!\!\includegraphics[scale=0.16,trim={1cm 2cm 0.5cm 1.5cm},clip]{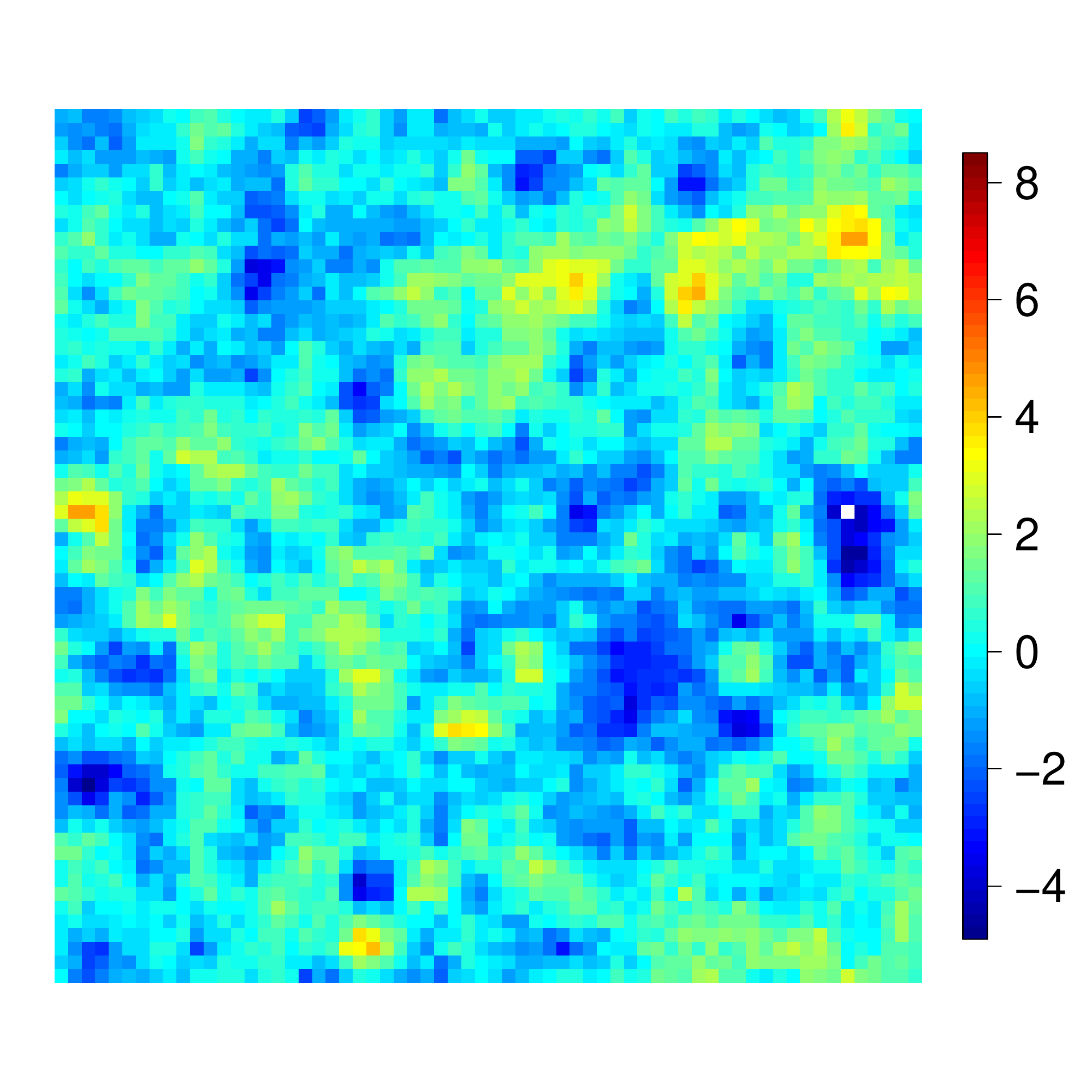}\\
\rotatebox{90}{$\quad \quad \phi=10$}~\includegraphics[scale=0.16,trim={1cm 2cm 0.5cm 1.5cm},clip]{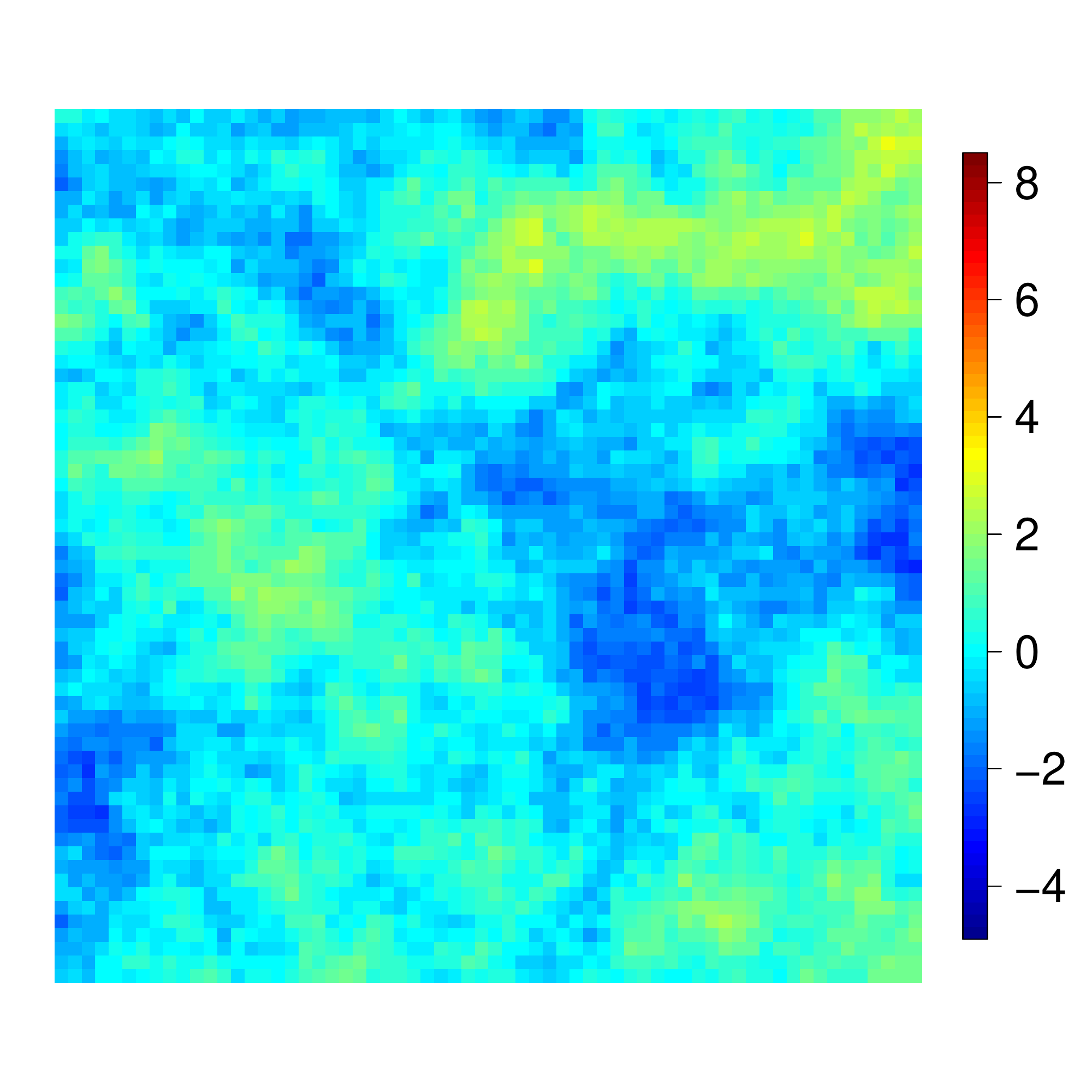} &
\!\!\!\includegraphics[scale=0.16,trim={1cm 2cm 0.5cm 1.5cm},clip]{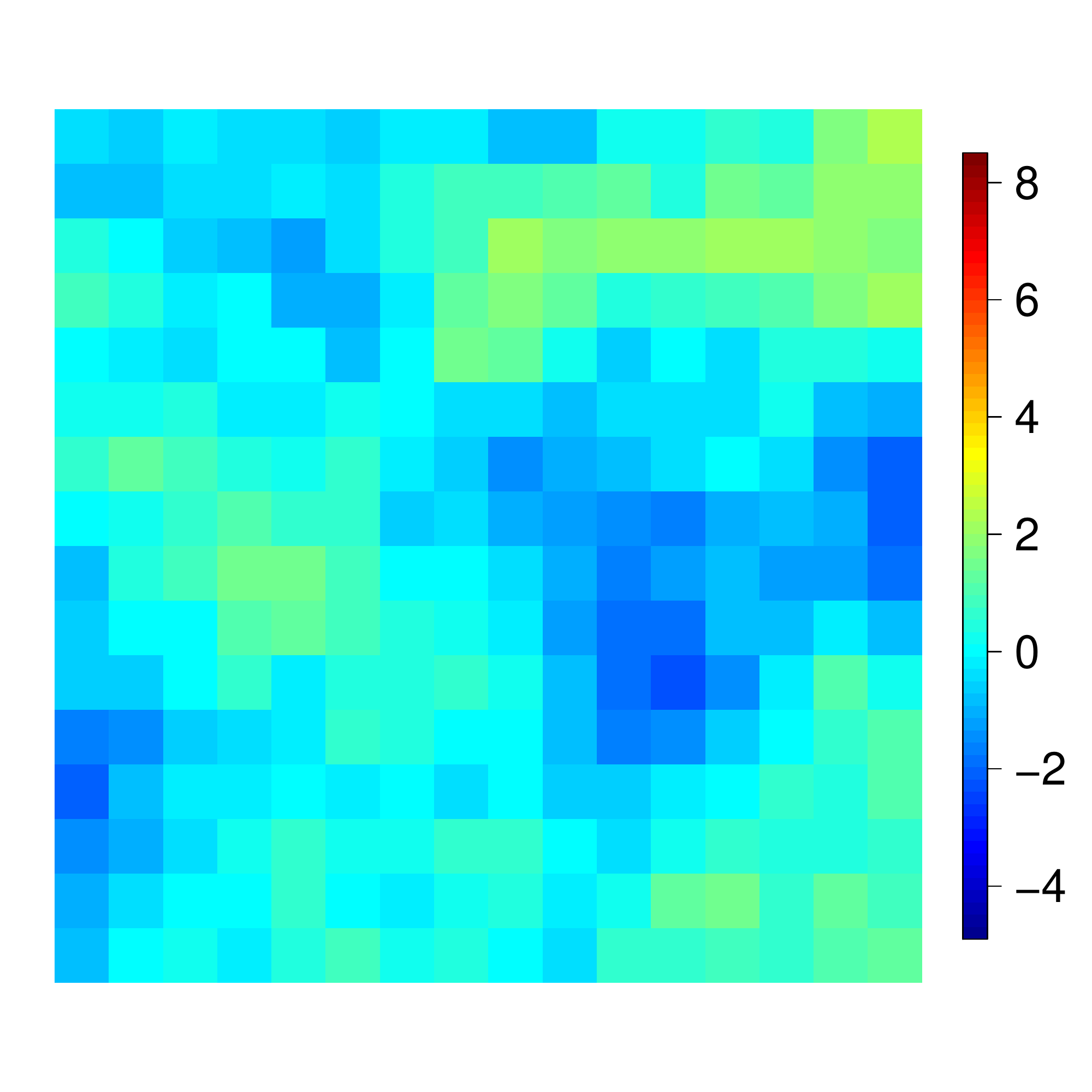}&
\!\!\!\includegraphics[scale=0.16,trim={1cm 2cm 0.5cm 1.5cm},clip]{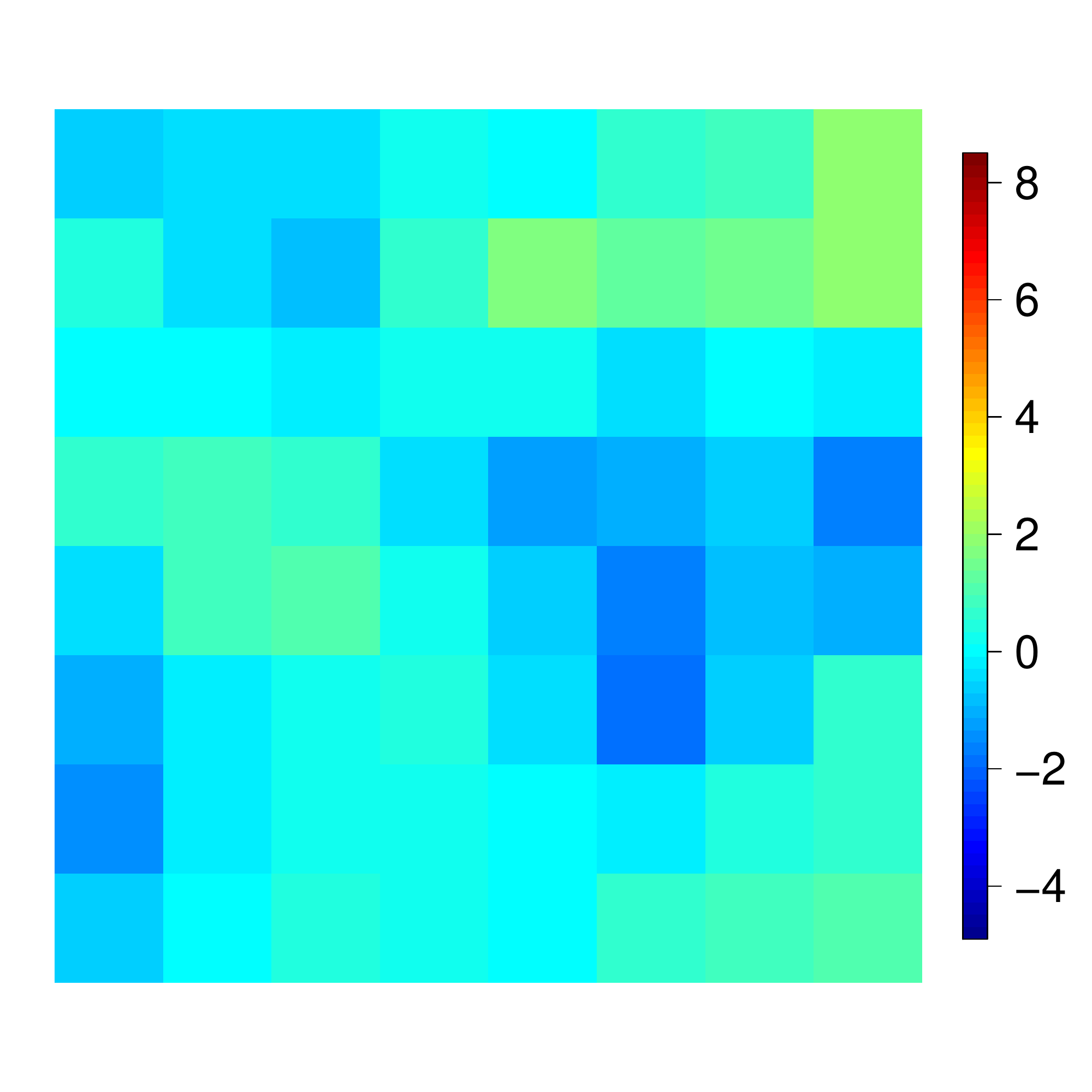}&
\!\!\!\includegraphics[scale=0.16,trim={1cm 2cm 0.5cm 1.5cm},clip]{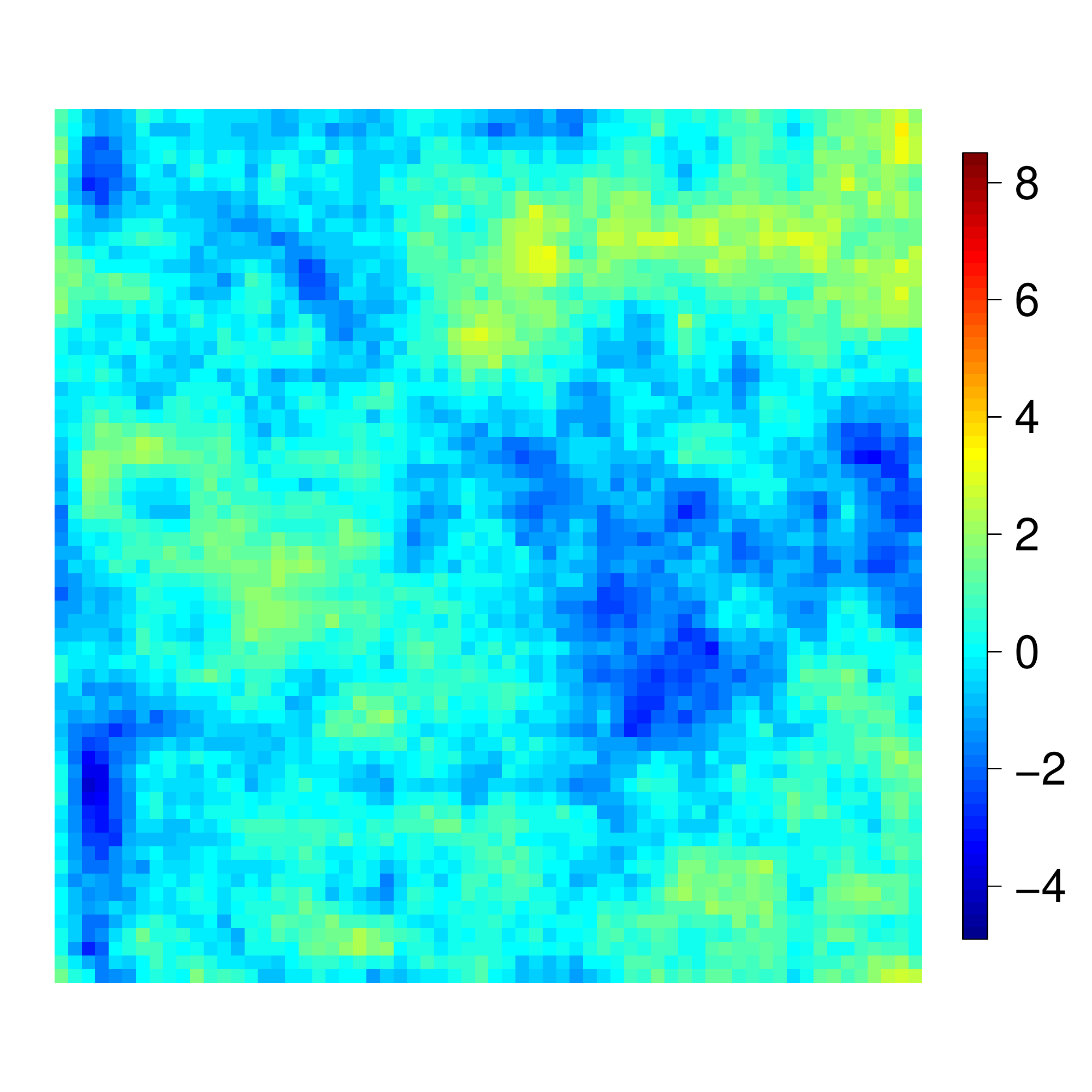}&
\!\!\!\includegraphics[scale=0.16,trim={1cm 2cm 0.5cm 1.5cm},clip]{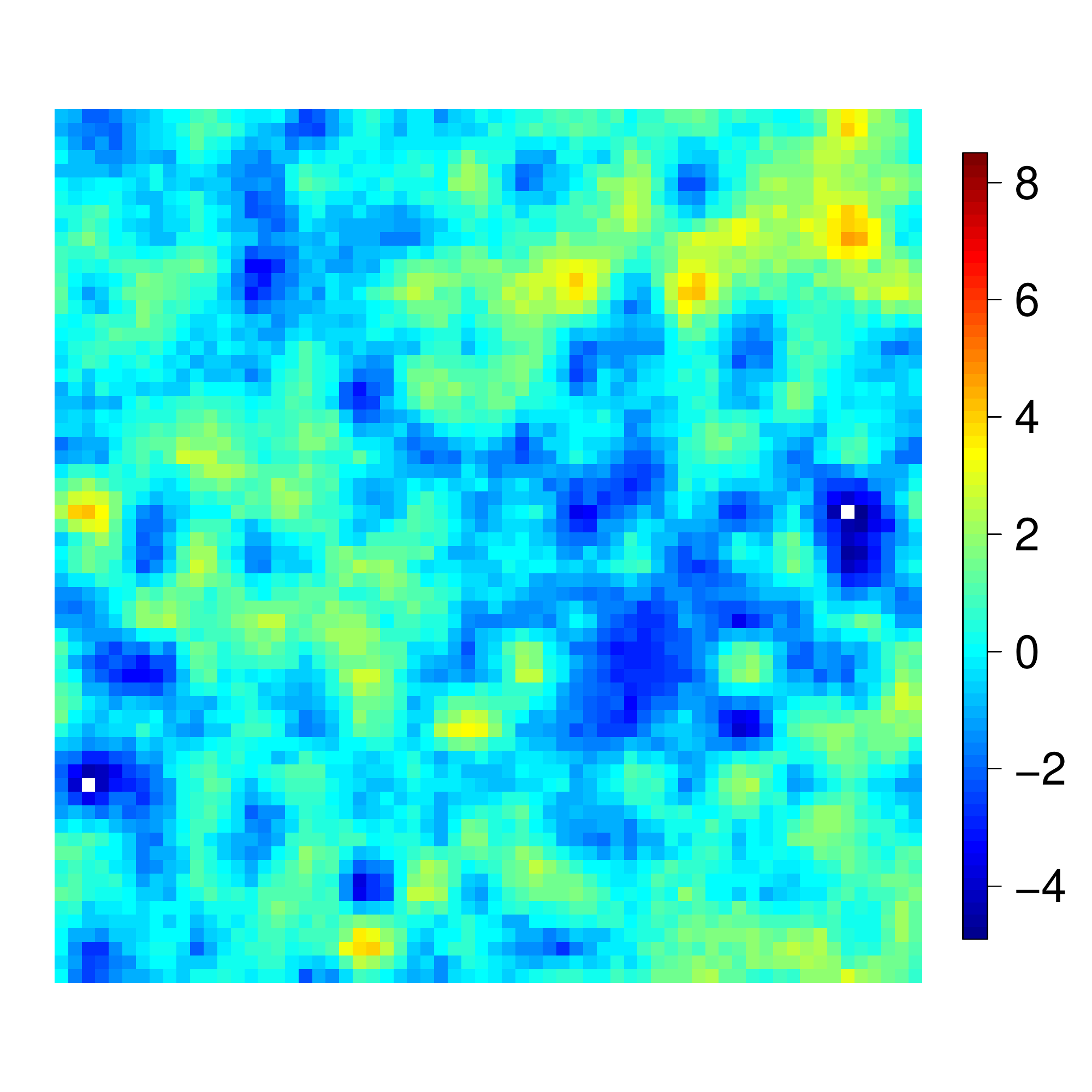}
\end{tabular}
\caption{Column (a): Three randomly generated images of $\tilde{\bm{Z}}_{64\times 64}\equiv\bm{Z}$ corresponding to $\phi\in\{0,5,10\}$
(respectively, from top to bottom);
column (b): The images of $\tilde{\bm{Z}}_{16\times 16}$ aggregated from $\tilde{\bm{Z}}_{64\times 64}$ in column (a);
column (c): The images of $\tilde{\bm{Z}}_{8\times 8}$ aggregated from $\tilde{\bm{Z}}_{64\times 64}$ in column (a);
column (d): A conditionally simulated image of $\bm{Z}$ conditional on $\tilde{\bm{Z}}_{16\times 16}$ given in column (b);
column (e): A conditionally simulated image of $\bm{Z}$ conditional on $\tilde{\bm{Z}}_{8\times 8}$ given in column (c).}
\label{fig:fourimage}
\end{figure}

\begin{figure}[tb]\centering
\begin{tabular}{ccc}
\includegraphics[scale=0.32,trim={1cm 2.5cm 2.5cm 2cm},clip]{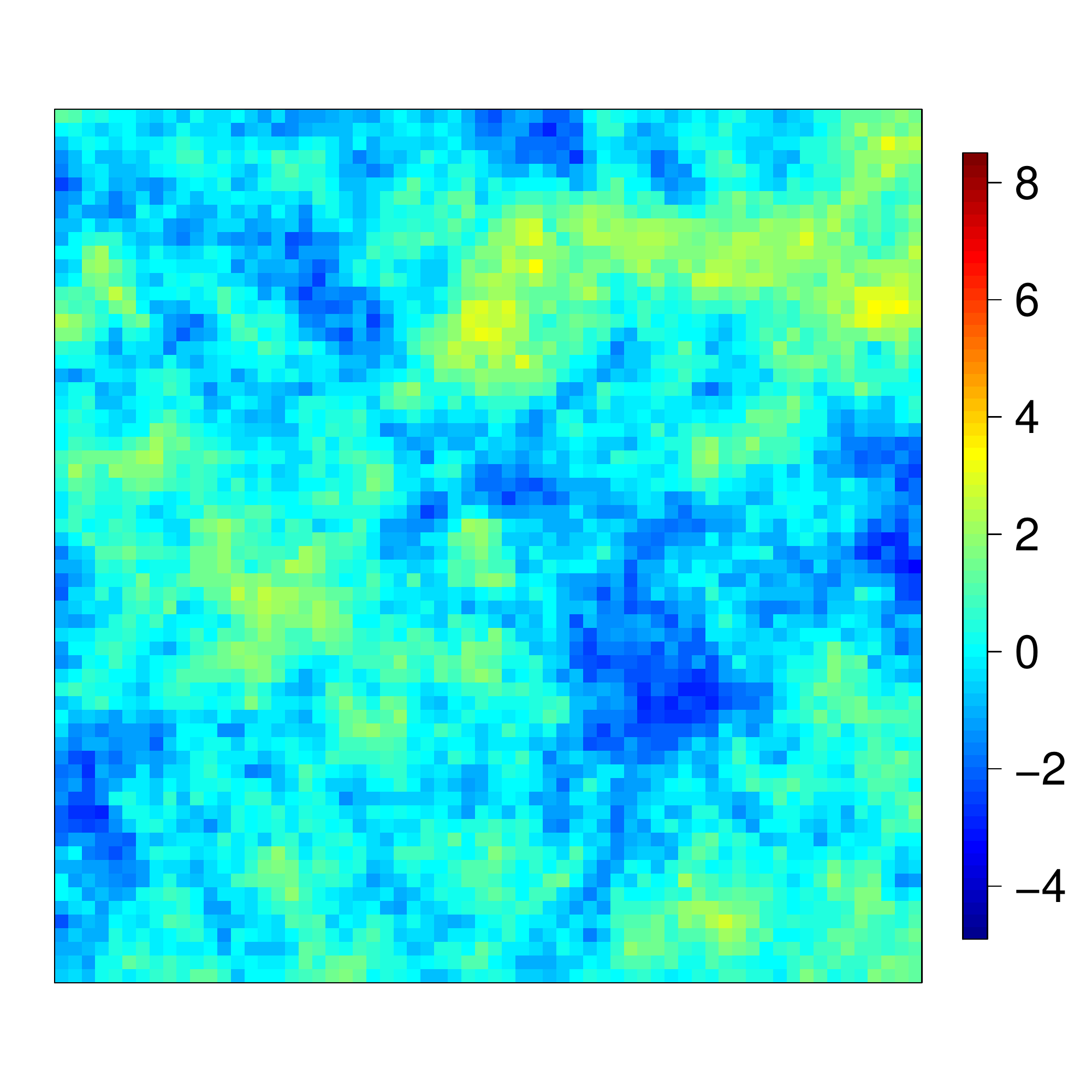} &
\!\!\!\!\!\includegraphics[scale=0.32,trim={1cm 2.5cm 2.5cm 2cm},clip]{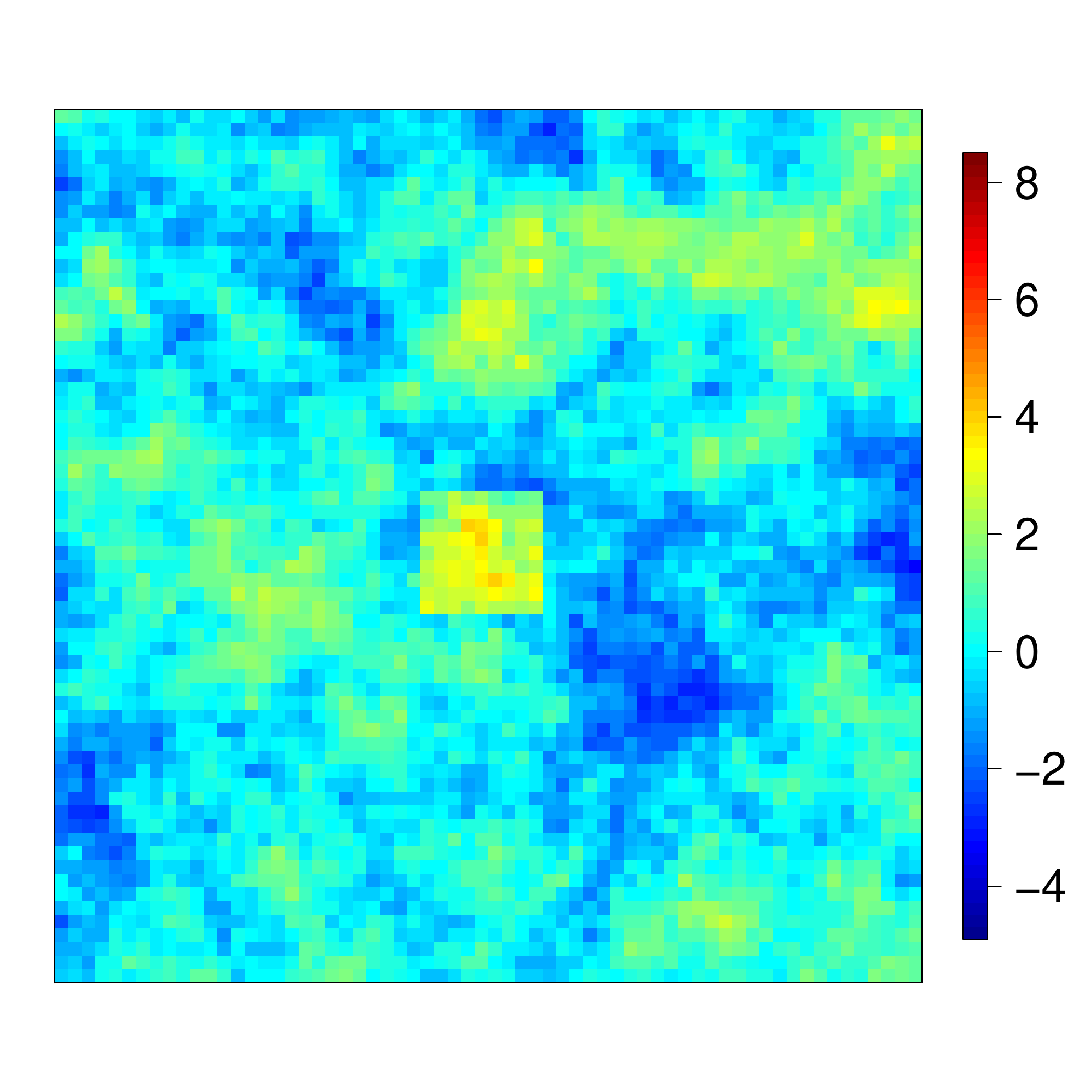} &
\!\!\!\!\!\includegraphics[scale=0.32,trim={1cm 2.5cm 2.5cm 2cm},clip]{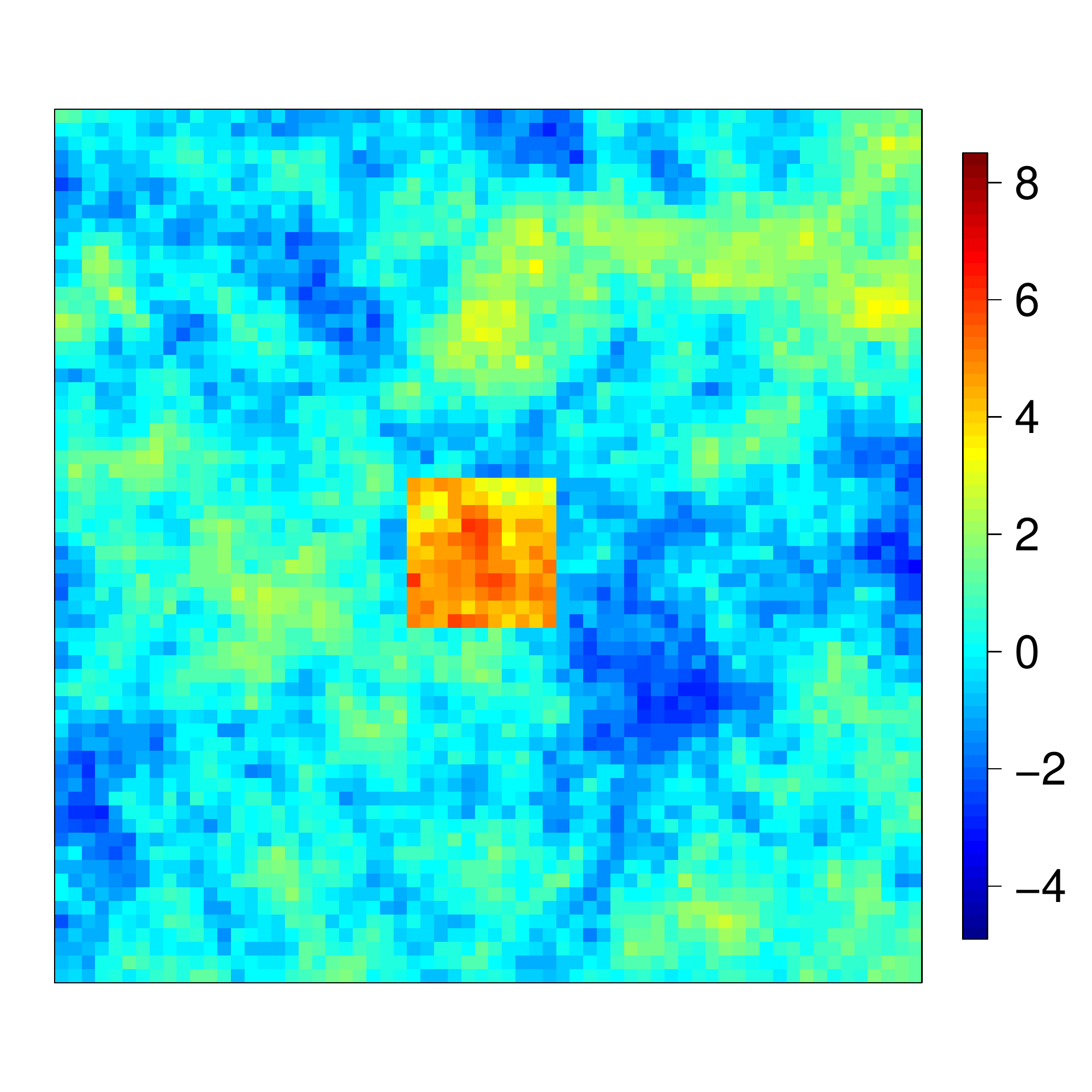} 
\end{tabular}
\caption{Images of $\bm{Z}$ with $\phi=5$ and signals of various extents and magnitudes
described in Section~\ref{sec:Exp1} and given, from left to right, by $(r,h)=(4,1)$, $(8,3)$, and $(10,5)$, respectively.}
\label{fig:signalonimageraw}
\end{figure}

\begin{figure}[tb]\centering
\begin{tabular}{ccc}
\includegraphics[scale=0.32,trim={1cm 2.5cm 2.5cm 2cm},clip]{./signal-r10-h5} &
\!\!\!\!\!\includegraphics[scale=0.32,trim={1cm 2.5cm 2.5cm 2cm},clip]{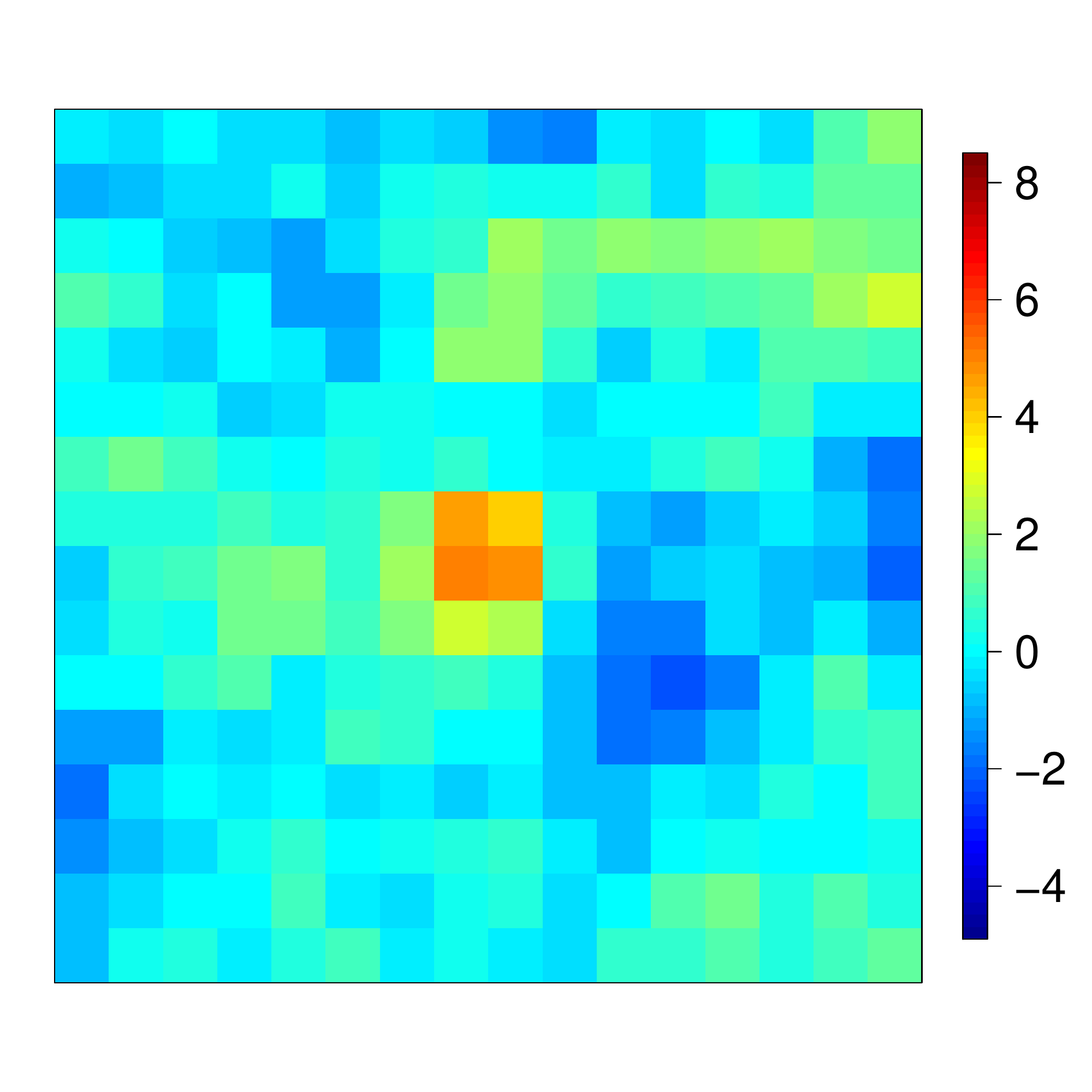} &
\!\!\!\!\!\includegraphics[scale=0.32,trim={1cm 2.5cm 2.5cm 2cm},clip]{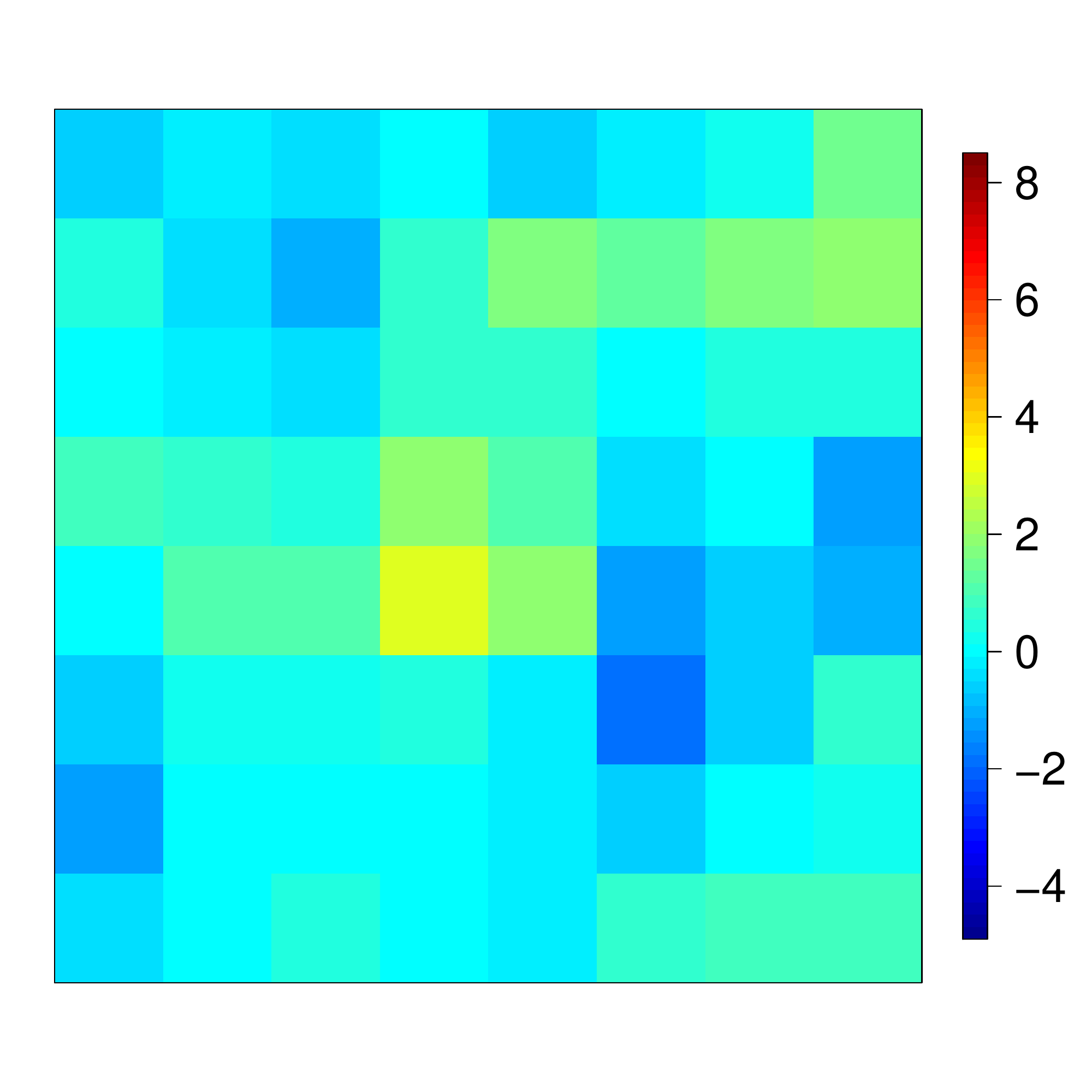} 
\end{tabular}
\caption{Aggregations of the image in Figure~\ref{fig:signalonimageraw}, right panel, where $(r,h)=(10,5)$; from left to right,
$\tilde{\bm{Z}}_{64\times 64}$, $\tilde{\bm{Z}}_{16\times 16}$, $\tilde{\bm{Z}}_{8\times 8}$.
The left panel is identical to the right panel of Figure~\ref{fig:signalonimageraw}.}
\label{fig:signalonimageraw2}
\end{figure}

Figure~\ref{fig:fourimage}(a) shows three randomly generated datasets with $h=0$ (i.e., no signal) and
strength of spatial dependence $\phi=0,5,10$, respectively.
Note that when $\phi$ is larger, it is more difficult to separate a signal from the spatially dependent noise,
since strong spatial dependence can take on the appearance of a non-zero mean vector of spatially coherent entries.
Data $\tilde{\bm{Z}}$ were generated by aggregating $\bm{Z}$ into $16\times 16$ and $8\times 8$ regular grid cells,
and they are denoted by $\tilde{\bm{Z}}_{16\times 16}$ and $\tilde{\bm{Z}}_{8\times 8}$, respectively.
Figures~\ref{fig:fourimage}(b) and \ref{fig:fourimage}(c) show the data, $\tilde{\bm{Z}}_{16\times 16}$ and $\tilde{\bm{Z}}_{8\times 8}$,
respectively, obtained through aggregation of the corresponding images in Figure~\ref{fig:fourimage}(a).
Figures~\ref{fig:fourimage}(d) and \ref{fig:fourimage}(e) show, for illustration, a single conditionally simulated $\bm{Z}$,
conditional on $\tilde{\bm{Z}}_{16 \times 16}$ and $\tilde{\bm{Z}}_{8\times 8}$, respectively.
Although we do not expect to reproduce the image $\bm{Z}$ shown in column (a) in Figure \ref{fig:fourimage} exactly,
the conditional simulations do produce patterns similar to $\bm{Z}$.
Figure \ref{fig:signalonimageraw} shows data $\bm{Z}$ generated from \eqref{eq:data}
with spatial signal $\bm{\mu}$ corresponding to $(r,h)\in\{(4,1),(8,3),(10,5)\}$ in \eqref{eq:signal} and $\phi=5$
(i.e., $C(\bm{u})=\exp(-\|\bm{u}\|/5)$).
These $64\times 64$ images are then aggregated, resulting in data, $\tilde{\bm{Z}}_{16\times 16}$ and $\tilde{\bm{Z}}_{8\times 8}$;
the case $(r,h)=(10,5)$ is shown in Figure \ref{fig:signalonimageraw2}.
Full sets of plots for all combinations of the factors with $\phi=5$ are
shown in Figures \ref{fig:signalonimageraw-all}--\ref{fig:signalonimageaggregated} in the Supplementary Material.

\begin{figure}[!tb]\centering
\begin{tabular}{ccc}
~~~~~$\phi=0$ & ~~$\phi=5$ & ~~$\phi=10$
\smallskip\\
\rotatebox{90}{$\quad \quad \quad r=6$}~~
\includegraphics[scale=0.22,trim={0cm 0.3cm 1cm 1cm},clip]{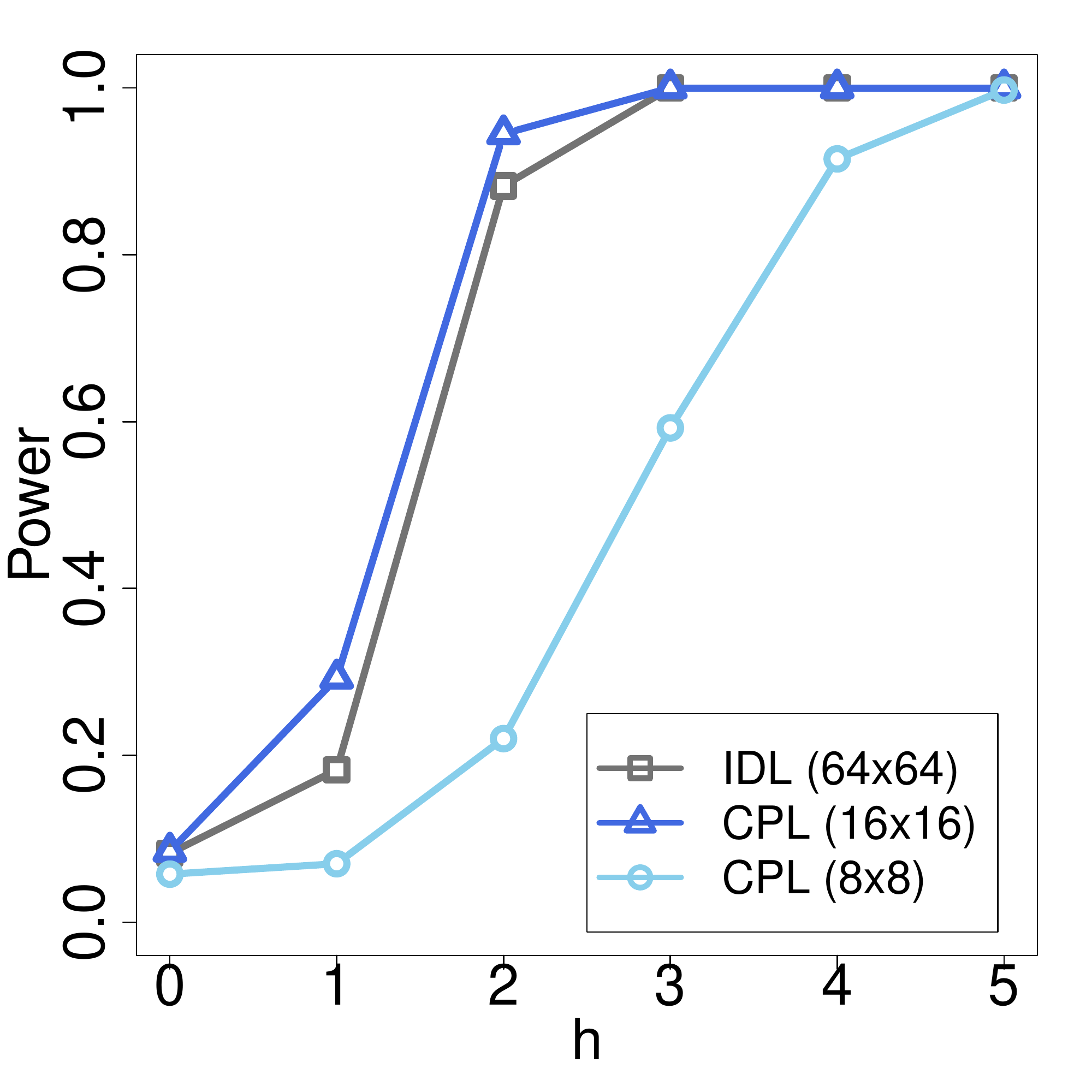} &
\includegraphics[scale=0.22,trim={0cm 0.3cm 1cm 1cm},clip]{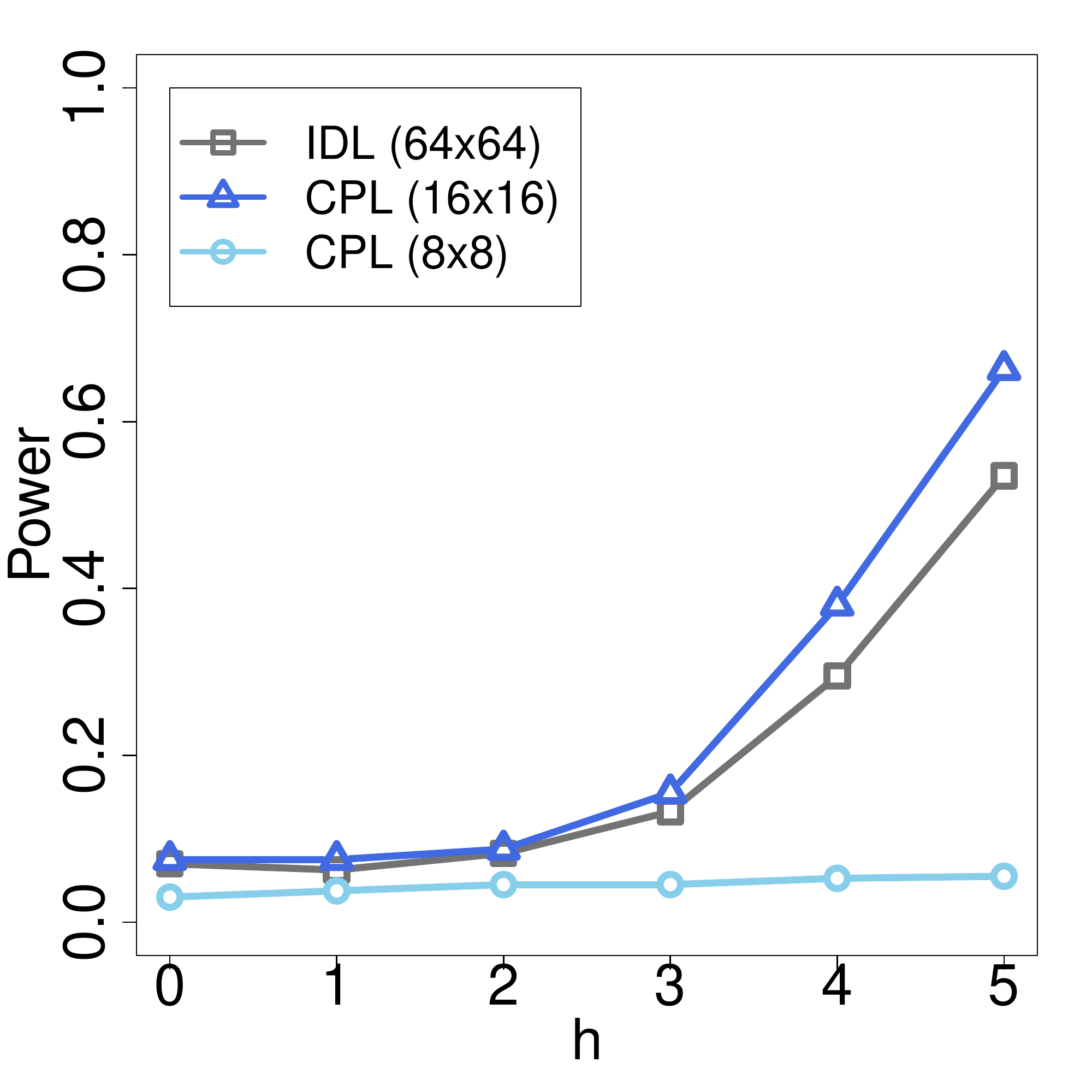} &
\includegraphics[scale=0.22,trim={0cm 0.3cm 1cm 1cm},clip]{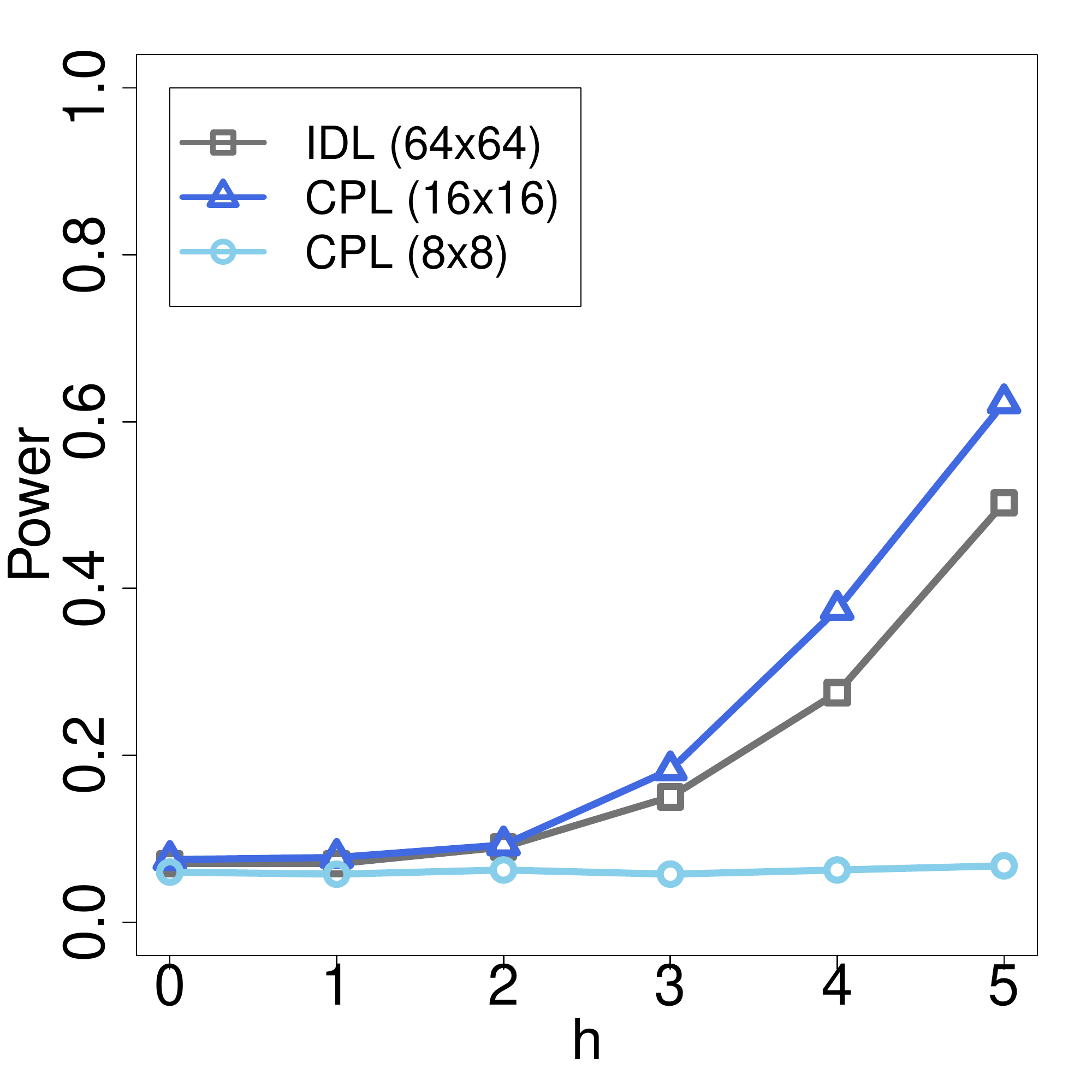} \\
\rotatebox{90}{$\quad \quad \quad r=10$}~~
\includegraphics[scale=0.22,trim={0cm 0.3cm 1cm 1cm},clip]{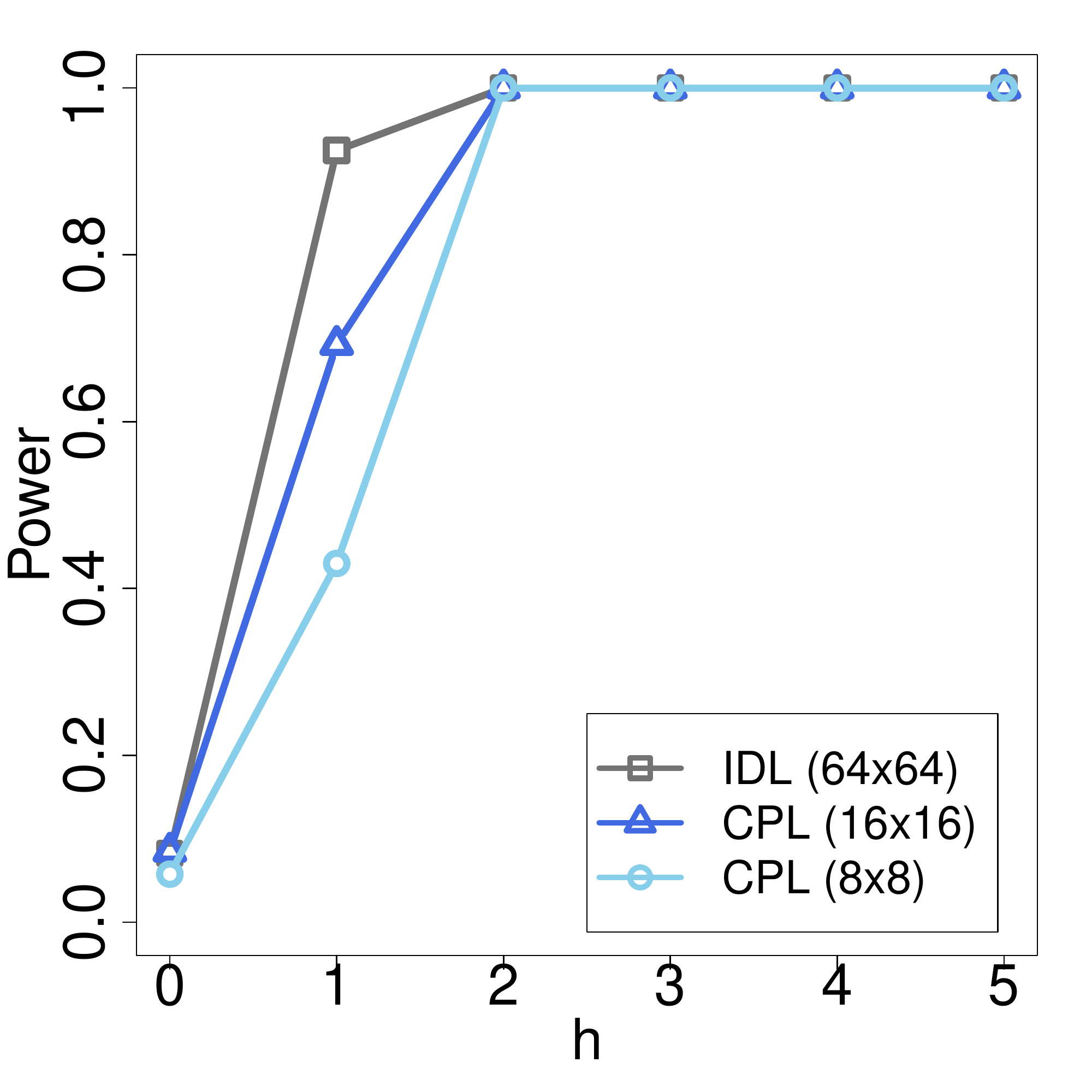} &
\includegraphics[scale=0.22,trim={0cm 0.3cm 1cm 1cm},clip]{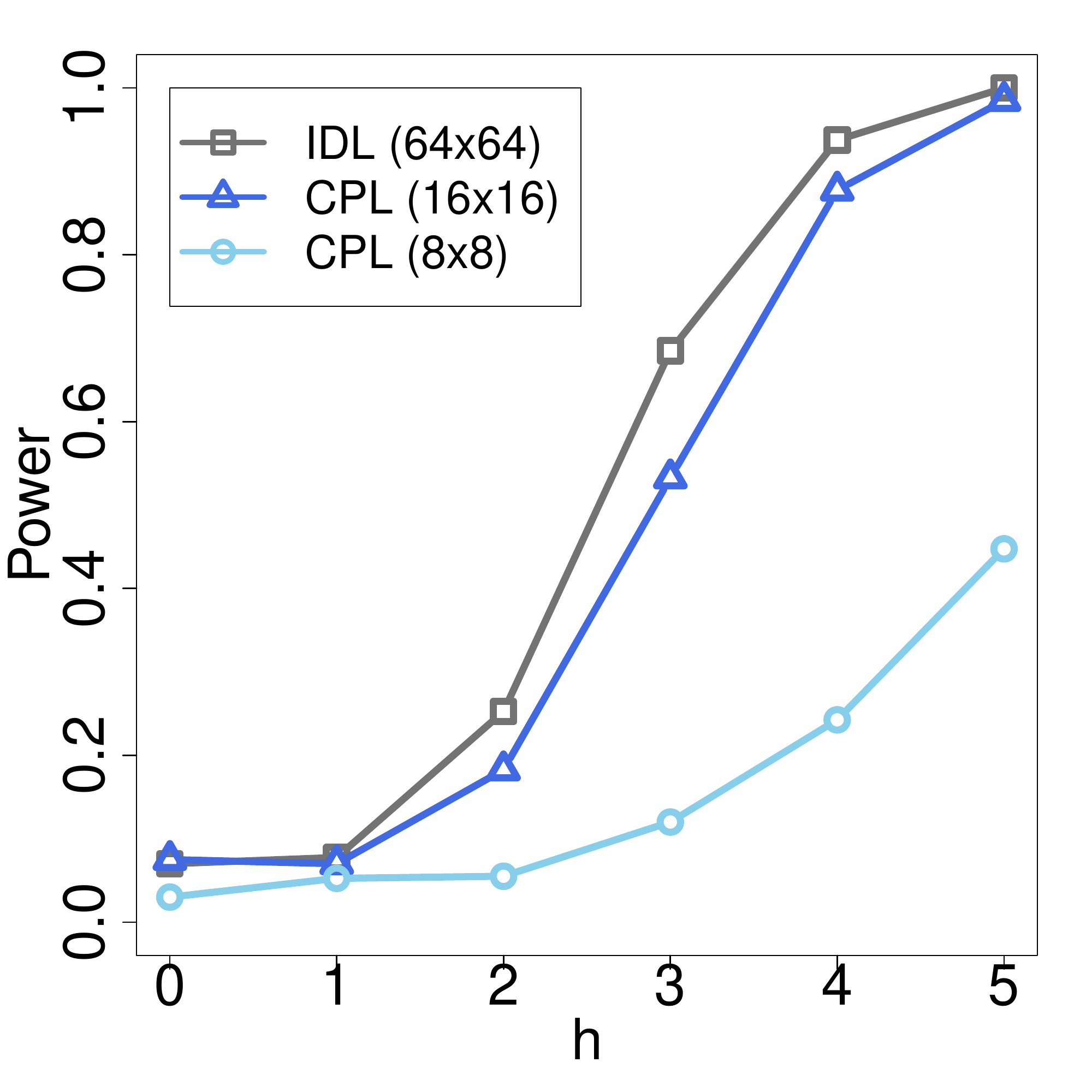} &
\includegraphics[scale=0.22,trim={0cm 0.3cm 1cm 1cm},clip]{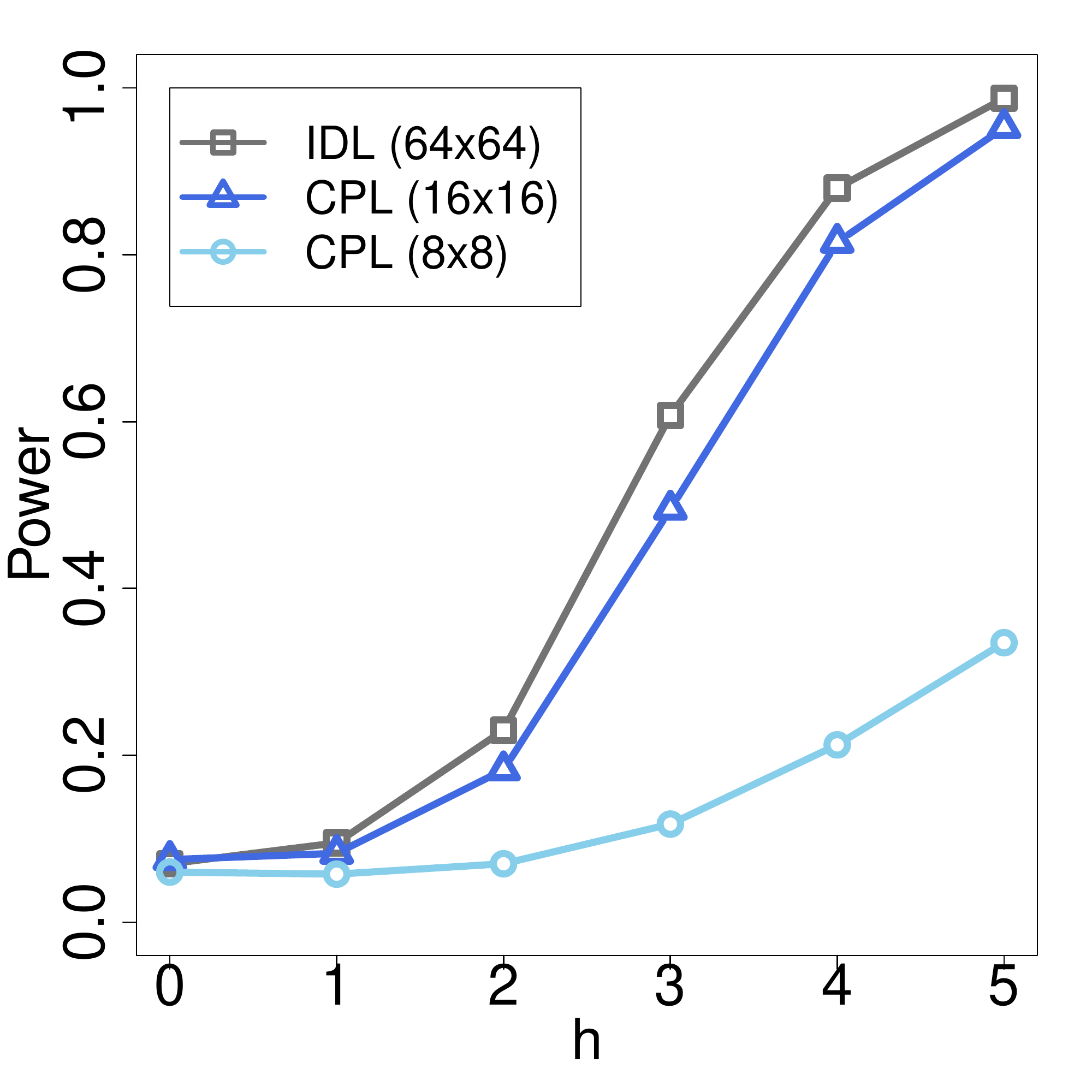} \\
\end{tabular}
\caption{Empirical power curves of IDL, CPL ($16\times 16$), and CPL ($8\times 8$) in Experiment 1
described in Section~\ref{sec:Exp1},
as a function of the signal's magnitude $h$, for $r\in\{6,10\}$ and $\phi\in\{0,5,10\}$.
Power curves for all values of $(r,\phi)$ and for all methods (IDL, CPL, MOM, and NVE) are shown in
Figure~\ref{fig:power for experiment 1-2} in the Supplementary Material.}
\label{fig:power for experiment 1}
\end{figure}

Tests were carried out at the commonly used $5\%$ significance level,
and empirical power curves as a function of $h$ for all values of $(r,\phi)$ and for all methods (IDL, CPL, MOM, and NVE) are shown in
Figure~\ref{fig:power for experiment 1-2} in the Supplementary Material; note that the empirical power curve at $h=0$ is equivalent to the empirical Type-I error rate.
These power curves suggest that our proposed procedure, CPL, is slightly more 
competitive than MOM and IDL (although not substantially),
so in Figure~\ref{fig:power for experiment 1} we show only a comparison of scenarios for
IDL and CPL as a function of signal magnitude $h$ for $r\in\{6,10\}$ in \eqref{eq:signal}, and $\phi\in\{0,5,10\}$.
Note that for a true power $\pi$, the Monte Carlo standard error of the estimated power is $\{\pi(1-\pi)/400\}^{1/2}$,
which is bounded above by $0.025$ (when $\pi=0.5$).

From the power-curve plots in Figure~\ref{fig:power for experiment 1},
we see that the Type-I error rates of CPL (and IDL) are under control,
reinforcing our conclusions from an initial simulation study; see Section \ref{sec:Type-I} in the Supplementary Material.
At worst, when $r=10$ and $\phi=10$, and from the nominal value of $0.05$, 
CPL has a Type-I error rate of $0.075$ for $\tilde{\bm{Z}}_{16\times 16}$ and $0.060$ for $\tilde{\bm{Z}}_{8\times 8}$
with Monte Carlo standard errors of $0.013$ and $0.012$, respectively.
It is clear that the power curve of CPL increases with the magnitude $h$ and the extent $r$ of the signal, as it does for IDL.
We can also see that signals can be detected much more easily for smaller $\phi$ (i.e., when the spatial dependence is weaker).
In particular, the power curves for $\phi=0$ are considerably larger than the corresponding powers for $\phi\in\{5,10\}$,
indicating that spatial dependence makes the signal-detection problem harder.
It is also not surprising to see that our proposed procedure CPL has more power when applied to $\tilde{\bm{Z}}_{16\times 16}$
than when applied to $\tilde{\bm{Z}}_{8\times 8}$,
and the empirical power curve increases more slowly with $h$ when the spatial dependence is stronger.

It is encouraging that IDL's and CPL's empirical power curves are close
after one coarsening of resolution to $\tilde{\bm{Z}}_{16\times 16}$.
Often CPL (and MOM) applied to $\tilde{\bm{Z}}_{16\times 16}$ outperformed IDL applied to $\bm{Z}$.
This is likely a consequence of basing the smoothed estimate $\hat{\bm{\Sigma}}$ on the exponential covariance function,
which is of the same form as that used to generate $\bm{\delta}$ in \eqref{eq:data2}.
IDL makes no such assumptions when estimating the parameters $\hat{\bm{\theta}}$.
In the presence of spatial dependence,
another coarsening of resolution to $\tilde{\bm{Z}}_{8\times 8}$ results in a substantial deterioration of the empirical power curve of CPL
(and that of MOM, not shown),
with power to find a signal occurring only when $r=10$. Again, as spatial dependence $\phi$ increases,
the power to detect a spatial signal weakens.
The full set of power curves is shown in Figure~\ref{fig:power for experiment 1-2} in the Supplementary Material.

\begin{figure}[!tb]\centering
\begin{tabular}{ccc}
~~IDL ($64\times 64$) & ~~~CPL ($16\times 16$) & ~~CPL ($8\times 8$) \\
\includegraphics[scale=0.25,trim={0.2cm 0.2cm 0.2cm 0.25cm},clip]{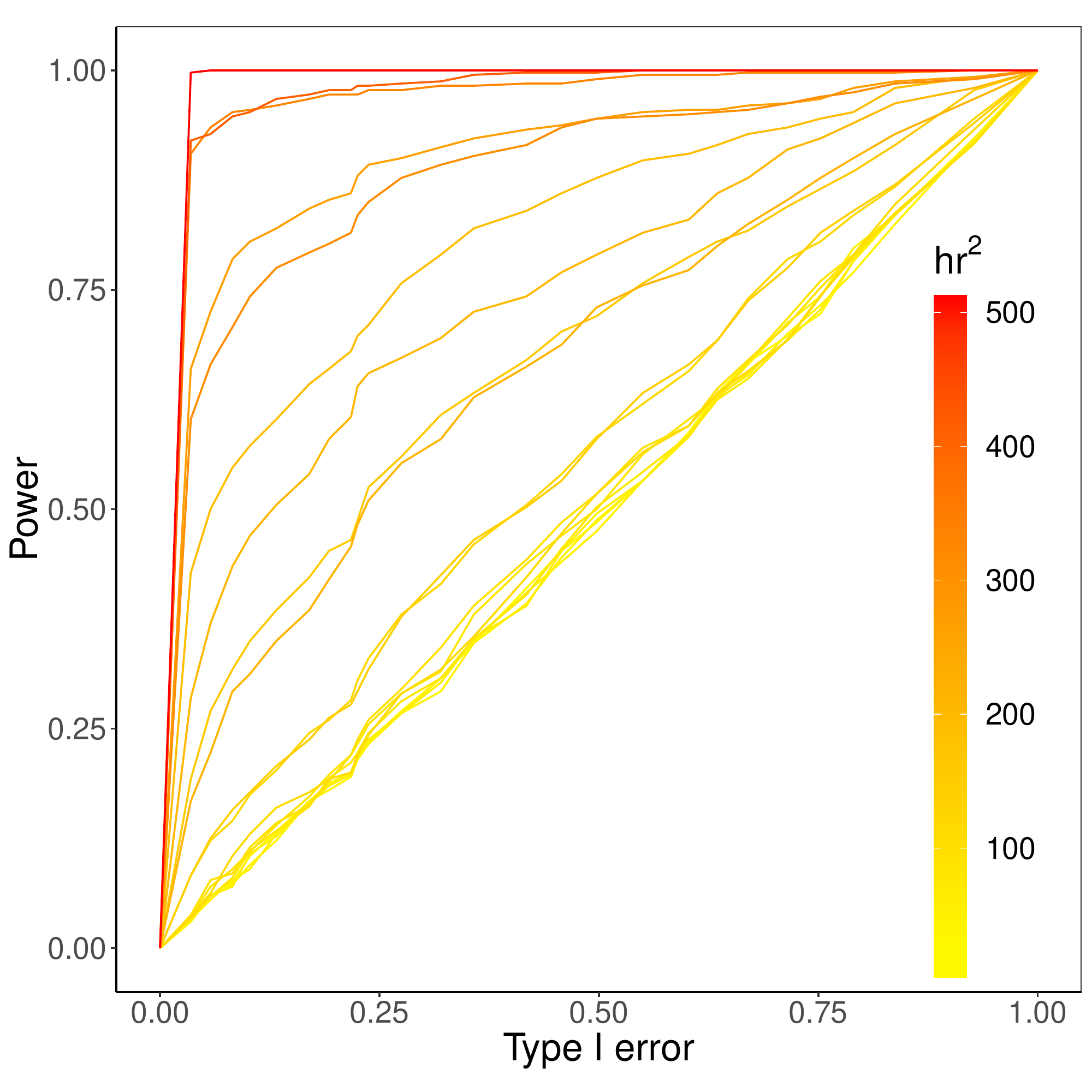} &
\includegraphics[scale=0.25,trim={0.2cm 0.2cm 0.2cm 0.25cm},clip]{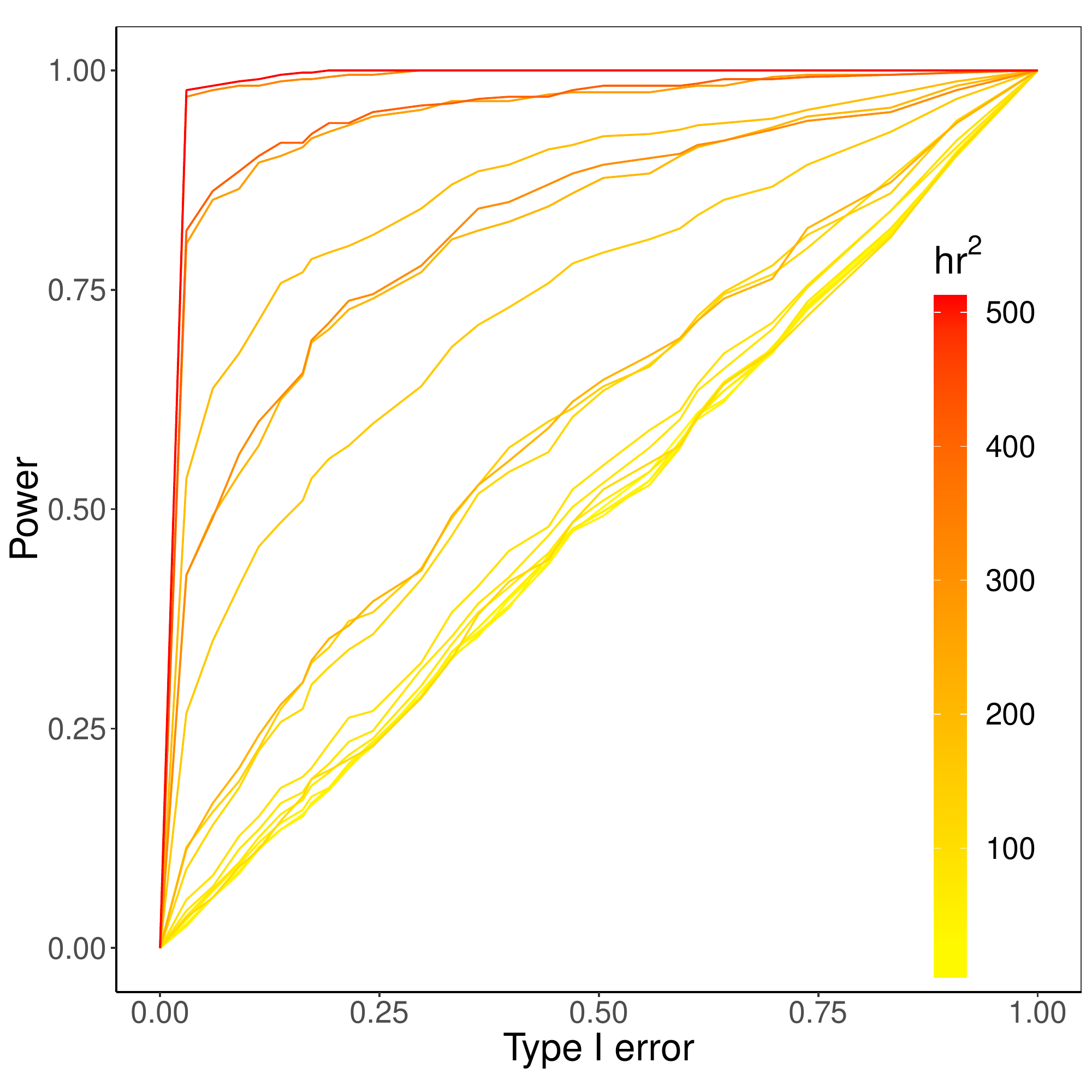} &
\includegraphics[scale=0.25,trim={0.2cm 0.2cm 0.2cm 0.25cm},clip]{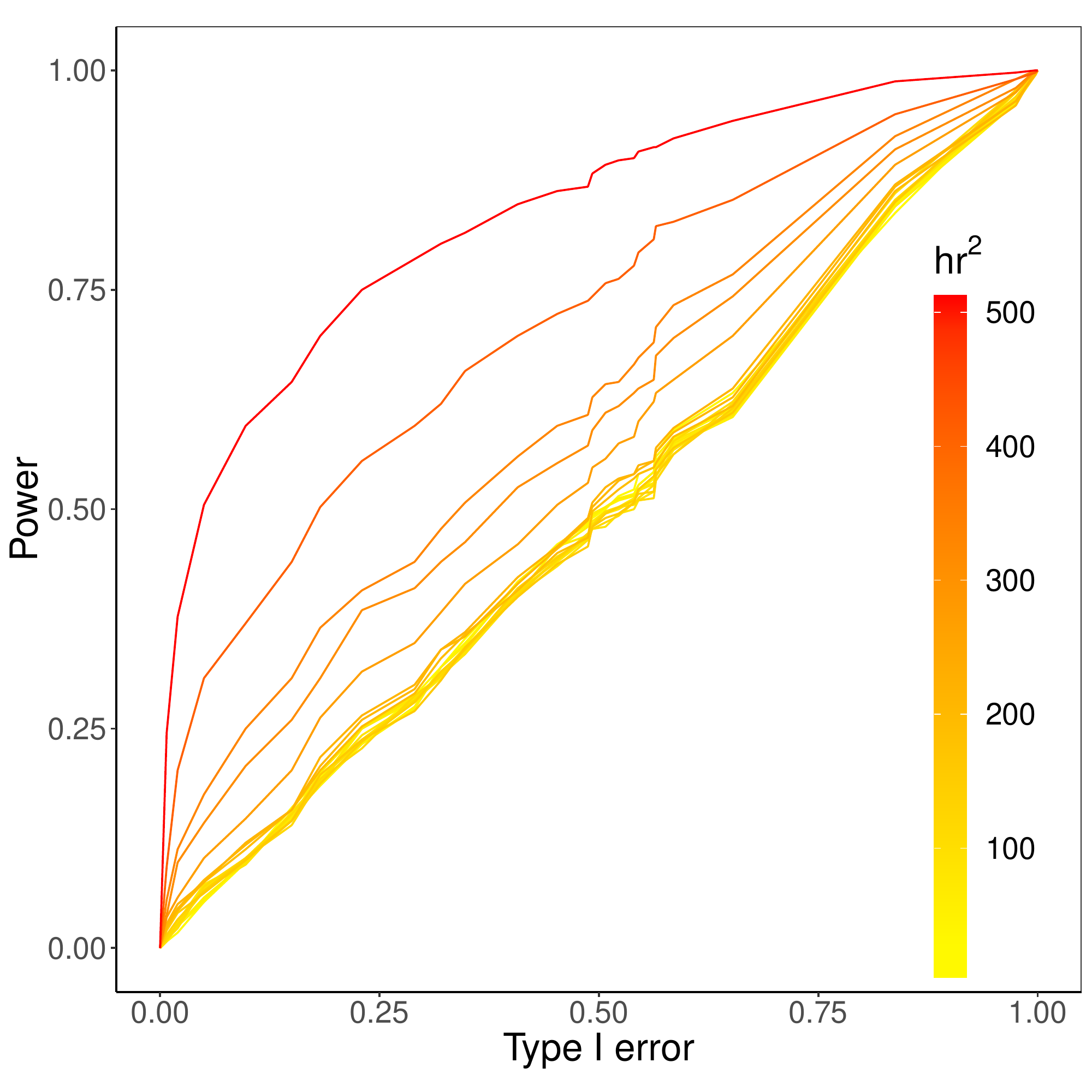} \\
\end{tabular}
\caption{Empirical ROC curves of IDL, CPL ($16\times 16$), and CPL ($8\times 8$) in Experiment 1
for $\phi=5$ and different scales of aggregations. Curves are
colored according to the volume $h r^2$ of the signal obtained from 24 combinations of $r\in\{4,6,8,10\}$ and $h\in\{0,1,2,3,4,5\}$ in \eqref{eq:signal}.
The full set of ROC curves are shown in Figure~\ref{fig:ROC for experiment 1-2} in the Supplementary Material.}
\label{fig:ROC for experiment 1}
\end{figure}

We also used the receiver operating characteristic (ROC) curve
(e.g., Egan, 1975) to facilitate comparison of the different methods.
Figure \ref{fig:ROC for experiment 1} shows empirical ROC curves with signals of various volumes,
$h r^2$, obtained from the 24 combinations of $r\in\{4,6,8,10\}$ and $h\in\{0,1,2,3,4,5\}$ in \eqref{eq:signal}, and $\phi=5$.
Each empirical ROC curve was computed based on 200 images of $\tilde{\bm{Z}}_{64\times 64}=\bm{Z}$,
100 of these images generated under the null hypothesis of no signal and the other 100 images generated under the alternative hypothesis
by adding a signal given by $r$ and $h$ in \eqref{eq:signal}.
Each curve was traced out by varying $\alpha$ on the right-hand side of \eqref{eq:test}.
From Figure \ref{fig:ROC for experiment 1},
we see that the area under the curve (AUC) tends to increase with the signal volume $h r^2$
and decrease with the amount of aggregation, as expected.
The full set of ROC curves for this experiment is shown in Figure~\ref{fig:ROC for experiment 1-2} in the Supplementary Material.

\subsection{Experiment 2: Missing data (in a contiguous block) at different scales of aggregation}
\label{sec:Exp2}

\begin{figure}[!tb]\centering
\begin{tabular}{ccc}
$\tilde{\bm{Z}}_{64\times64}$ & $\tilde{\bm{Z}}_{16\times 16}$~ & $\tilde{\bm{Z}}_{8\times 8}$~
\smallskip\\
\includegraphics[scale=0.32,trim={1cm 2.5cm 2.5cm 2cm},clip]{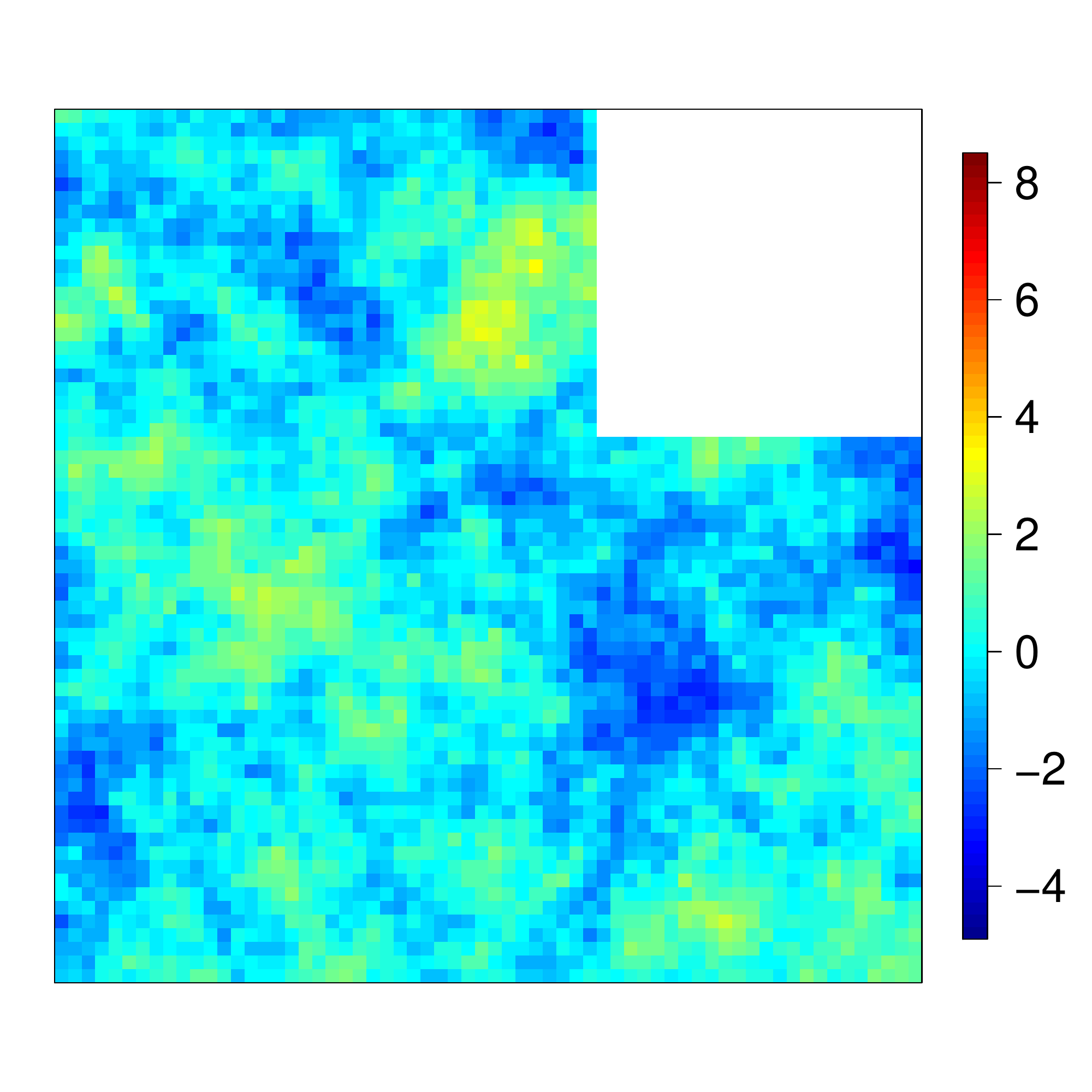} &
\!\!\!\!\!\!\includegraphics[scale=0.32,trim={1cm 2.5cm 2.5cm 2cm},clip]{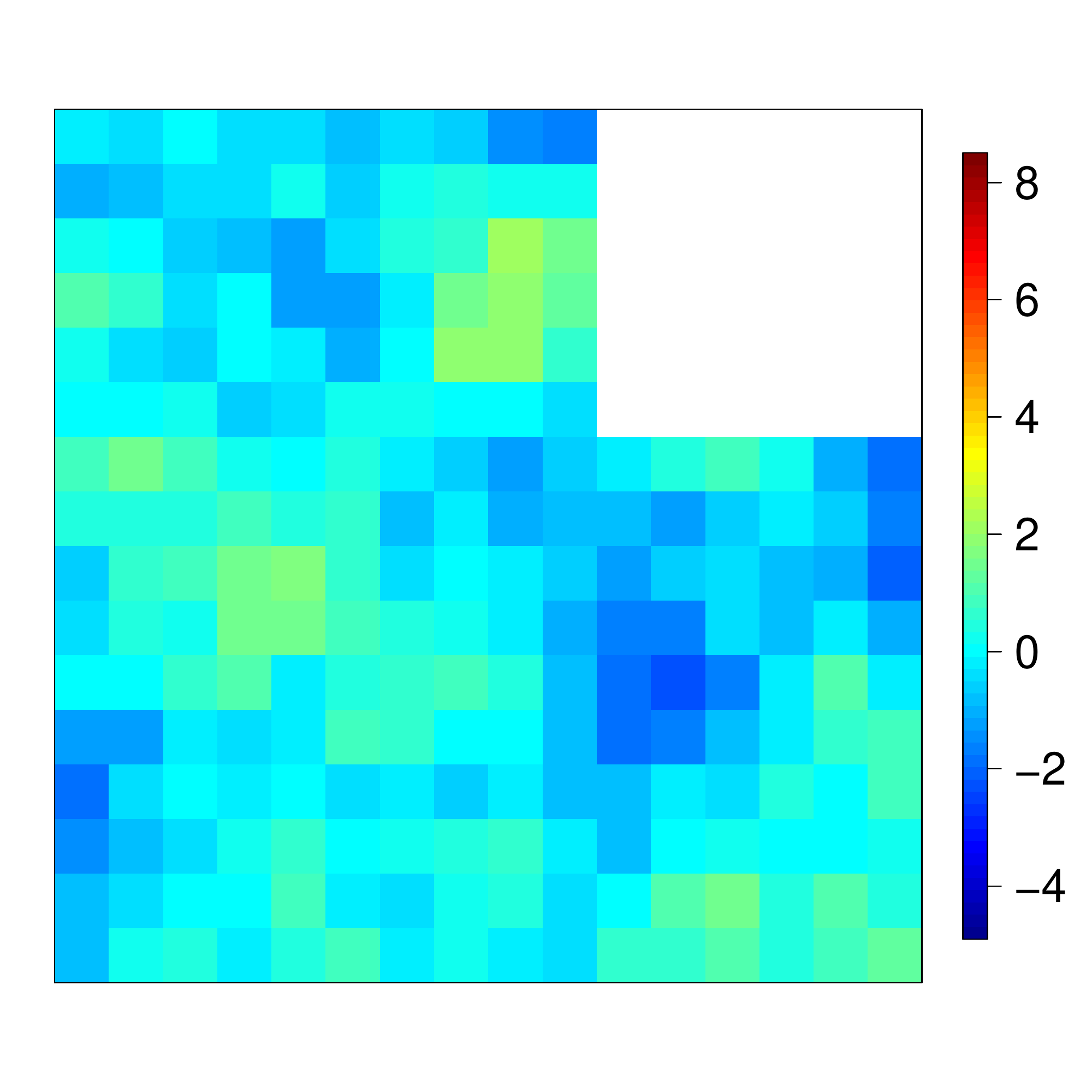} &
\!\!\!\!\!\!\includegraphics[scale=0.32,trim={1cm 2.5cm 2.5cm 2cm},clip]{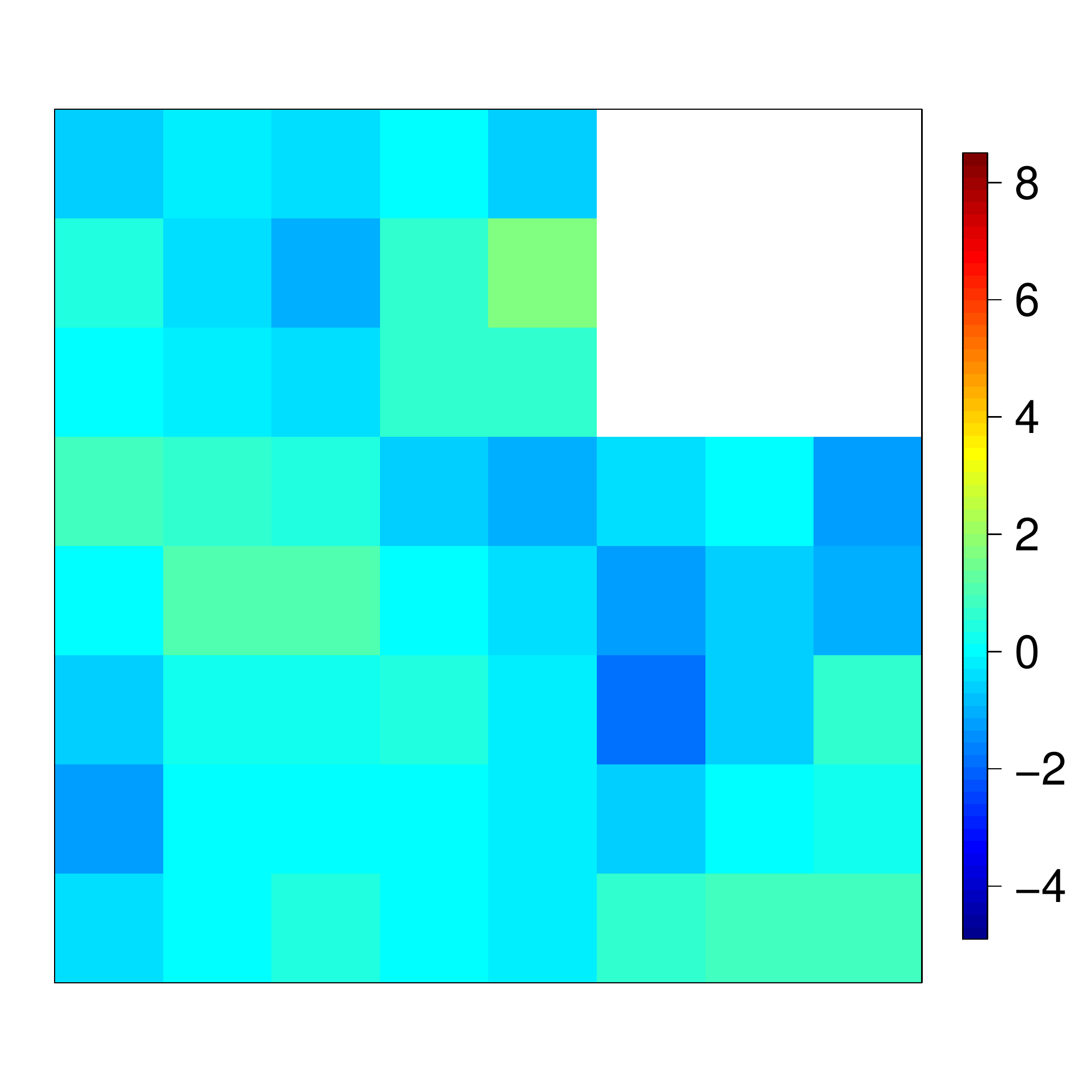} \\
\end{tabular}
\caption{Missing data (in a contiguous block) in Experiment 2 described in Section~\ref{sec:Exp2},
at various scales of aggregation; in all images, the missing fraction is 9/64, and
the spatial dependence $\phi=5$.}
\label{fig:fourimagewithmissingfixed}
\end{figure}

Experiment 2 is similar to Experiment 1, except that here we considered missing data in the upper-right corner of
$\tilde{\bm{Z}}_{64\times 64}$ ($=\bm{Z}$), $\tilde{\bm{Z}}_{16 \times 16}$, and $\tilde{\bm{Z}}_{8\times 8}$,
with the fraction of missing data fixed at $9/64$;
see Figure~\ref{fig:fourimagewithmissingfixed} for an illustration.
Tests were carried out at the usual $5\%$ significance level.
The empirical power curves and the empirical ROC curves obtained were similar to those in 
Figures~\ref{fig:power for experiment 1} and \ref{fig:ROC for experiment 1}, respectively;
for the full set of curves, see Figures~\ref{fig:power for experiment 2-2}
and \ref{fig:ROC for experiment 2-2} in the Supplementary Material.
The results demonstrate that our proposed
procedure is not likely to be affected by a large contiguous block of missing data
(as long as the block does not contain signal).

\subsection{Experiment 3: Missing data (at random) at different scales of aggregation}
\label{sec:Exp3}

\begin{figure}[!tb]\centering
\begin{tabular}{ccc}
$\tilde{\bm{Z}}_{64\times64}$ & $\tilde{\bm{Z}}_{16\times 16}$~ & $\tilde{\bm{Z}}_{8\times 8}$~
\smallskip\\
\includegraphics[scale=0.32,trim={1cm 2.5cm 2.5cm 2cm},clip]{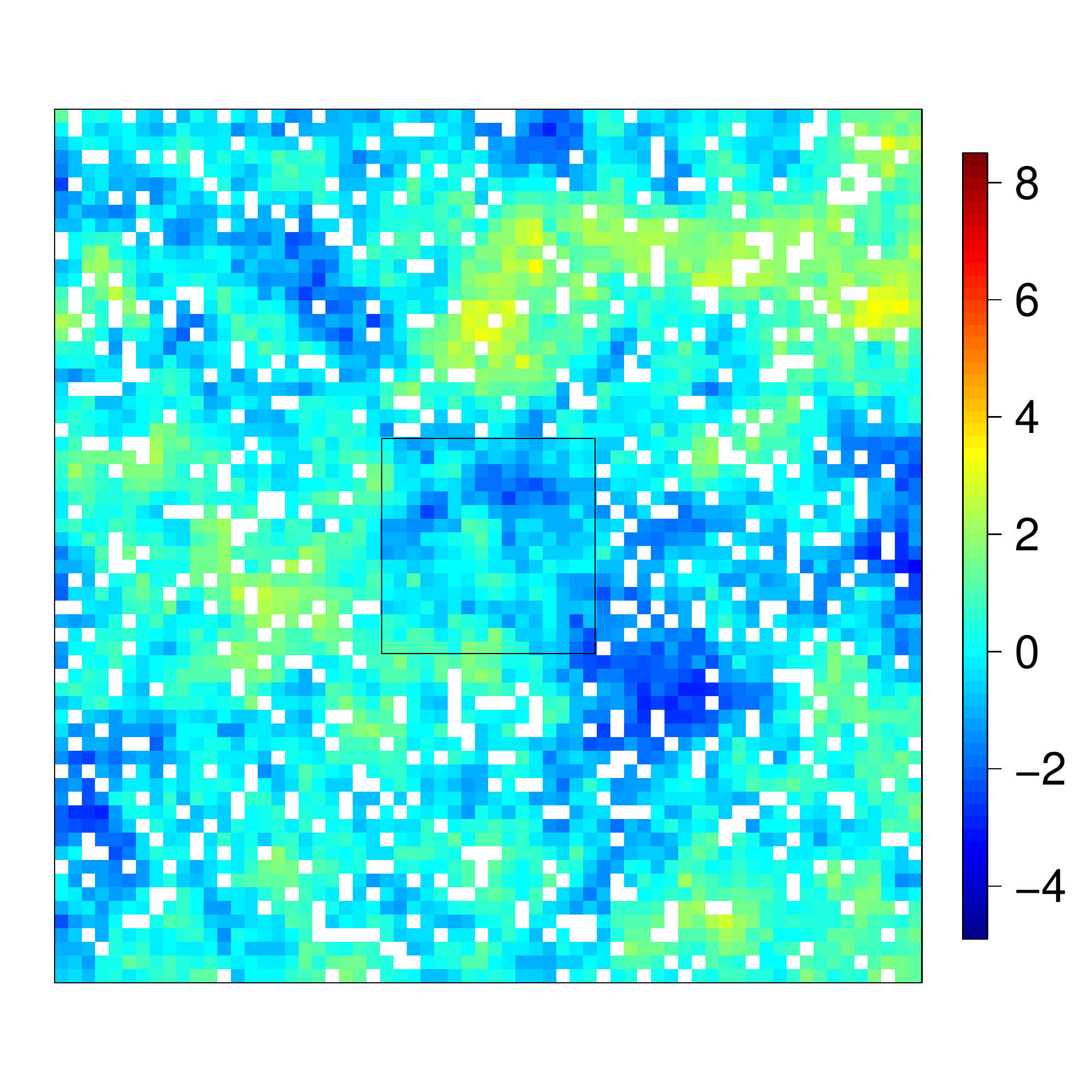} &
\!\!\!\!\!\!\includegraphics[scale=0.32,trim={1cm 2.5cm 2.5cm 2cm},clip]{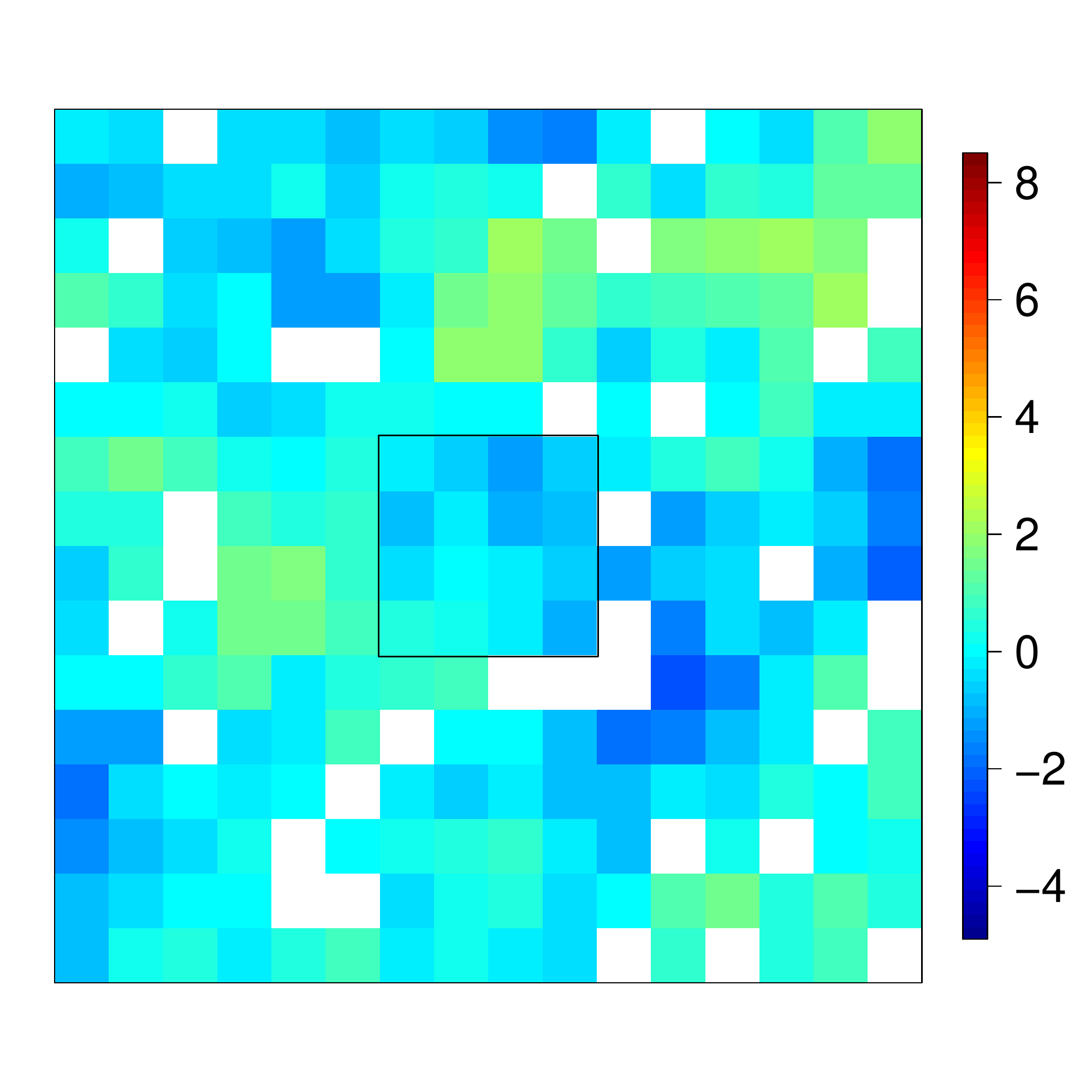} &
\!\!\!\!\!\!\includegraphics[scale=0.32,trim={1cm 2.5cm 2.5cm 2cm},clip]{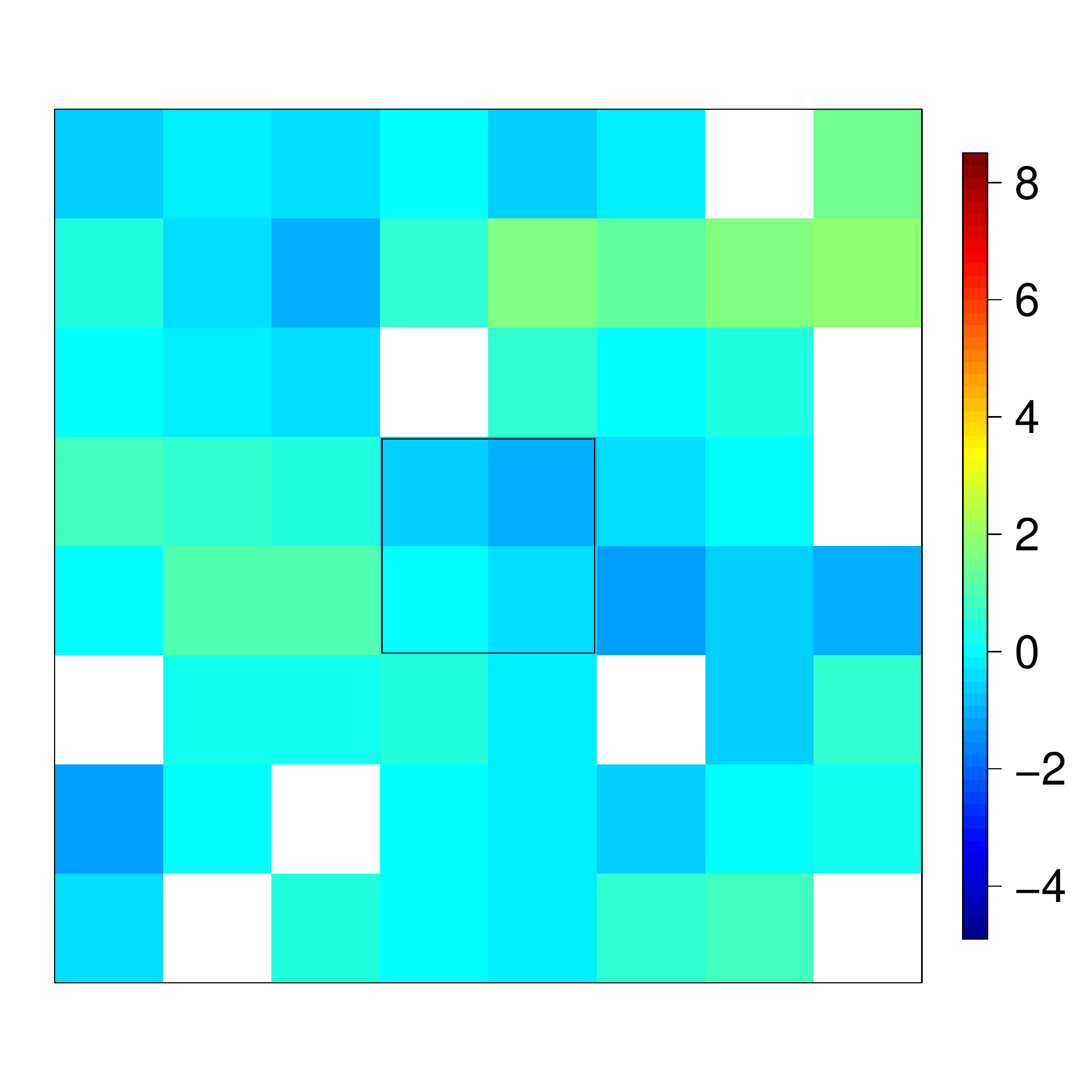} \\
\end{tabular}
\caption{Missing data (at random) in Experiment 3 described in Section~\ref{sec:Exp3},
at various scales of aggregation; in all images, the missing fraction is 9/64, and
the spatial dependence $\phi=5$.}
\label{fig:fourimagewithmissing}
\end{figure}

Experiment 3 is similar to Experiment 2, except that we considered small blocks missing at random, with the same fraction missing ($=9/64$);
see Figure~\ref{fig:fourimagewithmissing} for an illustration. These were
taken at random from all blocks except those in the central square region
(where the signal is located), shown with a black square outline in each panel of Figure~\ref{fig:fourimagewithmissing}.
The empirical power curves and the empirical ROC curves obtained were similar to those in Experiments 1 and 2;
for the full set of curves, see
Figures~\ref{fig:power for experiment 3-2} and \ref{fig:ROC for experiment 3-2} in the Supplementary Material.
A comparison of Experiments 2 and 3 shows that our proposed procedure, CPL, performs well
irrespective of whether the data are missing at random or in a contiguous block.
This is an illustration of how spatial modeling and its corresponding conditional simulation
can successfully borrow strength for two very different missing-data mechanisms.

\subsection{Experiment 4: Nonstationary noise}
\label{sec:Exp4}

Experiment 4 is similar to Experiment 1 except that data $\bm{Z}$ are generated with $\delta(\cdot)$ in \eqref{eq:signal+noise}
replaced by a nonstationary process $\delta^*(\cdot)$ obtained by deforming the coordinates of $\delta(\cdot)$:
\[
\delta^*(\bm{s})=\delta(\|\bm{s}\|^\kappa \bm{s});\quad \bm{s}\in D,
\]
where $\kappa$ is a parameter controlling the degree of nonstationarity with $\kappa=0$ corresponding to a non-deformed stationary process.
The spatial covariance function of $\delta^*(\cdot)$ is
\begin{align*}
\mathrm{cov}(\delta^*(\bm{s}),\delta^*(\bm{s^*}))
=\mathrm{cov}(\delta(\|\bm{s}\|^\kappa \bm{s}),\delta(\|\bm{s}^*\|^\kappa \bm{s}^*))
=\exp(-\|(\|\bm{s}\|^\kappa\bm{s}-\|\bm{s}^*\|^\kappa\bm{s}^*)\|/\phi);\quad\bm{s},\bm{s}^*\in D.
\end{align*}

\begin{figure}[tb]\centering
\begin{tabular}{ccccc}
$\kappa=-0.75$~~ & $\kappa=-0.5$~~~ & $\kappa=0$~~~~ & $\kappa=0.5$~~~ & $\kappa=2$~~~~
\smallskip\\
\!\!\!\includegraphics[scale=0.165,trim={1cm 2cm 0.5cm 1.5cm},clip]{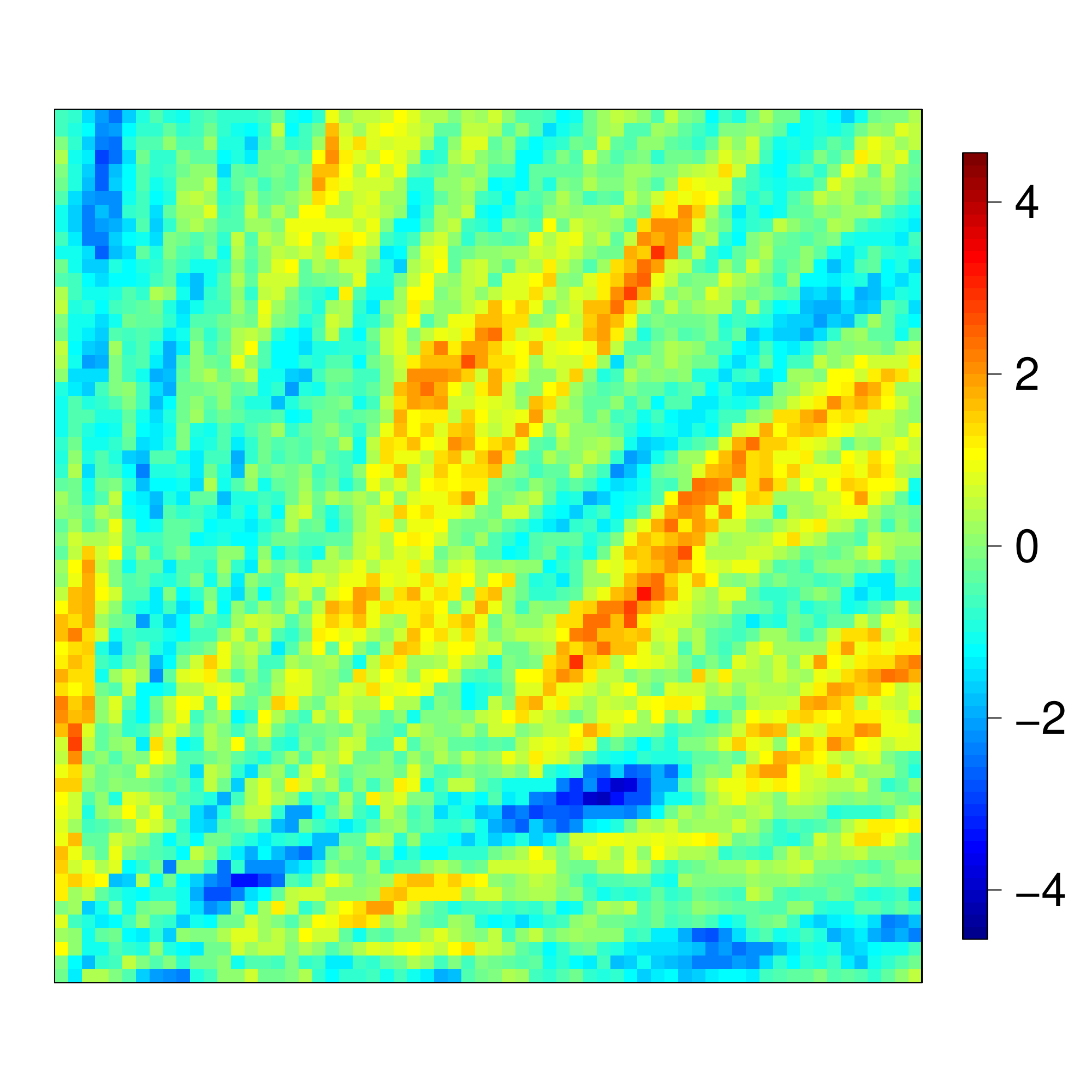} &
\!\!\!\includegraphics[scale=0.165,trim={1cm 2cm 0.5cm 1.5cm},clip]{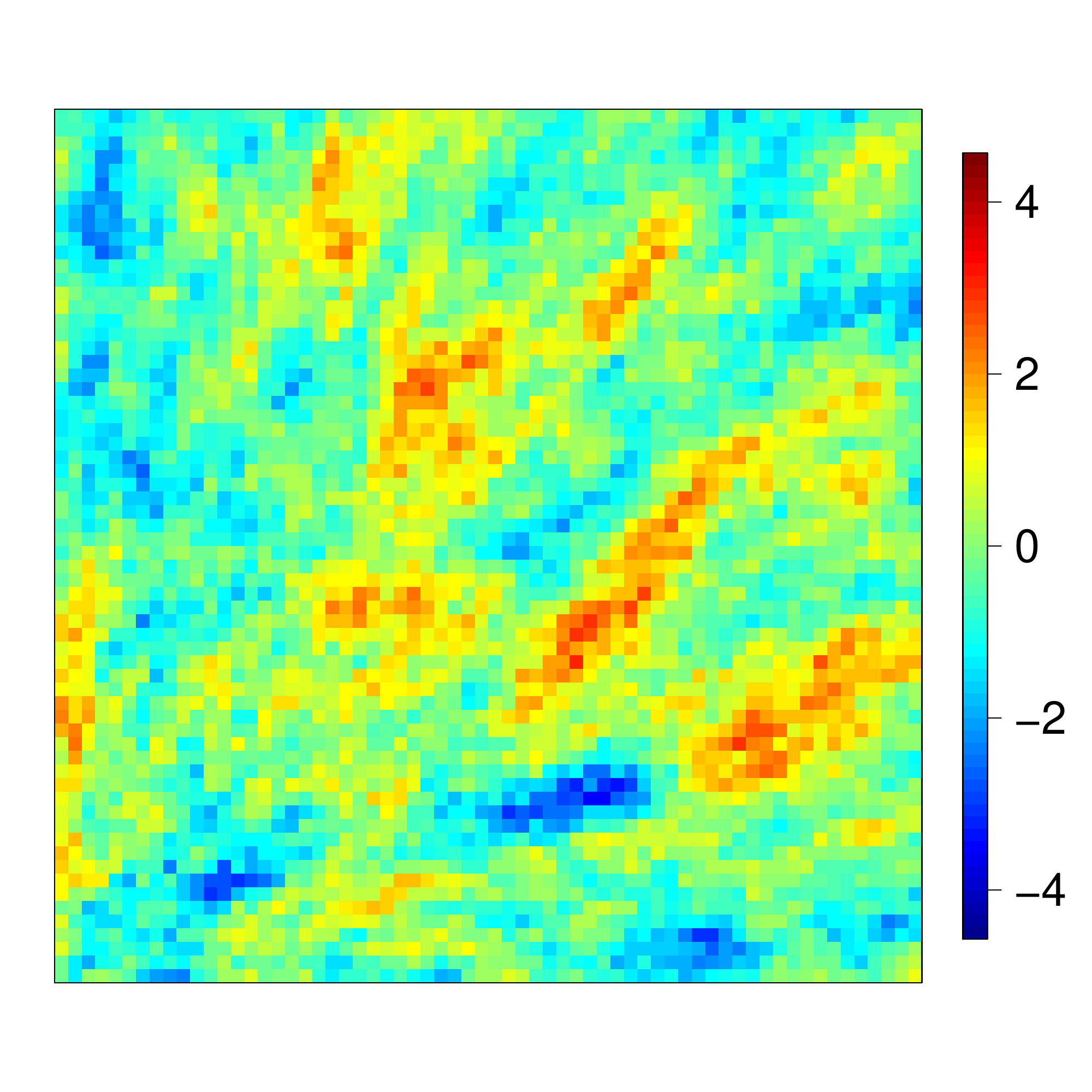} &
\!\!\!\includegraphics[scale=0.165,trim={1cm 2cm 0.5cm 1.5cm},clip]{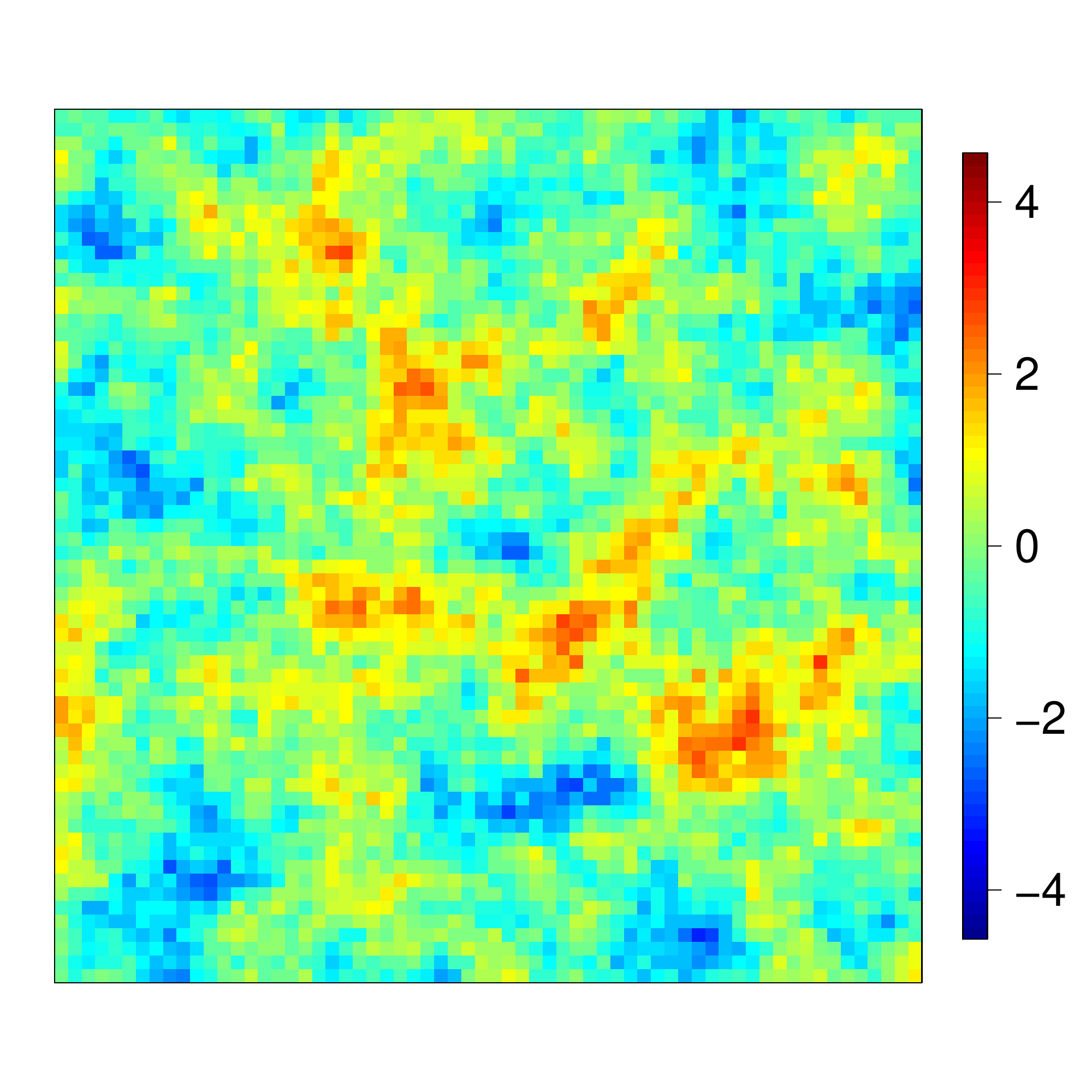}&
\!\!\!\includegraphics[scale=0.165,trim={1cm 2cm 0.5cm 1.5cm},clip]{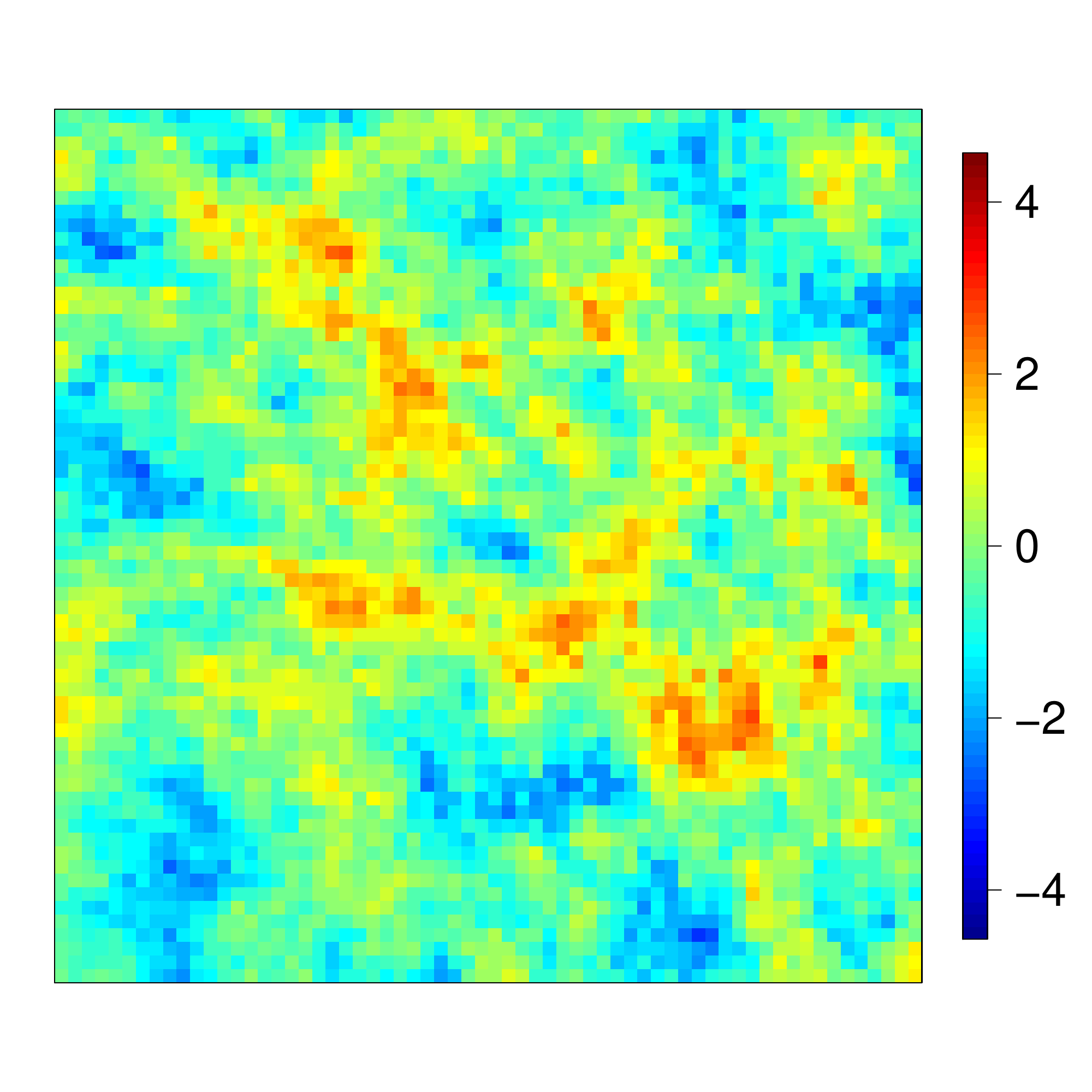}&
\!\!\!\includegraphics[scale=0.165,trim={1cm 2cm 0.5cm 1.5cm},clip]{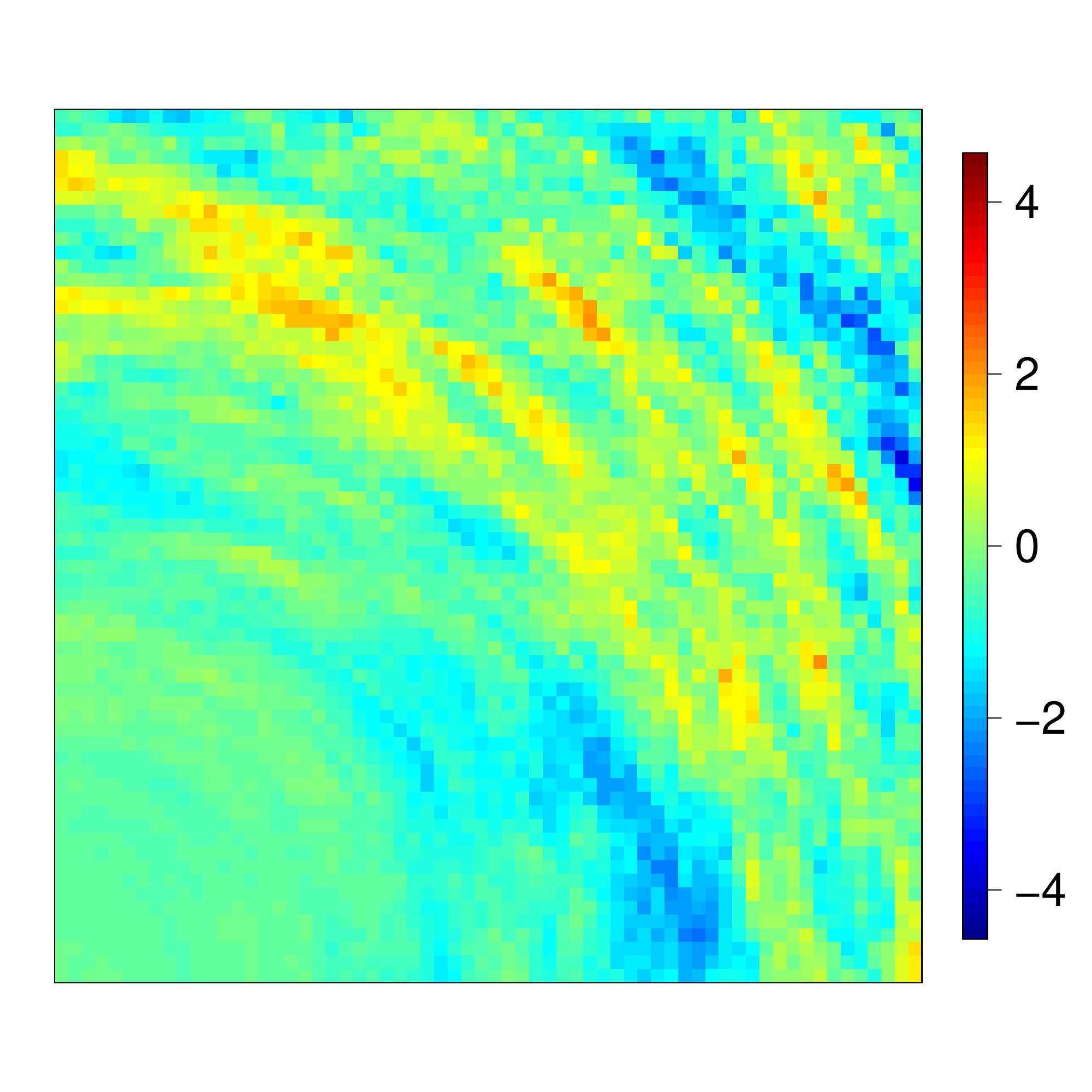}
\end{tabular}
\caption{Five randomly generated images of $\tilde{\bm{Z}}_{64\times 64}\equiv\bm{Z}$ corresponding to various degrees of nonstationarity
with $\kappa\in\{-0.75,-0.5,0,0.5,2\}$, where $\kappa=0$ corresponds to a stationary process, and signal magnitude $h=0$.}
\label{fig:noise}
\end{figure}

\noindent A larger departure of $\kappa$ from $0$ indicates a higher degree of nonstationarity.
In this experiment, we considered $r=6$, $\phi=5$, $h\in\{0,1,\dots,5\}$, and $\kappa\in\{-0.75,-0.5,0,0.5,2\}$.
Figure~\ref{fig:noise} shows five randomly generated datasets with signal magnitude $h=0$, for $\kappa\in\{-0.75,-0.5,0,0.5,2\}$.
We can see that the data show higher spatial dependence around the upper right-hand corner than the lower left-hand corner for $\kappa=-0.75$ and $\kappa=-0.5$.
In contrast, the data show higher spatial dependence around the lower left-hand corner than the upper right-hand corner
for $\kappa=0.5$ and $\kappa=2$.

\begin{figure}[!tb]\centering
\begin{tabular}{ccccc}
$\kappa=-0.75$ & $\kappa=-0.5$ & $\kappa=0$ & $\kappa=0.5$ & $\kappa=2$
\smallskip\\
\!\!\includegraphics[scale=0.16,trim={0cm 0.3cm 1cm 0.8cm},clip]{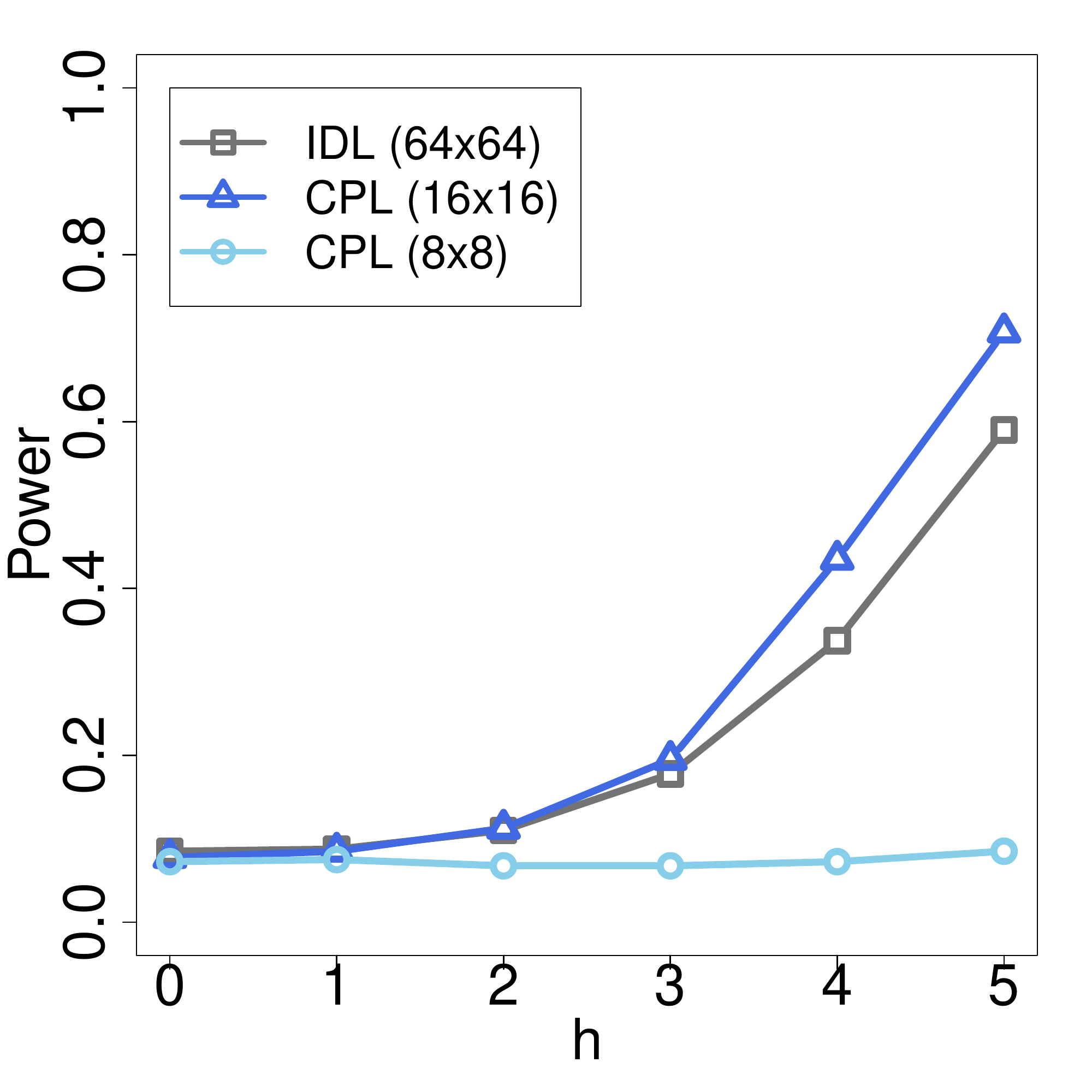} &
\!\!\!\includegraphics[scale=0.16,trim={0cm 0.3cm 1cm 0.8cm},clip]{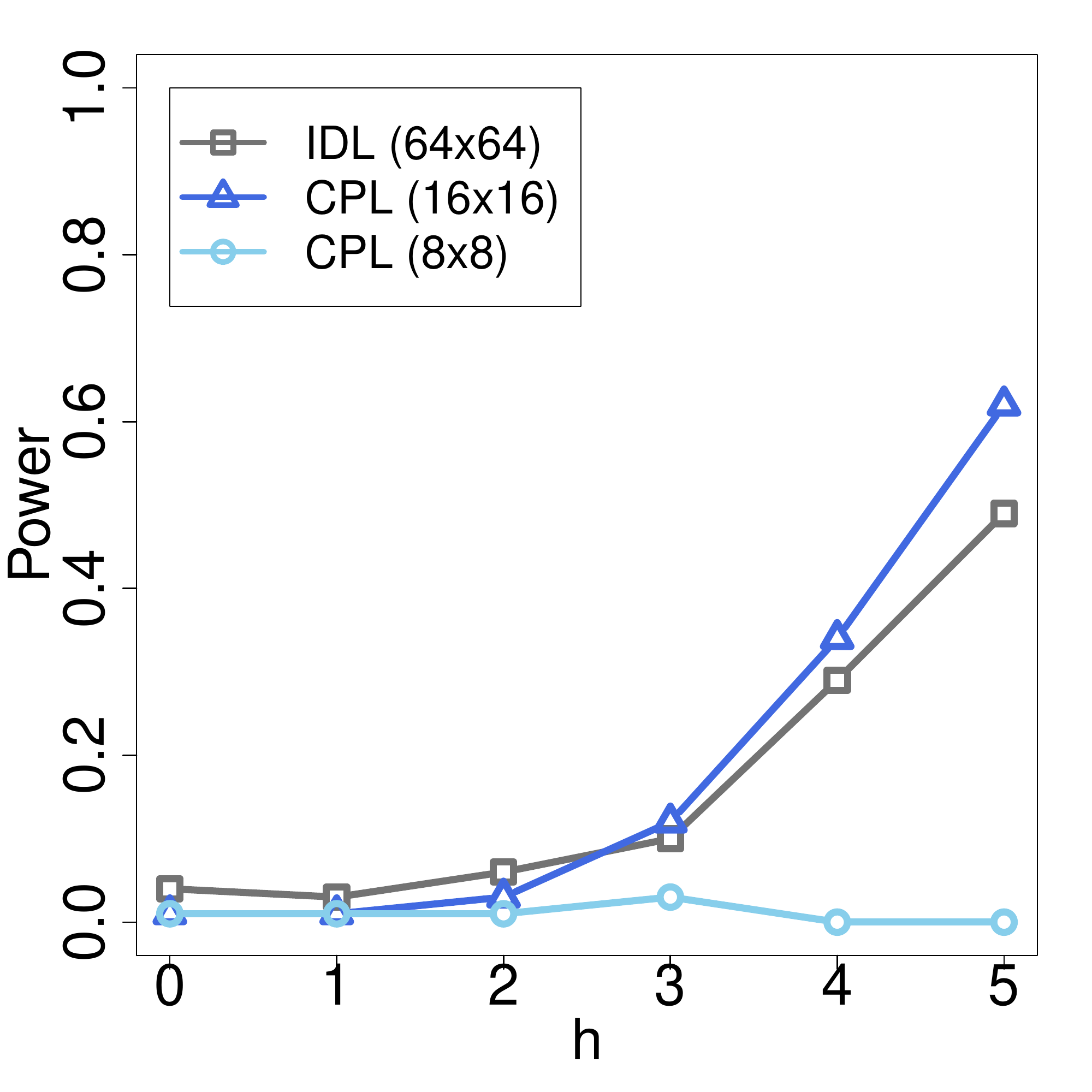} &
\!\!\!\includegraphics[scale=0.16,trim={0cm 0.3cm 1cm 0.8cm},clip]{./Exp1-power-phi5-r6-reduce} &
\!\!\!\includegraphics[scale=0.16,trim={0cm 0.3cm 1cm 0.8cm},clip]{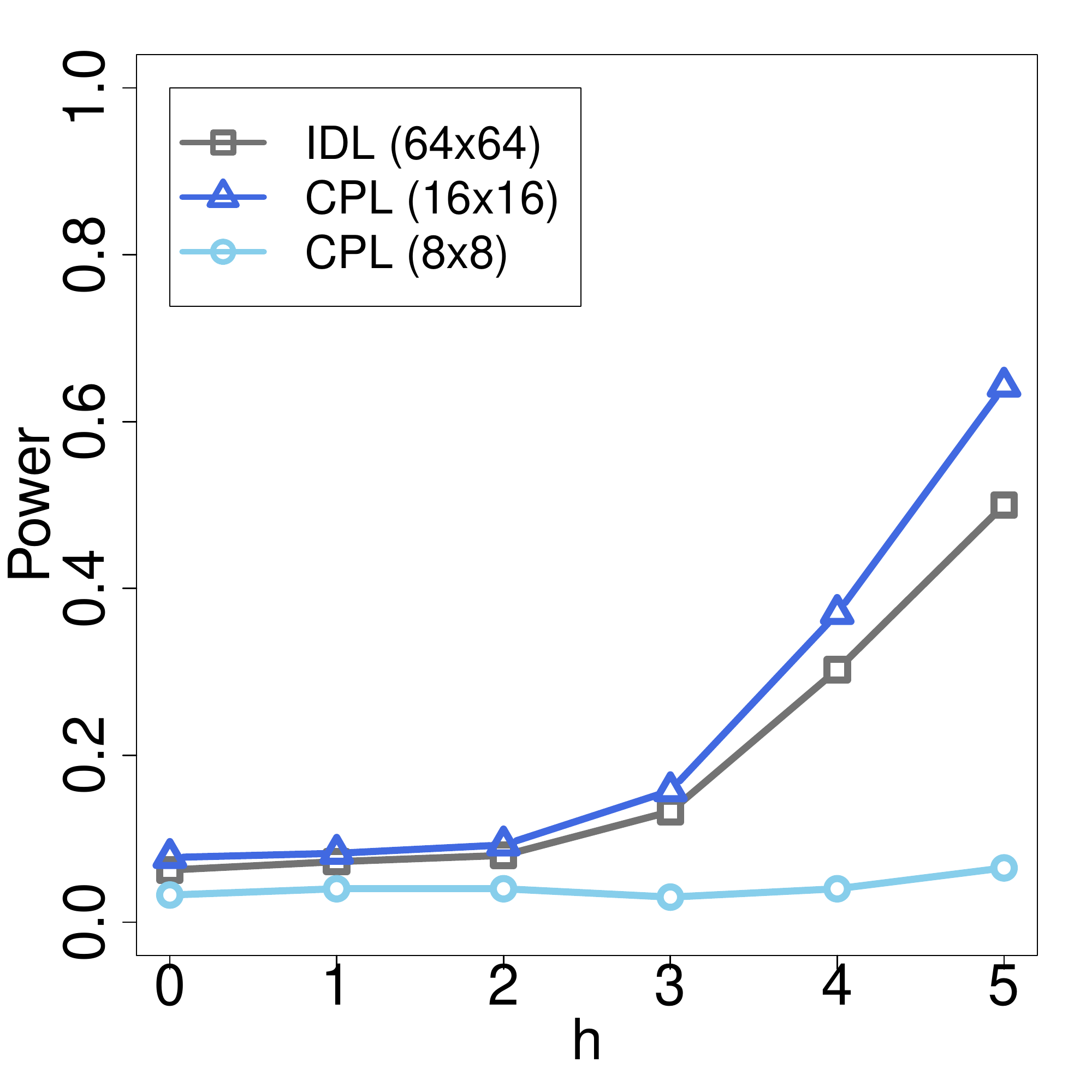} &
\!\!\!\includegraphics[scale=0.16,trim={0cm 0.3cm 1cm 0.8cm},clip]{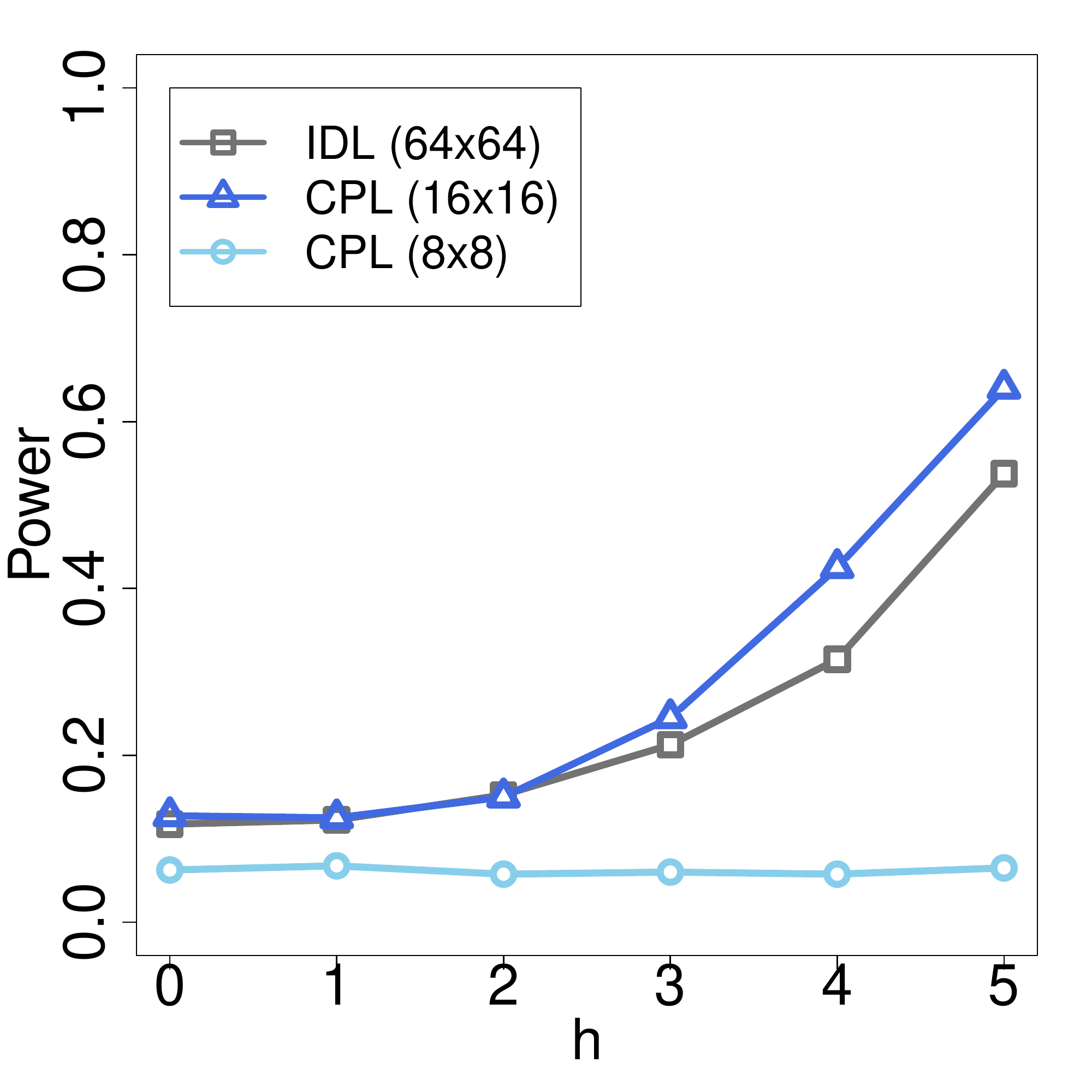} 
\end{tabular}
\caption{Empirical power curves of IDL, CPL ($16\times 16$), and CPL ($8\times 8$) in Experiment 4
described in Section~\ref{sec:Exp4},
as a function of the signal's magnitude $h$, for $r=6$, $\phi=5$ and $\kappa\in\{-0.75,-0.5,0,0.5,2\}$.
\label{fig:power for experiment 4}}
\end{figure}

The empirical power curves for $\kappa\in\{-0.75,-0.5,0,0.5,2\}$ are shown in Figure~\ref{fig:power for experiment 4},
where the curves are a function of $h\in\{0,1,\dots,5\}$, $r=6$ and $\phi=5$. Note that the plot for $\kappa=0$ is the same as the plot for $r=6$ and $\phi=5$ in Figure~\ref{fig:power for experiment 1}. 
Although both IDL and CPL (for $\tilde{\bm{Z}}_{16\times 16}$) show elevated Type-I error rates for $\kappa=-0.75$ and $\kappa=2$,
the five plots in Figure~\ref{fig:power for experiment 4} generally have similar power curves,
indicating that our proposed procedure is robust to mild nonstationarity.
Nevertheless, if the nonstationarity is strong
(e.g., the variance of $\delta(\bm{s})$ varies substantially as $\bm{s}$ ranges over $D$),
then it would not be possible to distinguish the mean from the random component without further prior knowledge on the covariance function.

\section{Applications}
\label{sec:application}
\bigskip

\subsection{An application to temperature data in the Asia-Pacific}
\label{sec:temperature}

\begin{figure}[!tb]\centering
\begin{tabular}{ccc}
\!\!\includegraphics[scale=0.285,trim={0.9cm 1.5cm 0.6cm 1.5cm},clip]{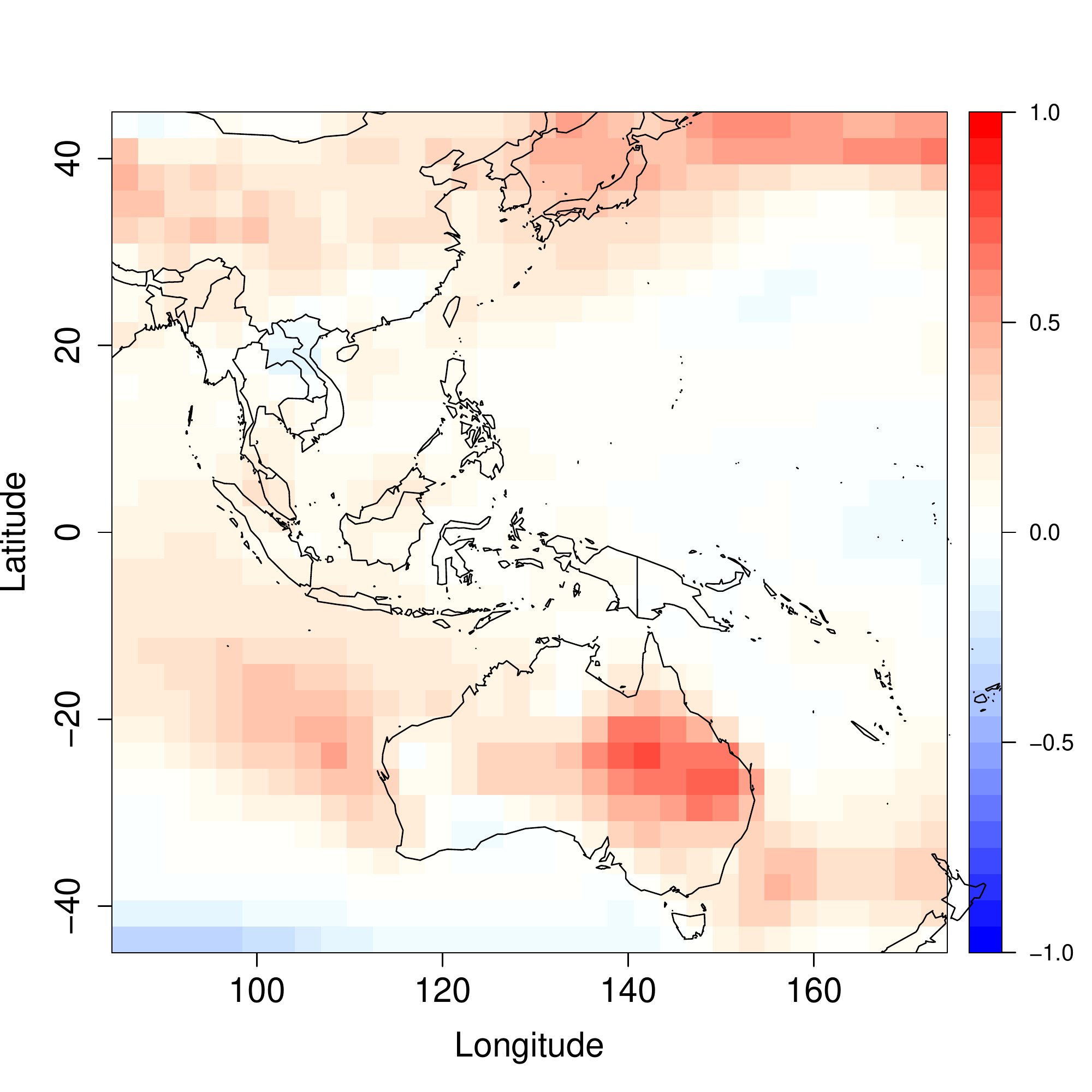} &
\!\!\!\!\includegraphics[scale=0.285,trim={0.9cm 1.5cm 0.6cm 1.5cm},clip]{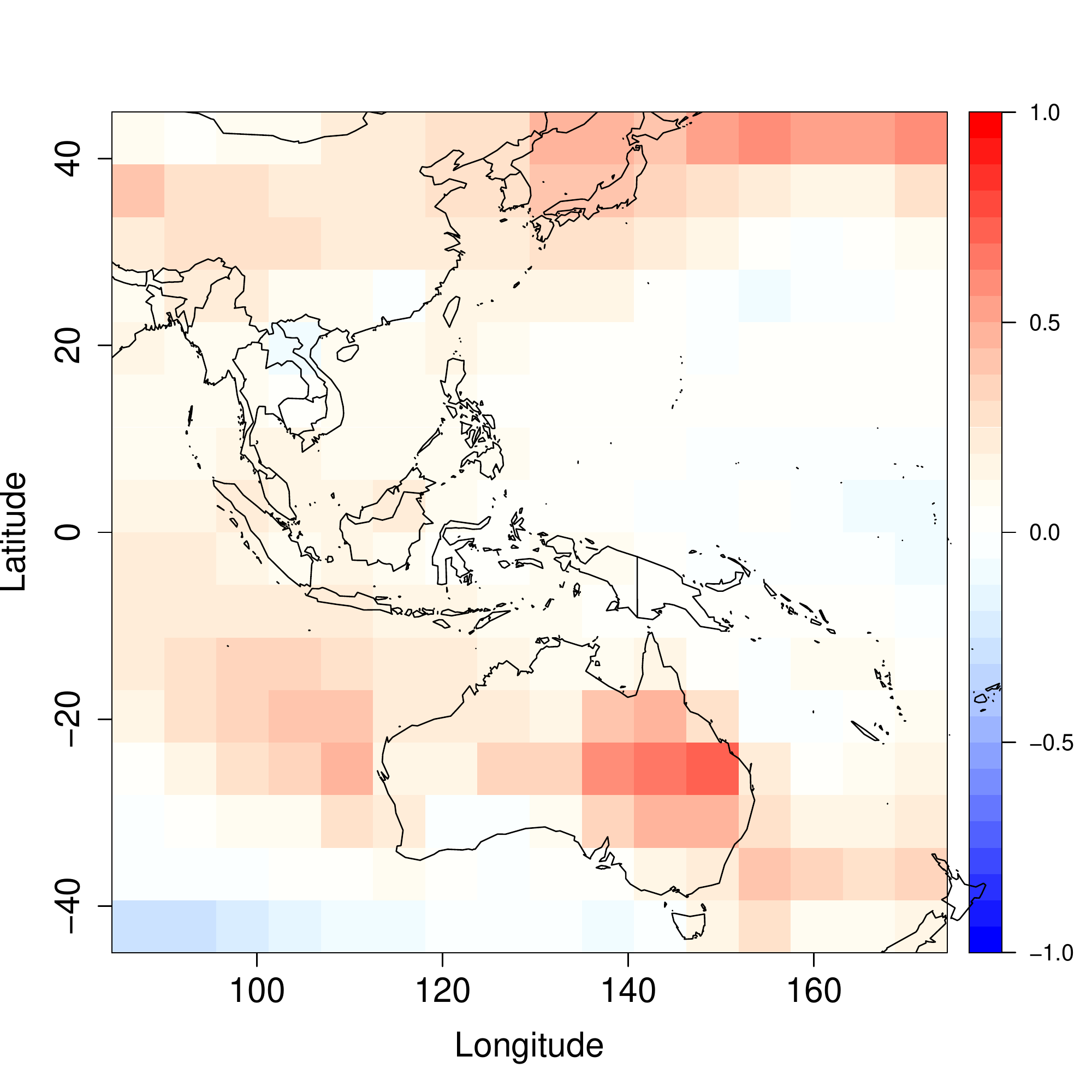} &
\!\!\!\!\includegraphics[scale=0.285,trim={0.9cm 1.5cm 0.6cm 1.5cm},clip]{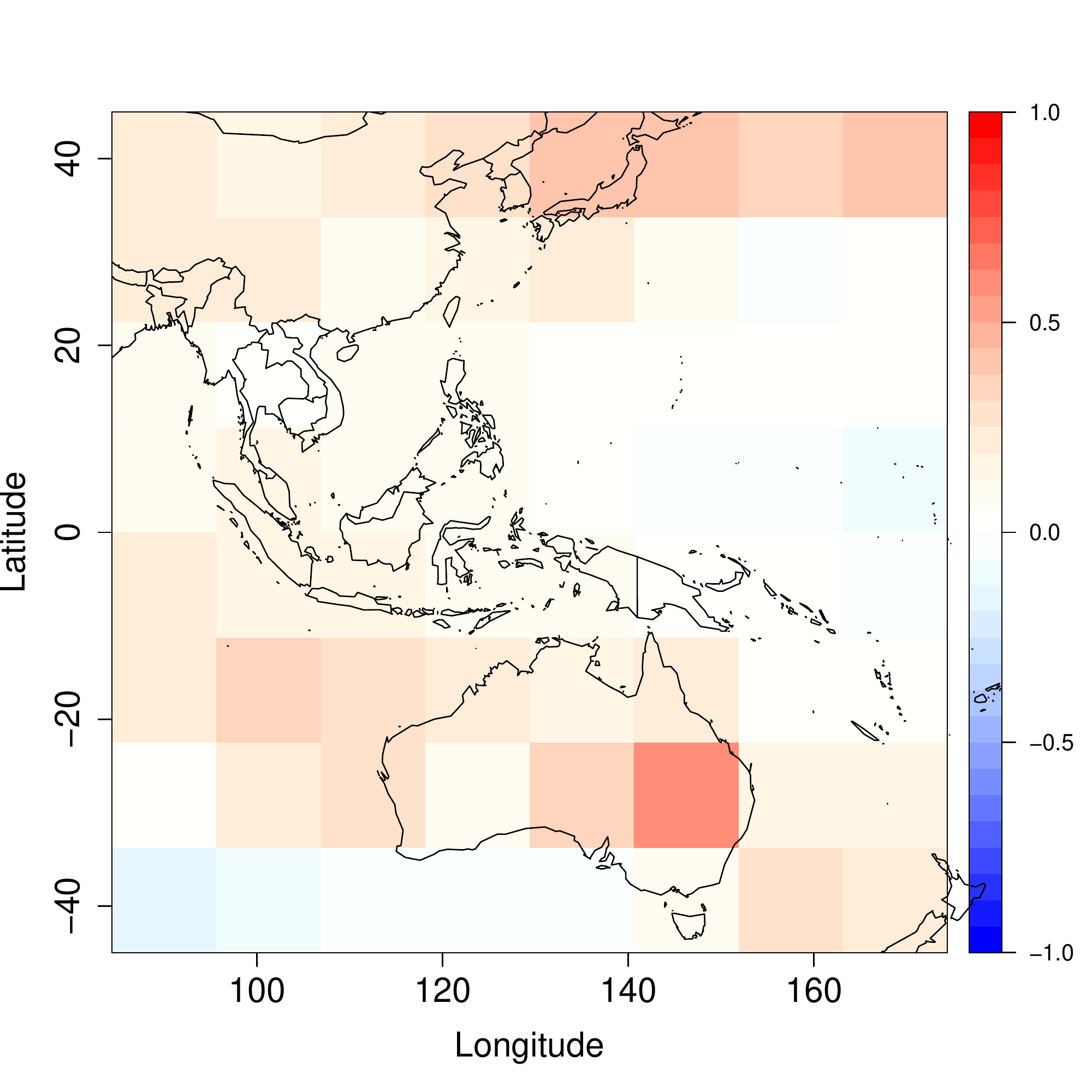} \\
(a)~~~ & (b)~~~ & (c)~~~
\bigskip\\
$\hat{p}=1.8\times 10^{-8}$~~ & $\hat{p}=7.4\times 10^{-7}$~~ & $\hat{p}=4.3\times 10^{-4}$~~
\vspace{-0.1cm}\\
\!\!\includegraphics[scale=0.285,trim={0.9cm 1.5cm 0.6cm 1.5cm},clip]{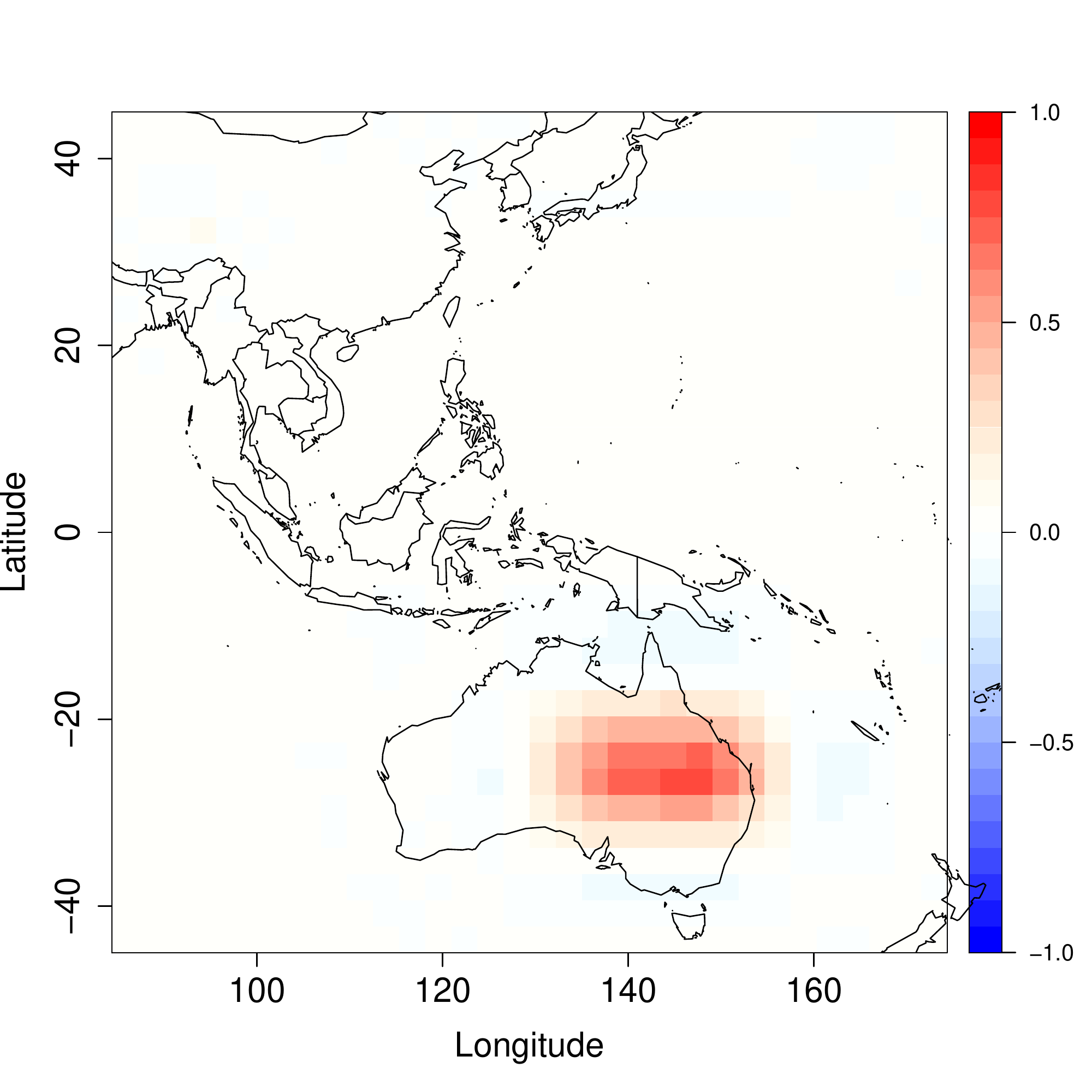} &
\!\!\!\!\includegraphics[scale=0.285,trim={0.9cm 1.5cm 0.6cm 1.5cm},clip]{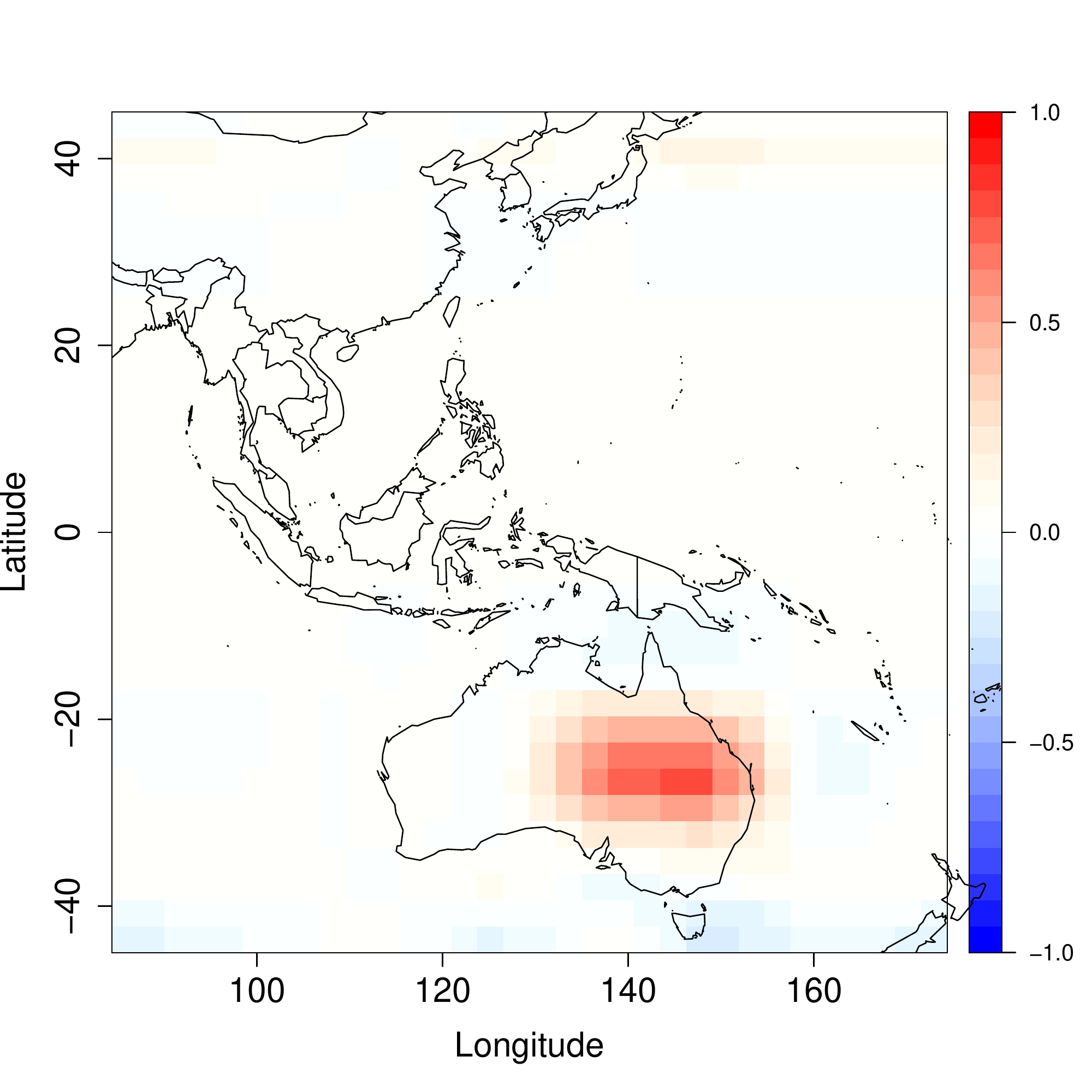} &
\!\!\!\!\includegraphics[scale=0.285,trim={0.9cm 1.5cm 0.6cm 1.5cm},clip]{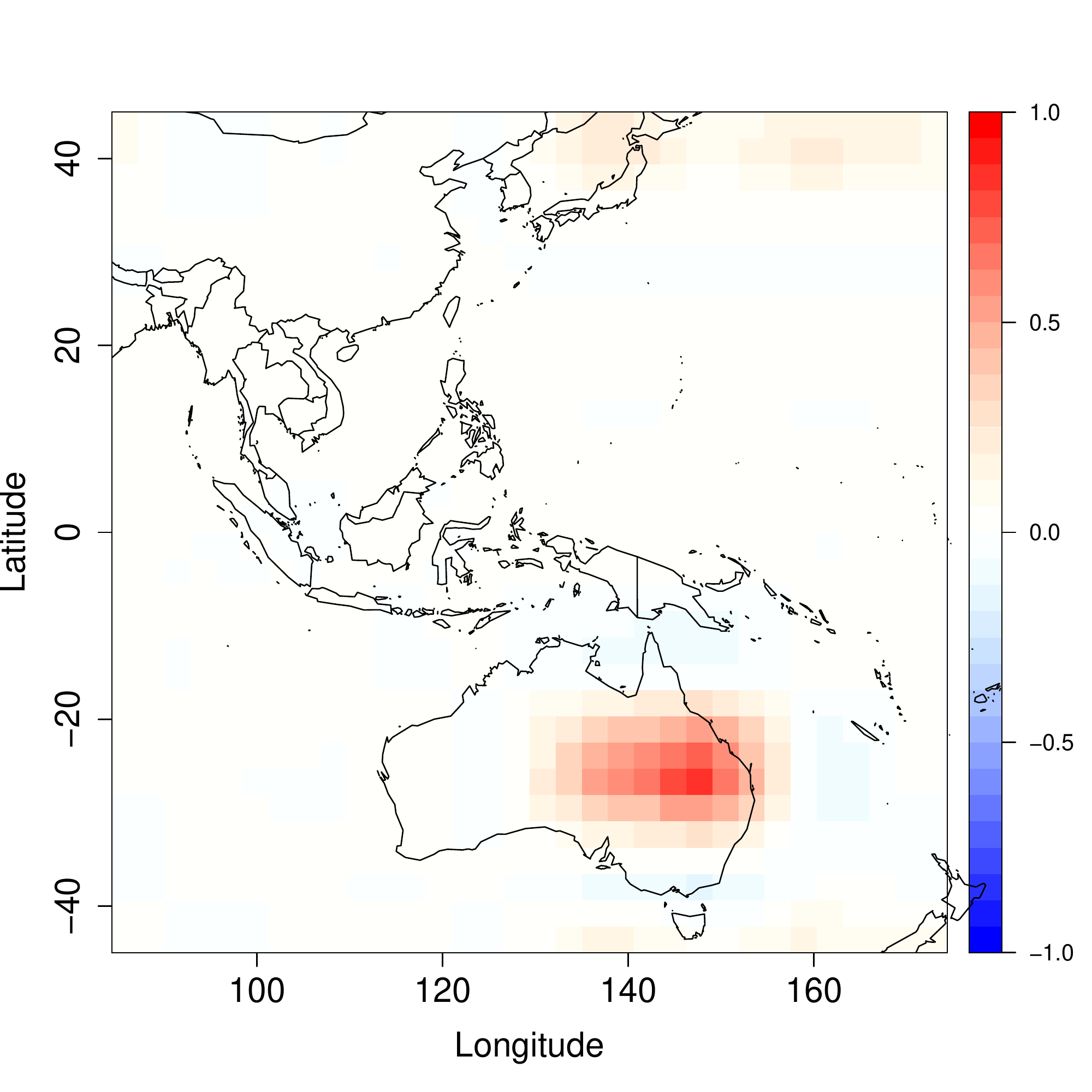} \\
(d)~~~ & (e)~~~ & (f)~~~ \\
\end{tabular}
\caption{(a) The average 2-meter air temperature differences between the 1990s and the 1980s on the Kelvin scale at the native resolution of $32\times 32$ pixels;
(b) The image aggregated from (a) into $16\times 16$ regular grid cells;
(c) The image aggregated from (a) into $8\times 8$ regular grid cells;
(d) The $32\times 32$ signal estimated by the EFDR-CS procedure and CPL based on pixel-level data in (a);
(e) The $32\times 32$ signal estimated by the EFDR-CS procedure and CPL based on the $16\times 16$ aggregated data in (b);
(f) The $32\times 32$ signal estimated by the EFDR-CS procedure and CPL based on the $8\times 8$ aggregated data in (c).
The $p$-values associated with (d), (e), and (f) are given at the top of the figures.}
\label{fig:temperature}
\end{figure}

Finding signals in climate data is critically important
for assessing the sustainability of Earth's ecosystems.
In this section, we apply our proposed procedure to a temperature dataset obtained from the National Center for Atmospheric Research (NCAR) climate system model.
The original data comprise monthly averages of 2-meter air temperatures on the Kelvin scale for the period 1980--1999
over the whole globe on $128\times 64$ equiangular longitude-latitude (about $2.8^\circ\times 2.8^\circ$) grid cells or pixels.
It is of interest to know if there is a decadal change in temperature (i.e., whether there is a possible signal) from the 1980s
to the 1990s and, if a change has occurred, to identify the magnitudes and locations of the change.
Here, we focus on a region $D$ of $32\times 32$ grid cells
containing most of the Asia-Pacific from $84^\circ$E to $174^\circ$E and from $45^\circ$S to $45^\circ$N.
We obtained the data $\bm{Z}$ by computing the average monthly temperature in the 1990s for each pixel in $D$,
from which we subtracted the corresponding average monthly temperature in the 1980s.
The resulting data, $\bm{Z}=\tilde{\bm{Z}}_{32\times 32}$, are shown in Figure~\ref{fig:temperature}(a).

\begin{figure}[!tb]\centering
\begin{tabular}{ccc}
\!\!\includegraphics[scale=0.285,trim={0.9cm 1.5cm 0.6cm 1.5cm},clip]{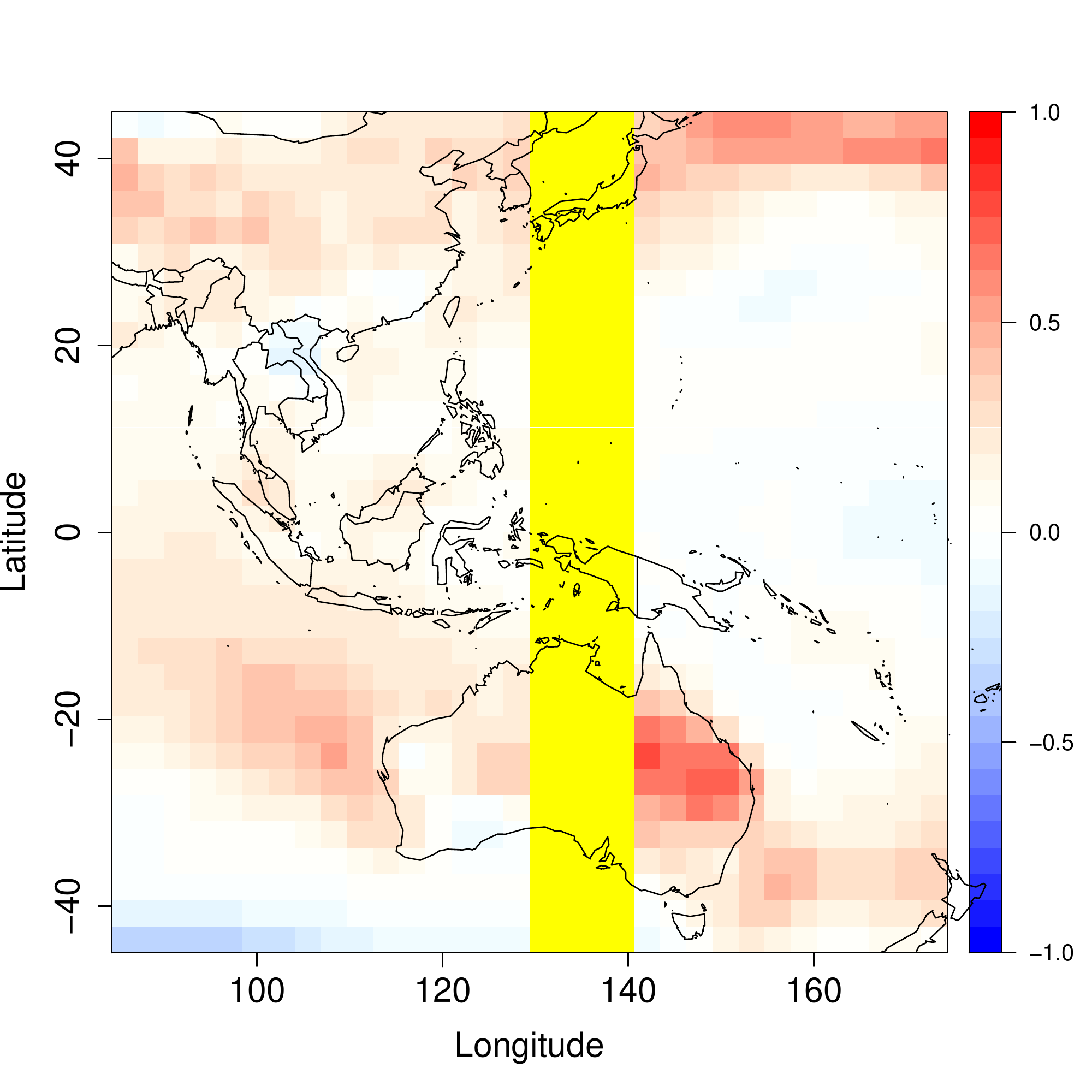} &
\!\!\!\!\includegraphics[scale=0.285,trim={0.9cm 1.5cm 0.6cm 1.5cm},clip]{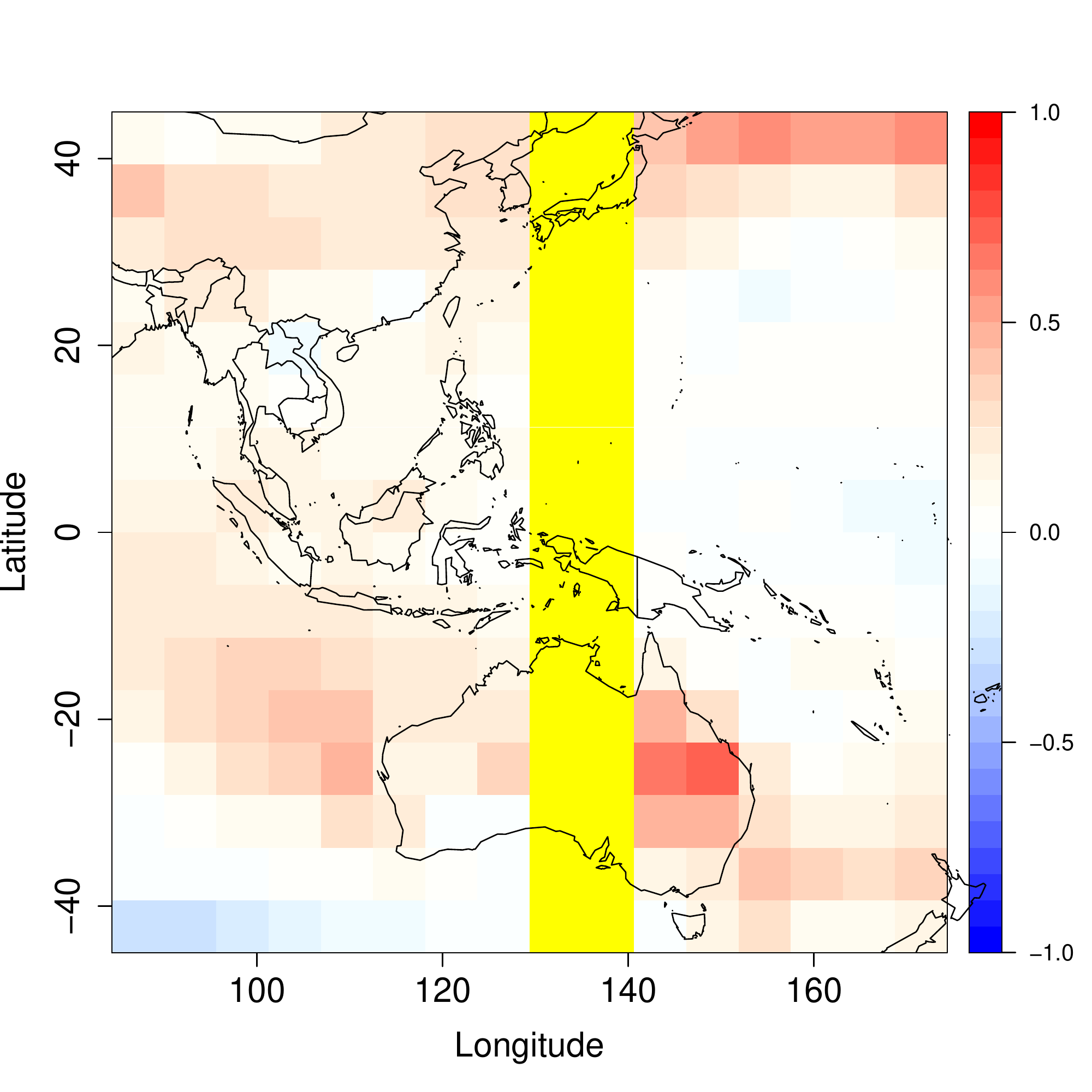} &
\!\!\!\!\includegraphics[scale=0.285,trim={0.9cm 1.5cm 0.6cm 1.5cm},clip]{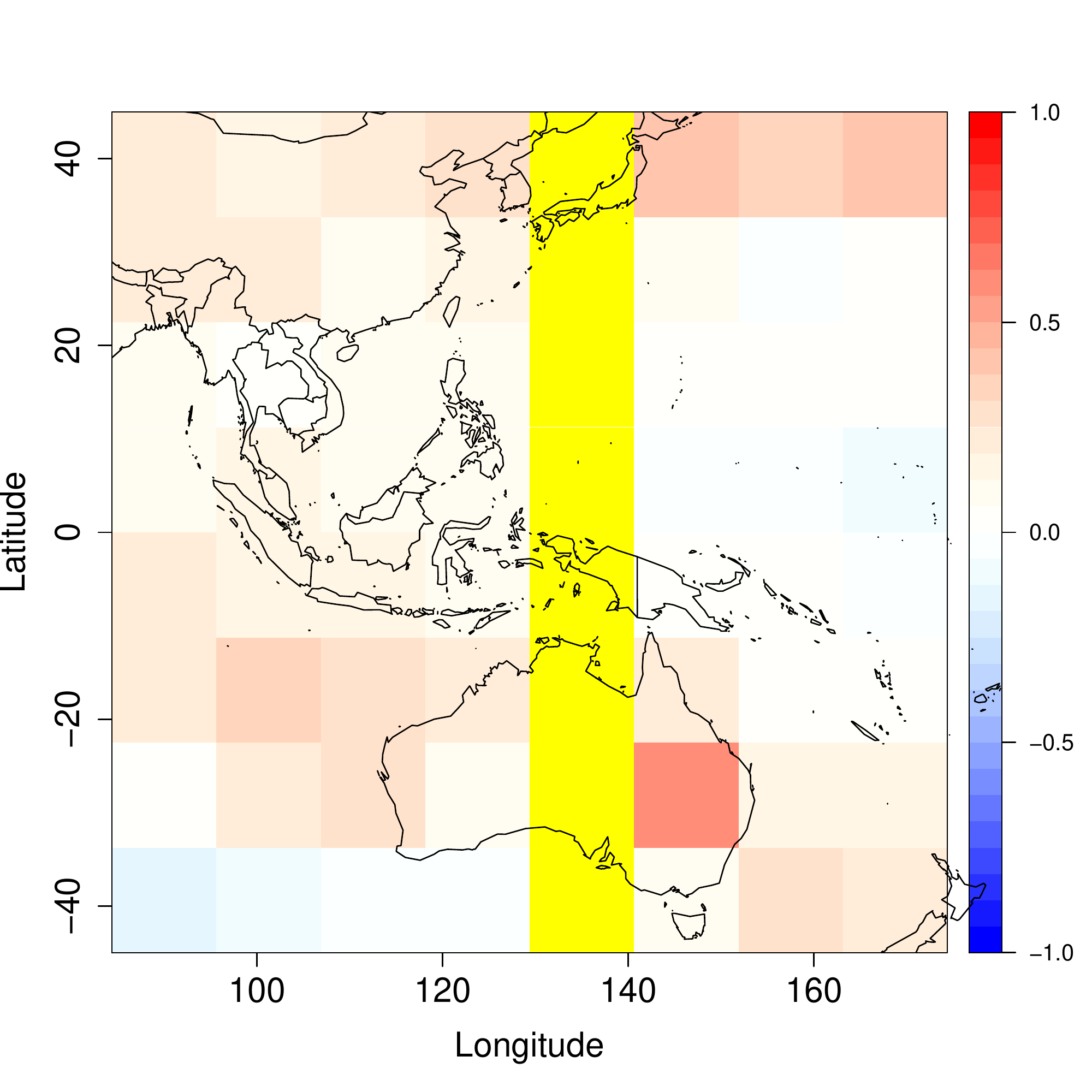} \\
(a)~~~ & (b)~~~ & (c)~~~
\bigskip\\
$\hat{p}=2.6\times 10^{-11}$~~ & $\hat{p}=1.2\times 10^{-5}$~~ & $\hat{p}=1.7\times 10^{-5}$~~
\vspace{-0.1cm}\\
\!\!\includegraphics[scale=0.285,trim={0.9cm 1.5cm 0.6cm 1.5cm},clip]{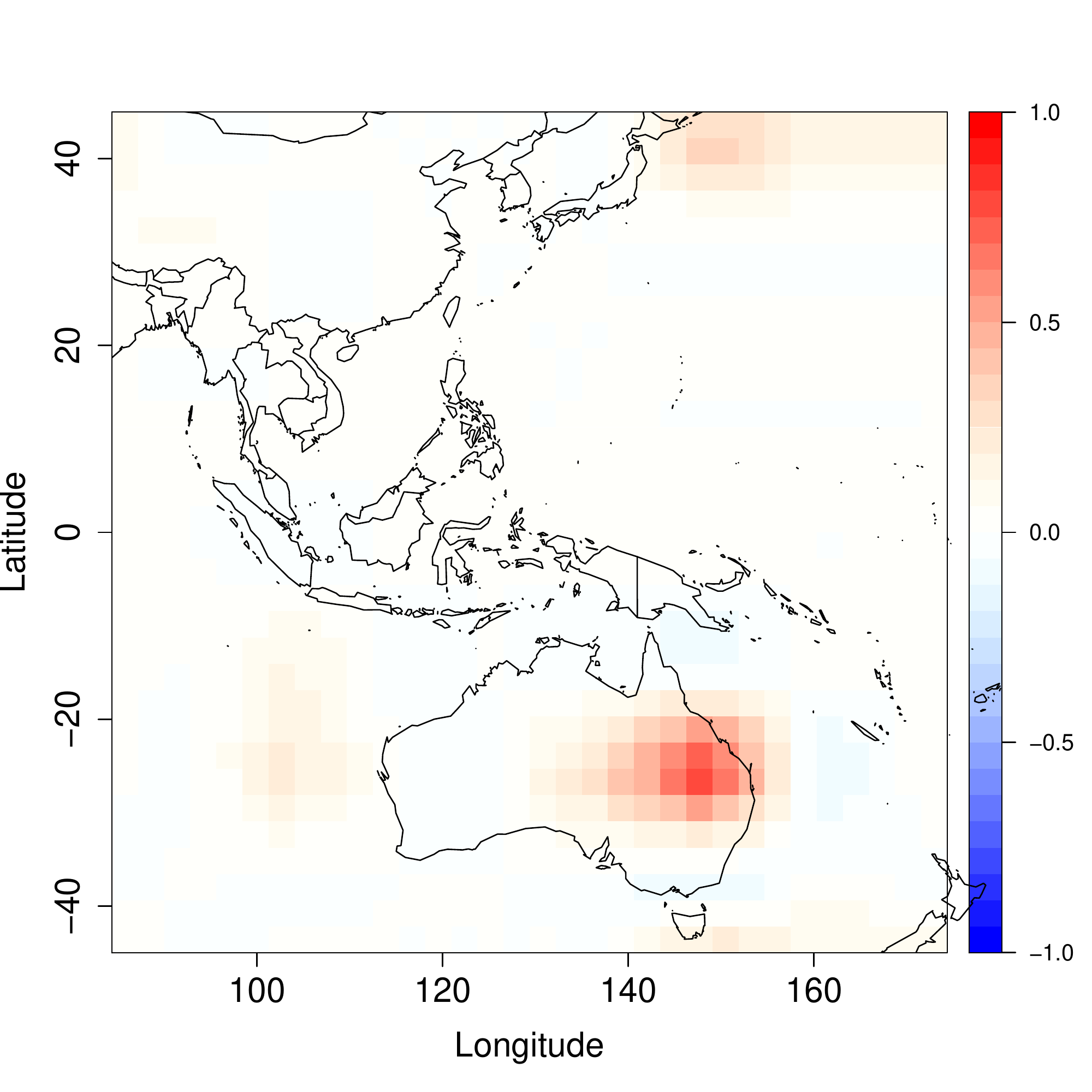} &
\!\!\!\!\includegraphics[scale=0.285,trim={0.9cm 1.5cm 0.6cm 1.5cm},clip]{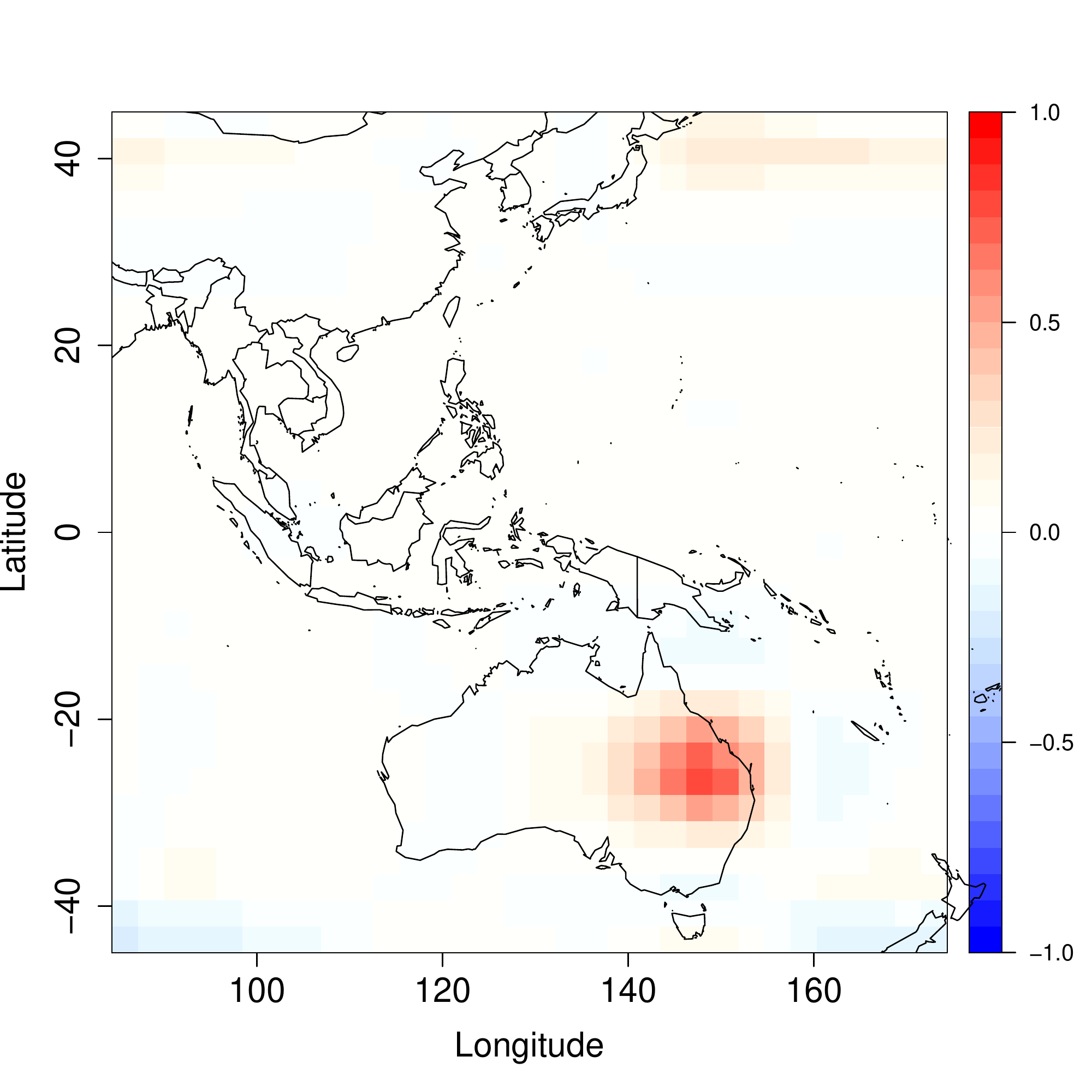} &
\!\!\!\!\includegraphics[scale=0.285,trim={0.9cm 1.5cm 0.6cm 1.5cm},clip]{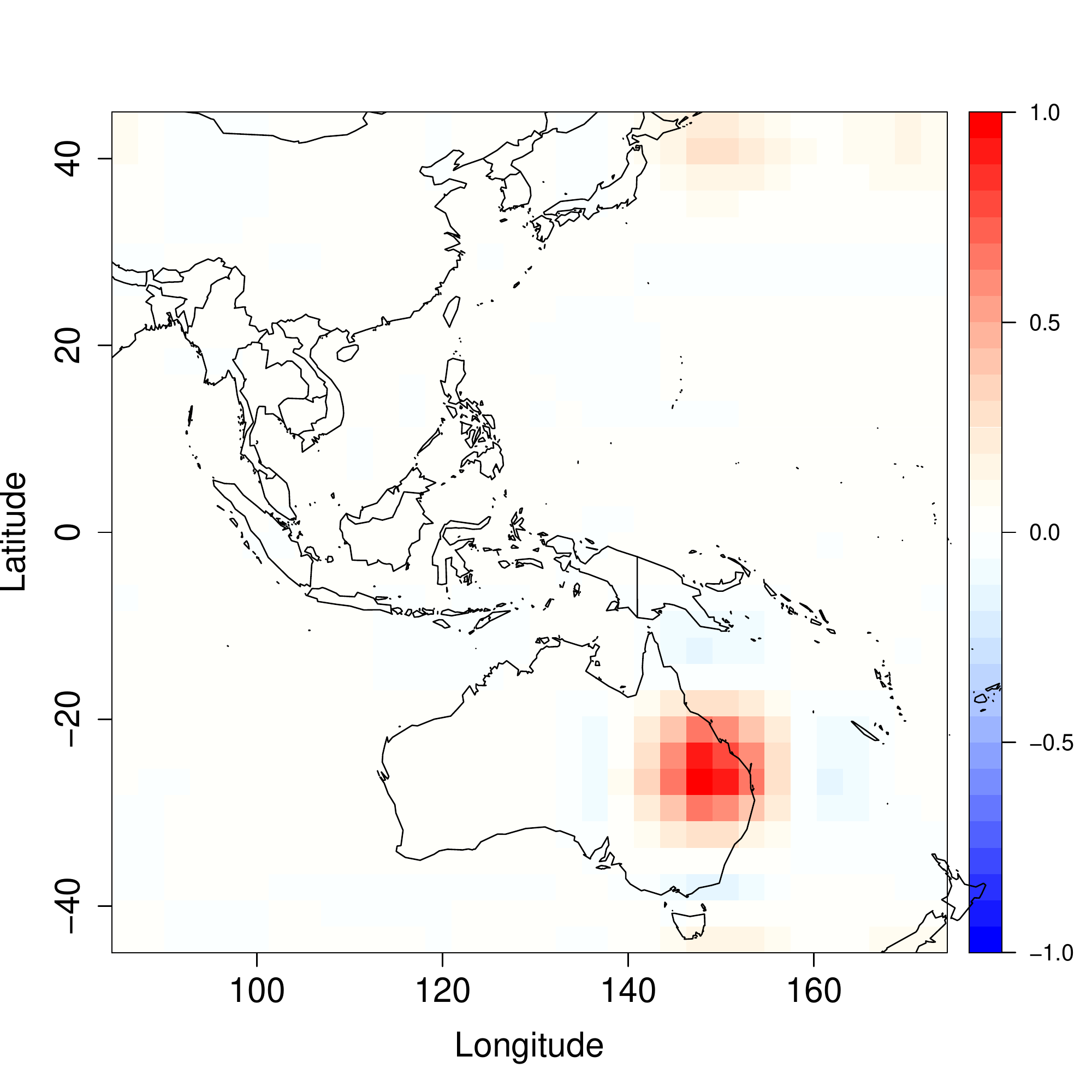} \\
(d)~~~ & (e)~~~ & (f)~~~ \\
\end{tabular}
\caption{Similar to Figure \ref{fig:temperature}, but with a missing strip of data (shown in yellow) in (a), (b), and (c).}
\label{fig:temperature2}
\end{figure}

\begin{figure}[!tbh]\centering
\begin{tabular}{ccc}
\!\!\includegraphics[scale=0.285,trim={0.9cm 1.5cm 0.6cm 1.5cm},clip]{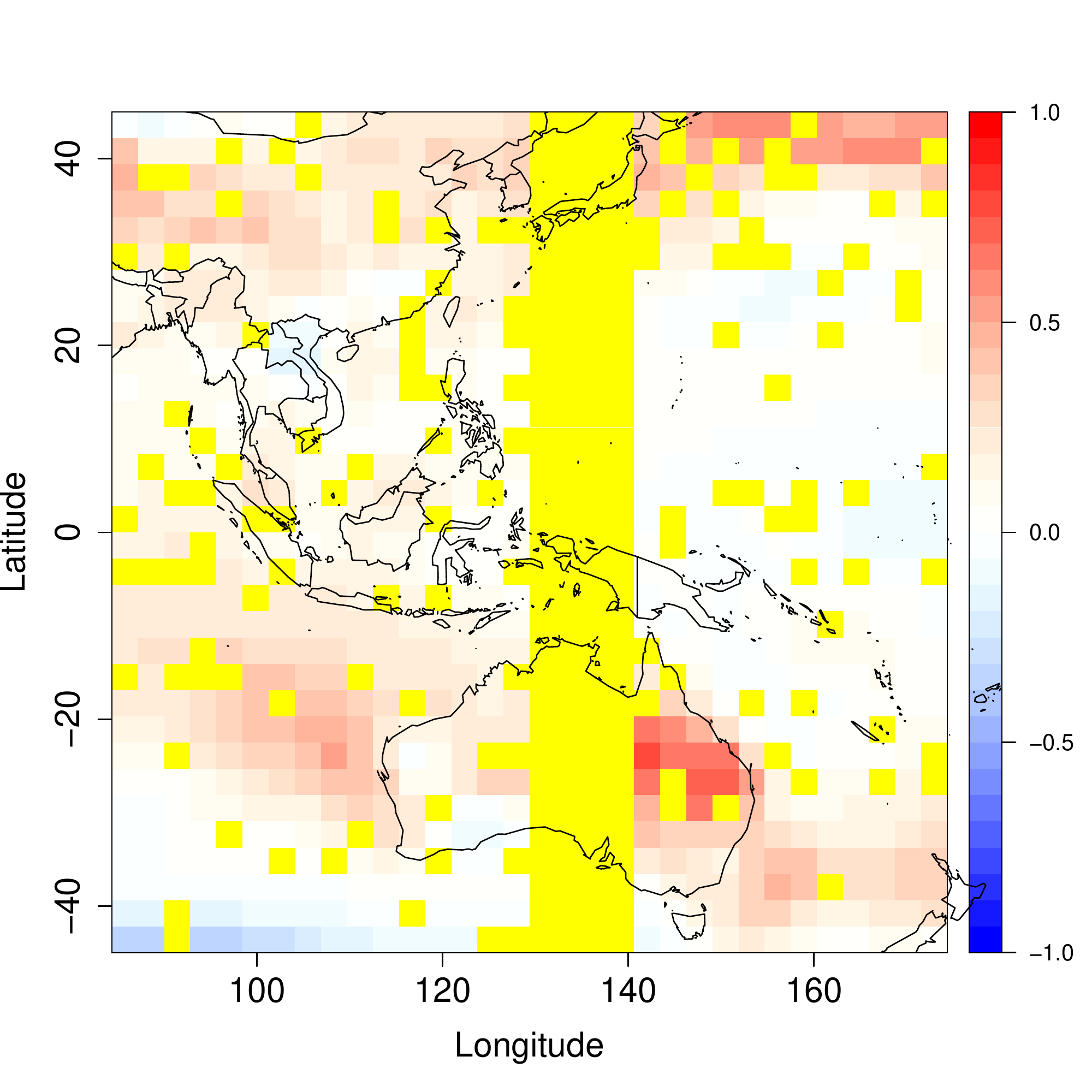} &
\!\!\!\!\includegraphics[scale=0.285,trim={0.9cm 1.5cm 0.6cm 1.5cm},clip]{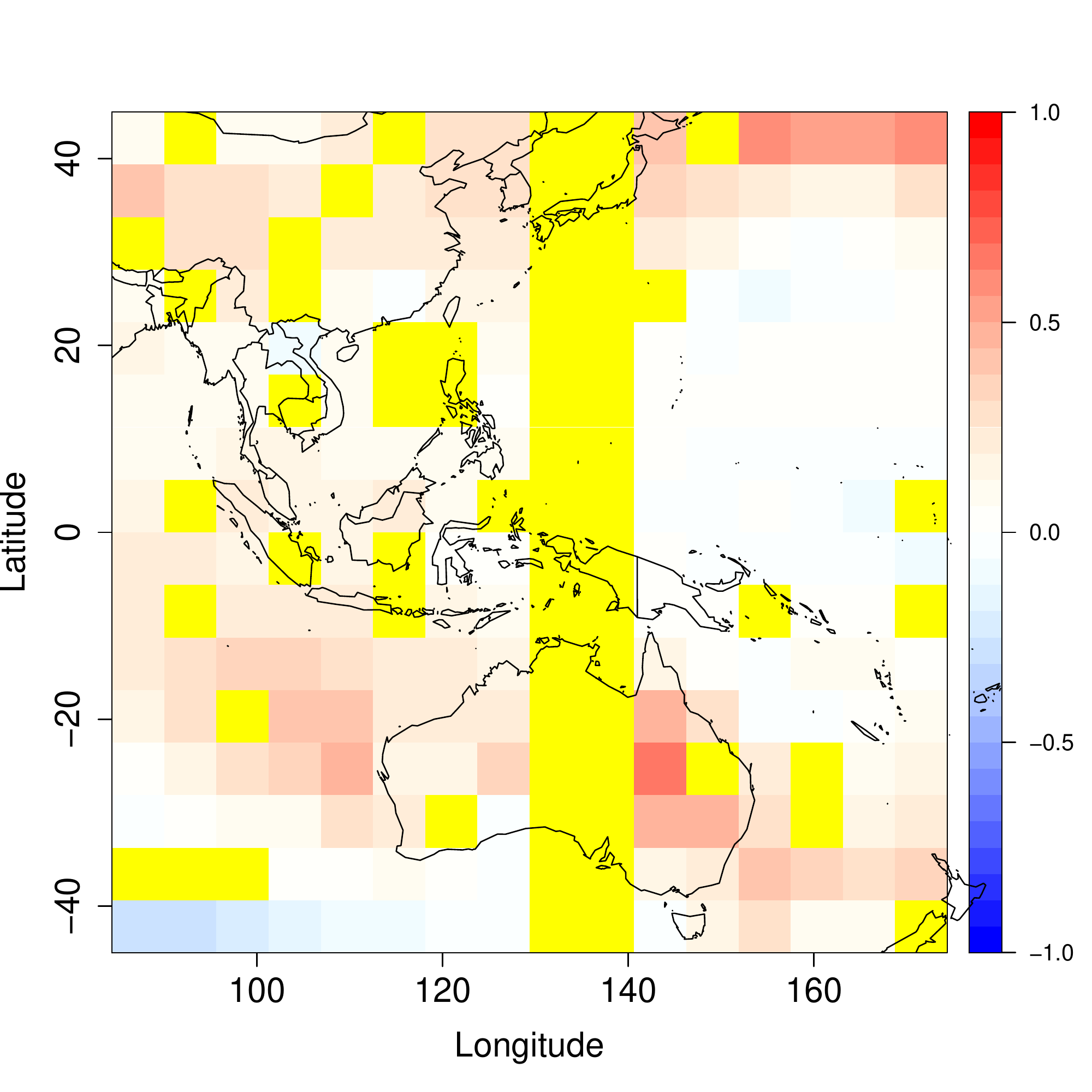} &
\!\!\!\!\includegraphics[scale=0.285,trim={0.9cm 1.5cm 0.6cm 1.5cm},clip]{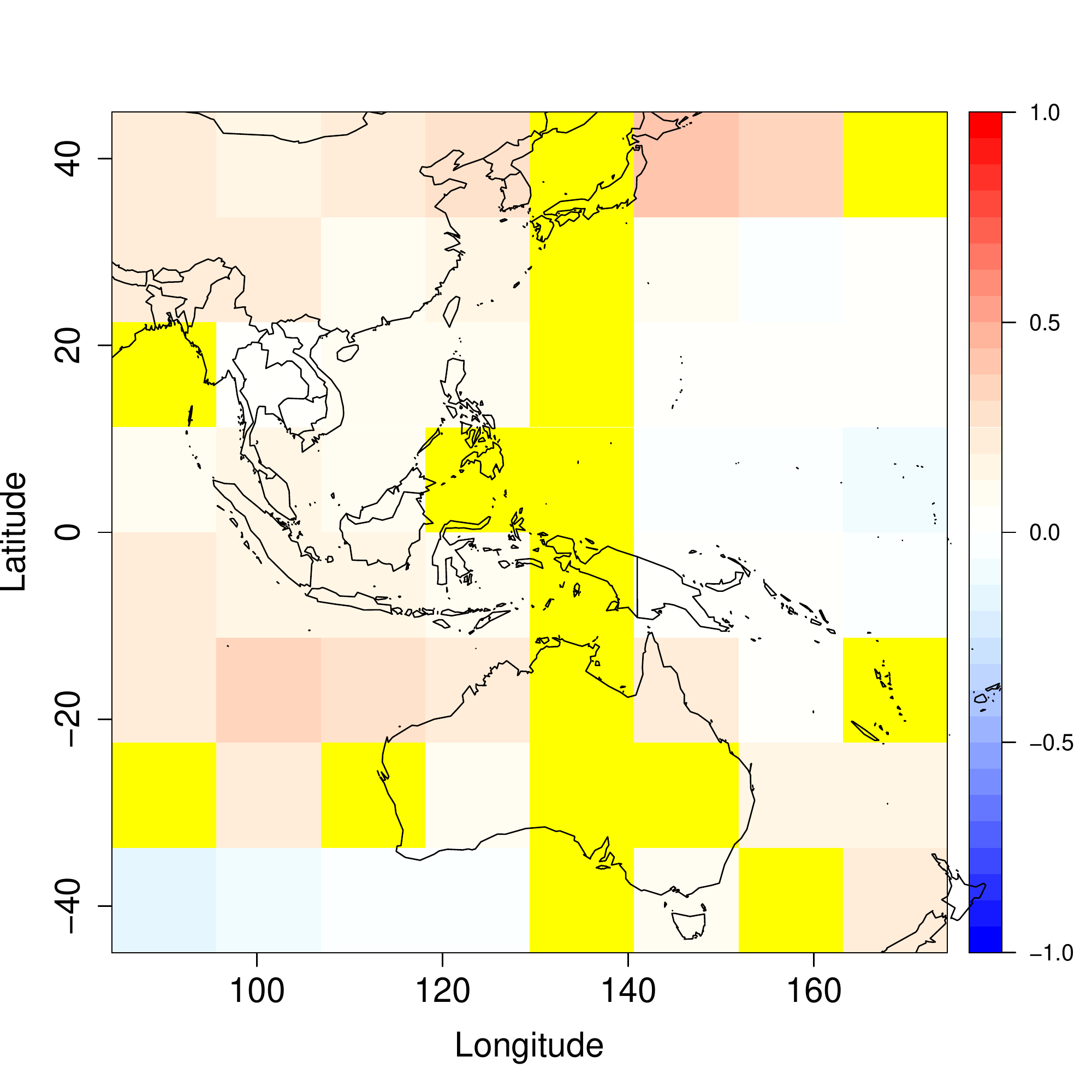} \\
(a)~~~ & (b)~~~ & (c)~~~ 
\bigskip\\
$\hat{p}=3.8\times 10^{-8}$~~ & $\hat{p}=1.9\times 10^{-4}$~~ & $\hat{p}=5.4\times 10^{-2}$~~
\vspace{-0.1cm}\\
\!\!\includegraphics[scale=0.285,trim={0.9cm 1.5cm 0.6cm 1.5cm},clip]{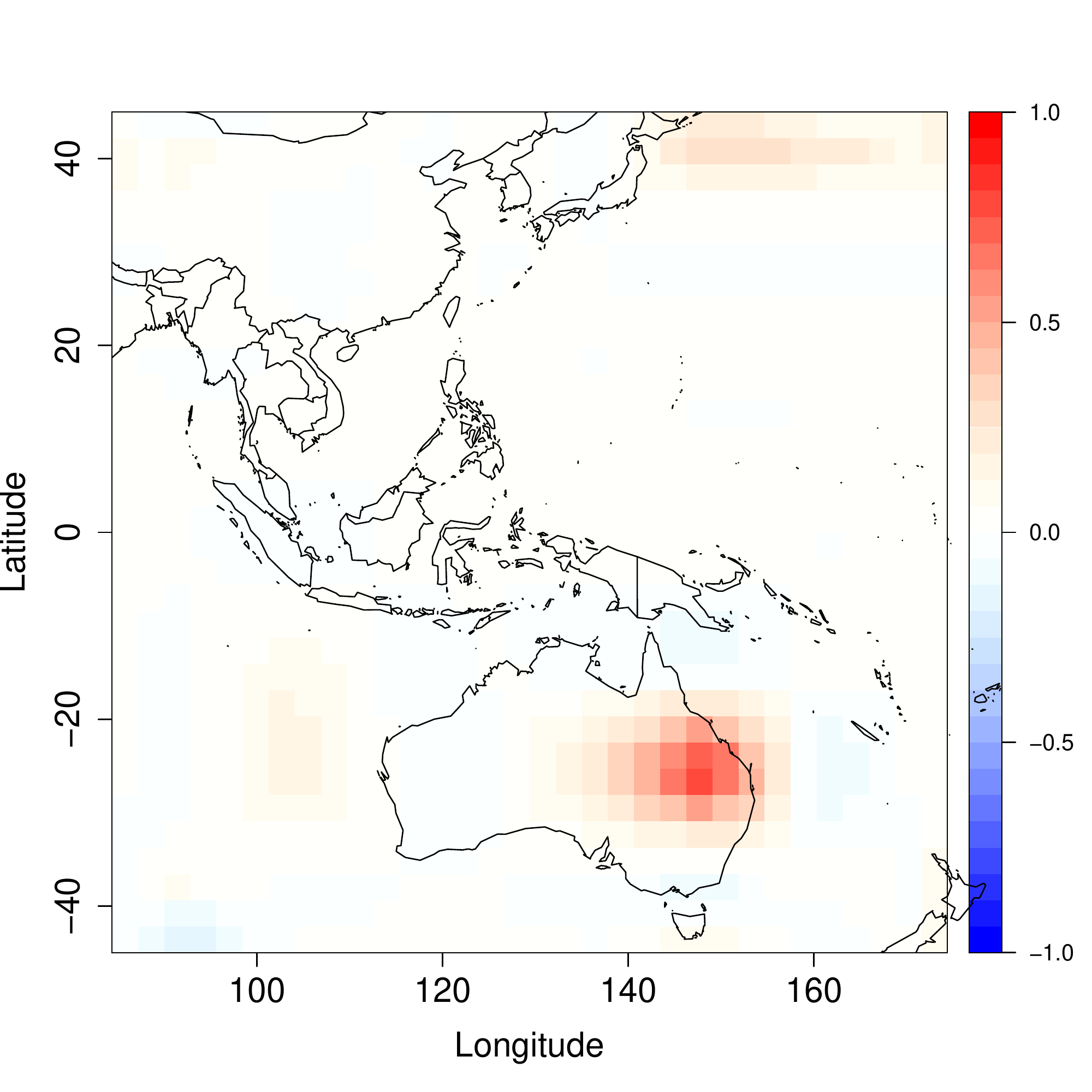} &
\!\!\!\!\includegraphics[scale=0.285,trim={0.9cm 1.5cm 0.6cm 1.5cm},clip]{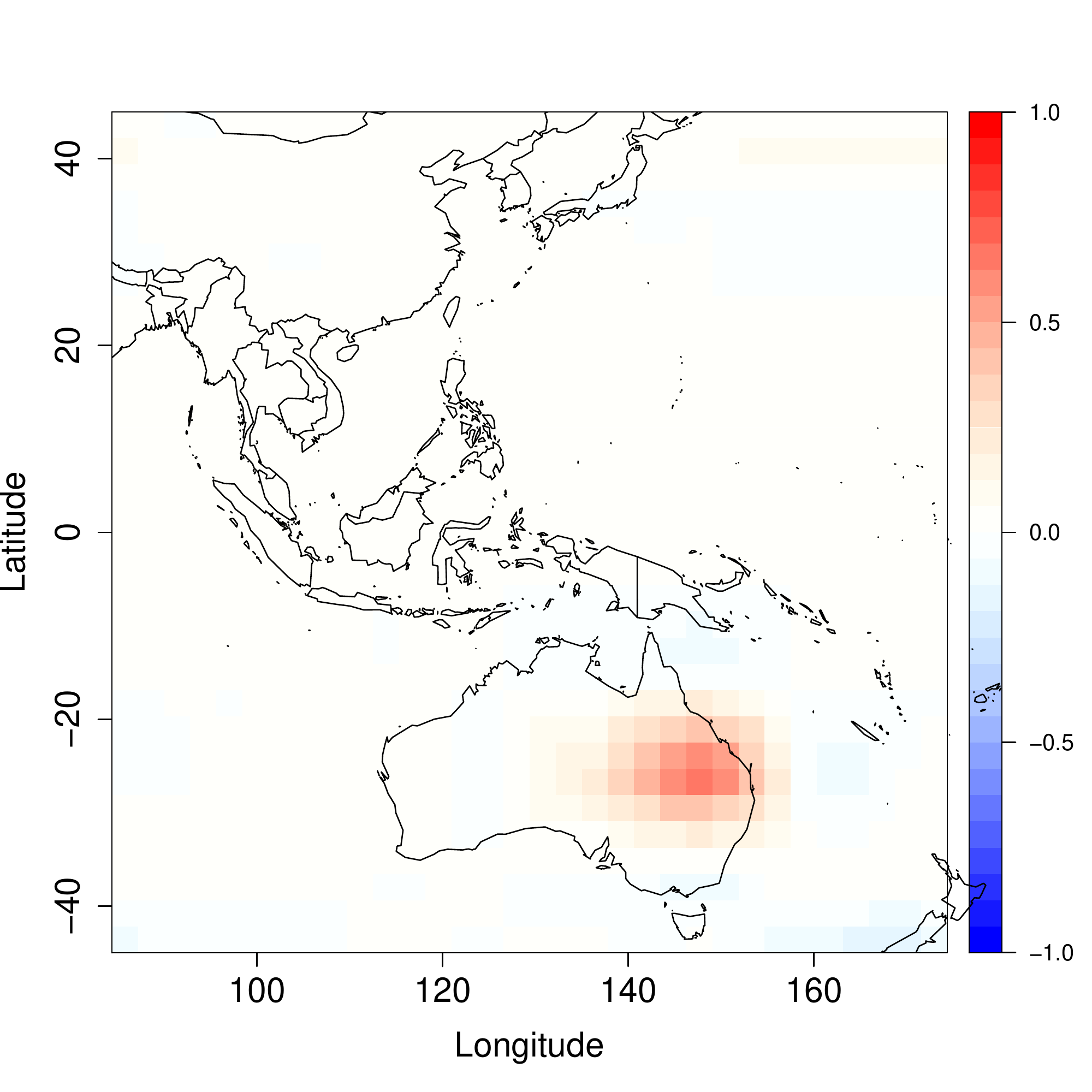} &
\!\!\!\!\includegraphics[scale=0.285,trim={0.9cm 1.5cm 0.6cm 1.5cm},clip]{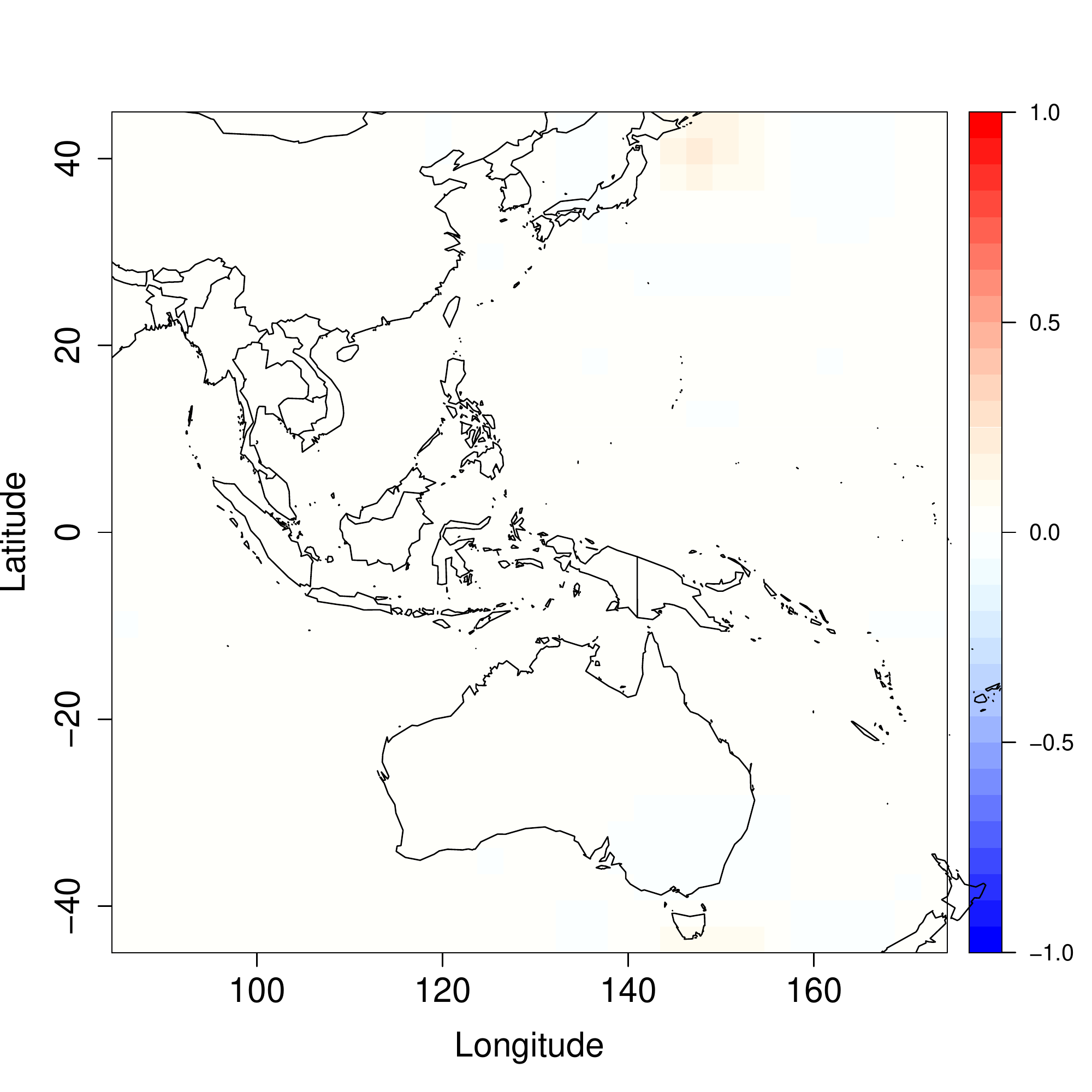} \\
(d)~~~ & (e)~~~ & (f)~~~ \\
\end{tabular}
\caption{Similar to Figure \ref{fig:temperature}, but with a missing strip and randomly missing pixels (shown in yellow) in (a), (b), and (c).}
\label{fig:temperature3}
\end{figure}

Just as for the simulation experiments in Section~\ref{sec:simulation},
we considered scenarios involving complete data at different resolutions and incomplete data at different resolutions.
Under the first of three scenarios, we aggregated $\bm{Z}$ into $16\times 16$ and $8\times 8$ regular grid cells,
denoted by $\tilde{\bm{Z}}_{16\times 16}$ and $\tilde{\bm{Z}}_{8\times 8}$
and shown in Figure~\ref{fig:temperature}(b) and Figure~\ref{fig:temperature}(c), respectively.
Under the next two scenarios, data missing in a contiguous block and data missing at random were considered:
Initially, we removed a strip of data from
$\tilde{\bm{Z}}_{32\times 32}$, $\tilde{\bm{Z}}_{16\times 16}$, and $\tilde{\bm{Z}}_{8\times 8}$ to mimic a missing swath
commonly seen in satellite data.
The resulting data are denoted by $\tilde{\bm{Z}}_{32\times 32}^{(1)}$, $\tilde{\bm{Z}}_{16\times 16}^{(1)}$, and $\tilde{\bm{Z}}_{8\times 8}^{(1)}$,
with the same fraction of fine-resolution pixels missing ($=1/8$), and they are shown in Figures~\ref{fig:temperature2}(a)--(c), respectively.
We then randomly removed a further 1/8 of the grid cells from the datasets $\tilde{\bm{Z}}_{32\times 32}^{(1)}$, $\tilde{\bm{Z}}_{16\times 16}^{(1)}$,
and $\tilde{\bm{Z}}_{8\times 8}^{(1)}$, respectively.
The resulting data, denoted by $\tilde{\bm{Z}}_{32\times 32}^{(2)}$, $\tilde{\bm{Z}}_{16\times 16}^{(2)}$, and $\tilde{\bm{Z}}_{8\times 8}^{(2)}$, respectively, are shown in Figures~\ref{fig:temperature3}(a)--(c), where it is seen that
after removing data on the strip and the random scattering of pixels, we obtain irregular lattice data.

We applied the EFDR-CS procedure (with CPL defined in Section~\ref{sec:method}), based on the significance test
\eqref{eq:test}, to the nine cases
of incomplete spatially aggregated data: $\tilde{\bm{Z}}_{32\times 32}$, $\tilde{\bm{Z}}_{16\times 16}$, $\tilde{\bm{Z}}_{8\times 8}$,
$\tilde{\bm{Z}}_{32\times 32}^{(1)}$, $\tilde{\bm{Z}}_{16\times 16}^{(1)}$, $\tilde{\bm{Z}}_{8\times 8}^{(1)}$,
$\tilde{\bm{Z}}_{32\times 32}^{(2)}$, $\tilde{\bm{Z}}_{16\times 16}^{(2)}$, and $\tilde{\bm{Z}}_{8\times 8}^{(2)}$.
Similar to the numerical experiments in Section \ref{sec:simulation},
we used the R package ``EFDR" (Zammit-Mangion and Huang, 2015) with its default setting of
the Daubechies least asymmetric wavelet filter of length $8$ and
the number of hypotheses to be tested in the wavelet space fixed at $100$.
As for the simulations in Section~\ref{sec:simulation}, we chose two wavelet scales
resulting in seven wavelet classes corresponding to different scales and orientations.
We estimated $\hat{\bm{\theta}}(\hat{\tau}^2,\hat{\phi})=\hat{\bm{\theta}}=(\hat{\theta}_1,\dots,\hat{\theta}_7)'$ through \eqref{eq:theta.hat}
in the Supplementary Material,
where $\hat{\tau}^2$ and $\hat{\phi}$ are the ML estimators based on the exponential covariance model, $C(\bm{u})=\tau^2\exp(\|\bm{u}\|/\phi)$.

Except for the last case, $\tilde{\bm{Z}}_{8\times 8}^{(2)}$, where one of the missing cells is coincident with the potential signal,
our proposed procedure rejected the null hypothesis of no decadal change in temperature at the $5\%$ significance level.
The final $p$-values (on the scale $t=-2\log p$) are shown in Table~\ref{table:app}.
As expected, for a given row of the Table, the values on the $t$-scale across rows tend to increase
(i.e., $p$-values tend to decrease) for data at finer-scale resolutions.
Comparison down columns supports our conclusion from the simulations in Section~\ref{sec:simulation},
that CPL is not greatly affected by incomplete data as long as the signal is observed.

The spatial patterns of temperature changes, given by $\hat{\bm{\mu}}$ in \eqref{eq:muhat} and
based on $\tilde{\bm{Z}}_{32\times 32}$, $\tilde{\bm{Z}}_{16\times 16}$, $\tilde{\bm{Z}}_{8\times 8}$,
are shown in Figures~\ref{fig:temperature}(d)--(f), respectively.
As we saw in the simulations in Section~\ref{sec:simulation},
our proposed procedure handles successive spatial aggregations well,
and very similar signals of temperature increase are observed over central eastern Australia.
For the incomplete data,
$\tilde{\bm{Z}}_{32\times 32}^{(1)}$, $\tilde{\bm{Z}}_{16\times 16}^{(1)}$, $\tilde{\bm{Z}}_{8\times 8}^{(1)}$,
$\tilde{\bm{Z}}_{32\times 32}^{(2)}$, and $\tilde{\bm{Z}}_{16\times 16}^{(2)}$,
our proposed procedure identified similar signals, albeit with smaller spatial extents.
The results are shown in Figures~\ref{fig:temperature2}(d)--(f) and Figures~\ref{fig:temperature3}(d)--(e), respectively.
However, Figure~\ref{fig:temperature3}(f) does not show a signal because
in $\tilde{\bm{Z}}_{8\times 8}^{(2)}$, one of the missing cells was coincident with the potential signal
(see Figure~\ref{fig:temperature3}(c)). This resulted in a failure to reject the null hypothesis and an almost blank image in
Figure~\ref{fig:temperature3}(f) with a final $p$-value of $5.4\%$ that indicates no signal is present.
When that cell was allowed to remain and a different cell removed from the $8\times 8$ dataset in Figure~\ref{fig:temperature3}(c),
$t=-2\log p$ went from $5.84$ to $15.76$,
and a spatial signal like those seen in Figures~\ref{fig:temperature3}(d) and \ref{fig:temperature3}(e) appeared.

\begin{table}[tb]\centering
\caption{The $p$-values (on the $t$-scale, where $t=-2\log p$) for our proposed procedure (CPL) applied to
incomplete spatially aggregated temperature datasets
in the Asia-Pacific under different levels of aggregation and incompleteness.}
\medskip
\begin{tabular}{c|rrr}
\hline
Missing Fraction & \multicolumn{3}{c}{Scales of Aggregation}\\
\cline{2-4}
          & $32\times 32$  & $16\times 16$ & $8\times 8$ \\\hline
$0$    & $35.67$           & $28.23$ & $15.51$ \\
$1/8$ & $48.75$           & $22.70$ & $21.94$ \\
$1/4$ & $34.18$           & $17.11$ & $5.84$ \\
\hline
\end{tabular}
\smallskip\\
\footnotesize Note:  $-2\log(0.05)=5.99$.~\hspace{4.5cm}~
\label{table:app}
\end{table}

\subsection{Total column CO$_2$ signal from remote sensing data}
\label{sec:co2}

\begin{figure}[!bt]\centering
\includegraphics[scale=0.58,trim={0cm 3cm 0cm 3cm},clip]{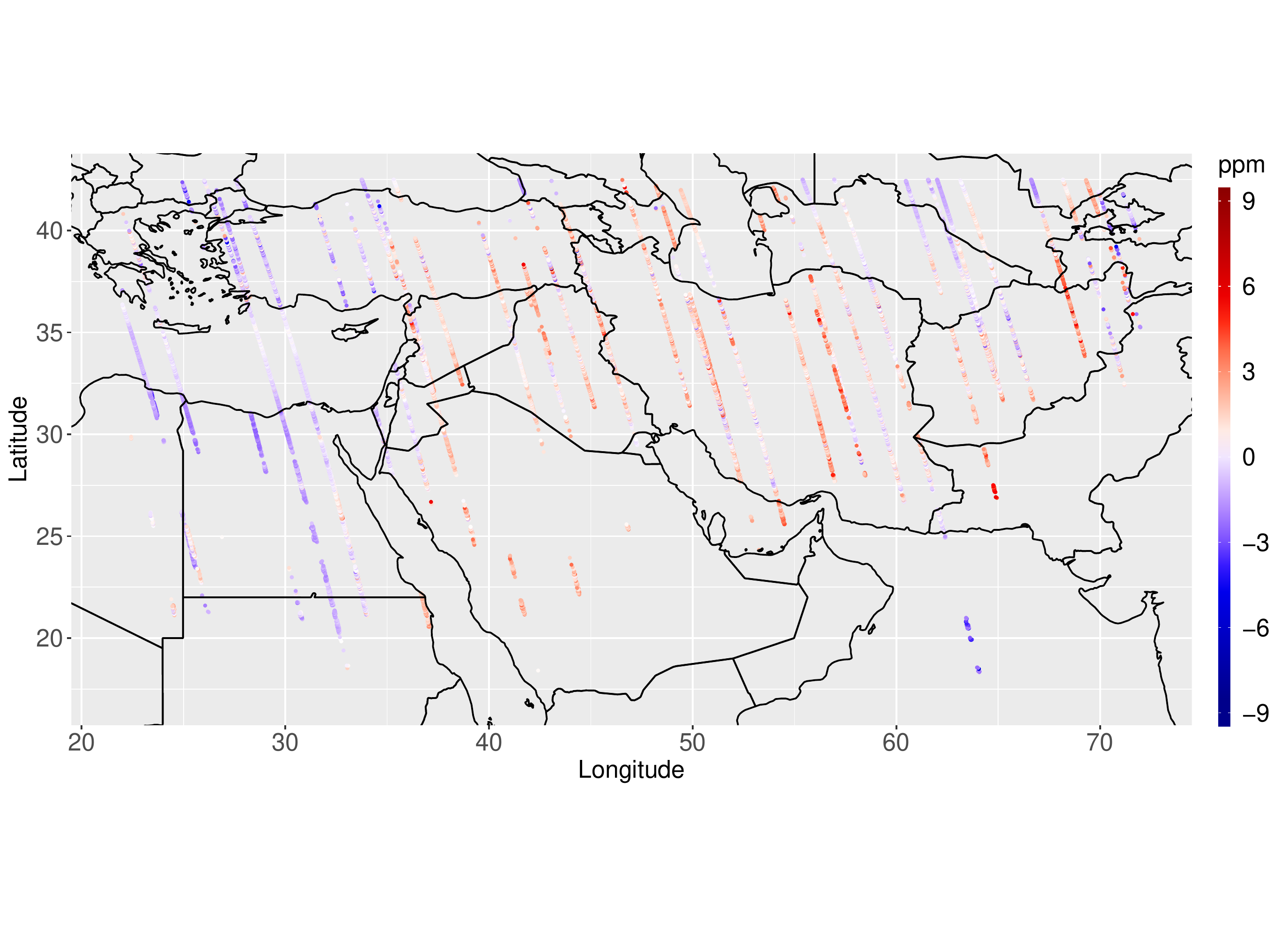}
\caption{Detrended column-average CO$_2$ data from NASA's OCO-2 satellite during 1--16 July 2018.}
\label{fig:R3}
\end{figure}

We applied the EFDR-CS procedure to a total-column carbon-dioxide (CO$_2$) dataset obtained from the Orbiting Carbon Observatory-2 (OCO-2) satellite of
the U.S.~National Aeronautics and Space Administration (NASA).
The data consist of $m=69903$ retrievals $\{Y(\bm{s}_i;t_i):i=1,\dots,m\}$ 
measured in parts per million (ppm) during 1--16 July 2018 in a region $D^*$ centered on the Middle East.
The retrieval locations $\{\bm{s}_i\}$ and $D^*$ are shown in Figure \ref{fig:R3},
and the daily retrieval times $\{t_i:i=1,\dots,m\}$ play a role, as we now explain.

First, we estimated a temporal trend:
\[
\hat{\mu}(t)\equiv\mathrm{ave}\{Y(\bm{s}_i;t_i):\,t_i\in\mbox{day}~t,\,i=1,\dots,m\};\quad t=1,\dots,16,
\]
and obtained the residuals,
\[
R_1(\bm{s}_i)~\equiv~Y(\bm{s}_i;t_i)-\hat{\mu}(t);\quad\mbox{for }t_i\in\mbox{day}~t, i=1,\dots,m,
\]
which we now treat as a purely spatial process. Second, a spatial trend that is linear in latitude, $\beta_0+\beta_1\,\mathrm{lat}(\bm{s})$,
was observed, estimated by ordinary least squares, and subtracted to obtain spatial residuals that now have mean zero:
\[
R_2(\bm{s}_i)~\equiv~R_1(\bm{s}_i)-\hat{\beta}_0-\hat{\beta}_1\,\mathrm{lat}(\bm{s}_i);\quad i=1,\dots,m.
\]

\begin{figure}[!tb]\centering
\begin{tabular}{ccc}
\includegraphics[scale=0.27,trim={1cm 1.2cm 0cm 1cm},clip]{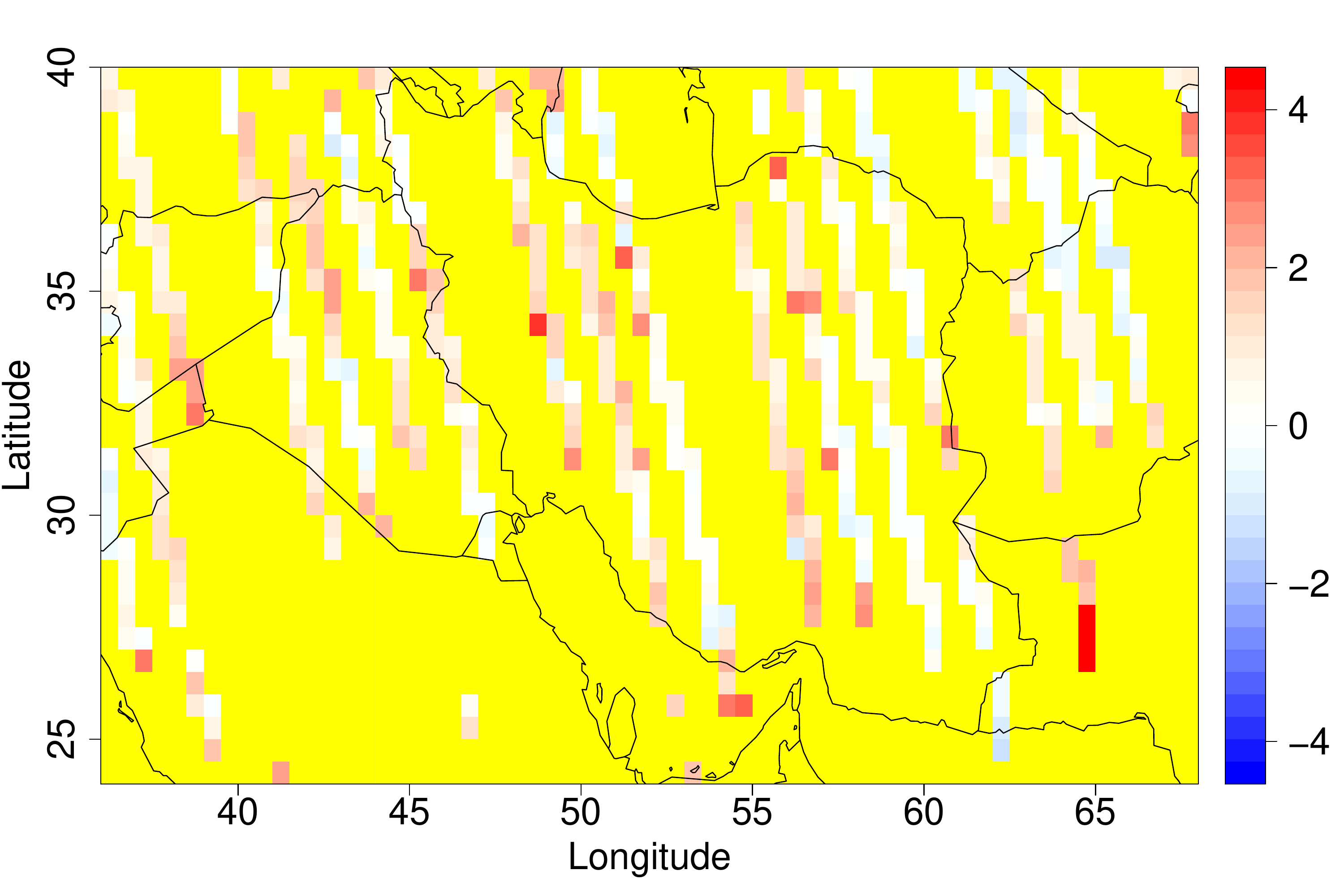} &
\includegraphics[scale=0.27,trim={1cm 1.2cm 0cm 1cm},clip]{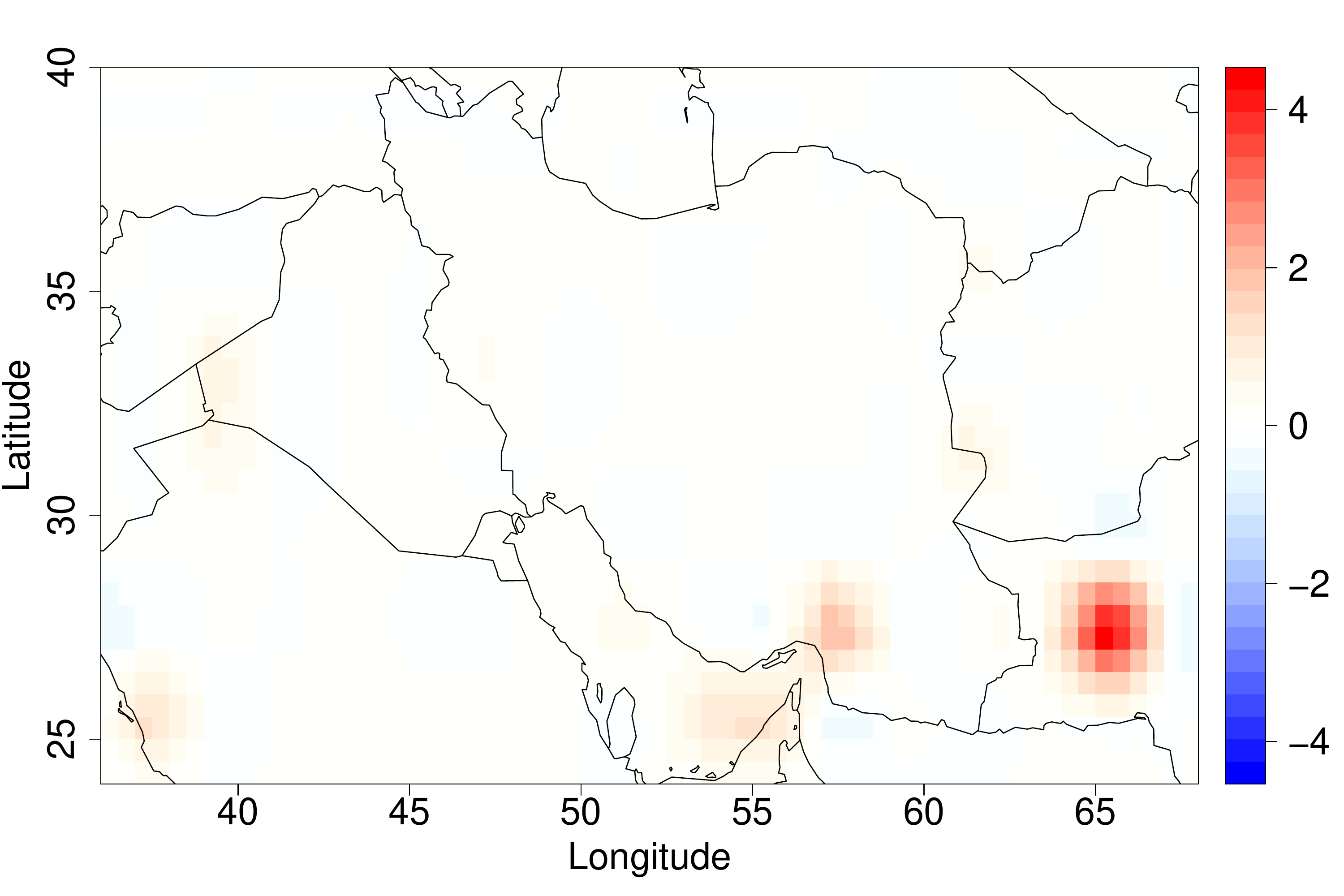} \\
(a)~~~ & (b)~~~ 
\end{tabular}
\caption{(a) Detrended column-average CO$_2$ data aggregated from Figure \ref{fig:R3} into $64\times 32$ ($0.5^\circ\times 0.5^\circ$) grid cells;
(b) The signal estimated by the EFDR-CS procedure and CPL based on the data in (a).}
\label{fig:co2}
\end{figure}

The residuals $\{R_2(\bm{s}_i)\}$ are shown in Figure \ref{fig:R3};
notice that there are large gaps in the map caused by the orbit tracks or no retrieval from the OCO-2 satellite.
Third, we obtained the data $\tilde{\bm{Z}}$ at the EFDR-CS procedure's finest resolution by aggregating $\{R_2(\bm{s}_i;t_i):i=1,\dots,m\}$ 
into $0.5^\circ\times 0.5^\circ$ longitude-latitude grid cells using simple averaging.
We focused on a region from $36^\circ$E to $68^\circ$E and from $24^\circ$N to $40^\circ$N
(a region of the Middle East, Afghanistan, and the western part of Pakistan), which consists of $64\times 32$ grid cells at $0.5^\circ\times 0.5^\circ$ resolution.
There were 1548 grid cells out of $64\times32=2048$ that contained no data.
The resulting data, $\tilde{\bm{Z}}_{64\times 32}$,
consisting of $N=2048-1548=500$ observations, are shown in Figure \ref{fig:co2}(a),
where the 1548 missing data are shown in yellow, and the missing fraction is $1548/(64\times 32)\approx 76\%$.

We applied our EFDR-CS procedure with CPL (defined in Section~\ref{sec:method}) based on the significance test
\eqref{eq:test} in the same way as in Section \ref{sec:temperature}.
We estimated $\hat{\bm{\theta}}(\hat{\tau}^2,\hat{\phi},\hat{\lambda})=\hat{\bm{\theta}}=(\hat{\theta}_1,\dots,\hat{\theta}_7)'$ through \eqref{eq:theta.hat} in the Supplementary Material,
but now $\hat{\tau}^2$, $\hat{\phi}$, and $\hat{\lambda}$ are ML estimators based on the exponential covariance model with
a \textit{nugget effect} (i.e., $C(\bm{u})=\tau^2\{\exp(\|\bm{u}\|/\phi)+\lambda I(\bm{u}=\bm{0})\}$) to account for fine-scale-process variability.
The estimated map of the signal $\bm{\mu}$ is displayed in Figure \ref{fig:co2}(b), which shows a large hotspot in west Pakistan in the vicinity of Karachi,
of about +4ppm.
Our inference on $\bm{\mu}$ rejected the null hypothesis $H_0$: $\bm{\mu}=\bm{0}$ at the 5$\%$ significance level with a $p$-value of $7\times 10^{-6}$,
indicating that the signal is real.

\baselineskip=22pt

\section{Discussion and conclusions}
\label{sec:discussion}

In this article, we have proposed a spatial inference procedure to detect signals from possibly incomplete, area-aggregated data,
in the presence of spatially correlated noise.
The procedure, which we call EFDR-CS, is based on using the Enhanced False Discovery Rate and
Conditional Simulations to infer the fine-scale spatial signal from incomplete spatially aggregated data.

A critical component of the research presented in this article is a novel
methodology to combine exchangeable $p$-values into a single $p$-value
using a bivariate Gaussian copula and a composite likelihood.
In Section~\ref{sec:simulation} and the Supplementary Material, we show that
the methodology is able to properly control the Type-I error rate, even when the $p$-values are strongly correlated.
Further, we can extend the rectangular dimensions $n_1$ and $n_2$ to their next powers of two,
to deal with any boundary effects caused by applying a DWT to a rectangular image.

While we consider that the image $\bm{Z}$ at the original pixel resolution follows a multivariate Gaussian distribution,
it is possible to extend the CS approach for images generated from non-Gaussian distributions
(or even discrete distributions).
For example, one might use a generalized linear mixed model for non-Gaussian data,
with the random effects derived from a latent Gaussian spatial process
(e.g., Sengupta \textit{et al.}, 2016; Wilson and Wakefield, 2020).
This would be useful when, for example,
$\bm{Z}$ are counts of events aggregated over a number of regions from a log-Gaussian Cox process.
It would be straightforward to adapt the procedure proposed in this paper to these models by defining the signal $\bm{\mu}$ to be the mean of the latent Gaussian process.
The null hypothesis, $H_0$: $\bm{\mu}=\bm{\mu}_0$, could be tested by
conditionally simulating the hidden Gaussian process $M$ times, conditional on the data.
Then EFDR would be applied to each simulated process and the resulting $p$-values combined into a single $p$-value,
as we have done for Gaussian data.

\section*{Acknowledgments}

Hsin-Cheng Huang's research was supported by ROC Ministry of Science and Technology grants MOST 105-2119-M-007-032 and MOST 106-2118-M-001-002-MY3.
Noel Cressie's research was supported by
Australian Research Council Discovery Projects DP150104576 and DP190100180,
and by NSF grant SES-1132031 funded through the NSF-Census Research Network (NCRN) program.
Andrew Zammit-Mangion's research was supported by Australian Research Council Discovery Early Career Research Award (DECRA) DE180100203
and Discovery Project DP190100180.
Cressie's and Zammit-Mangion's research was also supported by NASA ROSES grant 17-OCO2-17-0012.
The authors would like to thank Chris Wikle, Scott Holan, Vineet Yadav, and Mike Gunson for comments on parts of this research,
and Yi Cao for assistance with Section~\ref{sec:co2}.

\section*{Supplementary Material}

The supplementary material consists of three sections.
Section \ref{sec:estimation} gives estimation of
the spatial covariance parameters $\bm{\theta}$ introduced in Section \ref{sec:testing}.
Section \ref{sec:Type-I} provides an initial simulation study to investigate the Type-I error rates obtained using the testing rule \eqref{eq:test}.
Section \ref{sec:figure} contains complete figures for the simulations in Section~\ref{sec:simulation}.
These include: $\tilde{\bm{Z}}_{16\times 16}$ and $\tilde{\bm{Z}}_{8\times 8}$,
corresponding to $\bm{Z}$ in Figure \ref{fig:signalonimageraw};
empirical power curves as a function of the signal's magnitude $h$, for various procedures (NVE, IDL, our proposed procedure CPL,
and its variant MOM);
and empirical ROC curves for CPL, where the signal volume $h r^2$ is varied.

\section*{References}

\begin{description}
\item Benjamini, Y.~and Heller, R.~(2007). False discovery rates for spatial signals.
  \textit{Journal of the American Statistical Association}, \textbf{102}, 1272-1281.
\item Benjamini, Y.~and Hochberg, Y.~(1995). Controlling the false discovery rate: A practical and powerful approach to multiple testing,
  \textit{Journal of the Royal Statistical Society, Series B}, \textbf{57}, 289-300.
\item Brown, M.~(1975). A method for combining non-independent, one-sided tests of significance, 
  \textit{Biometrics}, \textbf{31}, 987-992.
\item Cressie, N.~(1993). Statistics for Spatial Data, rev.~edn, Wiley, New York, NY.
\item Daubechies, I.~(1992). \textit{Ten Lectures on Wavelets},
  CBMS-NSF Regional Conference Series in Applied Mathematics, SIAM, Philadelphia, PA.
\item Egan, J.~P.~(1975). \textit{Signal Detection Theory and ROC Analysis}, Academic Press, New York, NY.
\item Fisher, R.~A.~(1925). \textit{Statistical Methods for Research Workers}, Oliver and Boyd, Edinburgh, UK.
\item Gilleland, E.~(2013). Testing competing precipitation forecasts accurately and efficiently:
  The spatial prediction comparison test, \textit{Monthly Weather Review}, \textbf{141}, 340-355.
\item Hering, A.~S.~and Genton, M.~G.~(2011). Comparing spatial predictions,
  \textit{Technometrics}, \textbf{53}, 414-425.
\item Lei, L., Ramdas, A., and Fithian, W.~(2017).
  STAR: A general interactive framework for FDR control under structural constraints,
  \textit{arXiv preprint arXiv:1710.02776}.
\item Littell, R.~C.~and Folks, J.~L.~(1971). Asymptotic optimality of Fisher's method of combining independent tests,
  \textit{Journal of the American Statistical Association}, \textbf{66}, 802-806.
\item Little, R.~J.~A.~and Rubin, D.~B.~(2002). \textit{Statistical Analysis with Missing Data}, 2nd edn, Wiley, New York, NY.
\item Martinez, J.~G., Bohn, K.~M., Carroll, R.~J., and Morris, J.~S.~(2013). A study of Mexican free-tailed bat chirp syllables: Bayesian functional mixed models for nonstationary acoustic time series,
 \textit{Journal of the American Statistical Association}, \textbf{108}, 514-526.
\item Nguyen, H., Cressie, N., and Braverman, A.~(2012). Spatial statistical data fusion for remote sensing applications,
  \textit{Journal of the American Statistical Association}, \textbf{107}, 1004-1018.
\item Risser, M.~D., Paciorek, C.~J., and Stone, D.~A.~(2019). Spatially-dependent multiple testing under model misspecification, with application to detection of anthropogenic influence on extreme climate events,
\textit{Journal of the American Statistical
Association}, \textbf{114}, 61-78.
\item Schott, J.~R.~(2017). \textit{Matrix Analysis for Statistics}, 3rd edn, Wiley, Hoboken, NJ.
\item Sengupta, A., Cressie, N., Kahn, B.~H., and Frey, R.~(2016).
  Predictive inference for big, spatial, non-Gaussian data: MODIS cloud data and its change-of-support,
  \textit{Australian and New Zealand Journal of Statistics}, \textbf{58}, 15-45.
\item Shen, X., Huang, H.-C., and Cressie, N.~(2002). Nonparametric hypothesis testing for a spatial signal,
  \textit{Journal of the American Statistical Association}, \textbf{97}, 1122-1140.
\item Song, P.~X.-K.~(2000). Multivariate dispersion models generated from Gaussian copula,
  \textit{Scandinavian Journal of Statistics}, \textbf{27}, 305-320.
\item Spiegelhalter, D.~J., Abrams, K.~R., and Myles, J.~P.~(2004).
  \textit{Bayesian Approaches to Clinical Trials and Health-Care Evaluation}, Wiley, New York, NY.
\item Sun, W., Reich, B.J., Cai, T.T., Guindani, M., and Schwartzman, A.~(2015).
  False discovery control in large-scale spatial multiple testing, \textit{Journal of the Royal Statistical Society, Series B},
  \textbf{77}, 59-83.
\item Wasserstein, R.~L., Schirm, A.~L., and Lazar, N.~A.~(2019). Moving to a world beyond ``$p<0.05$,"
  \textit{The American Statistician}, \textbf{73}, 1-19.
\item Wilson, K.~and Wakefield, J.~(2020). Pointless spatial modeling, \textit{Biostatistics}, \textbf{21}, e17-e32.
\item Ye, J.~(1998). On measuring and correcting the effects of data mining and model selection,
  \textit{Journal of the American Statistical Association}, \textbf{93}, 120-131.
\item Yekutieli, D and Benjamini, Y.~(1999). Resampling-based false discovery rate controlling multiple test procedures
  for correlated test statistics, \textit{Journal of Statistical Planning and Inference}, \textbf{82}, 171-196.
\item Yun, S., Zhang, X., and Li, B.~(2018). Detection of local differences between two spatiotemporal random fields, Manuscript.
\item Zammit-Mangion, A.~and Huang, H.-C.~(2015). EFDR: Wavelet-based enhanced FDR for signal detection in noisy images,
  R package version 0.1.1. URL \url{https://CRAN.R-project.org/package=EFDR}.
\end{description}

\newpage
\beginsupplement
\setcounter{section}{0}
\setcounter{equation}{0}
\setcounter{table}{0}

\section*{Supplementary Material for ``False Discovery Rates to Detect Signals from Incomplete Spatially Aggregated Data"}

The supplementary material consists of three sections.
Section \ref{sec:estimation} gives estimation of $\bm{\theta}$ introduced in Section \ref{sec:testing}.
Section \ref{sec:Type-I} provides an initial simulation study to investigate the Type-I error rates obtained using the significance test \eqref{eq:test}.
Section \ref{sec:figure} contains nine figures for the simulations in Section~\ref{sec:simulation}.

\section{Estimation of spatial dependence from incomplete spatially aggregated data}
\label{sec:estimation}

To obtain $\bm{\Sigma}(\bm{\theta})$ in Section \ref{sec:testing},
we start with the null hypothesis $H_0$ in \eqref{eq:hypothesis} and a model in the wavelet domain for $\mathcal{W}\bm{\delta}$ defined by \eqref{eq:dwt delta}.
As in Shen \textit{et al.}~(2002), the wavelet coefficients, $\mathcal{W}\bm{Z}$, are modeled independently as
\begin{equation}
\bm{d}_{-j}^{(m)}\sim \mathrm{Gau}(\bm{0},\theta_{3(j-1)+m}\bm{I});\,m=1,2,3;\,j=1,\dots,J,\quad\mbox{and}\quad
\bm{c}_{-J}\sim \mathrm{Gau}(\bm{0},\theta_{3J+1}\bm{I}),
\label{eq:wavelet model}
\end{equation}

\noindent and from \eqref{eq:dwt delta}, $\bm{\delta}=\mathcal{W}^{-1}(\bm{d}'_{-1},\dots,\bm{d}'_{-J}, \bm{c}'_{-J})'$. Hence,
from \eqref{eq:wavelet model}, the vector of the variance components
$\bm{\theta}\,\equiv\,(\theta_1,\dots,\theta_{3J+1})'$ parameterizes
$\mathrm{var}(\bm{\delta})=\bm{\Sigma}(\bm{\theta})$.

The covariance matrix in the wavelet domain,
$\bm{V}(\bm{\theta})\equiv\mathrm{var}(\mathcal{W}\bm{\delta})=\mathcal{W}\bm{\Sigma}(\bm{\theta})\mathcal{W}'$ is a diagonal matrix
whose $k$-th diagonal block is given by $\theta_k\bm{I}$, for $k=1,\dots,(3J+1)$.
We estimate $\bm{\theta}$ through
\begin{equation}
\hat{\bm{\theta}}\,\equiv\,\mathop{\arg\min}_{\bm{\theta}\in(0,\infty)^{3J+1}}\|\mathcal{W}'\bm{V}(\bm{\theta})\mathcal{W}-\hat{\bm{\Sigma}}\|_F=
\mathop{\arg\min}_{\bm{\theta}\in(0,\infty)^{3J+1}}\|\bm{V}(\bm{\theta})-\mathcal{W}\hat{\bm{\Sigma}}\mathcal{W}'\|_F\:,
\label{eq:theta.hat0}
\end{equation}

\noindent since $\mathcal{W}$ is orthogonal. In \eqref{eq:theta.hat0},
$\|\cdot\|_F$ is the Frobenius norm and
$\hat{\bm{\Sigma}}$ is a smooth empirical estimator, here the maximum likelihood (ML) estimator of $\bm{\Sigma}$ under a parametric covariance model,
$C(\cdot)$, and $\bm{\mu}=\bm{0}$.
Note that it is generally not possible to define the usual method-of-moments estimator of $\bm{\Sigma}$
from the incomplete spatially aggregated data $\tilde{\bm{Z}}$,
for which there are no replications available.

As an example, consider the Mat\'{e}rn covariance model for $\mathrm{cov}(Z(\bm{s}),Z(\bm{s}^*))$ given by,
\begin{equation}
C(\bm{u})=\tau^2\bigg\{\frac{1}{2^{\nu-1}\Gamma(\nu)}\bigg(\frac{\sqrt{2\nu}}{\phi}\|\bm{u}\|\bigg)^{\nu}\mathcal{K}_{\nu}
\bigg(\frac{\sqrt{2\nu}}{\phi}\|\bm{u}\|\bigg)+\lambda I(\bm{u}=\bm{0})\bigg\};\quad\bm{u}=\bm{s}^*-\bm{s}\in\mathbb{R}^2,
\label{eq:Matern}
\end{equation}

\noindent where $\mathcal{K}_\nu(\cdot)$ is the modified Bessel function of the second kind of order $\nu$,
$\tau^2$ is a variance parameter,
and $\bm{\gamma}\,\equiv\,(\phi,\nu,\lambda)'$ consists of a spatial-scale parameter $\phi$, a smoothness parameter $\nu$,
and a nugget-effect parameter $\lambda$.
Under $H_0$, the ML estimator of  $\bm{\gamma}$ can be obtained from the data, $\tilde{\bm{Z}}$, by minimizing the negative log profile likelihood, as follows:
\[
\hat{\bm{\gamma}}\,\equiv\,\big(\hat{\phi},\hat{\nu},\hat{\lambda}\big)'\,\equiv\,
\mathop{\arg\min}_{\bm{\gamma}}\bigg\{\frac{1}{2}\log|\bm{H}\bm{\Omega}(\bm{\gamma})\bm{H}'|
+\frac{K}{2}\log\big\{\tilde{\bm{Z}}'(\bm{H}\bm{\Omega}(\bm{\gamma})\bm{H}')^{-1}\tilde{\bm{Z}}\big\}+\mathrm{constant}\bigg\},
\]
where $\bm{\Omega}(\bm{\gamma})$ is an $n\times n$ correlation matrix whose $(i,j)$-th entry is
$C(\bm{s}_i-\bm{s}_j)/\tau^2$.
Then the ML estimator $\hat{\tau}^2$ of the stationary variance $\tau^2$ is:
\[
\hat{\tau}^2\,\equiv\,\frac{1}{K}\tilde{\bm{Z}}'(\bm{H\Omega}(\hat{\bm{\gamma}})\bm{H}')^{-1}\tilde{\bm{Z}}.
\]
Consequently, a smooth empirical estimator of $\bm{\Sigma}$ for use in \eqref{eq:theta.hat0} is given by $\hat{\bm{\Sigma}}\,\equiv\,\hat{\tau}^2\bm{\Omega}(\hat{\bm{\gamma}})$.

Let $\mathcal{W}_k$ be a sub-matrix of $\mathcal{W}$, consisting of the rows corresponding to the $k$-th wavelet component.
Then $\mathcal{W}'=(\mathcal{W}'_1,\dots,\mathcal{W}'_{3J+1})$. It follows from \eqref{eq:theta.hat0} that
$\hat{\bm{\theta}}\,\equiv\,\big(\hat{\theta}_1,\dots,\hat{\theta}_{3J+1}\big)'$ is given by,
\begin{equation}
\hat{\theta}_k\,=\,\frac{\hat{\tau}^2}{n_k}\mathrm{tr}\big(\mathcal{W}_k\bm{\Omega}(\hat{\bm{\gamma}})\mathcal{W}'_k\big);\quad
k=1,\dots,3J+1,
\label{eq:theta.hat}
\end{equation}

\noindent where $n_k$ is the number of rows of $\mathcal{W}_k$, and $\mathrm{tr}(\bm{A})$ denotes the trace of a square matrix $\bm{A}$.

\section{Observed Type-I error rates using $p$-values from correlated $z$-tests}
\label{sec:Type-I}

We conducted an initial simulation study to investigate the Type-I error rates obtained using the
significance test \eqref{eq:test}
when the level of exchangeability $\rho$ was estimated using copulas (CPL)
and the method-of-moments (MOM).
The significance test where the $p$-values are combined na\"{i}vely
through their simple average (NVE) was also considered for comparison.
In this study, the $p$-values $\{p_i\}$ came from a two-sided $z$-test for the mean of a Gaussian distribution with known variance;
the purpose of the study is to assess the validity of our proposed procedure in a very simple, non-spatial setting where there is exchangeability.

Let the set $\{x_1,\dots,x_{100}\}$ be made up of elements $x_i $
distributed independently as $\mathrm{Gau}(\mu,1)$, for $i=1,\dots,100$.
We put $\mu=0$ and drew $\{x_1,\dots,x_{100}\}$ from $\mathrm{Gau}(0,1)$, and then
we randomly drew $M$ subsamples of size $N\leq 100$ without replacement from $\{x_1,\dots,x_{100}\}$.
For $k=1,\dots,M$, let $\{x^*_{k,1},\dots,x^*_{k,N}\}$ be the $k$-th subsample
and $z_k=(1 / N) \sum_{i = 1}^{N}{x^*_{k,i}}$ be a statistic for testing the hypotheses, $H_0: \mu = 0$ versus $H_1: \mu \ne 0$.
Under $H_0$, it is easy to see that $\mathrm{E}(z_k)=0$ and $\mathrm{var}(z_k)=1/N$.
However, the $\{z_k\}$ are dependent;  indeed, they are exchangeable due to the sampling-without-replacement from $\{x_1,\dots,x_{100}\}$.

The individual $z$-test, based on the statistic $z_k$, rejects $H_0$ if
\begin{equation}
p_k\,\equiv\, 2(1 - \Phi(\sqrt{N}|z_k|))<\alpha,
\label{eq:test2}
\end{equation}

\noindent where $0<\alpha<1$ is a pre-specified significance level.
Because $\{z_k:k=1,\dots,M\}$ are exchangeable, so too are the $p$-values $\{p_k:k=1,\dots,M\}$.
We are interested in knowing how well they can be combined into a single $p$-value using
the na\"{i}ve procedure of averaging (NVE), our proposed copula-based procedure (CPL),
and its method-of-moments variant (MOM). That is, we compare
\begin{enumerate}
\item NVE: A na\"{i}ve procedure where the final $p$-value is $\sum_{k=1}^M p_k/M$\,.
\item CPL: The final $p$-value is given by \eqref{eq:hat_pvalue}.
\item MOM: The final $p$-value is given by \eqref{eq:tilde_pvalue}.
\end{enumerate}

\noindent Note that a Type-I error occurs if, under $H_0$, the resulting $p$-value is smaller than $\alpha$.

In this experimental set-up, the level of exchangeability $\rho$ in the intra-class correlation model \eqref{eq:IC} is
higher when $N$ (i.e., the subsample size) is closer to the full sample size of $100$.
So we considered only $N\in\{80,85,90,95\}$.
In addition, we considered the three significance levels, $\alpha\in\{0.01,0.05,0.1\}$, commonly used in practice.
The resulting empirical Type-I error rates for the three methods
(NVE, CPL, MOM) under the 12 different combinations of $N$ and $\alpha$,
based on 50,000 simulation replicates, are shown in Table~\ref{table:simulation}.

\begin{table}[tb]\centering
\caption{Empirical Type-I error rates under different values of $(\alpha,N)$ defined in Section~\ref{sec:Type-I}
based on 50,000 simulation replicates,
where values given in parentheses are the Monte Carlo standard errors.}
\medskip
\begin{tabular}{cc|ccc}
\hline
$\alpha$ & $N$ & NVE & CPL & MOM \\\hline
0.01 & 80 & 0.0016 (0.0002) & 0.0078 (0.0004) & 0.0063 (0.0004) \\
        & 85 & 0.0030 (0.0002) & 0.0089 (0.0004) & 0.0073 (0.0004) \\
        & 90 & 0.0049 (0.0003) & 0.0098 (0.0004) & 0.0087 (0.0004) \\
        & 95 & 0.0065 (0.0004) & 0.0098 (0.0004) & 0.0090 (0.0004) \\\hline
0.05 & 80 & 0.0161 (0.0006) & 0.0454 (0.0009) & 0.0446 (0.0009) \\
        & 85 & 0.0233 (0.0007) & 0.0488 (0.0010) & 0.0481 (0.0010) \\
        & 90 & 0.0305 (0.0008) & 0.0482 (0.0010) & 0.0478 (0.0010)\\
        & 95 & 0.0389 (0.0009) & 0.0485 (0.0010) & 0.0482 (0.0010) \\\hline
0.10 & 80 & 0.0443 (0.0009) & 0.0958 (0.0013) & 0.1033 (0.0014) \\
        & 85 & 0.0574 (0.0010) & 0.0968 (0.0013) & 0.1044 (0.0014) \\
        & 90 & 0.0686 (0.0011) & 0.0949 (0.0013) & 0.1001 (0.0013) \\
        & 95 & 0.0828 (0.0012) & 0.0971 (0.0013) & 0.0997 (0.0013) \\
\hline
\end{tabular}
\label{table:simulation}
\end{table}

NVE consistently gives Type-I error rates that are too small,
which was expected since the sample average of the $p$-values
results in a combined $p$-value that tends to be too large,
causing the null hypothesis to be rejected less often than it should.
The effect is less pronounced when the $p$-values are more correlated;
that is, as the size of the subsample $N$ increases, the Type-I error rate of NVE improves.
Our proposed procedure, whether it is CPL or MOM, is adaptive to the amount of dependence, and the
Type-I error rates are very close to the nominal levels for all cases.

This initial study is encouraging and indicates that our proposal is valid in the presence of exchangeably dependent $p$-values.
\pagebreak

\section{Figures for the simulations in Section~\ref{sec:simulation}}
\label{sec:figure}

\begin{figure}[tb]\centering
\begin{tabular}{cccccc}
~~~~~$h=0$ & $h=1$~~ & $h=2$~~ & $h=3$~~ & $h=4$~~ & $h=5$~~
\smallskip\\
\rotatebox{90}{$\quad \quad r=4$}~~\includegraphics[scale=0.155,trim={1cm 2.5cm 2.5cm 2cm},clip]{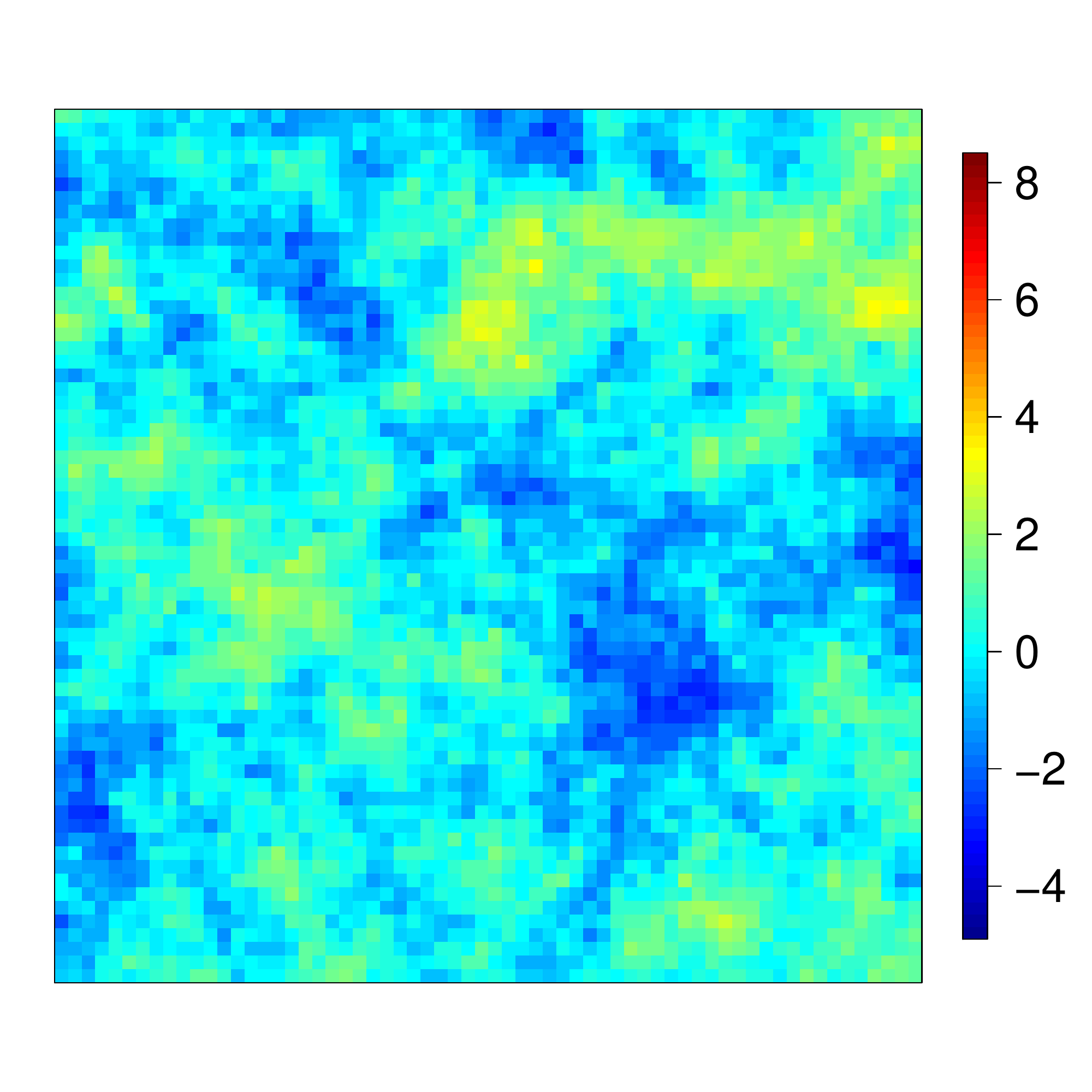} &
\!\!\!\!\!\!\includegraphics[scale=0.155,trim={1cm 2.5cm 2.5cm 2cm},clip]{./signal-r4-h1} &
\!\!\!\!\!\!\includegraphics[scale=0.155,trim={1cm 2.5cm 2.5cm 2cm},clip]{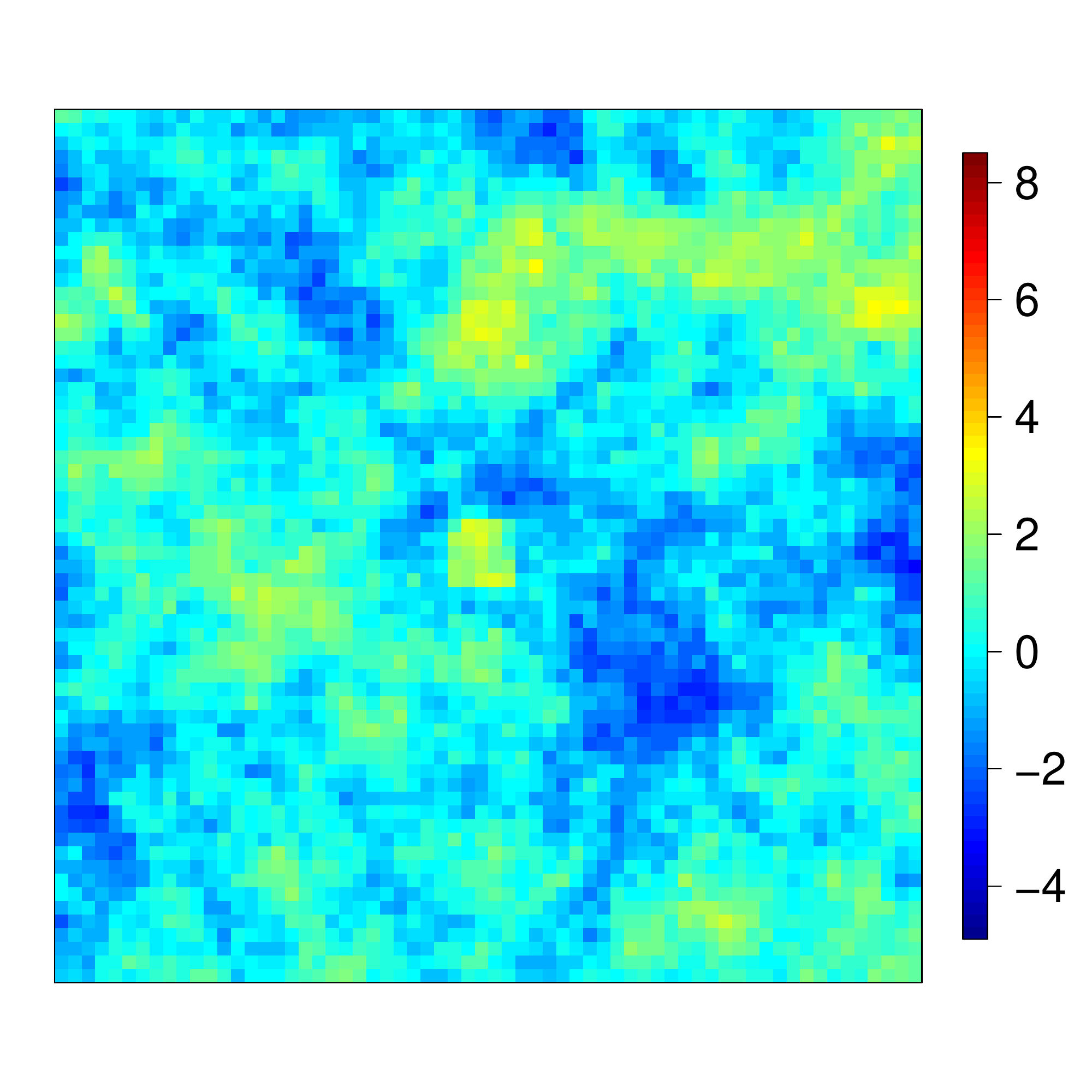} &
\!\!\!\!\!\!\includegraphics[scale=0.155,trim={1cm 2.5cm 2.5cm 2cm},clip]{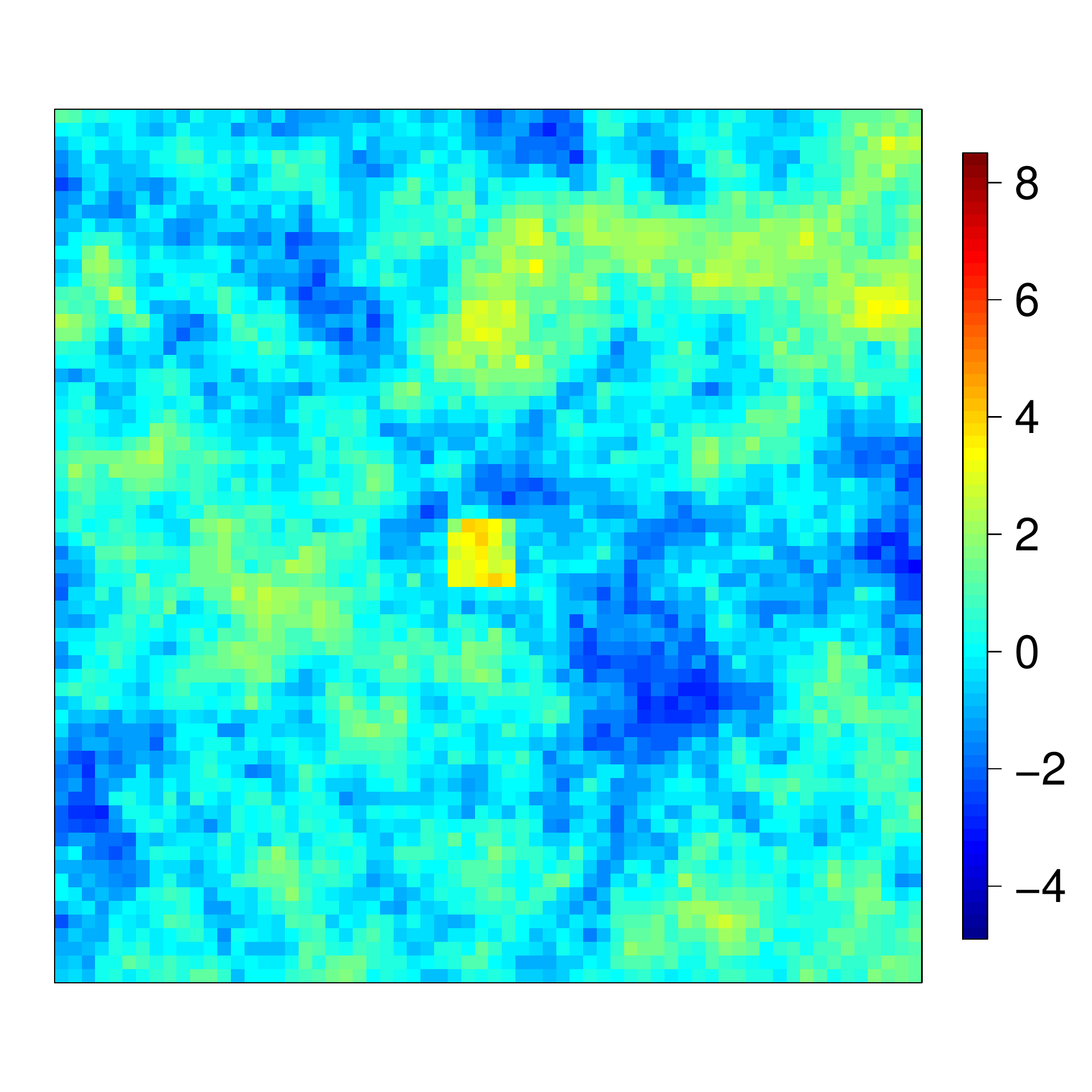} &
\!\!\!\!\!\!\includegraphics[scale=0.155,trim={1cm 2.5cm 2.5cm 2cm},clip]{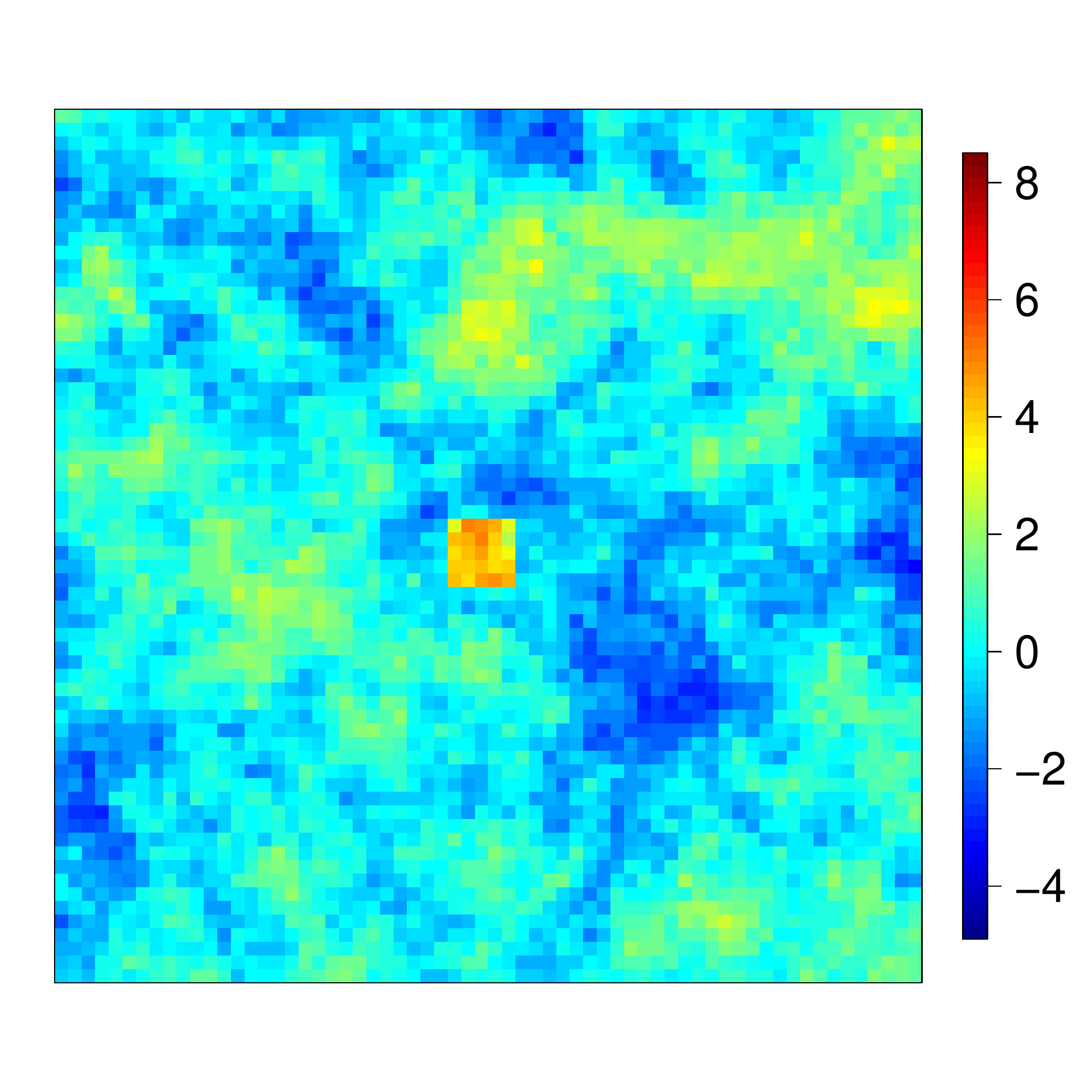} &
\!\!\!\!\!\!\includegraphics[scale=0.155,trim={1cm 2.5cm 2.5cm 2cm},clip]{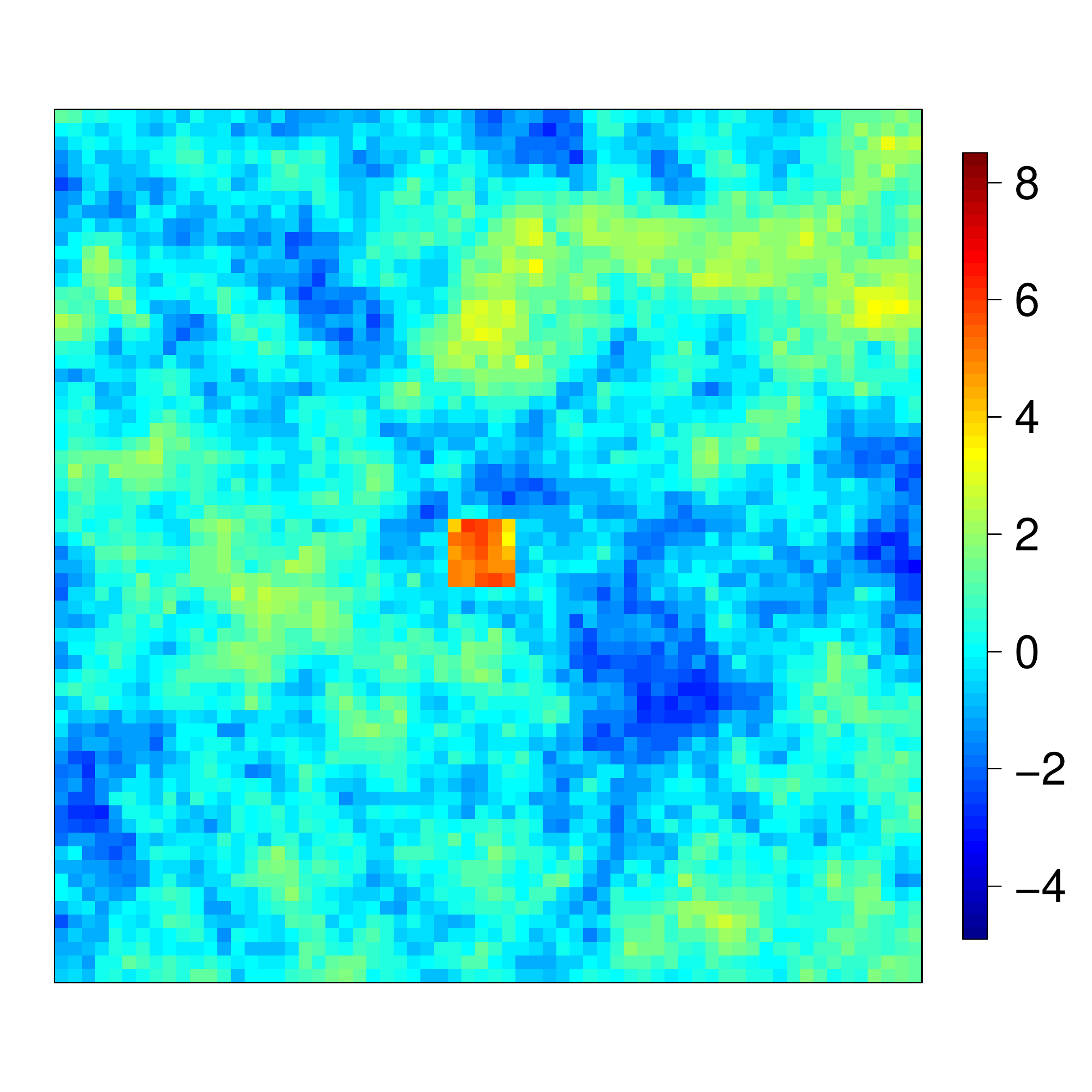} \\
\rotatebox{90}{$\quad \quad r=6$}~~\includegraphics[scale=0.155,trim={1cm 2.5cm 2.5cm 2cm},clip]{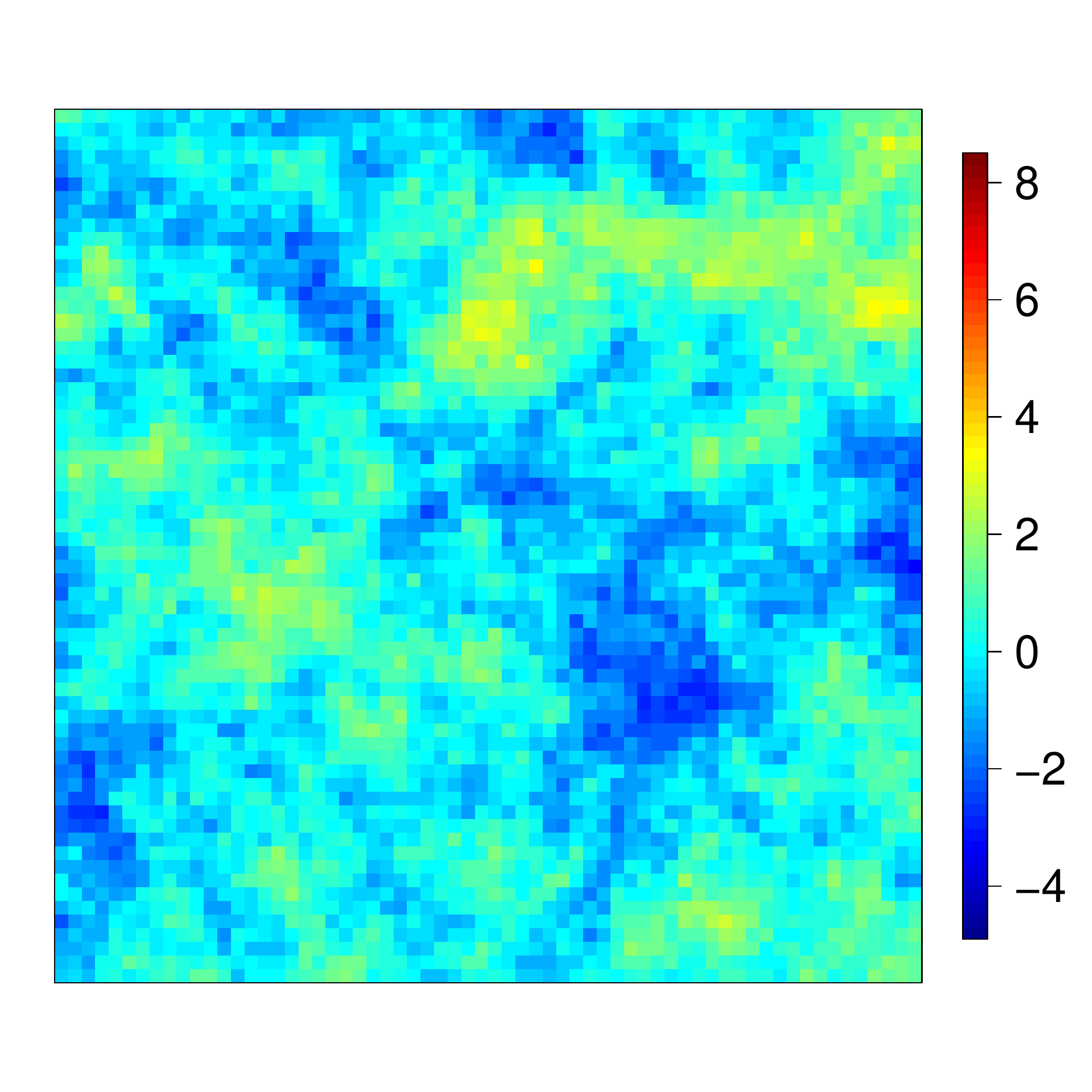} &
\!\!\!\!\!\!\includegraphics[scale=0.155,trim={1cm 2.5cm 2.5cm 2cm},clip]{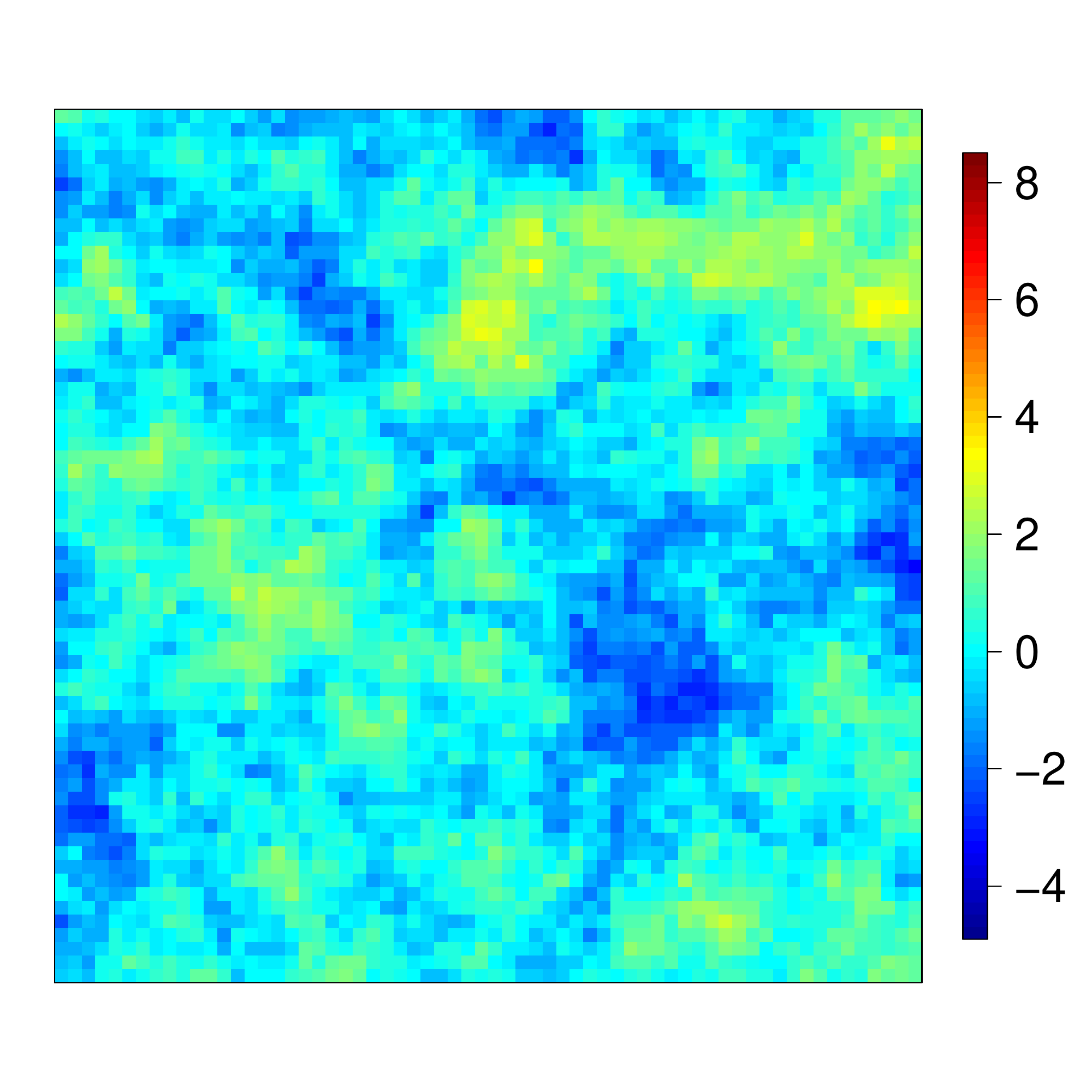} &
\!\!\!\!\!\!\includegraphics[scale=0.155,trim={1cm 2.5cm 2.5cm 2cm},clip]{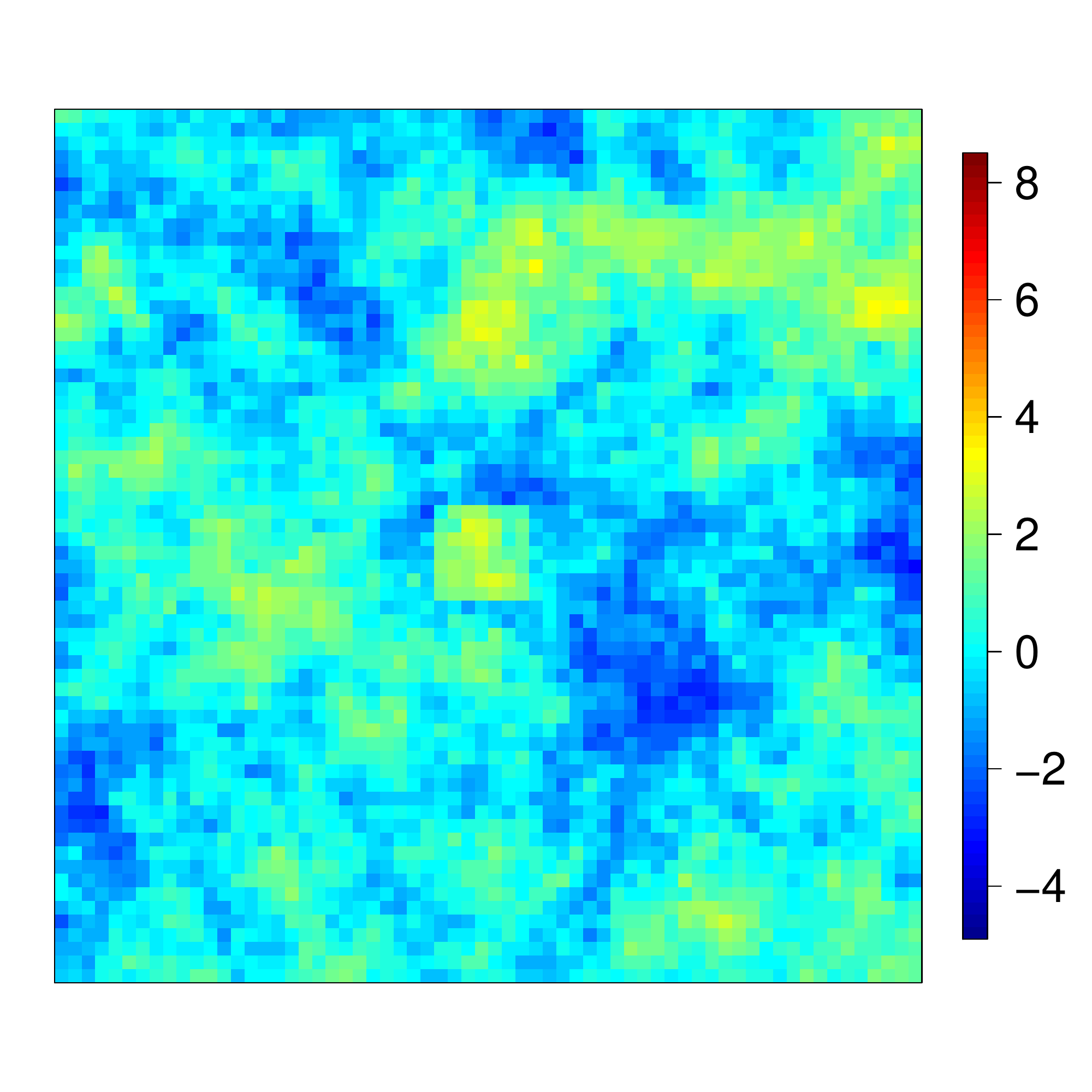} &
\!\!\!\!\!\!\includegraphics[scale=0.155,trim={1cm 2.5cm 2.5cm 2cm},clip]{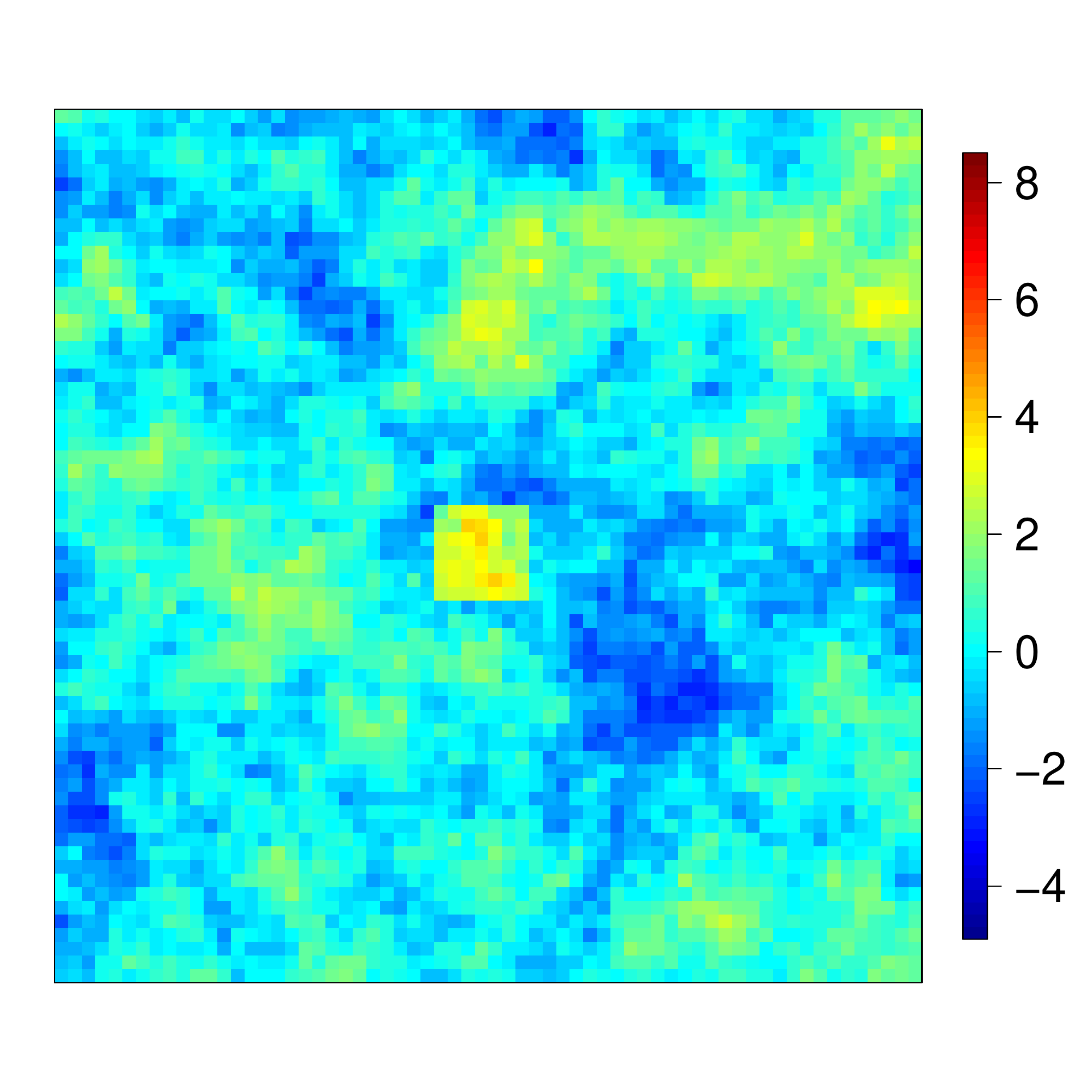} &
\!\!\!\!\!\!\includegraphics[scale=0.155,trim={1cm 2.5cm 2.5cm 2cm},clip]{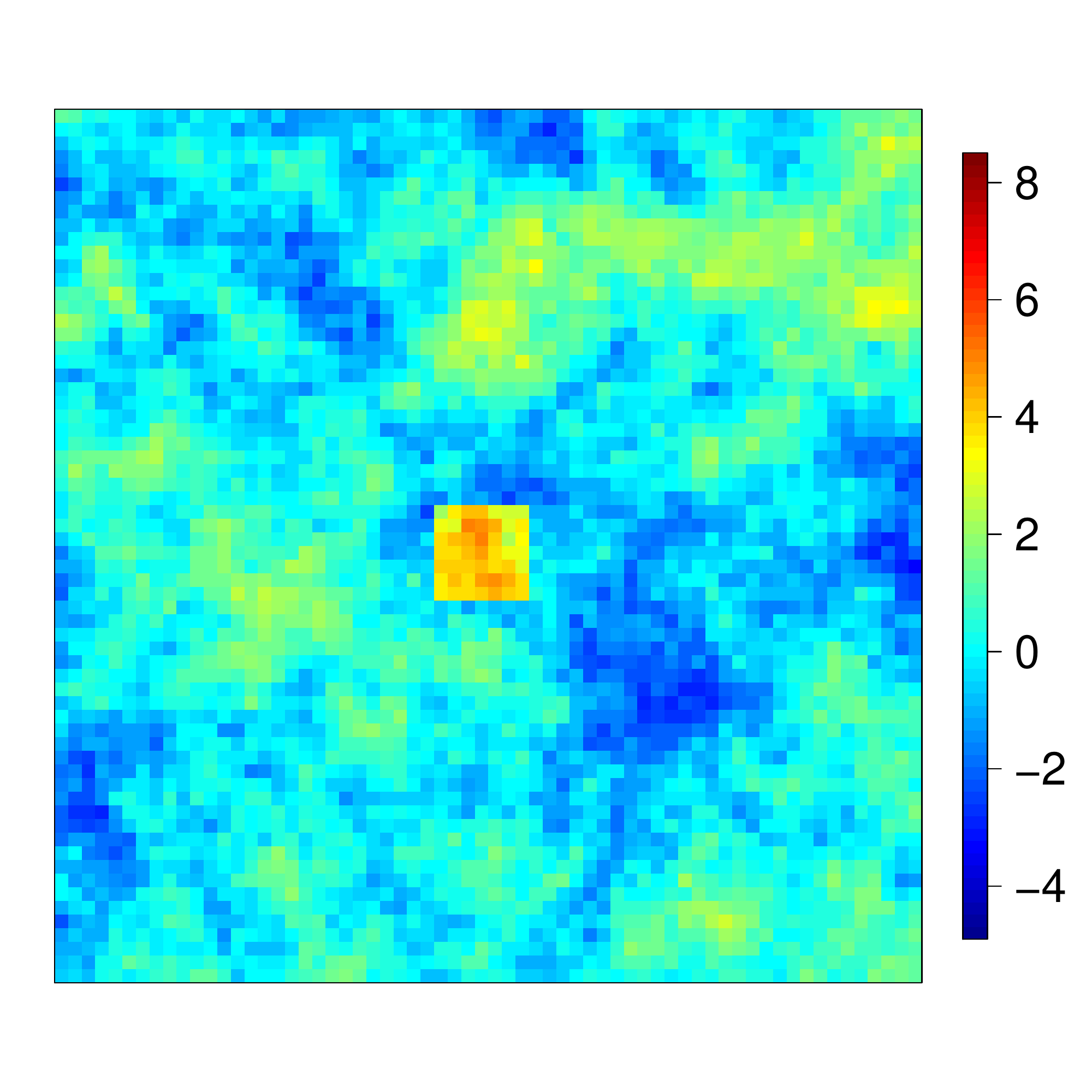} &
\!\!\!\!\!\!\includegraphics[scale=0.155,trim={1cm 2.5cm 2.5cm 2cm},clip]{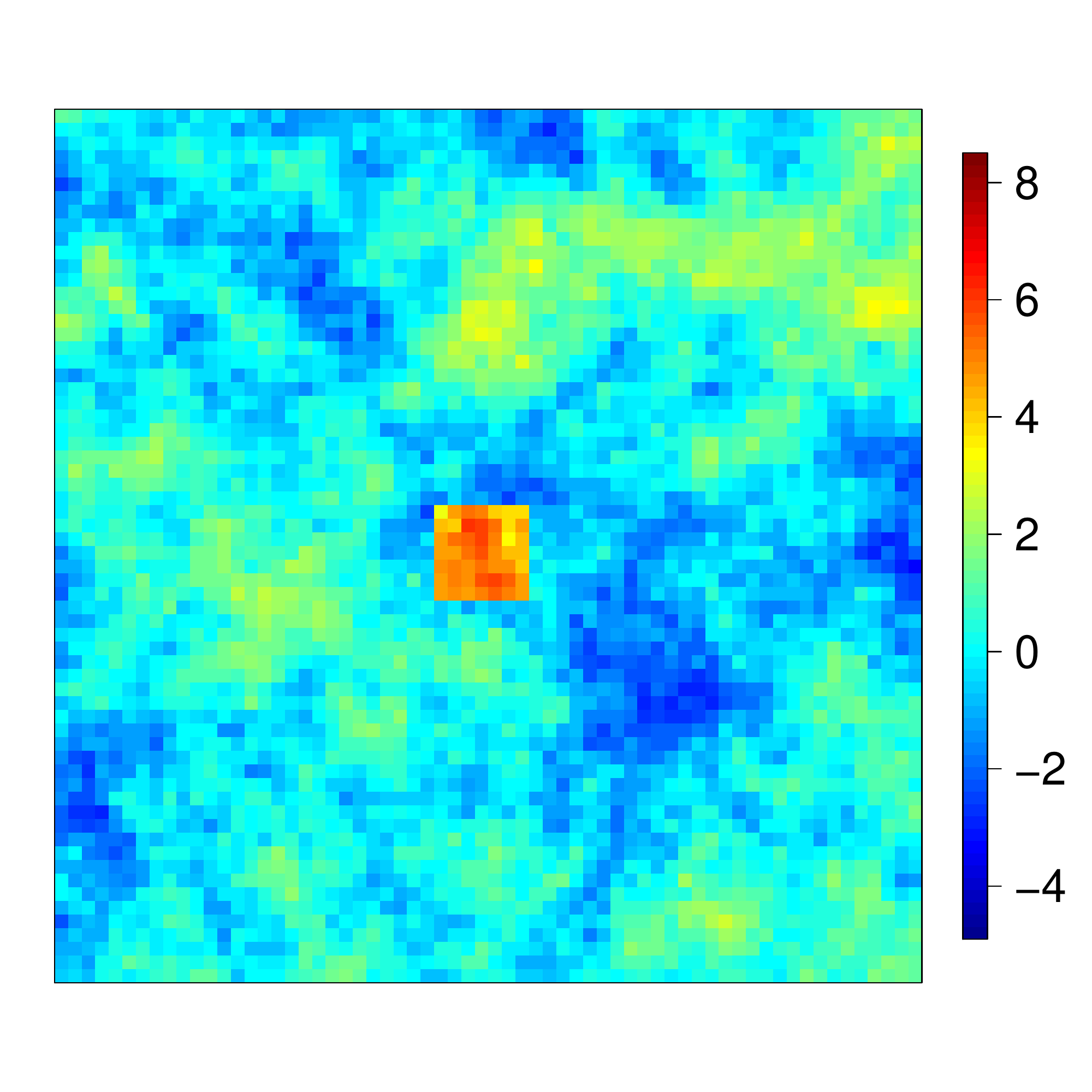} \\
\rotatebox{90}{$\quad \quad r=8$}~~\includegraphics[scale=0.155,trim={1cm 2.5cm 2.5cm 2cm},clip]{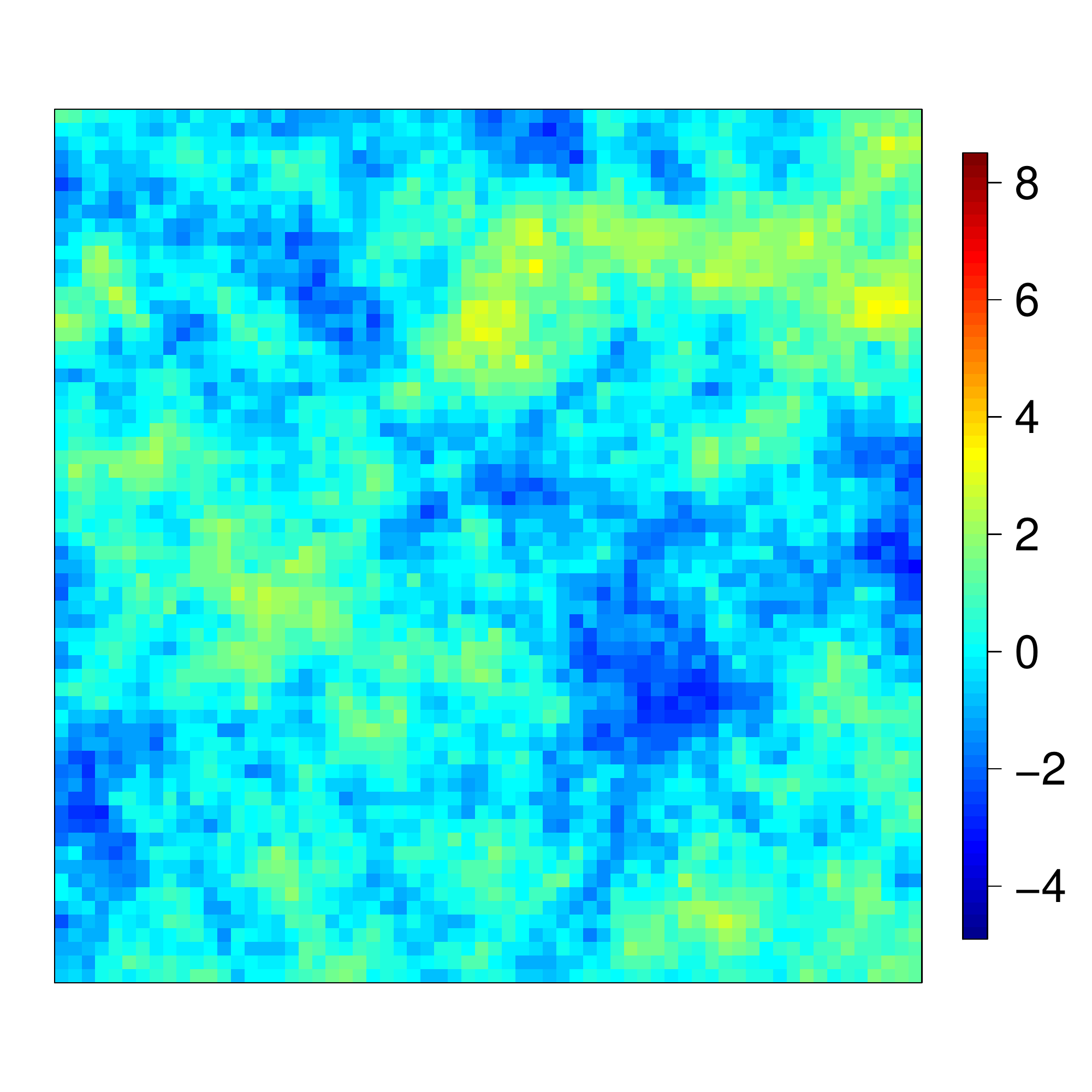} &
\!\!\!\!\!\!\includegraphics[scale=0.155,trim={1cm 2.5cm 2.5cm 2cm},clip]{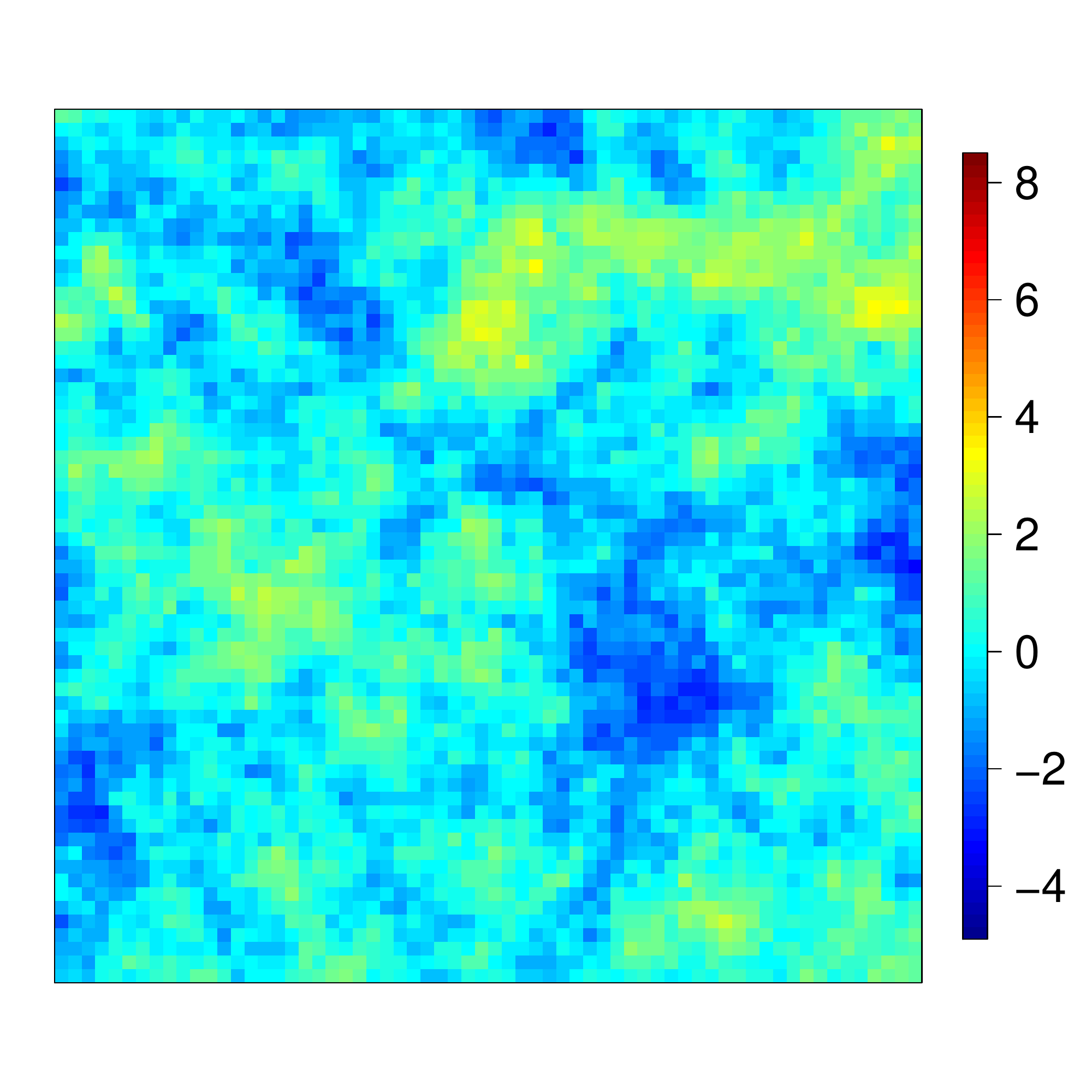} &
\!\!\!\!\!\!\includegraphics[scale=0.155,trim={1cm 2.5cm 2.5cm 2cm},clip]{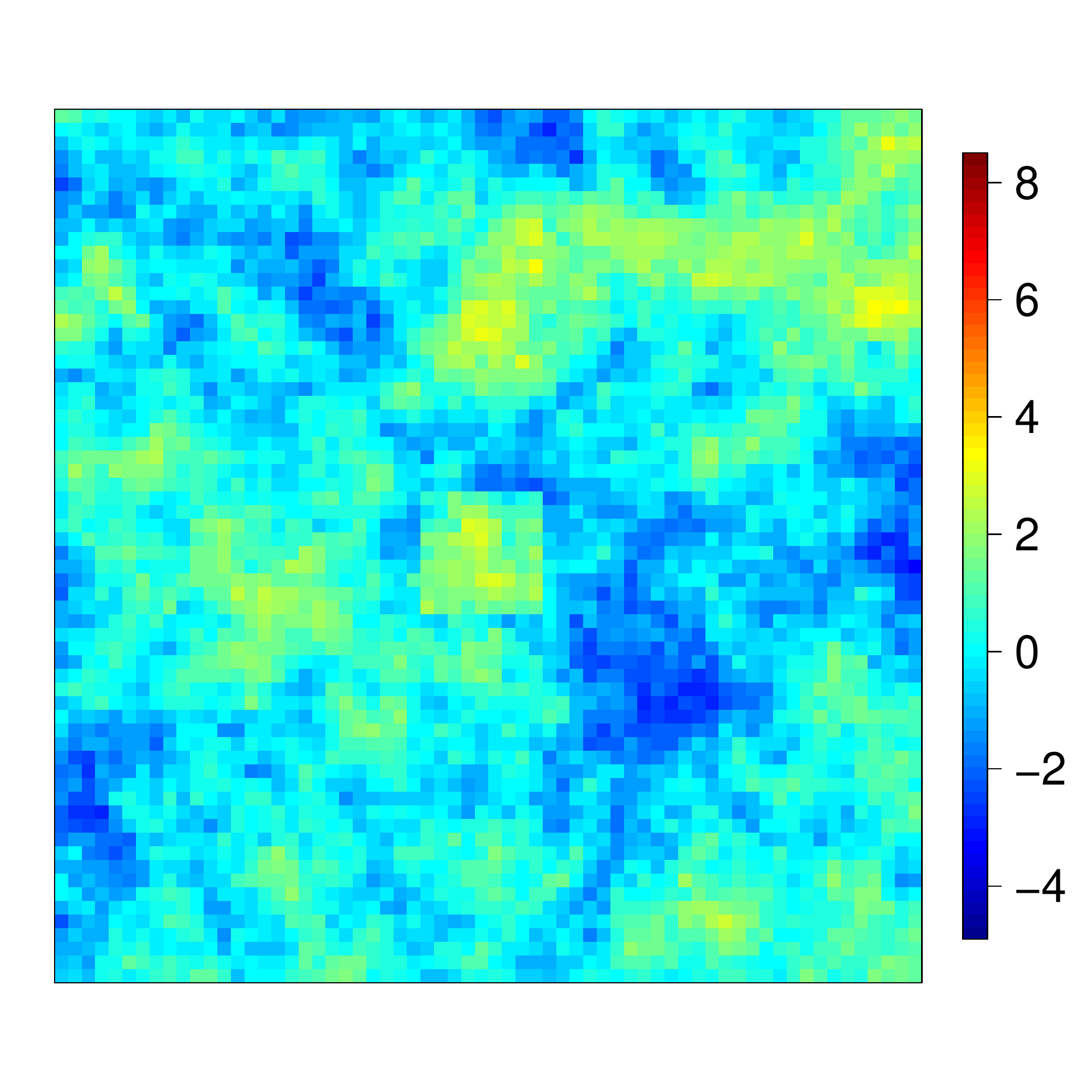} &
\!\!\!\!\!\!\includegraphics[scale=0.155,trim={1cm 2.5cm 2.5cm 2cm},clip]{./signal-r8-h3} &
\!\!\!\!\!\!\includegraphics[scale=0.155,trim={1cm 2.5cm 2.5cm 2cm},clip]{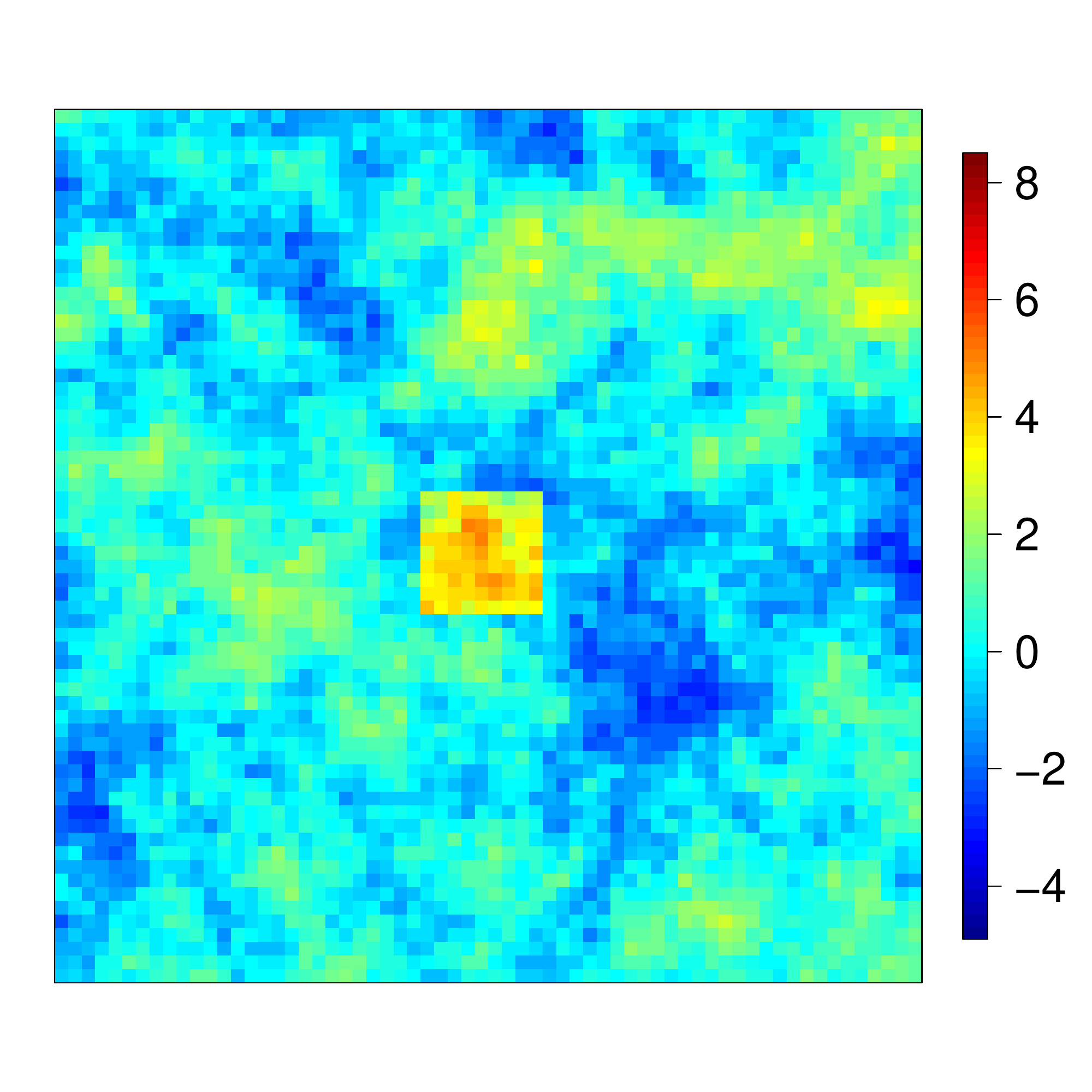} &
\!\!\!\!\!\!\includegraphics[scale=0.155,trim={1cm 2.5cm 2.5cm 2cm},clip]{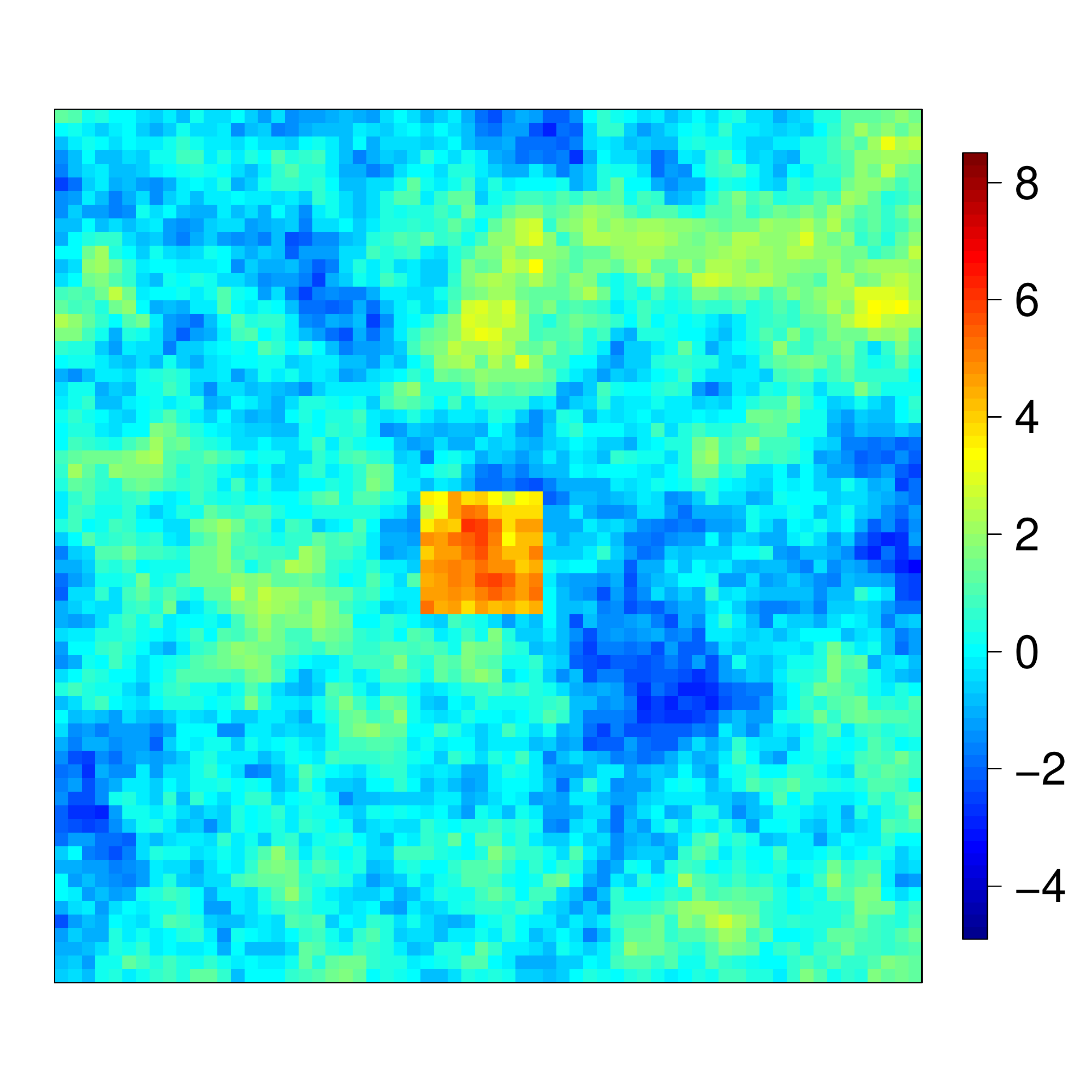} \\
\rotatebox{90}{$\quad \quad r=10$}~~\includegraphics[scale=0.155,trim={1cm 2.5cm 2.5cm 2cm},clip]{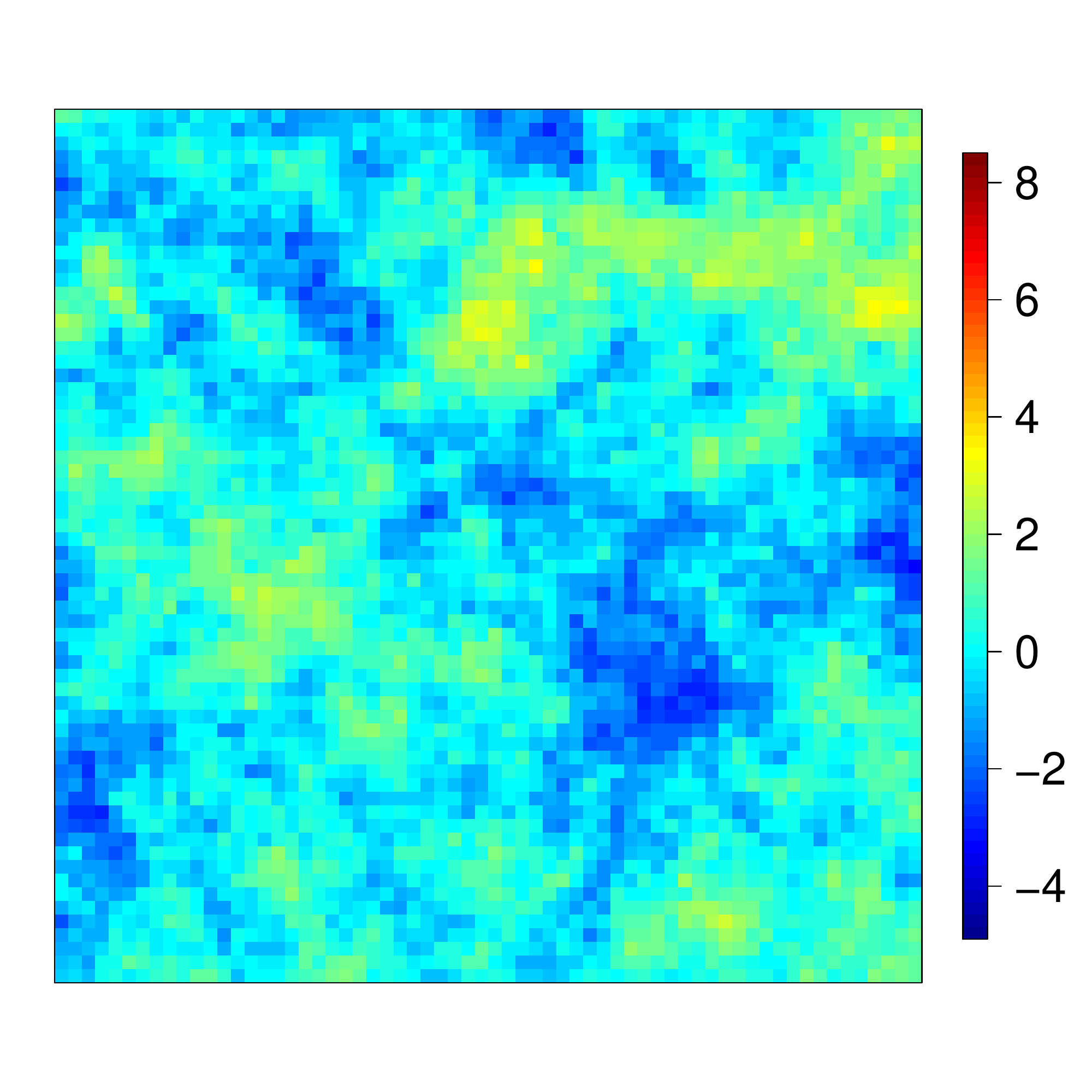} &
\!\!\!\!\!\!\includegraphics[scale=0.155,trim={1cm 2.5cm 2.5cm 2cm},clip]{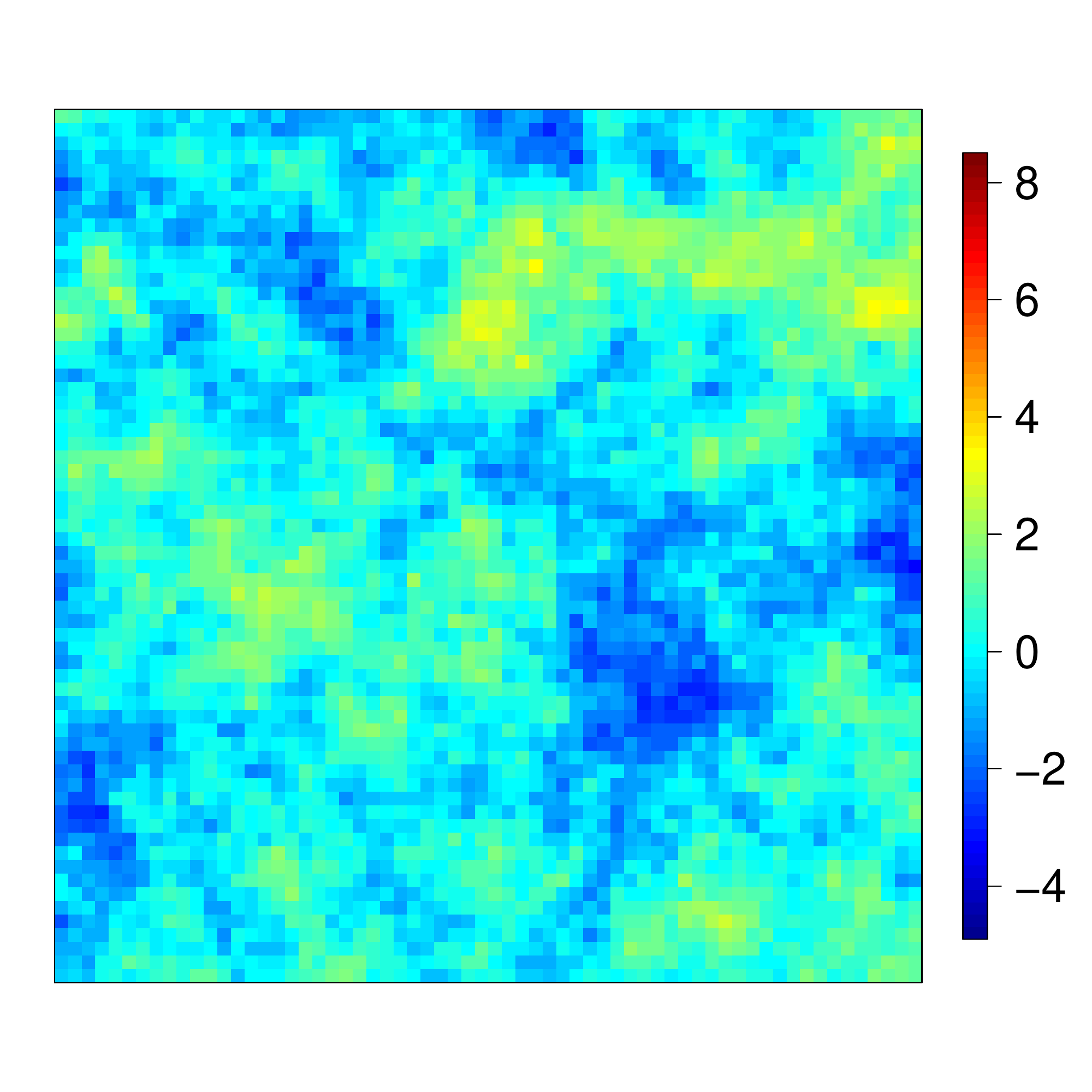} &
\!\!\!\!\!\!\includegraphics[scale=0.155,trim={1cm 2.5cm 2.5cm 2cm},clip]{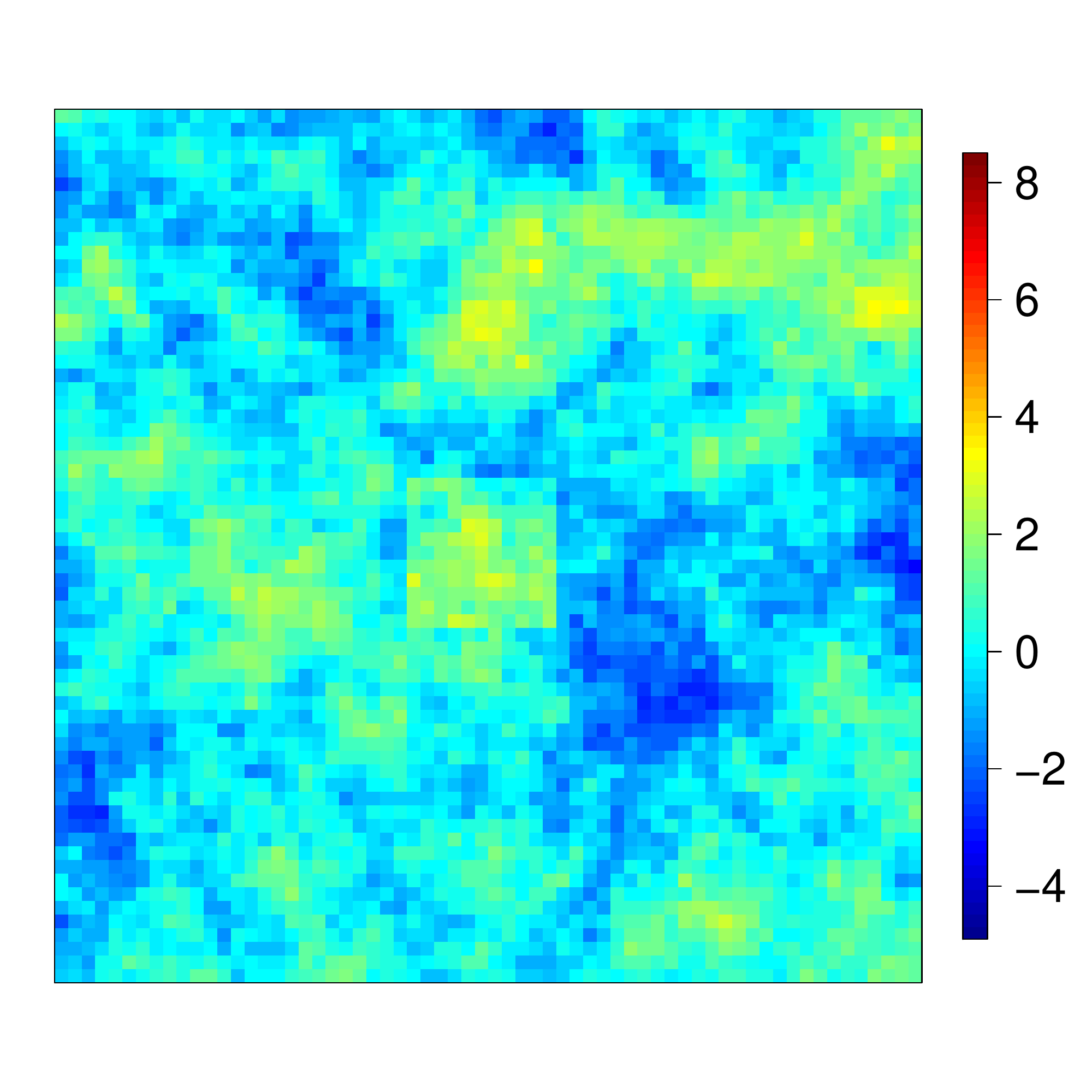} &
\!\!\!\!\!\!\includegraphics[scale=0.155,trim={1cm 2.5cm 2.5cm 2cm},clip]{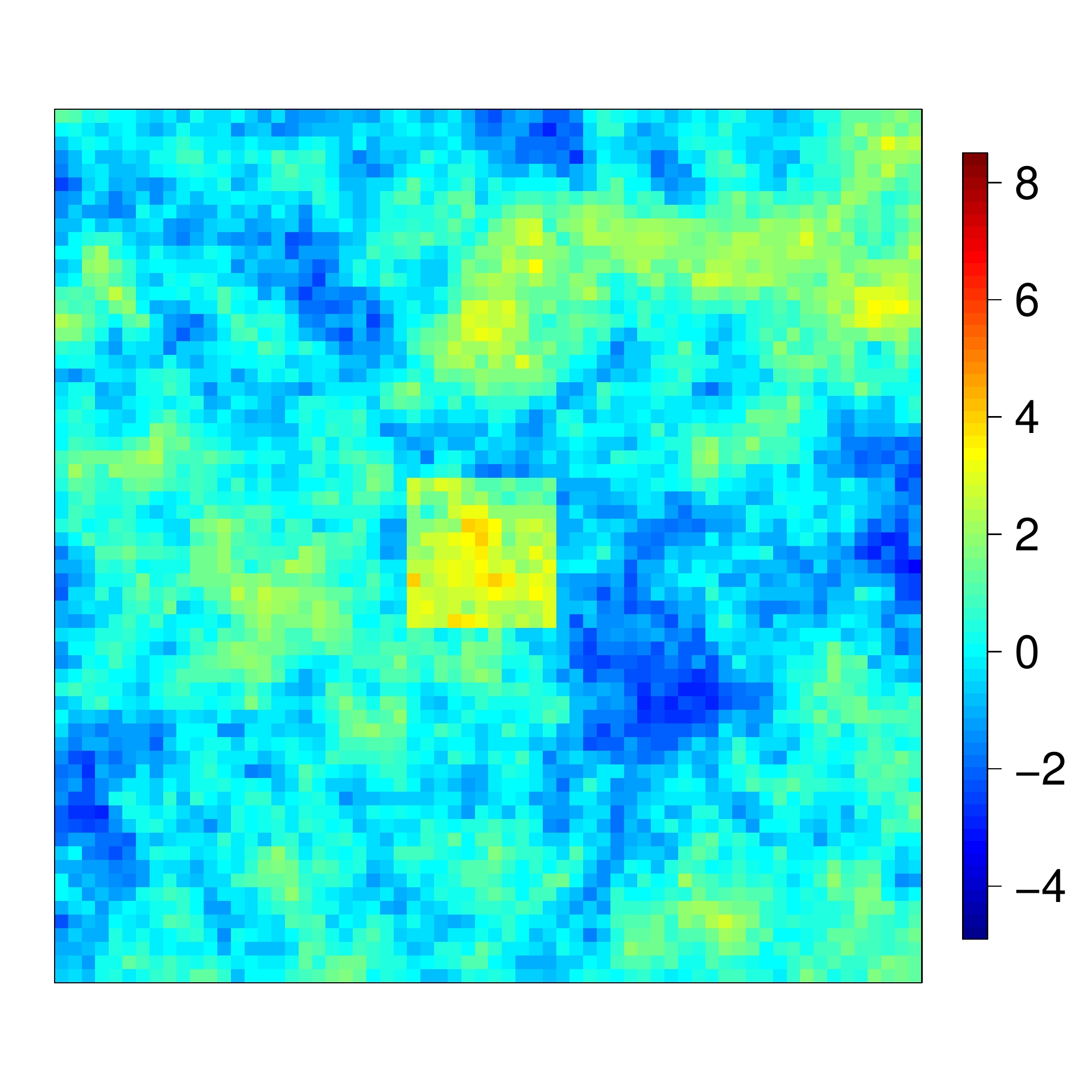} &
\!\!\!\!\!\!\includegraphics[scale=0.155,trim={1cm 2.5cm 2.5cm 2cm},clip]{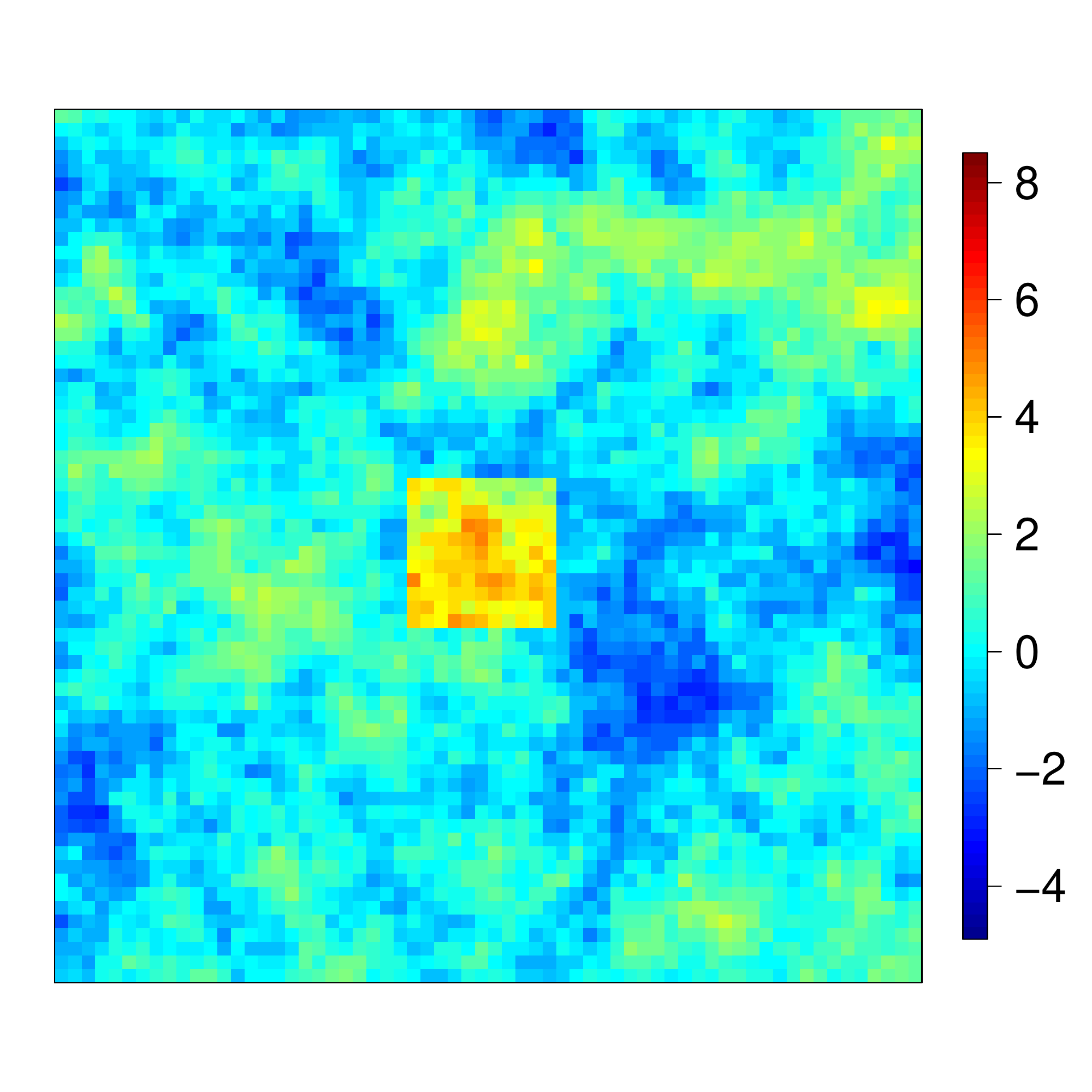} &
\!\!\!\!\!\!\includegraphics[scale=0.155,trim={1cm 2.5cm 2.5cm 2cm},clip]{./signal-r10-h5} 
\end{tabular}
\caption{Images of $\bm{Z}$ with $\phi=5$ and signals from \eqref{eq:signal} of various extents $r\in\{4,6,8,10\}$ down rows
and various magnitudes $h\in\{0,1,2,3,4,5\}$ across columns.}
\label{fig:signalonimageraw-all}
\end{figure}

This section contains the following nine figures:

\begin{itemize}
\item[S1] Images of $\bm{Z}$ with $\phi=5$ and signals from \eqref{eq:signal} of various extents $r\in\{4,6,8,10\}$ down rows and various magnitudes $h\in\{0,1,2,3,4,5\}$ across columns.
\item[S2] Images of $\bm{Z}_{16\times 16}$ obtained by aggregating $\bm{Z}$ in Figure~\ref{fig:signalonimageraw-all}
  into $4\times 4$ blocks resulting in $16\times 16$ grid cells.
\item[S3] Images of $\bm{Z}_{8\times 8}$ obtained by aggregating $\bm{Z}$ in Figure~\ref{fig:signalonimageraw-all}
  into $8\times 8$ blocks resulting in $8\times 8$ grid cells.
\item[S4--S6] Empirical power curves as a function of the signal's magnitude $h$, for various procedures for testing of $H_0$ in Experiments 1--3 in Section \ref{sec:simulation}, respectively.
\item[S7--S9] Empirical ROC curves for IDL and the proposed procedure, CPL, in Experiments 1--3 in Section \ref{sec:simulation}, respectively.
\end{itemize}

\begin{figure}[!tbh]\centering
\begin{tabular}{cccccc}
~~~~~$h=0$ & $h=1$~~ & $h=2$~~ & $h=3$~~ & $h=4$~~ & $h=5$~~
\smallskip\\
\rotatebox{90}{$\quad \quad r=4$}~~\includegraphics[scale=0.155,trim={1cm 2.5cm 2.5cm 2cm},clip]{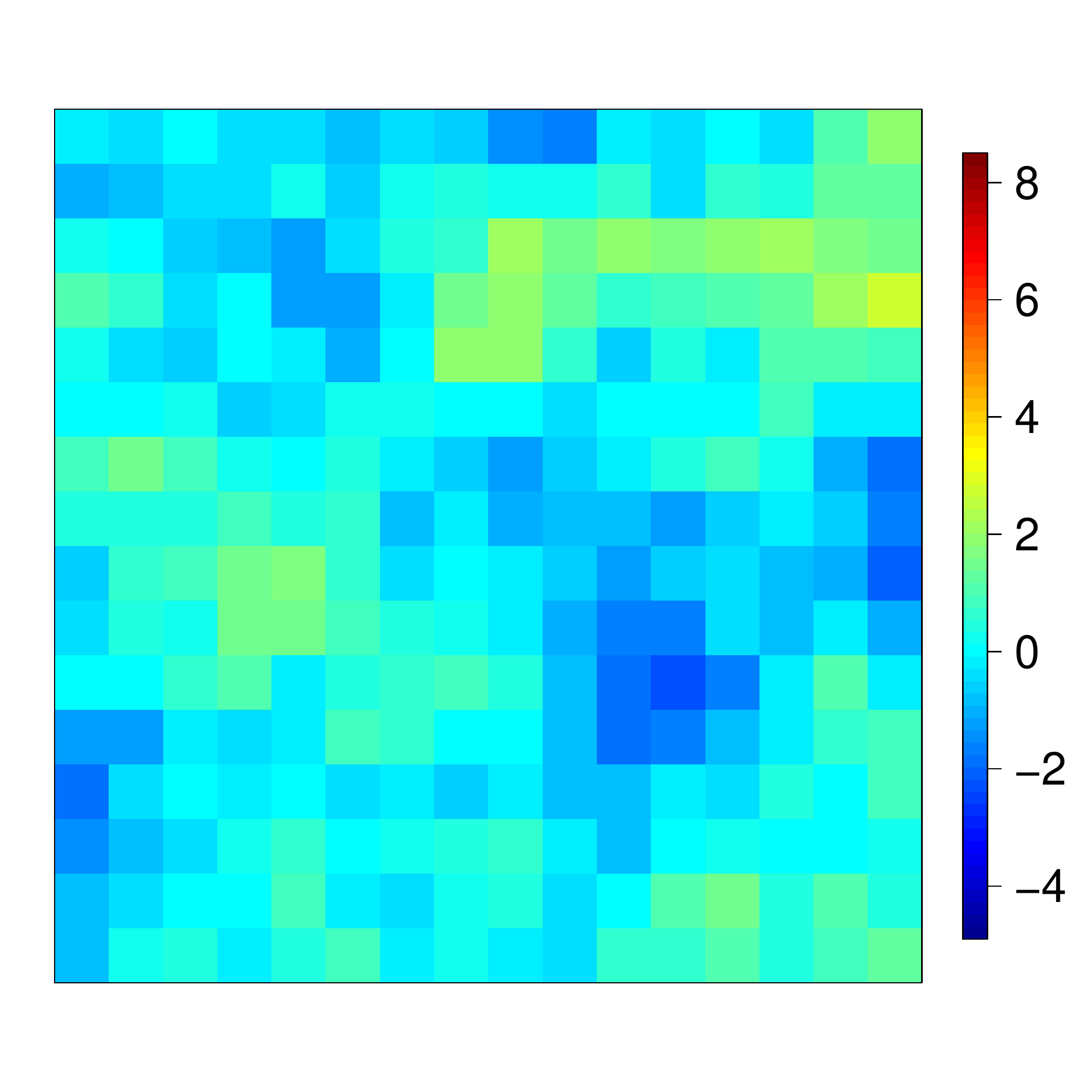} &
\!\!\!\!\!\!\includegraphics[scale=0.155,trim={1cm 2.5cm 2.5cm 2cm},clip]{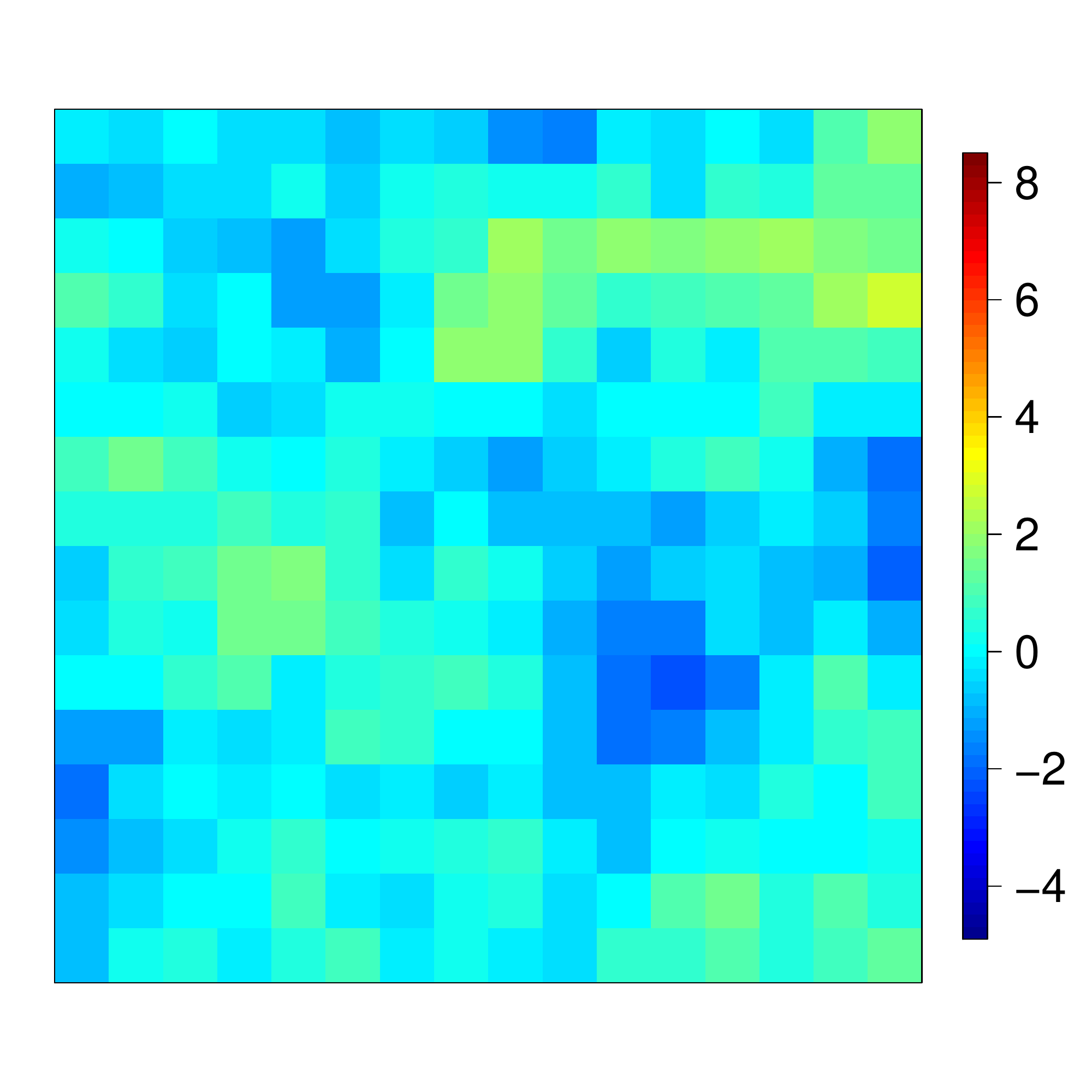} &
\!\!\!\!\!\!\includegraphics[scale=0.155,trim={1cm 2.5cm 2.5cm 2cm},clip]{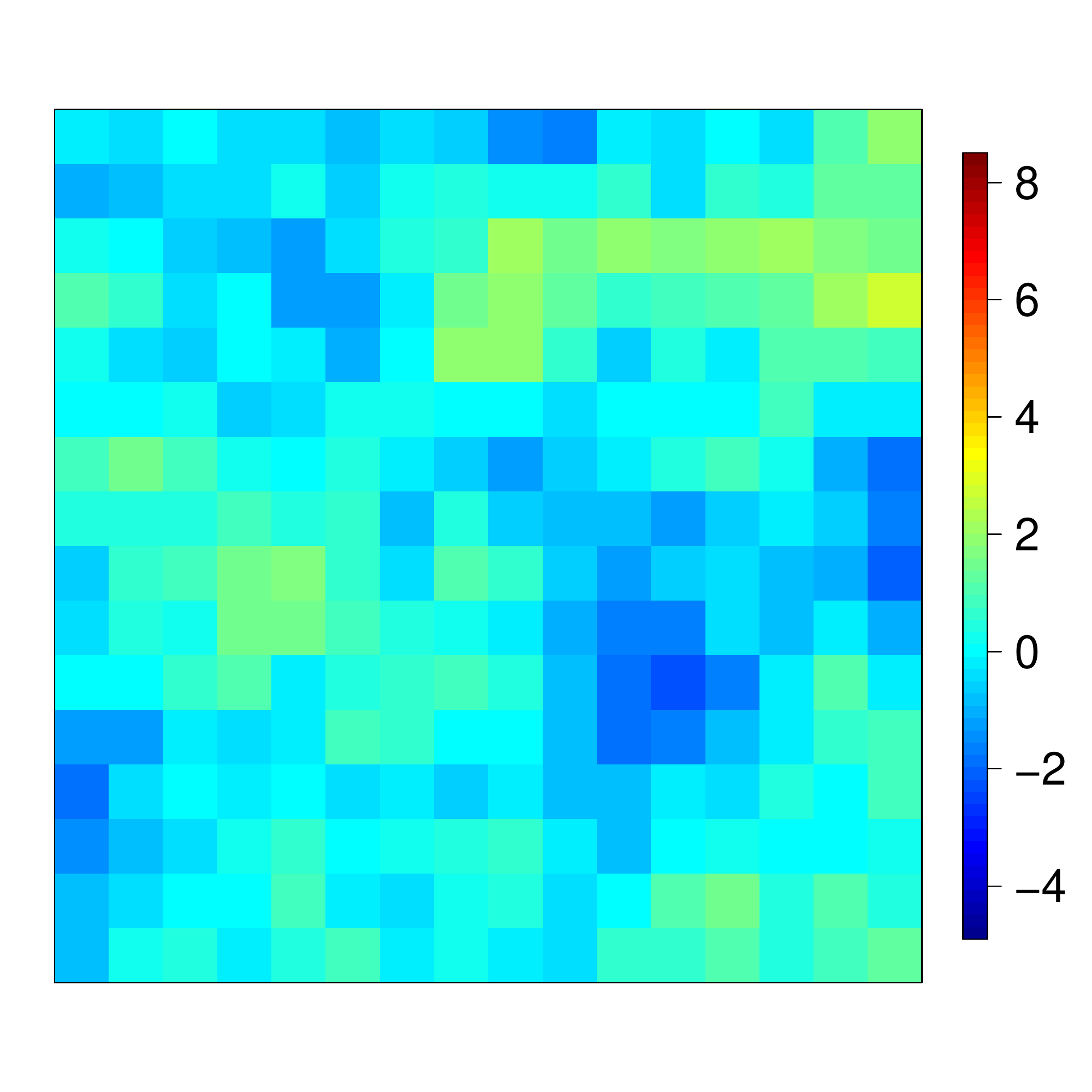} &
\!\!\!\!\!\!\includegraphics[scale=0.155,trim={1cm 2.5cm 2.5cm 2cm},clip]{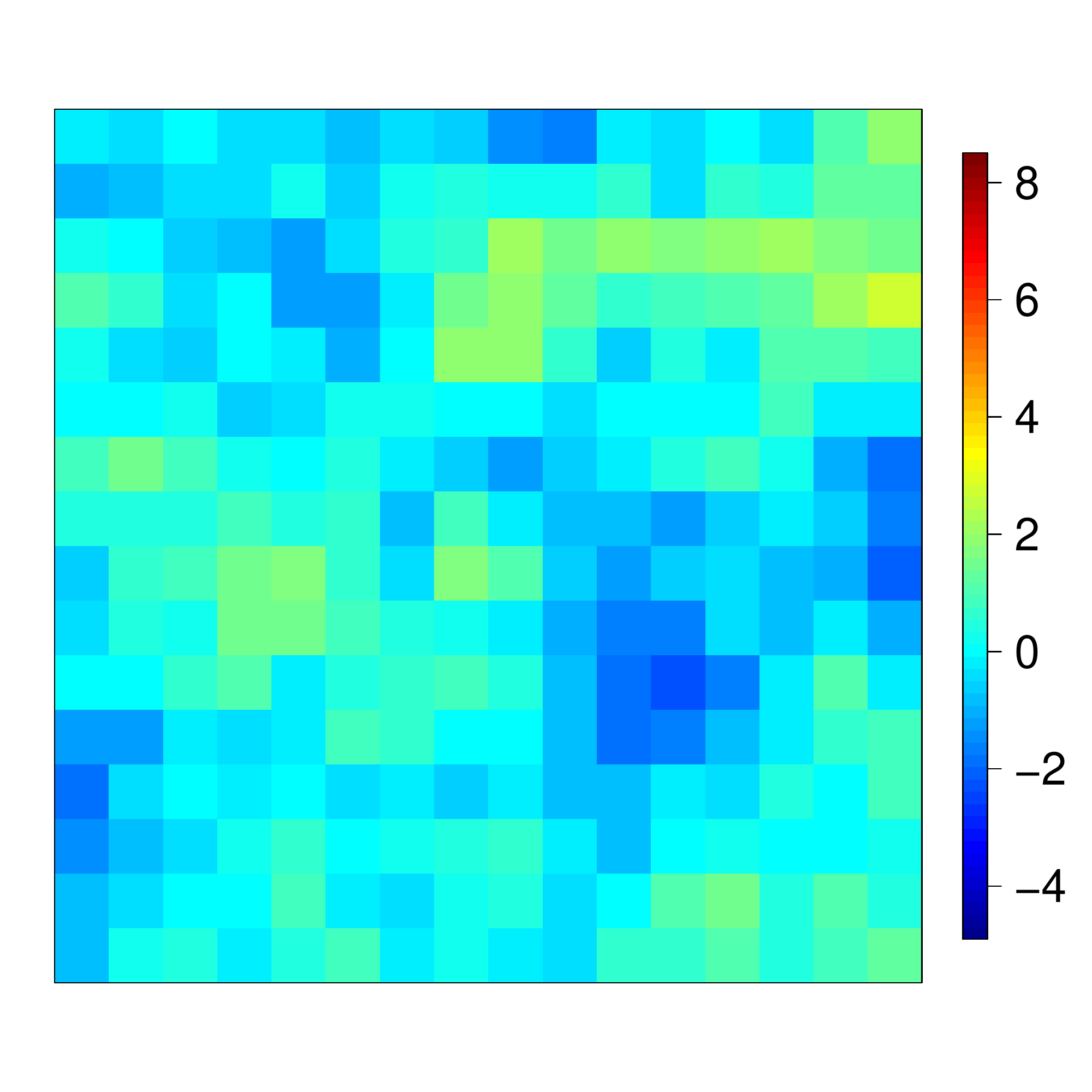} &
\!\!\!\!\!\!\includegraphics[scale=0.155,trim={1cm 2.5cm 2.5cm 2cm},clip]{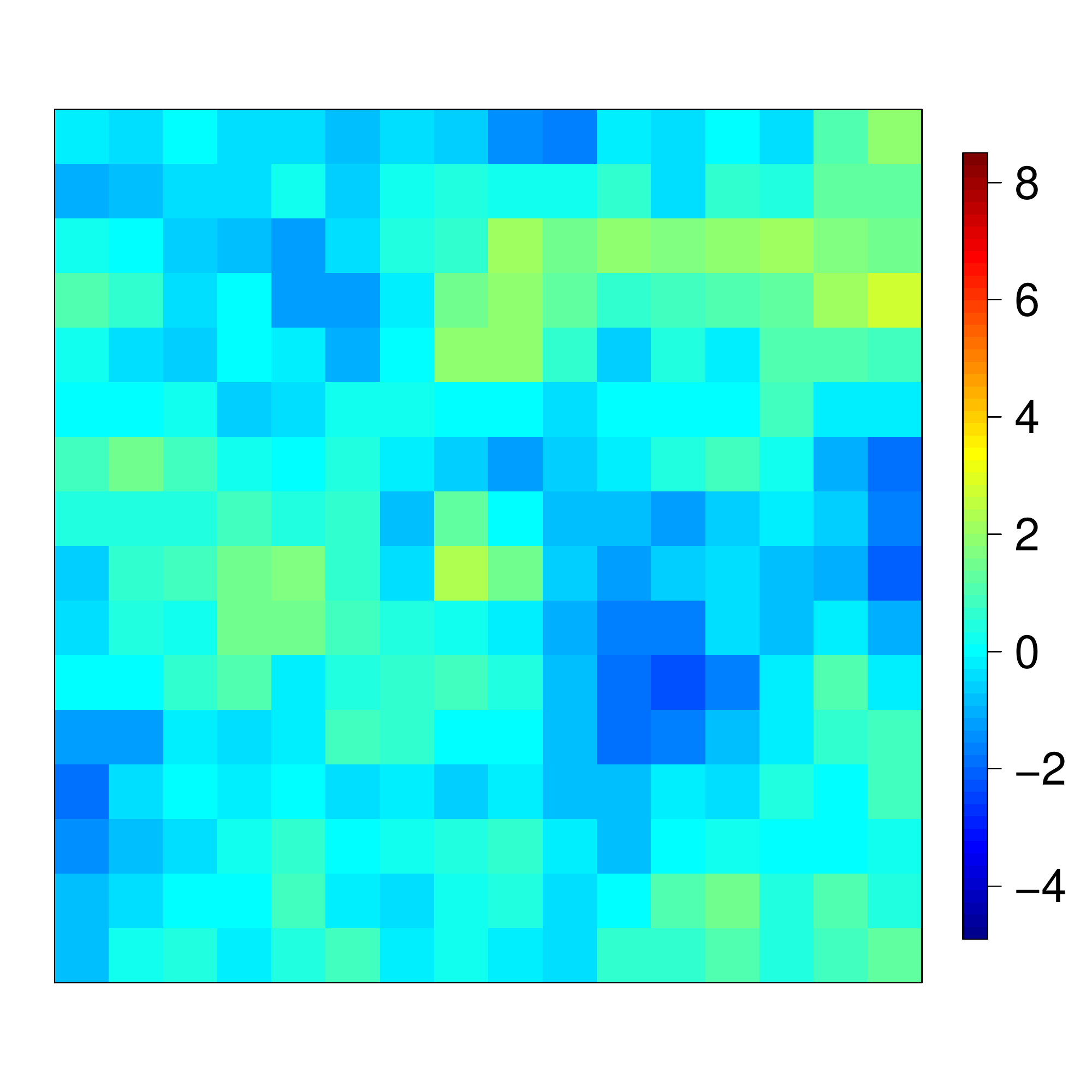} &
\!\!\!\!\!\!\includegraphics[scale=0.155,trim={1cm 2.5cm 2.5cm 2cm},clip]{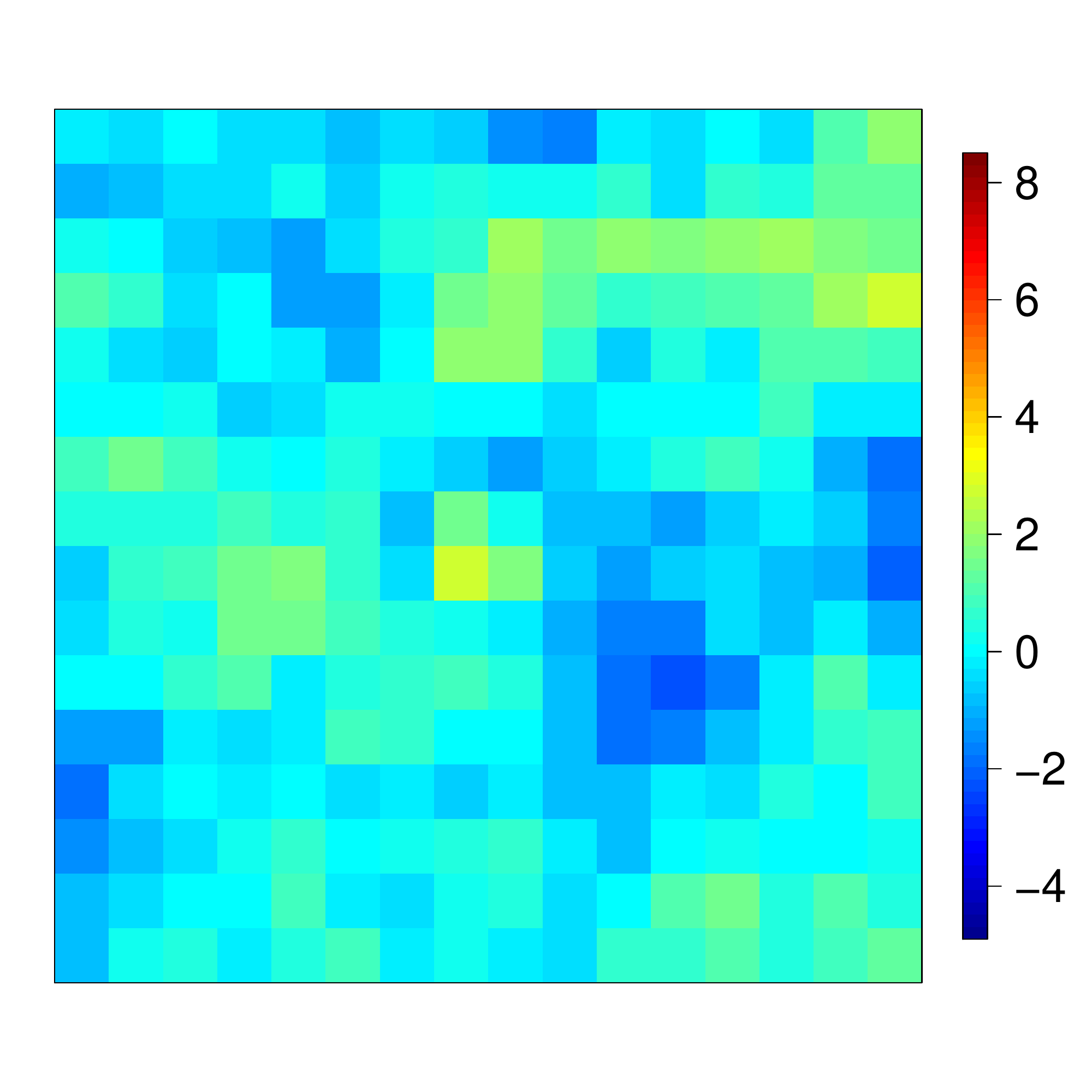} \\
\rotatebox{90}{$\quad \quad r=6$}~~\includegraphics[scale=0.155,trim={1cm 2.5cm 2.5cm 2cm},clip]{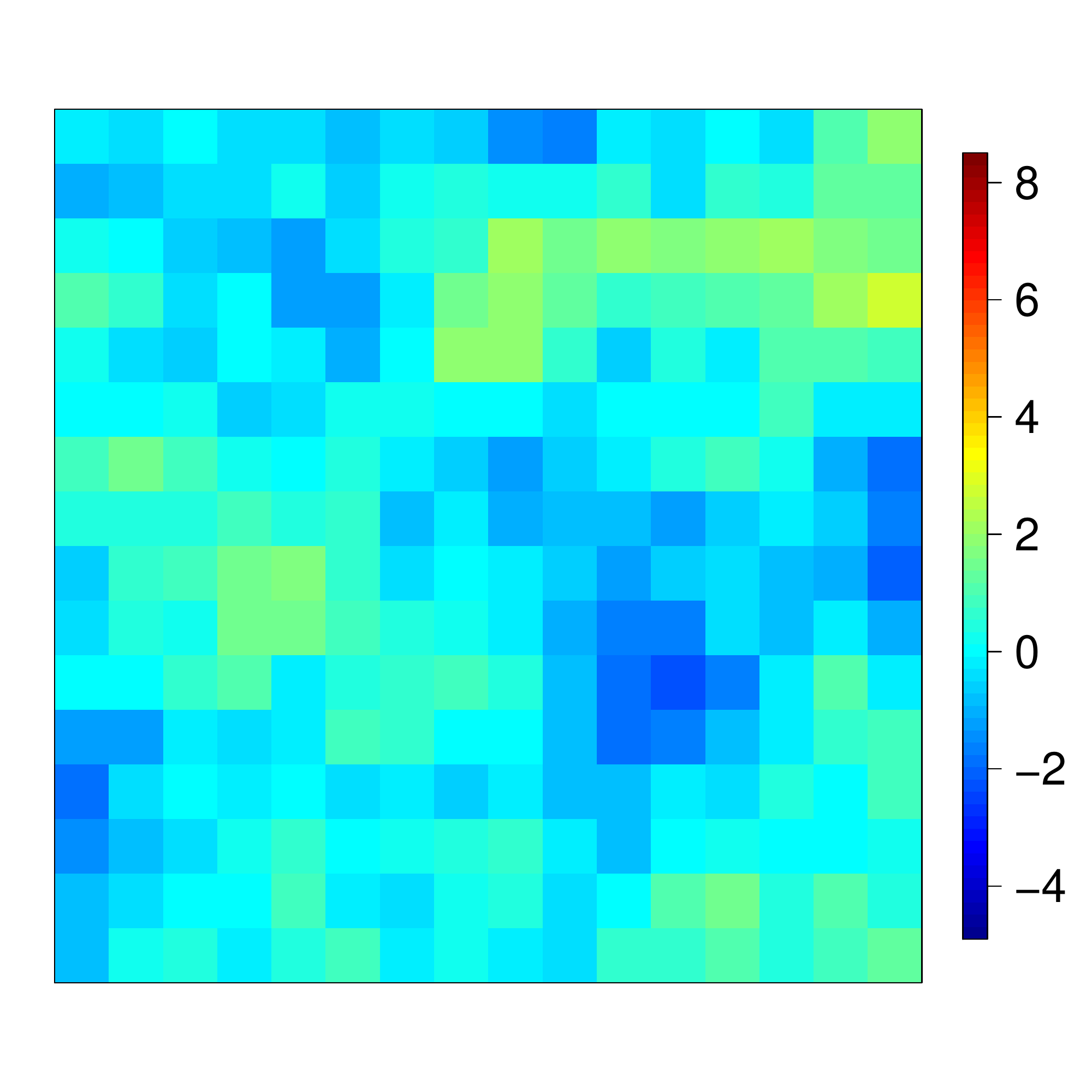} &
\!\!\!\!\!\!\includegraphics[scale=0.155,trim={1cm 2.5cm 2.5cm 2cm},clip]{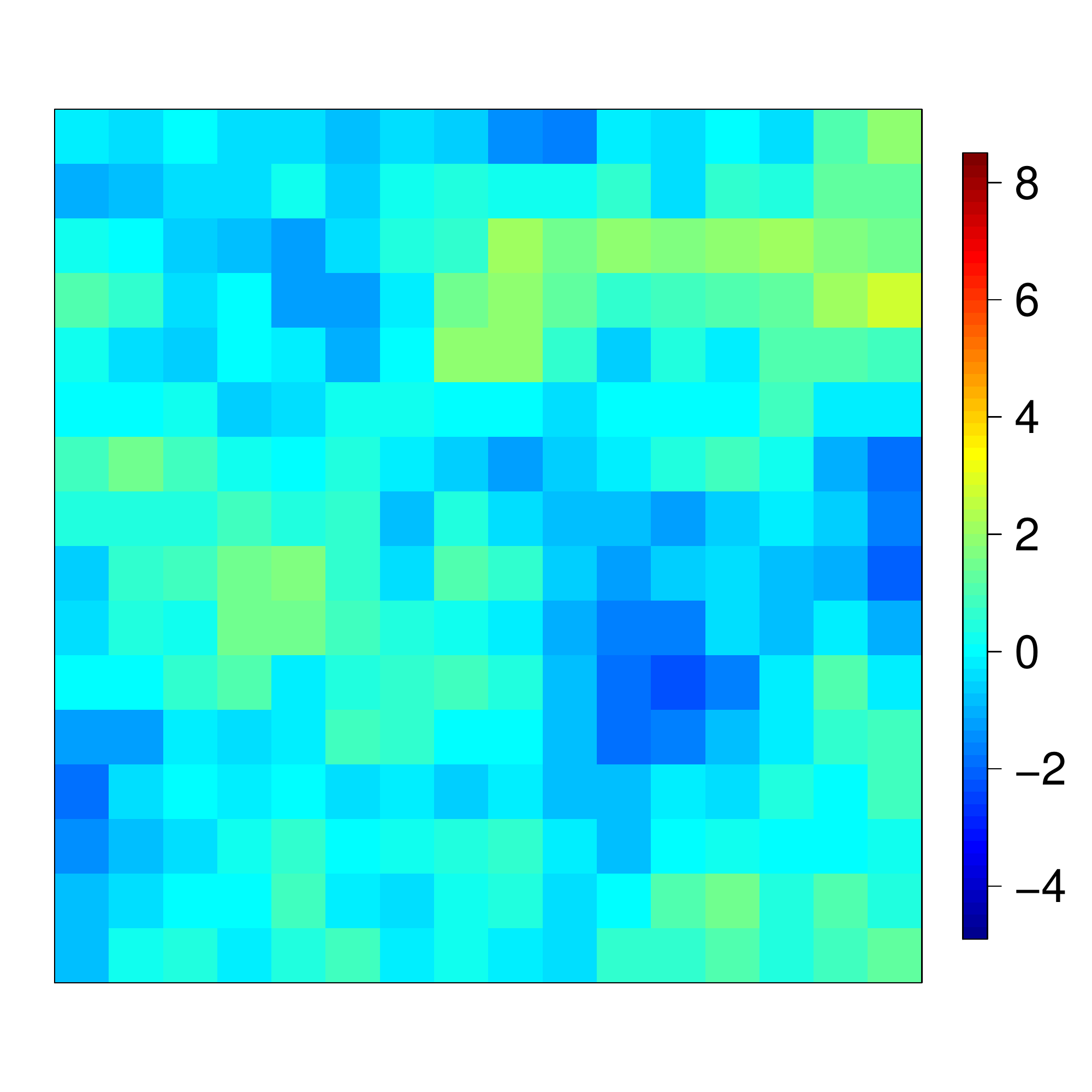} &
\!\!\!\!\!\!\includegraphics[scale=0.155,trim={1cm 2.5cm 2.5cm 2cm},clip]{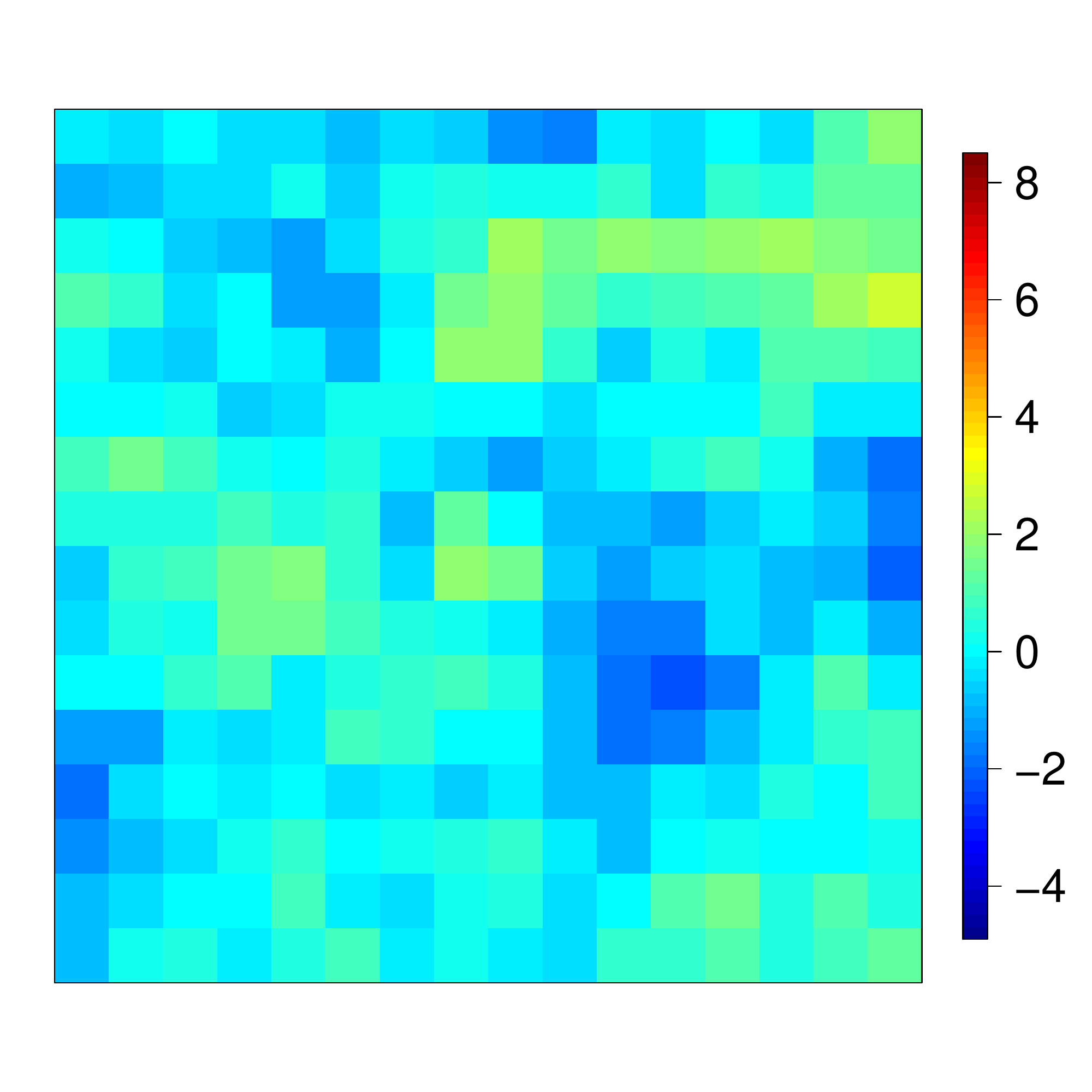} &
\!\!\!\!\!\!\includegraphics[scale=0.155,trim={1cm 2.5cm 2.5cm 2cm},clip]{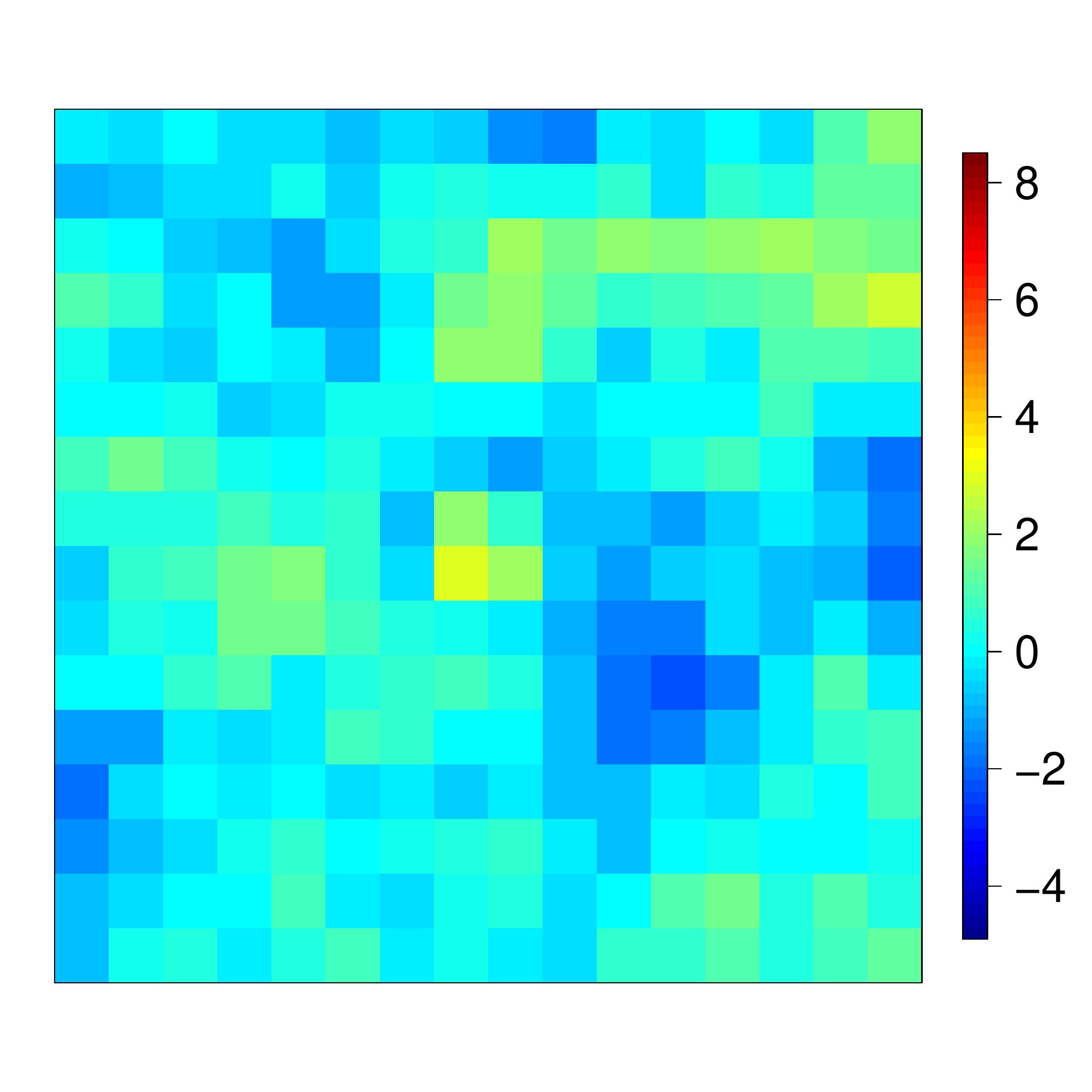} &
\!\!\!\!\!\!\includegraphics[scale=0.155,trim={1cm 2.5cm 2.5cm 2cm},clip]{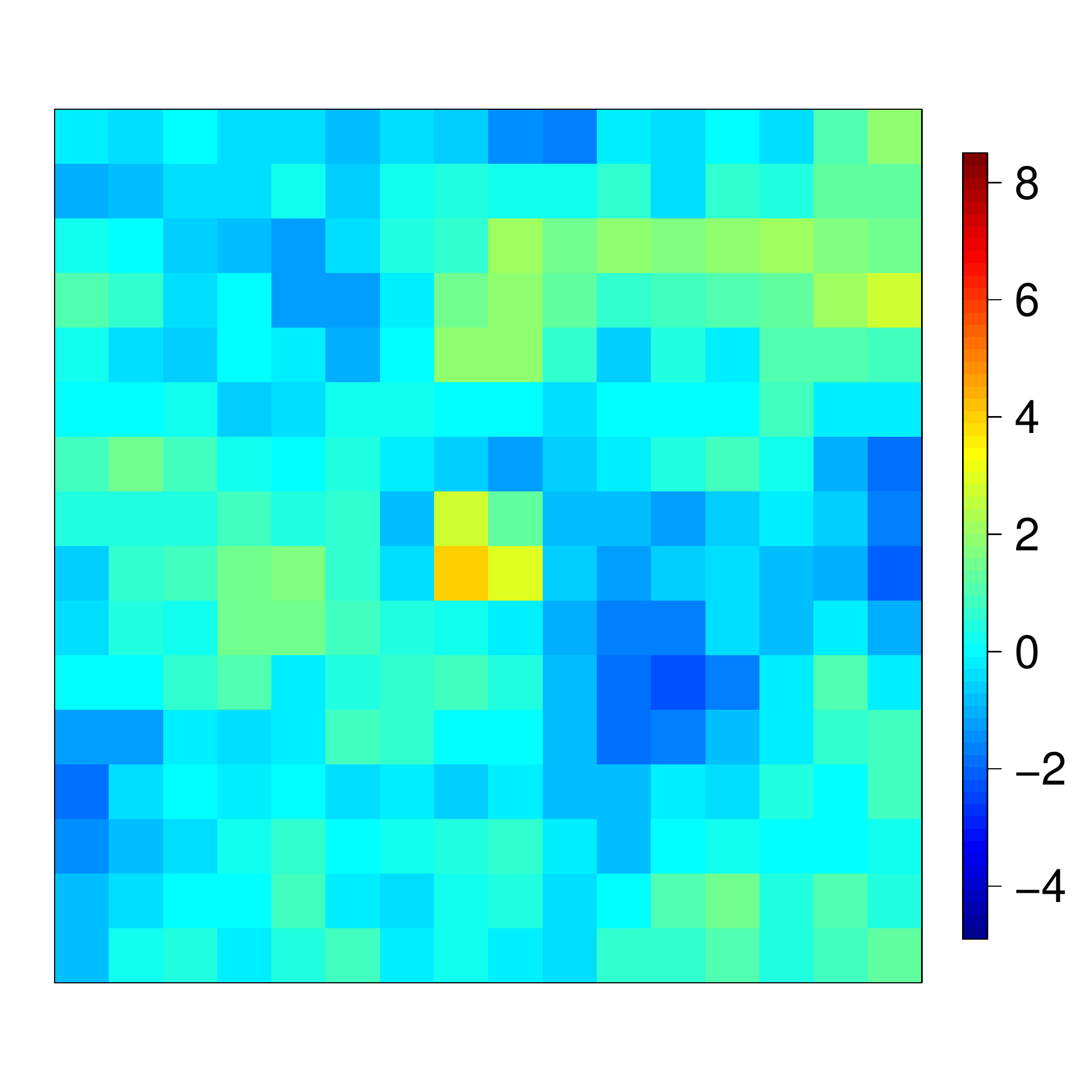} &
\!\!\!\!\!\!\includegraphics[scale=0.155,trim={1cm 2.5cm 2.5cm 2cm},clip]{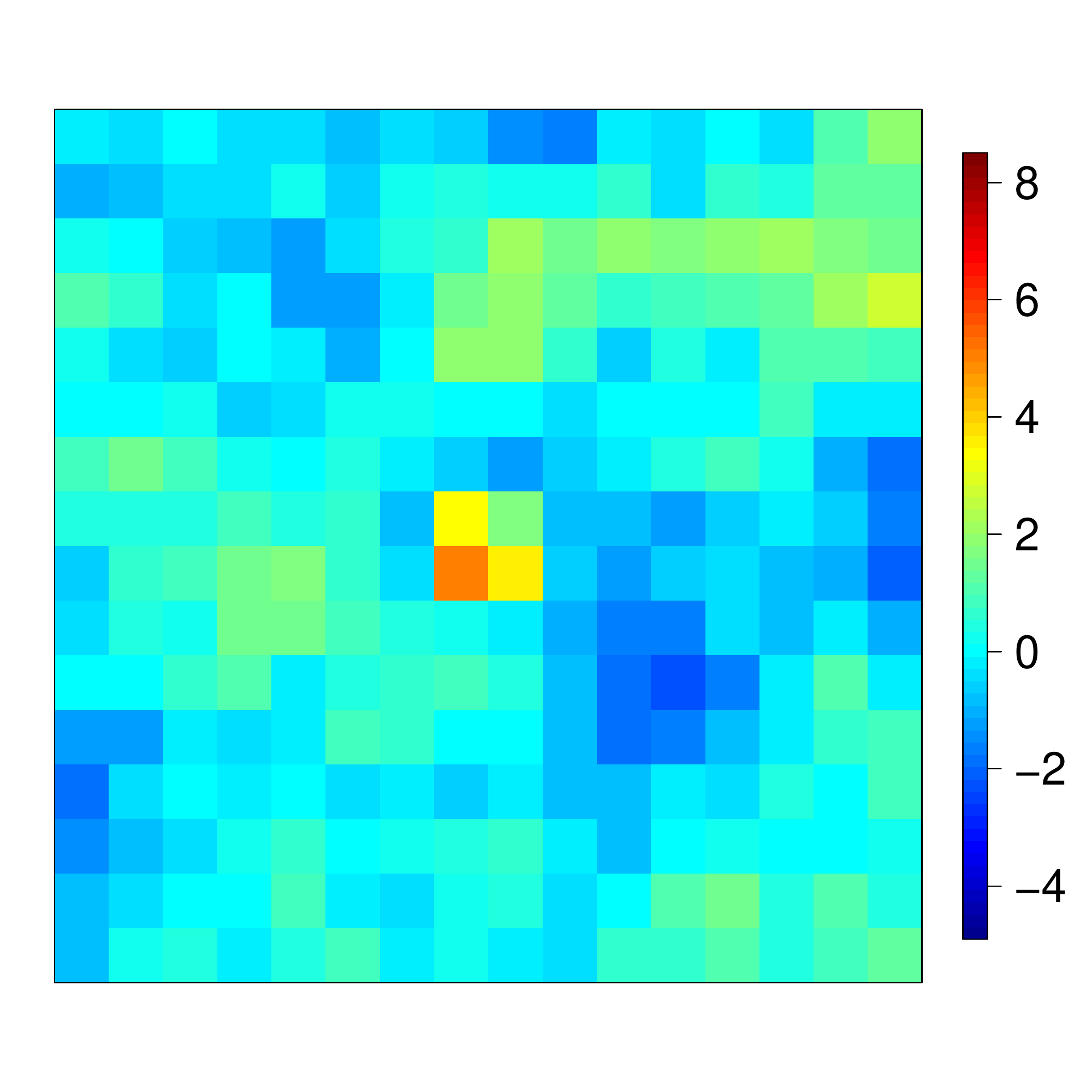} \\
\rotatebox{90}{$\quad \quad r=8$}~~\includegraphics[scale=0.155,trim={1cm 2.5cm 2.5cm 2cm},clip]{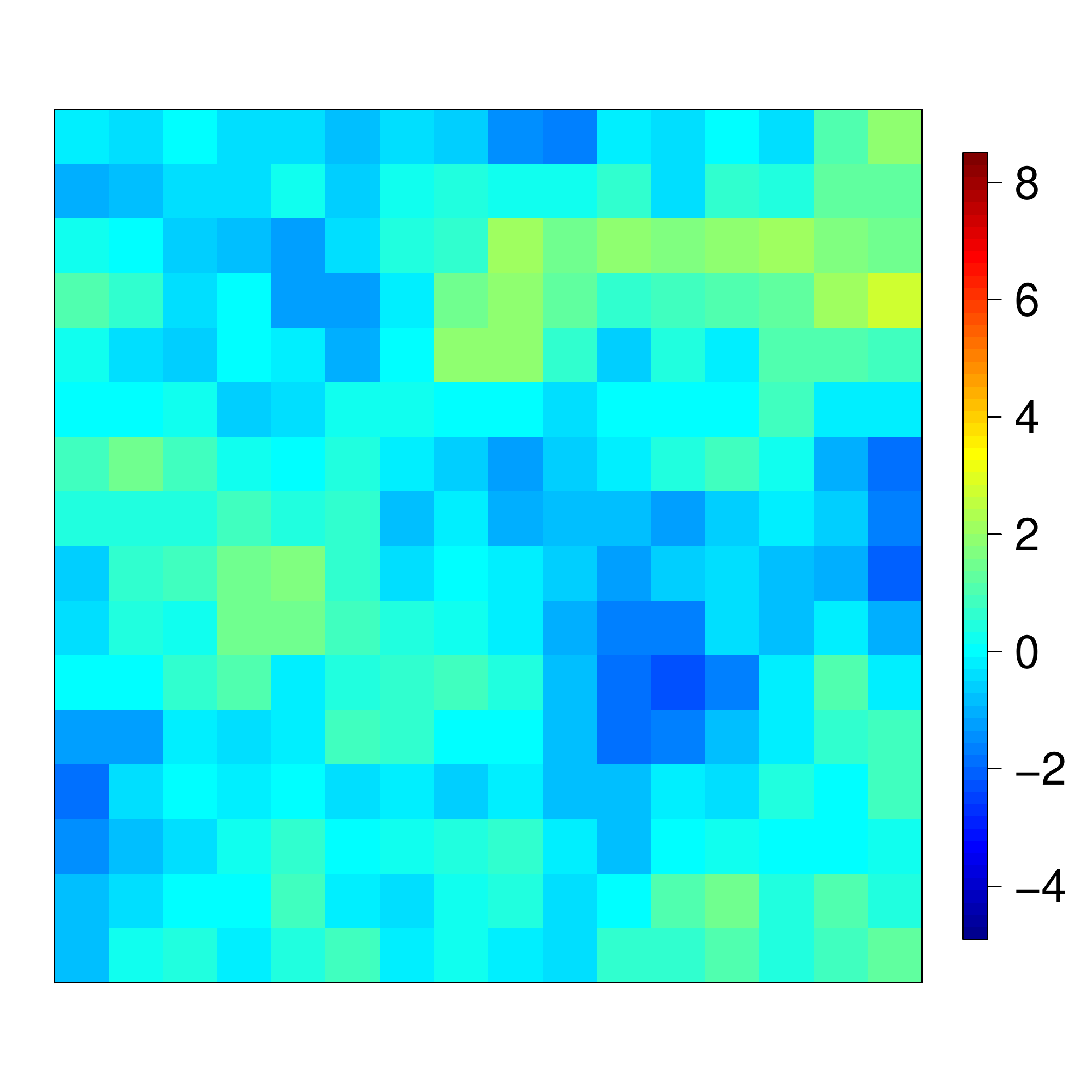} &
\!\!\!\!\!\!\includegraphics[scale=0.155,trim={1cm 2.5cm 2.5cm 2cm},clip]{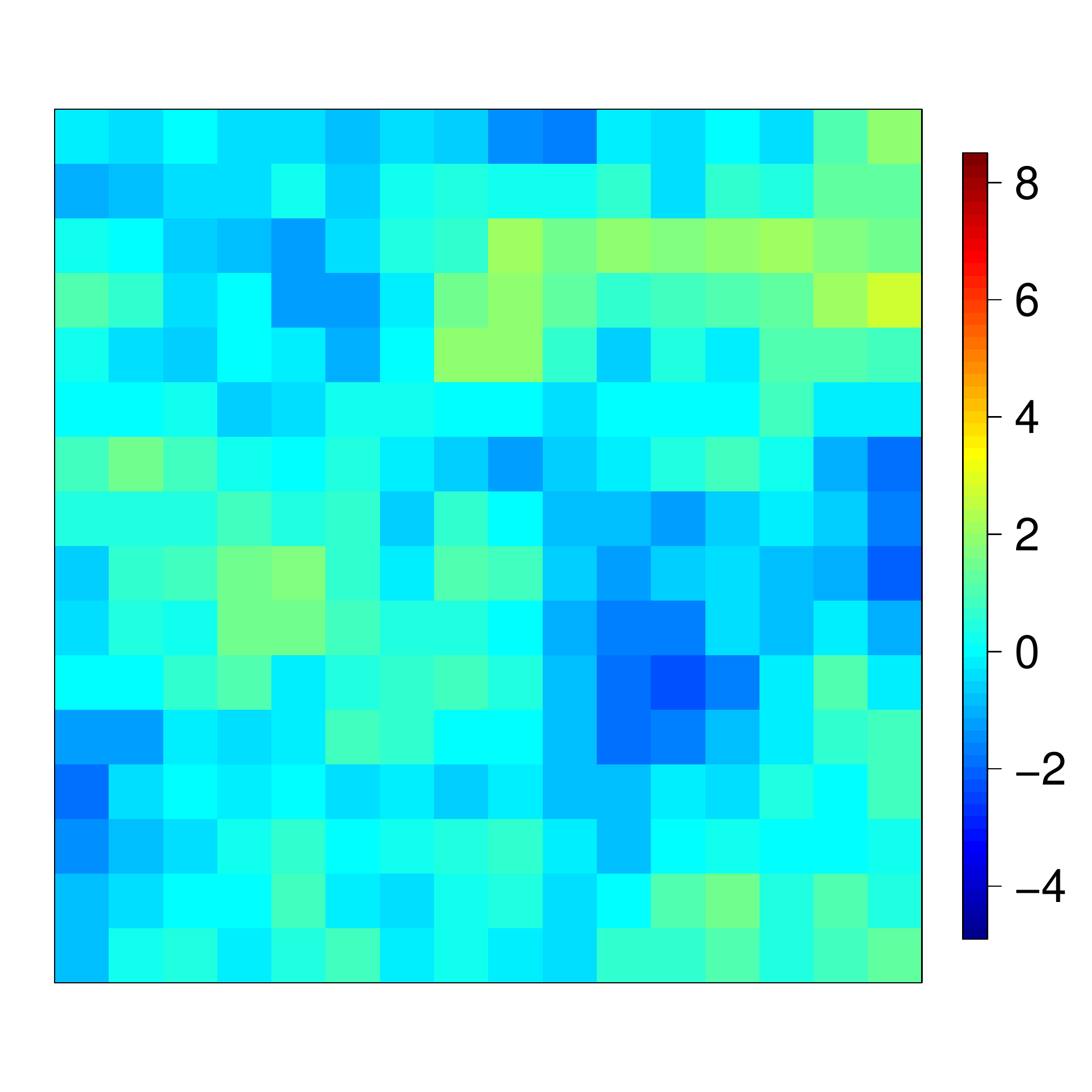} &
\!\!\!\!\!\!\includegraphics[scale=0.155,trim={1cm 2.5cm 2.5cm 2cm},clip]{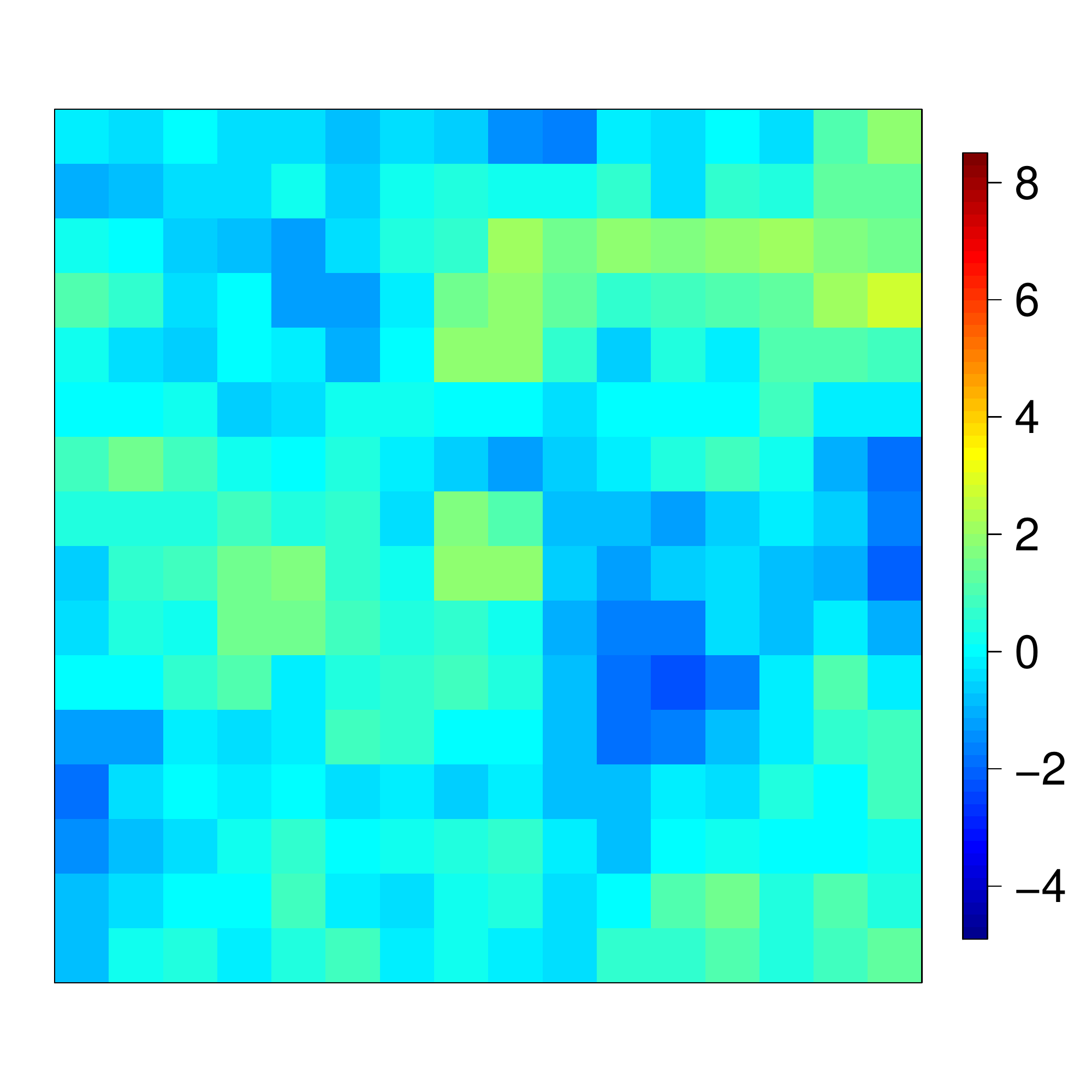} &
\!\!\!\!\!\!\includegraphics[scale=0.155,trim={1cm 2.5cm 2.5cm 2cm},clip]{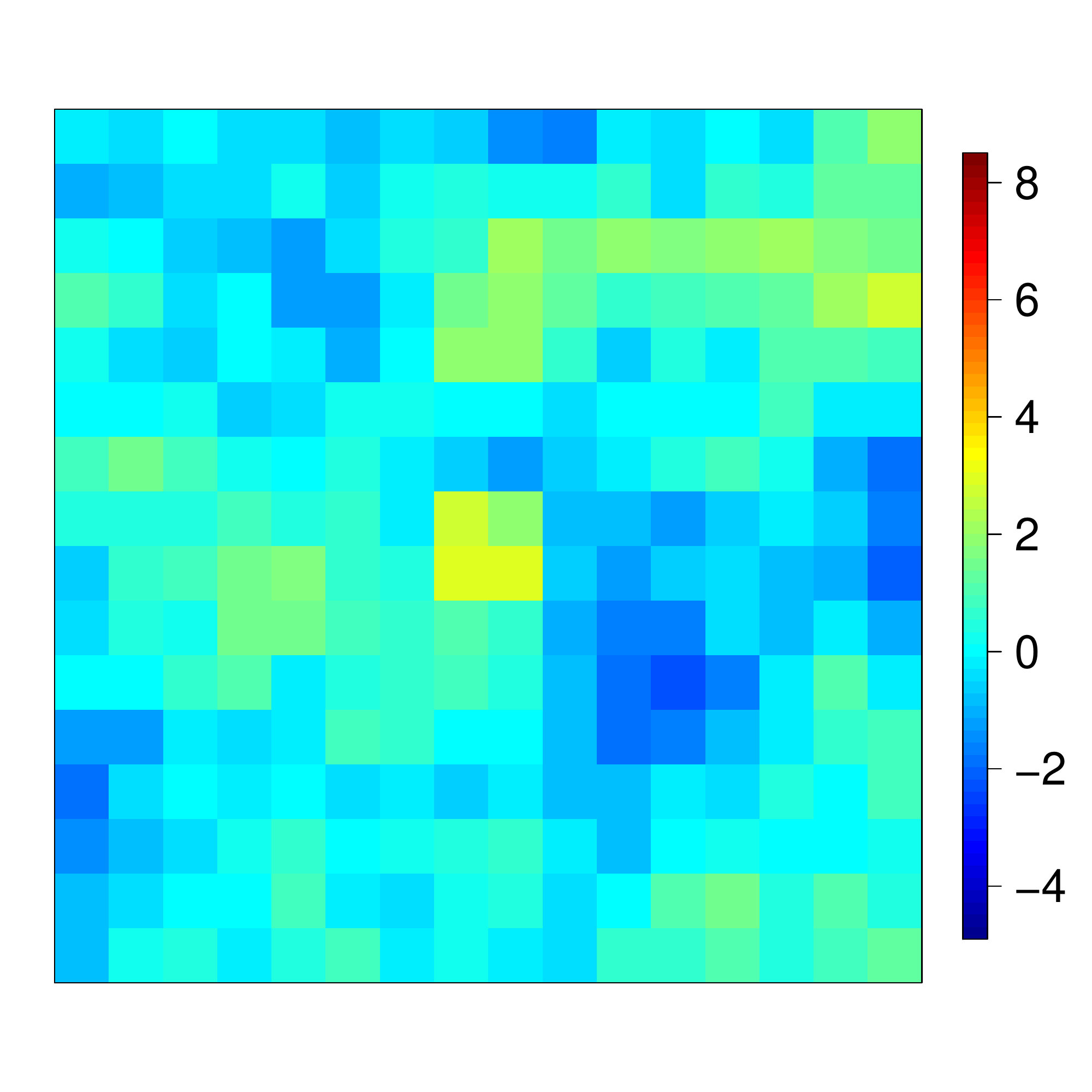} &
\!\!\!\!\!\!\includegraphics[scale=0.155,trim={1cm 2.5cm 2.5cm 2cm},clip]{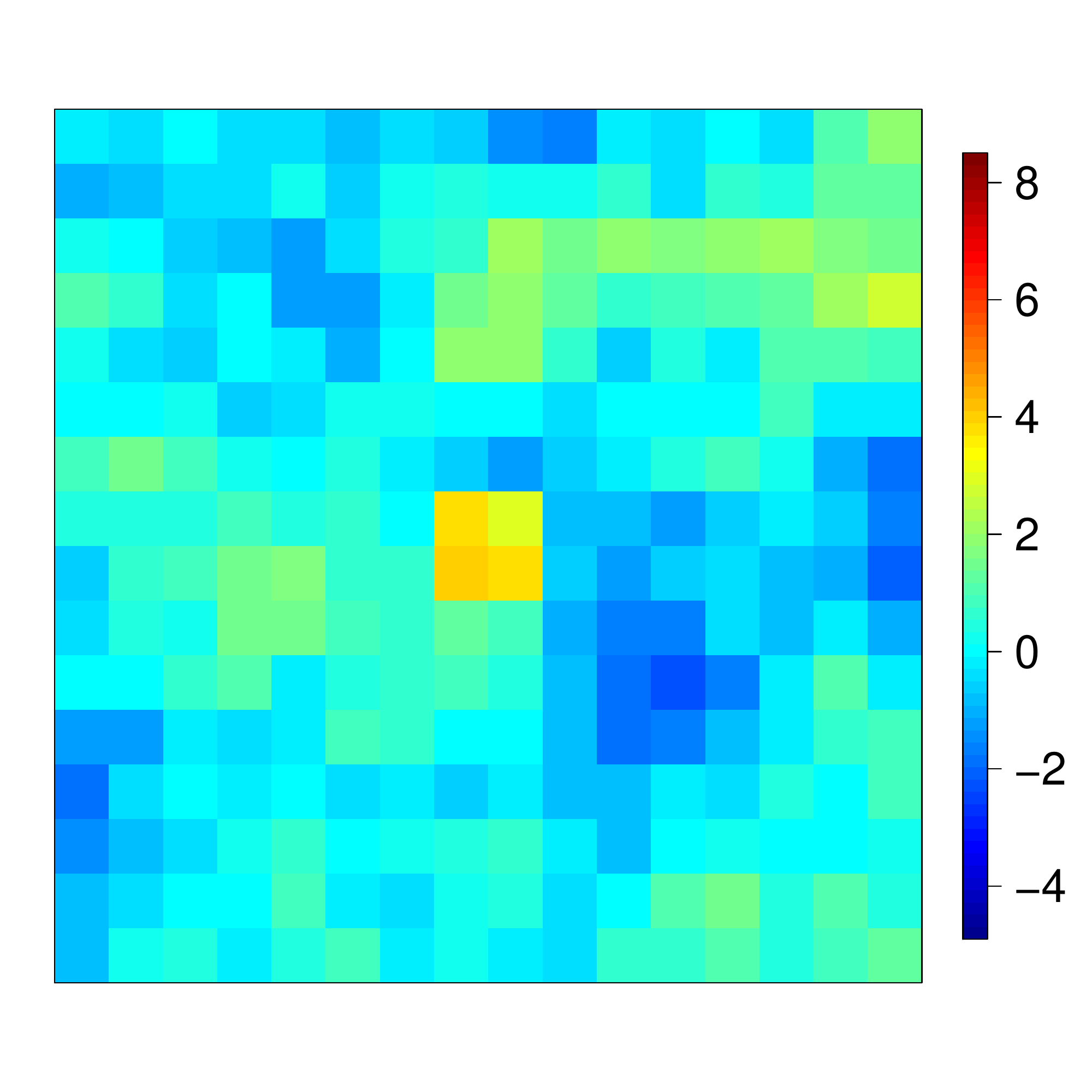} &
\!\!\!\!\!\!\includegraphics[scale=0.155,trim={1cm 2.5cm 2.5cm 2cm},clip]{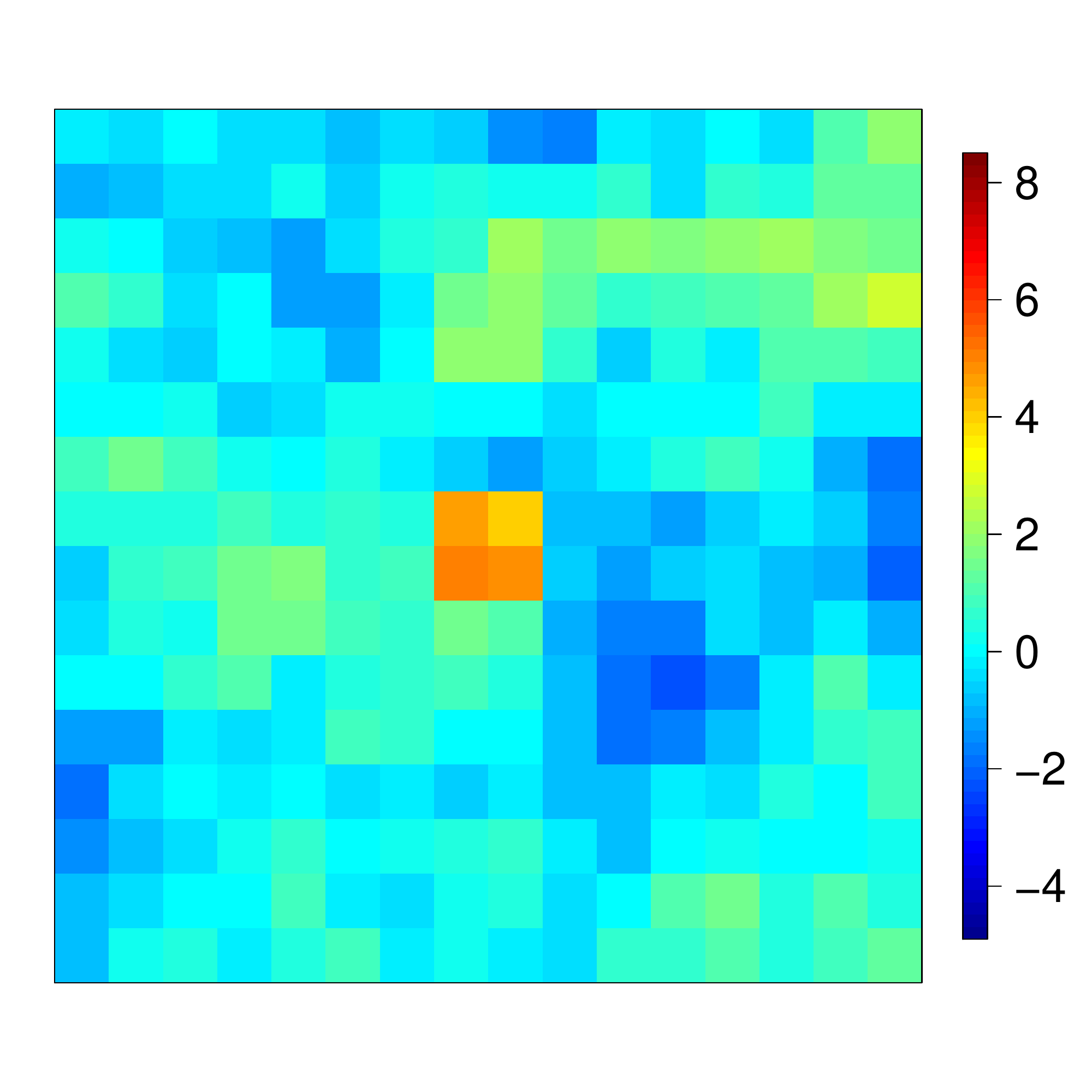} \\
\rotatebox{90}{$\quad \quad r=10$}~~\includegraphics[scale=0.155,trim={1cm 2.5cm 2.5cm 2cm},clip]{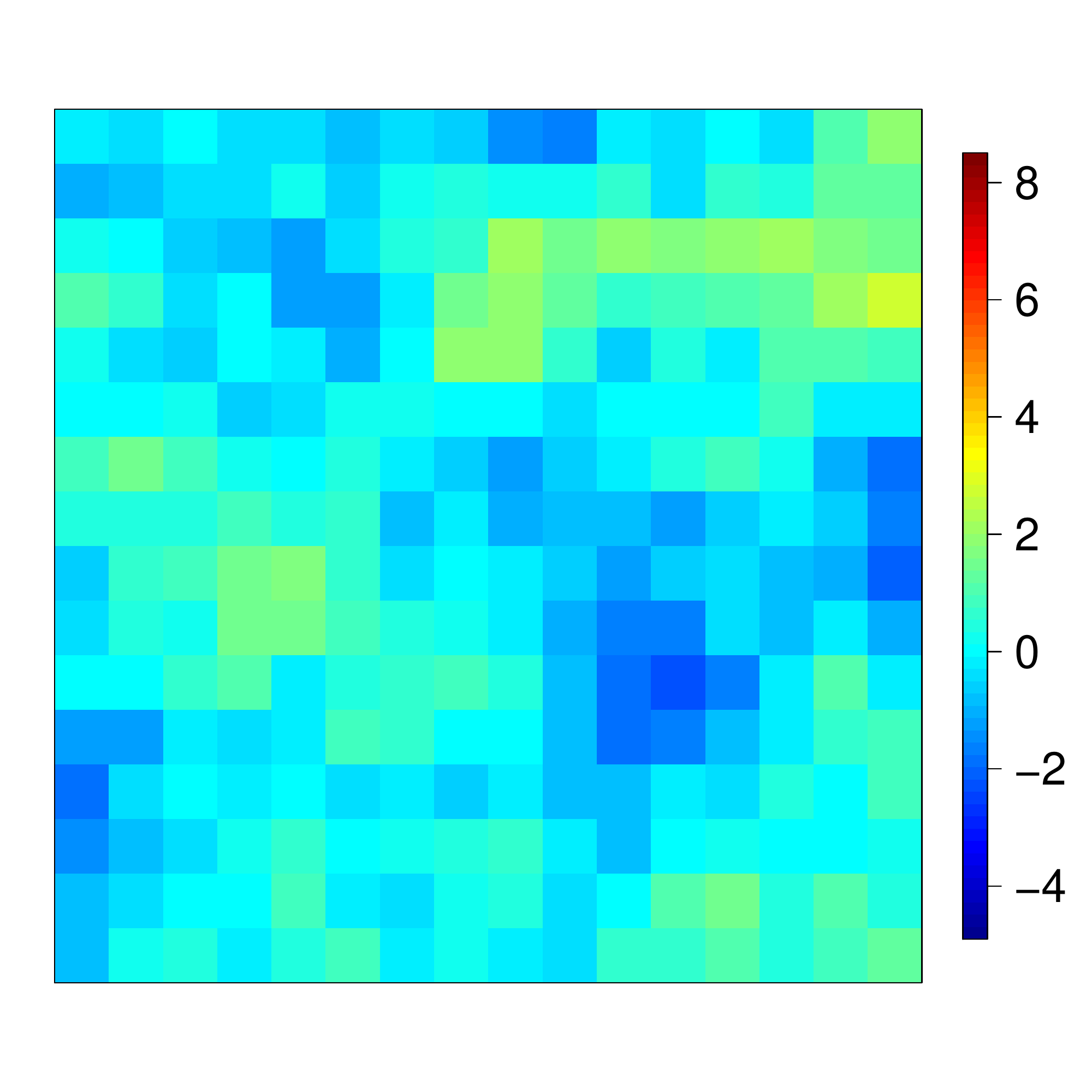} &
\!\!\!\!\!\!\includegraphics[scale=0.155,trim={1cm 2.5cm 2.5cm 2cm},clip]{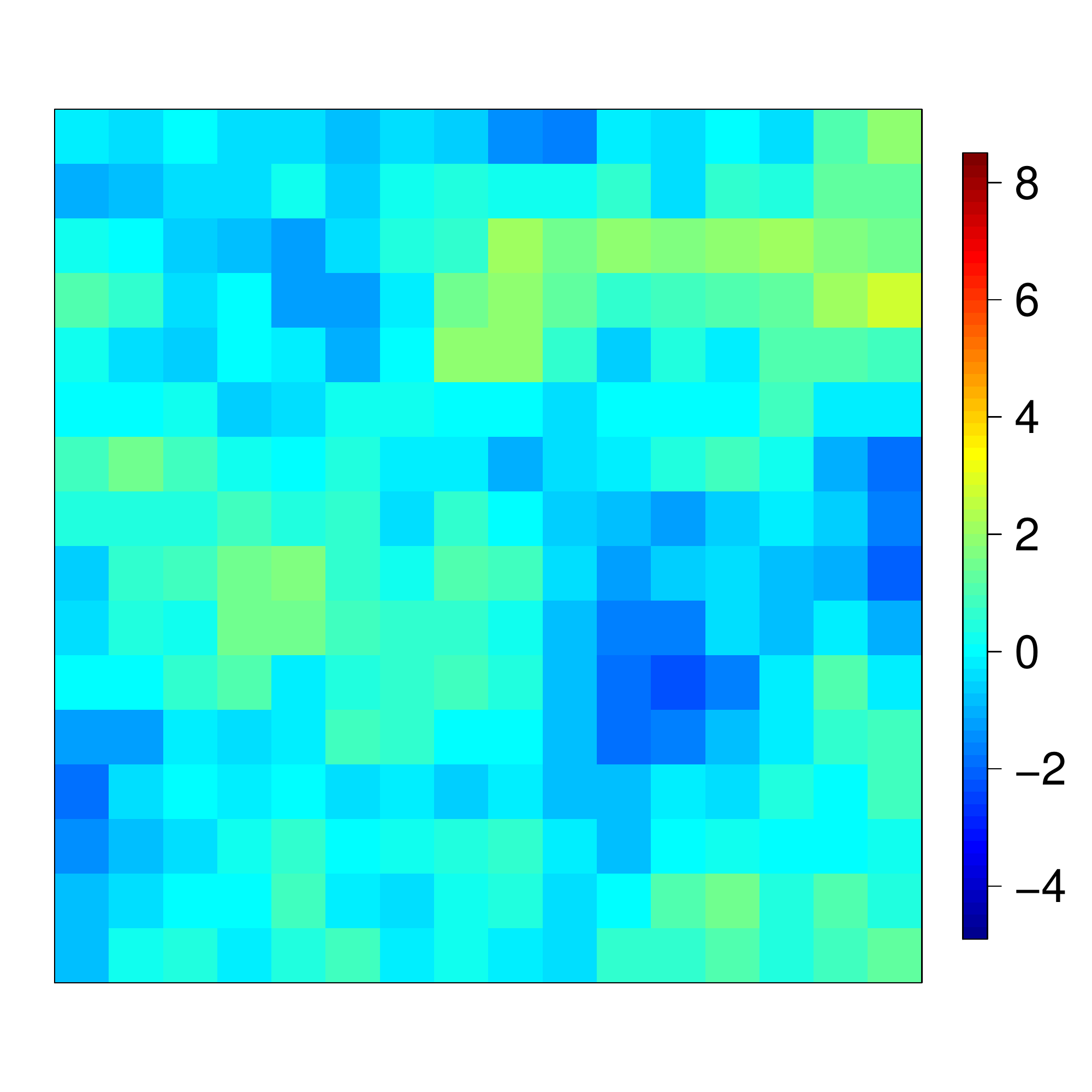} &
\!\!\!\!\!\!\includegraphics[scale=0.155,trim={1cm 2.5cm 2.5cm 2cm},clip]{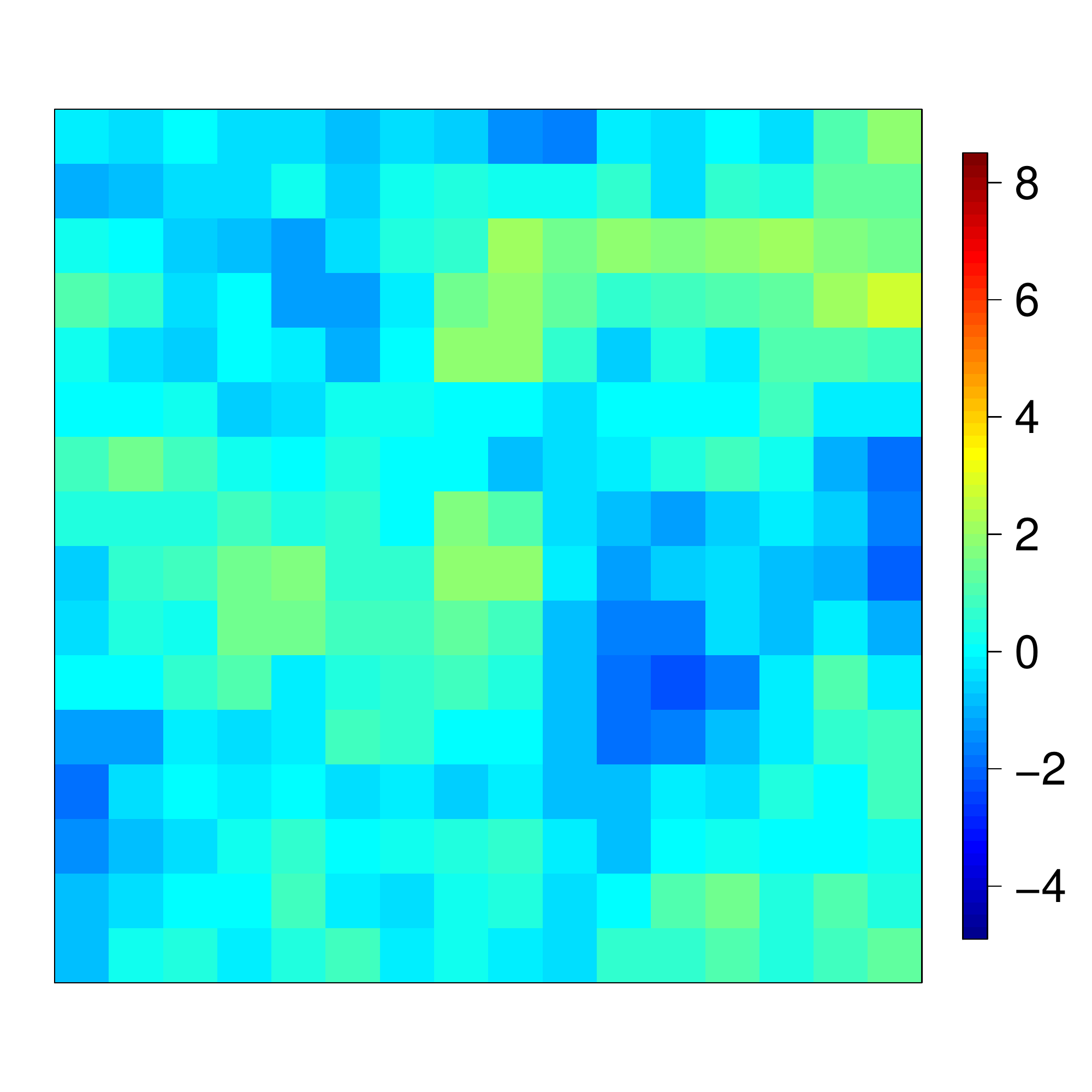} &
\!\!\!\!\!\!\includegraphics[scale=0.155,trim={1cm 2.5cm 2.5cm 2cm},clip]{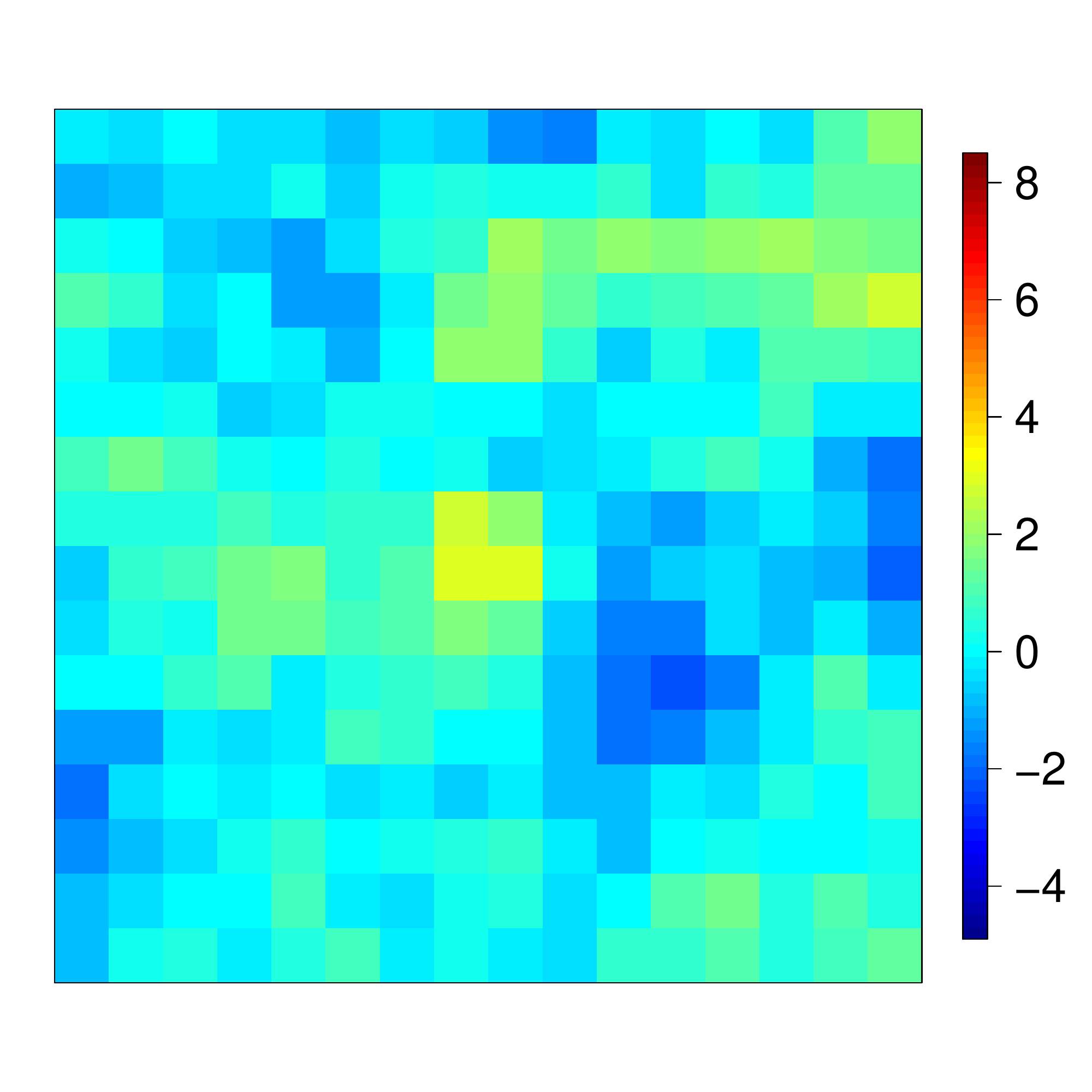} &
\!\!\!\!\!\!\includegraphics[scale=0.155,trim={1cm 2.5cm 2.5cm 2cm},clip]{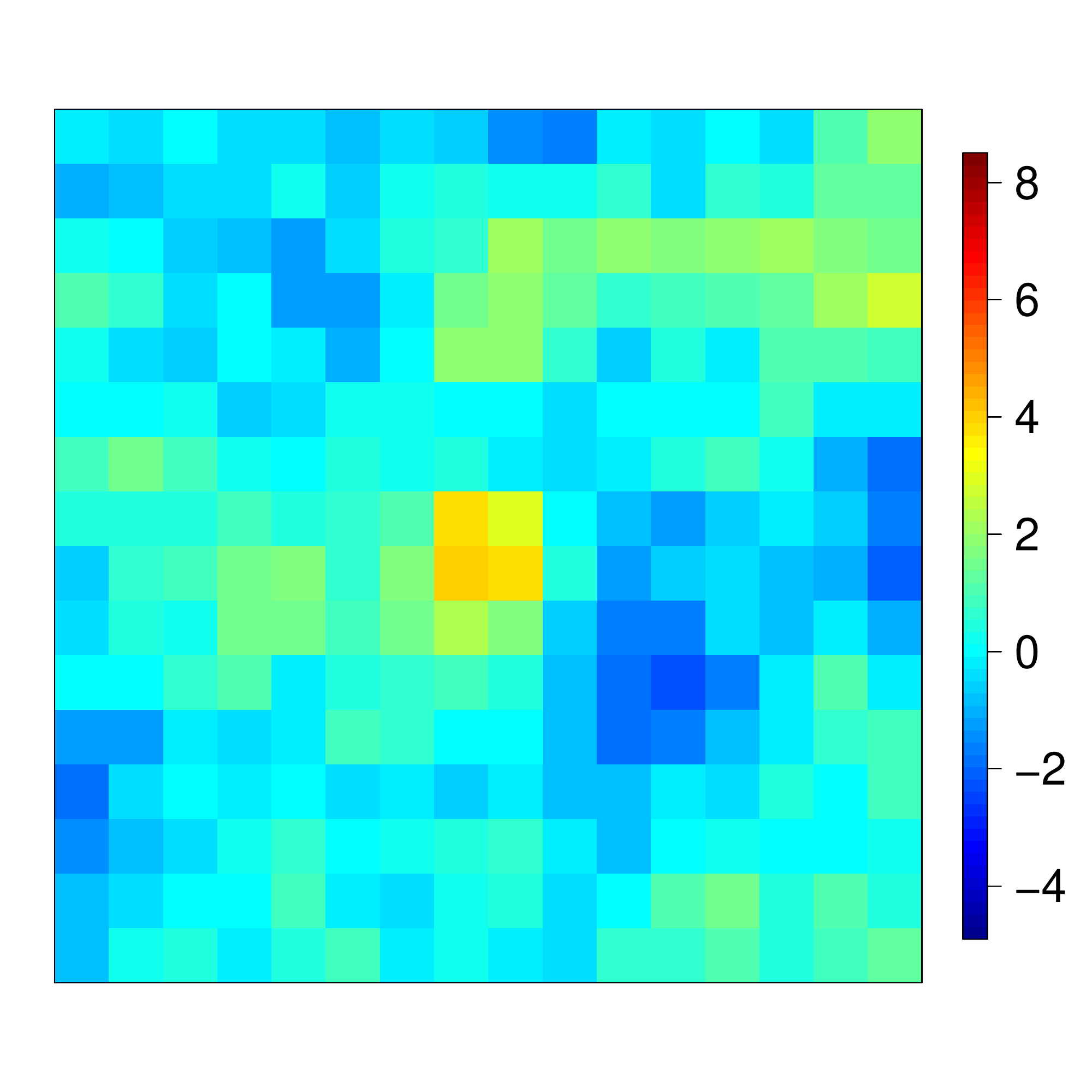} &
\!\!\!\!\!\!\includegraphics[scale=0.155,trim={1cm 2.5cm 2.5cm 2cm},clip]{./signal-r10-h5-grid16} 
\end{tabular}
\caption{Images of $\tilde{\bm{Z}}_{16 \times 16}$ obtained by aggregating $\bm{Z}$ in Figure~\ref{fig:signalonimageraw-all}
into $4\times 4$ blocks resulting in $16\times 16$ grid cells.}
\label{fig:signalonimageaggregated16by16}
\end{figure}

\begin{figure}[!tbh]\centering
\begin{tabular}{cccccc}
~~~~~$h=0$ & $h=1$~~ & $h=2$~~ & $h=3$~~ & $h=4$~~ & $h=5$~~
\smallskip\\
\rotatebox{90}{$\quad \quad r=4$}~~\includegraphics[scale=0.155,trim={1cm 2.5cm 2.5cm 2cm},clip]{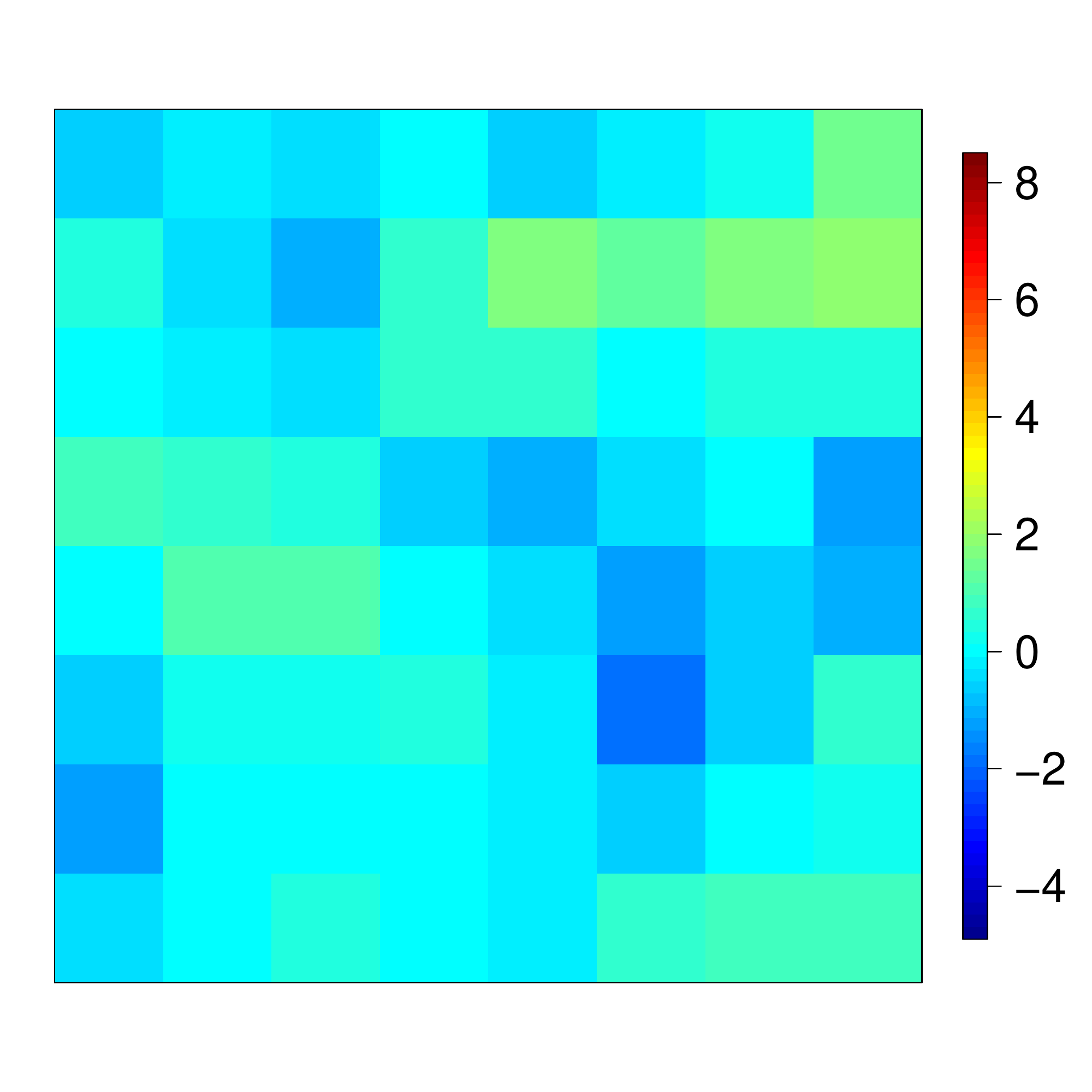} &
\!\!\!\!\!\!\includegraphics[scale=0.155,trim={1cm 2.5cm 2.5cm 2cm},clip]{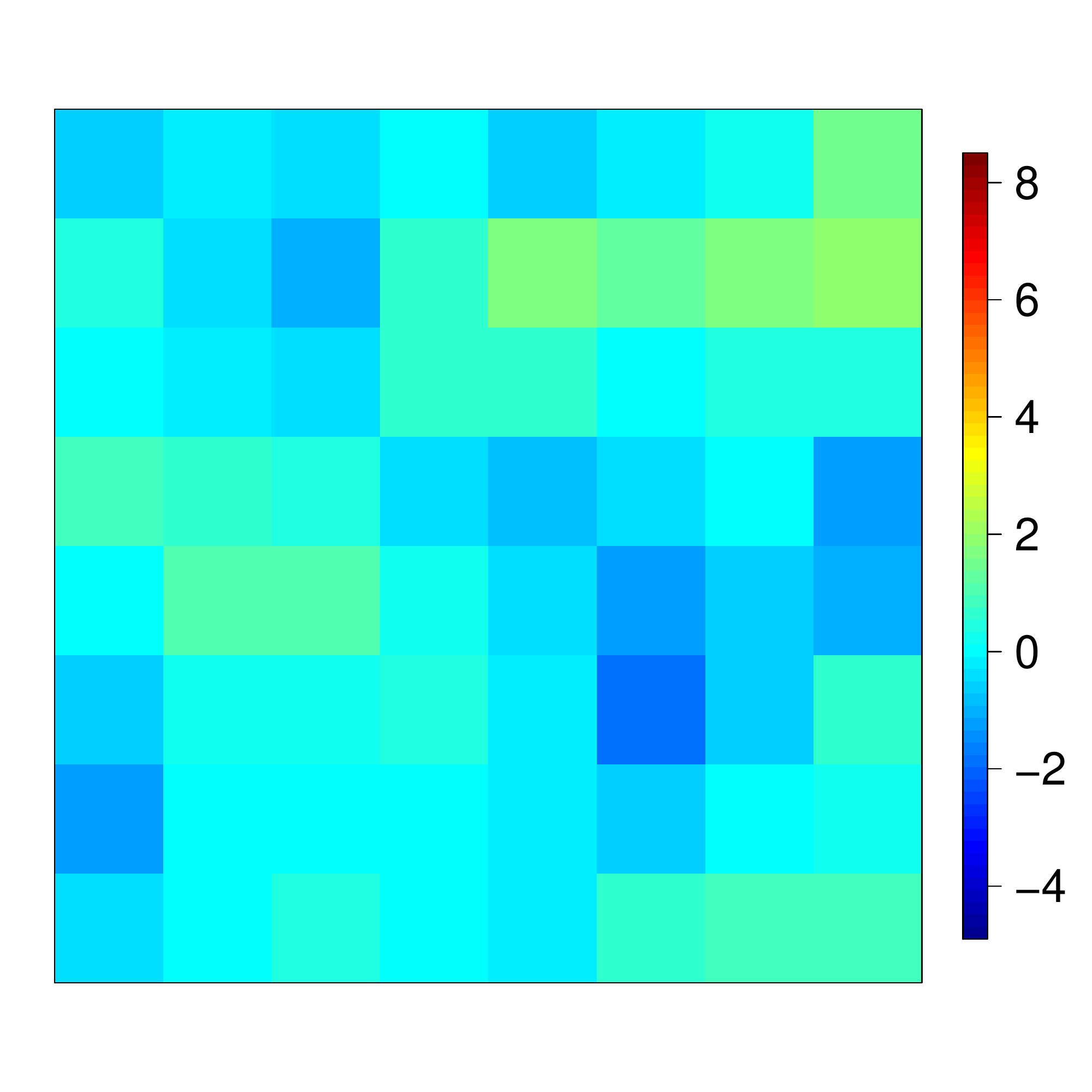} &
\!\!\!\!\!\!\includegraphics[scale=0.155,trim={1cm 2.5cm 2.5cm 2cm},clip]{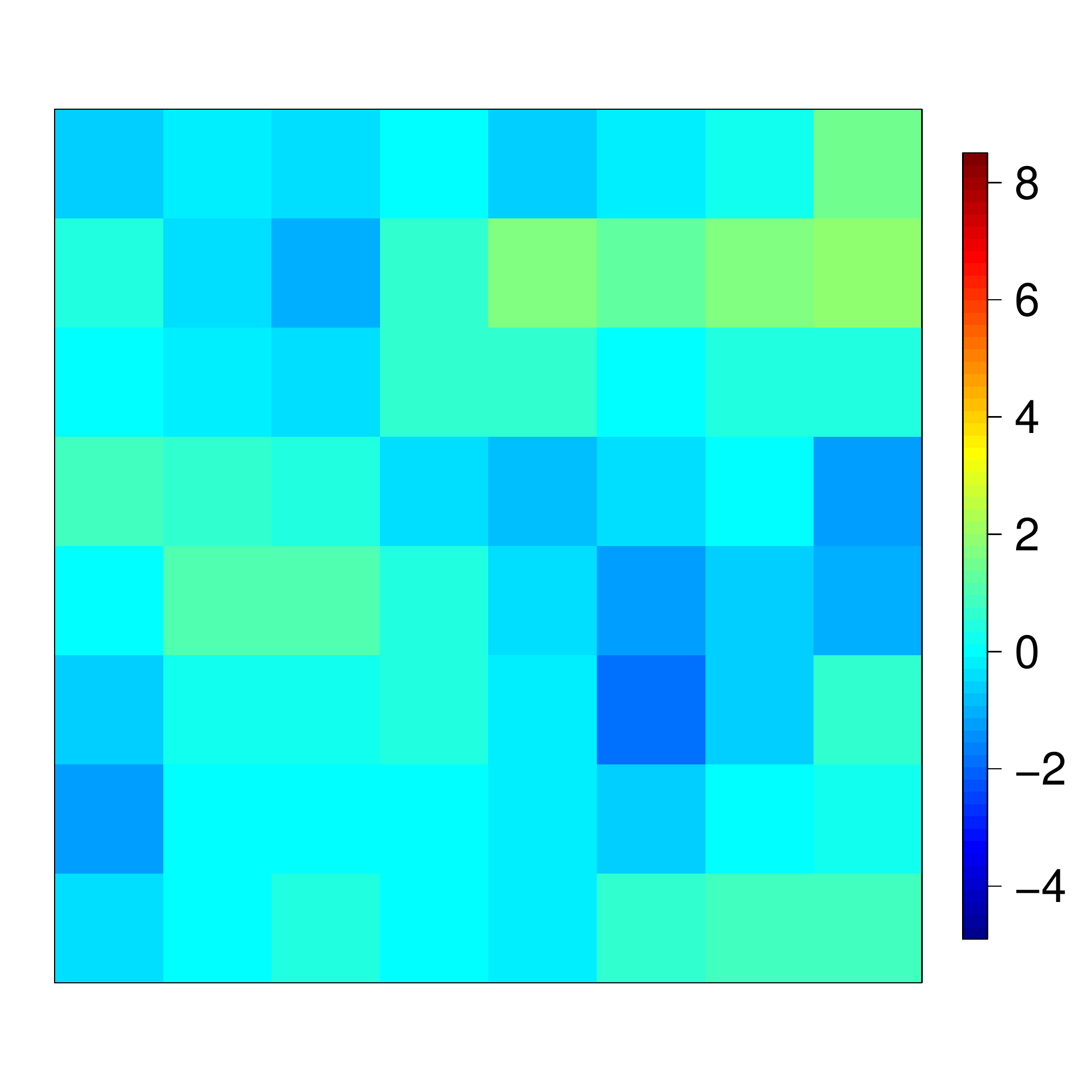} &
\!\!\!\!\!\!\includegraphics[scale=0.155,trim={1cm 2.5cm 2.5cm 2cm},clip]{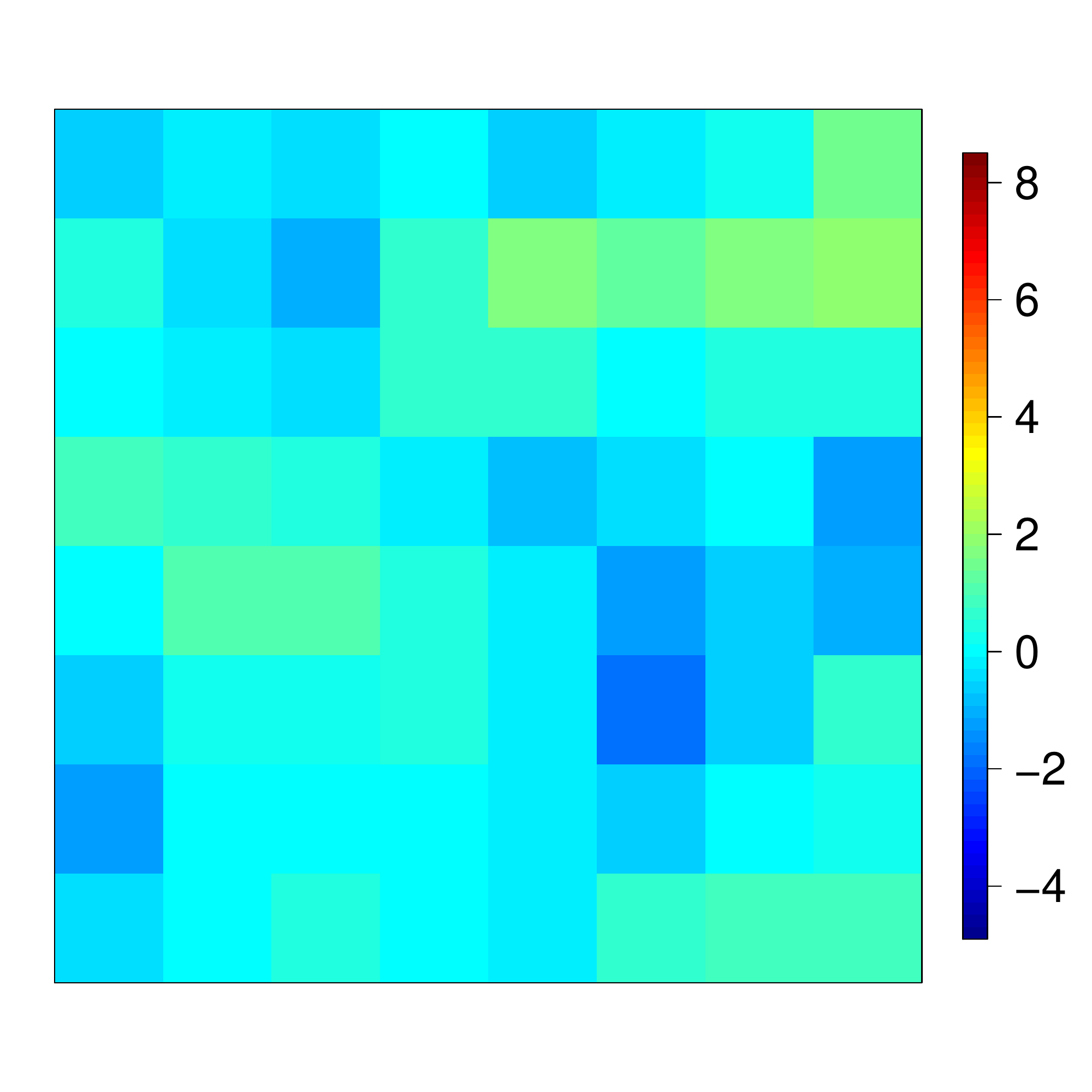} &
\!\!\!\!\!\!\includegraphics[scale=0.155,trim={1cm 2.5cm 2.5cm 2cm},clip]{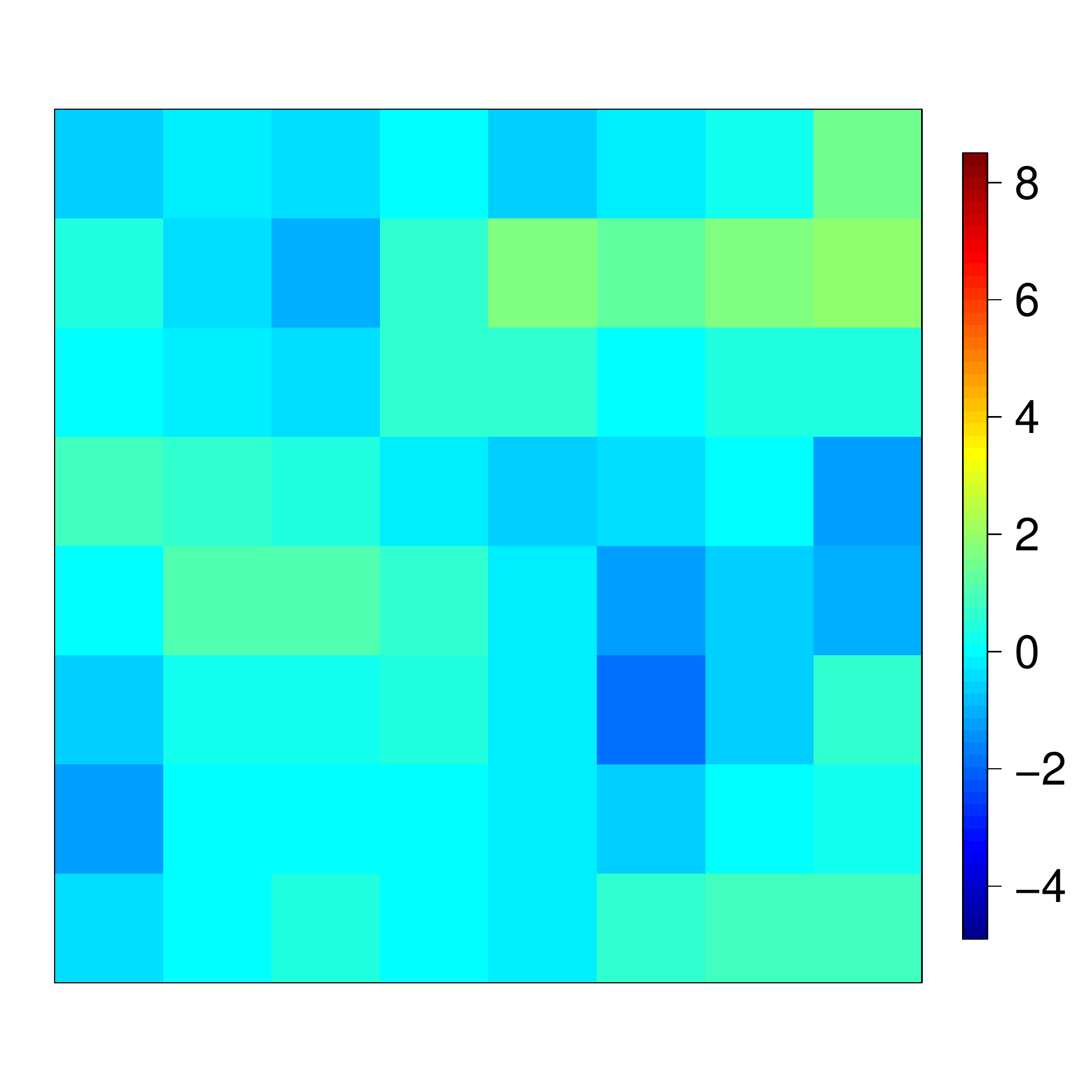} &
\!\!\!\!\!\!\includegraphics[scale=0.155,trim={1cm 2.5cm 2.5cm 2cm},clip]{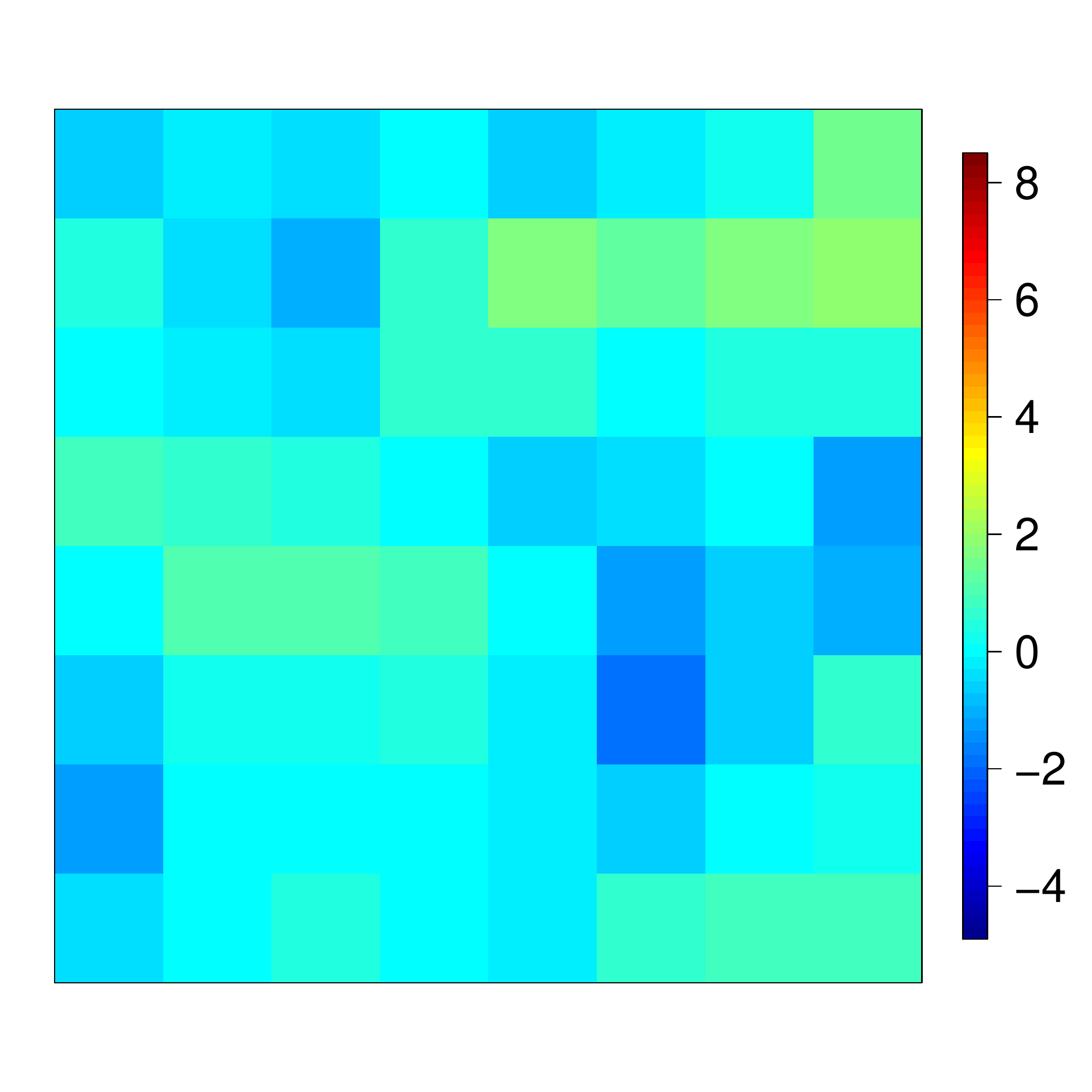} \\
\rotatebox{90}{$\quad \quad r=6$}~~\includegraphics[scale=0.155,trim={1cm 2.5cm 2.5cm 2cm},clip]{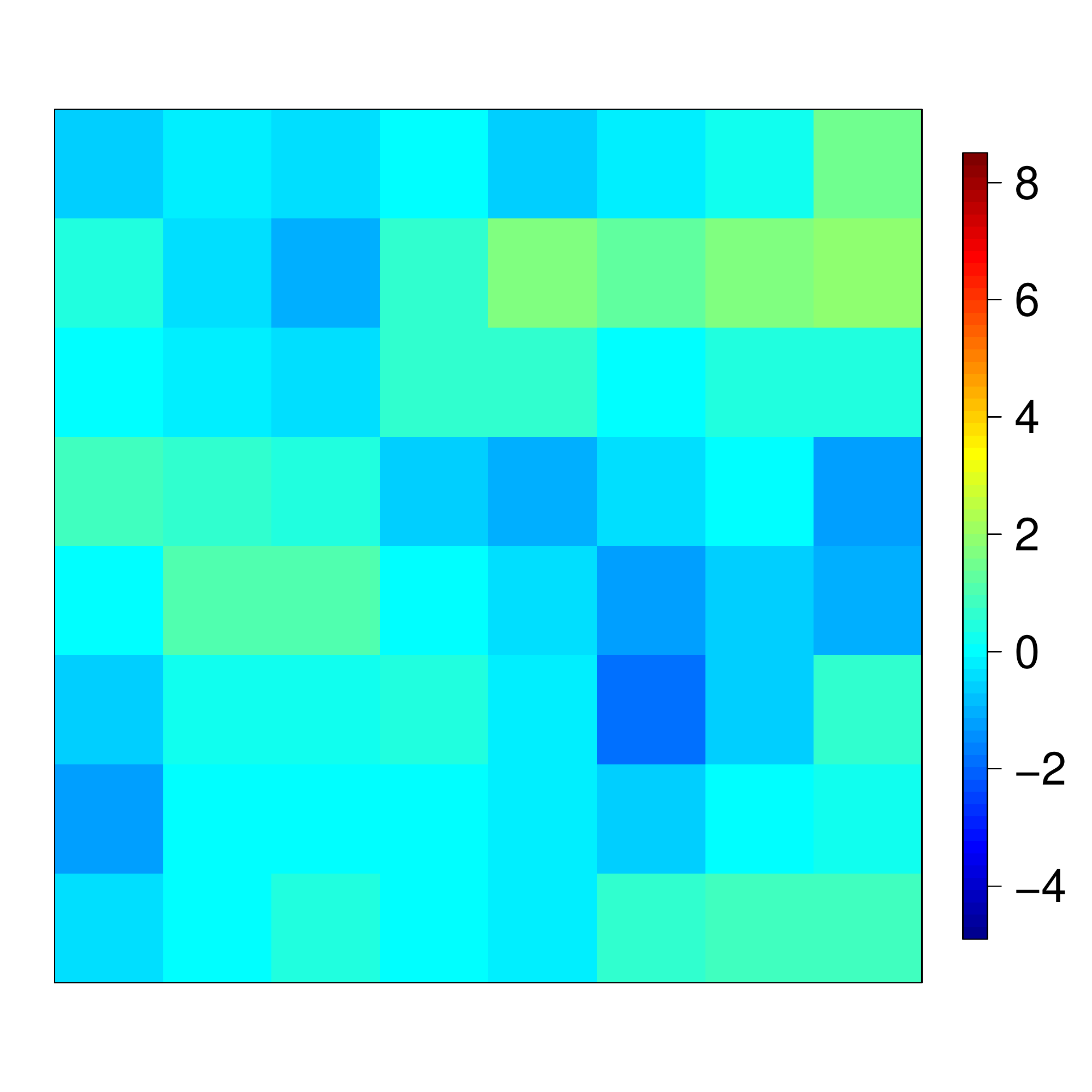} &
\!\!\!\!\!\!\includegraphics[scale=0.155,trim={1cm 2.5cm 2.5cm 2cm},clip]{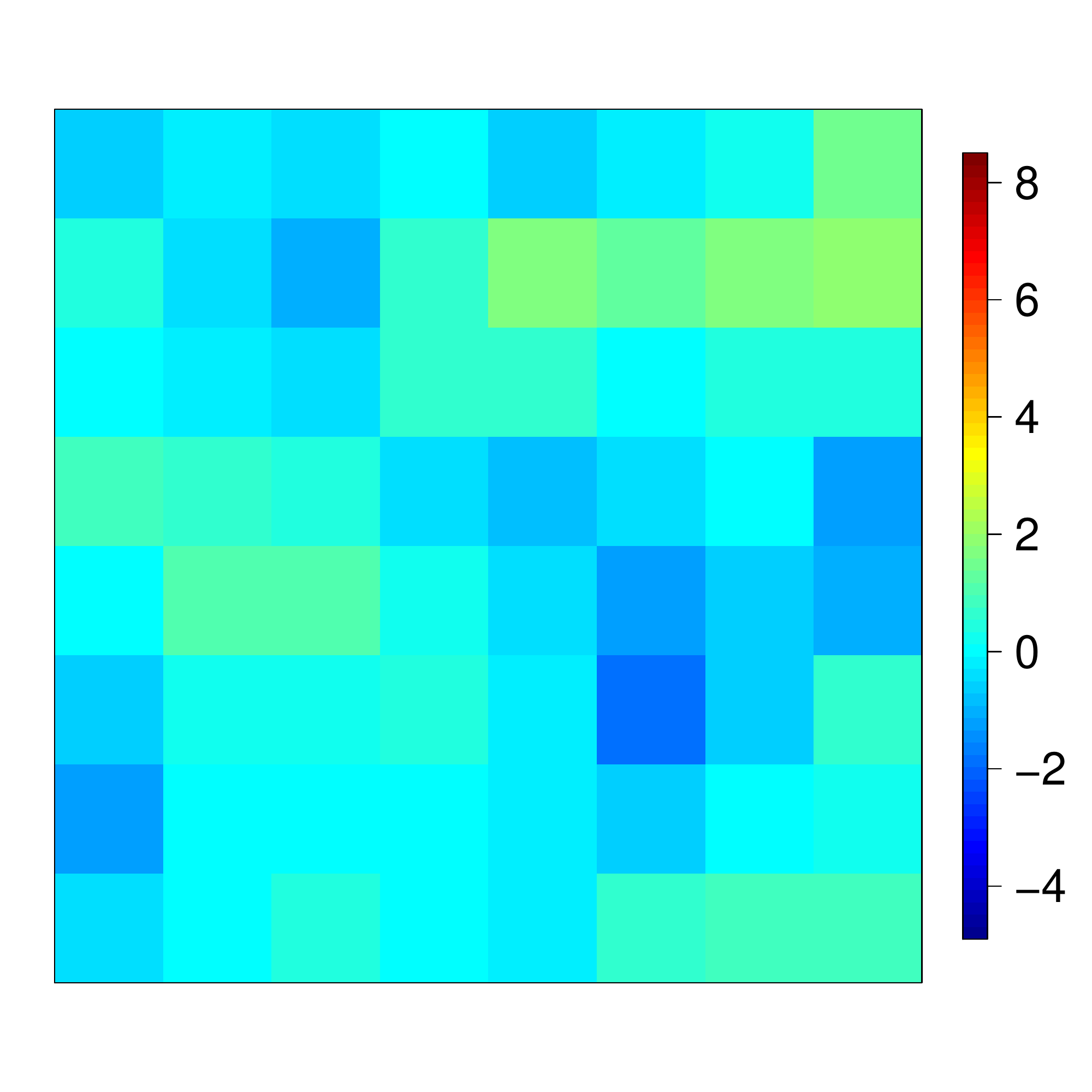} &
\!\!\!\!\!\!\includegraphics[scale=0.155,trim={1cm 2.5cm 2.5cm 2cm},clip]{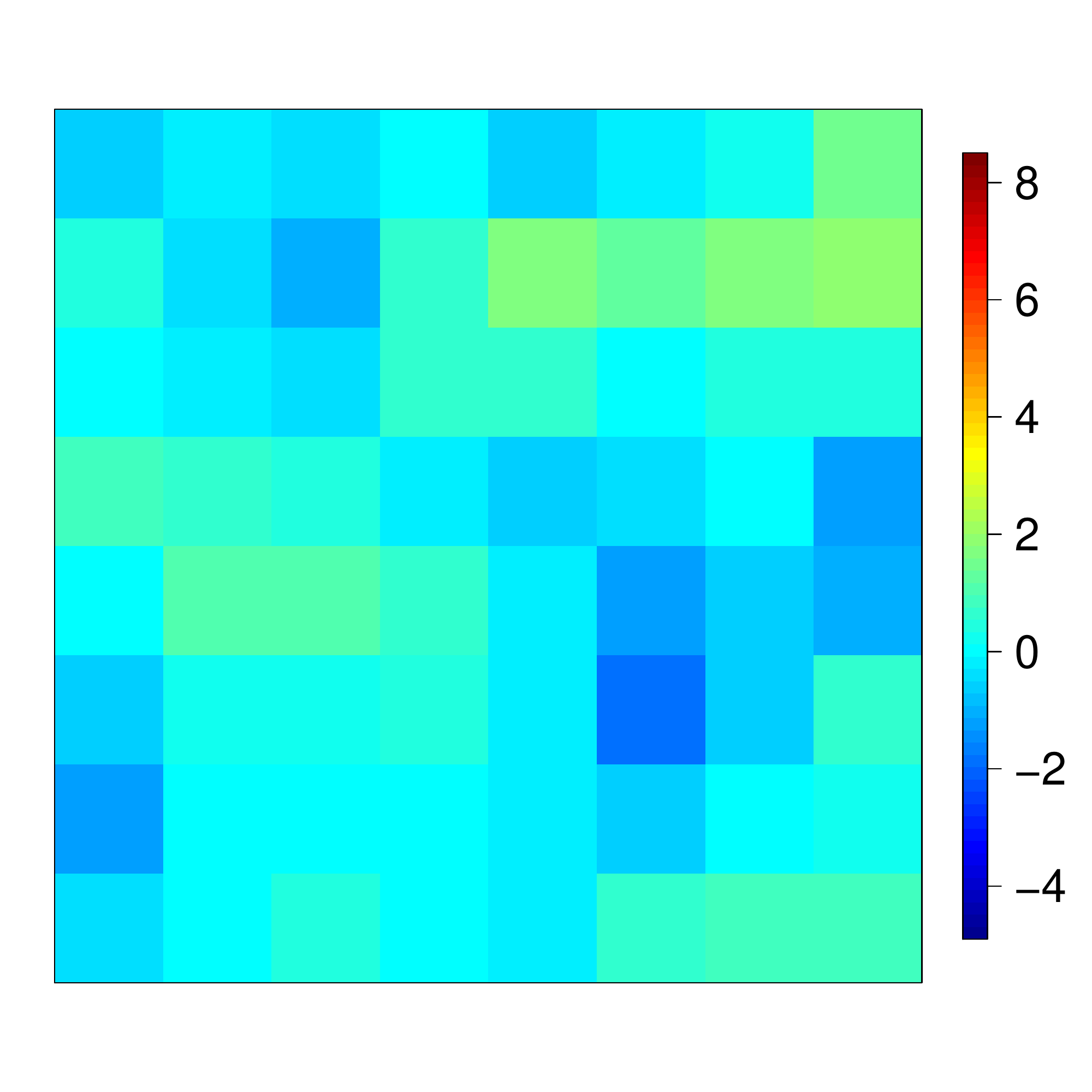} &
\!\!\!\!\!\!\includegraphics[scale=0.155,trim={1cm 2.5cm 2.5cm 2cm},clip]{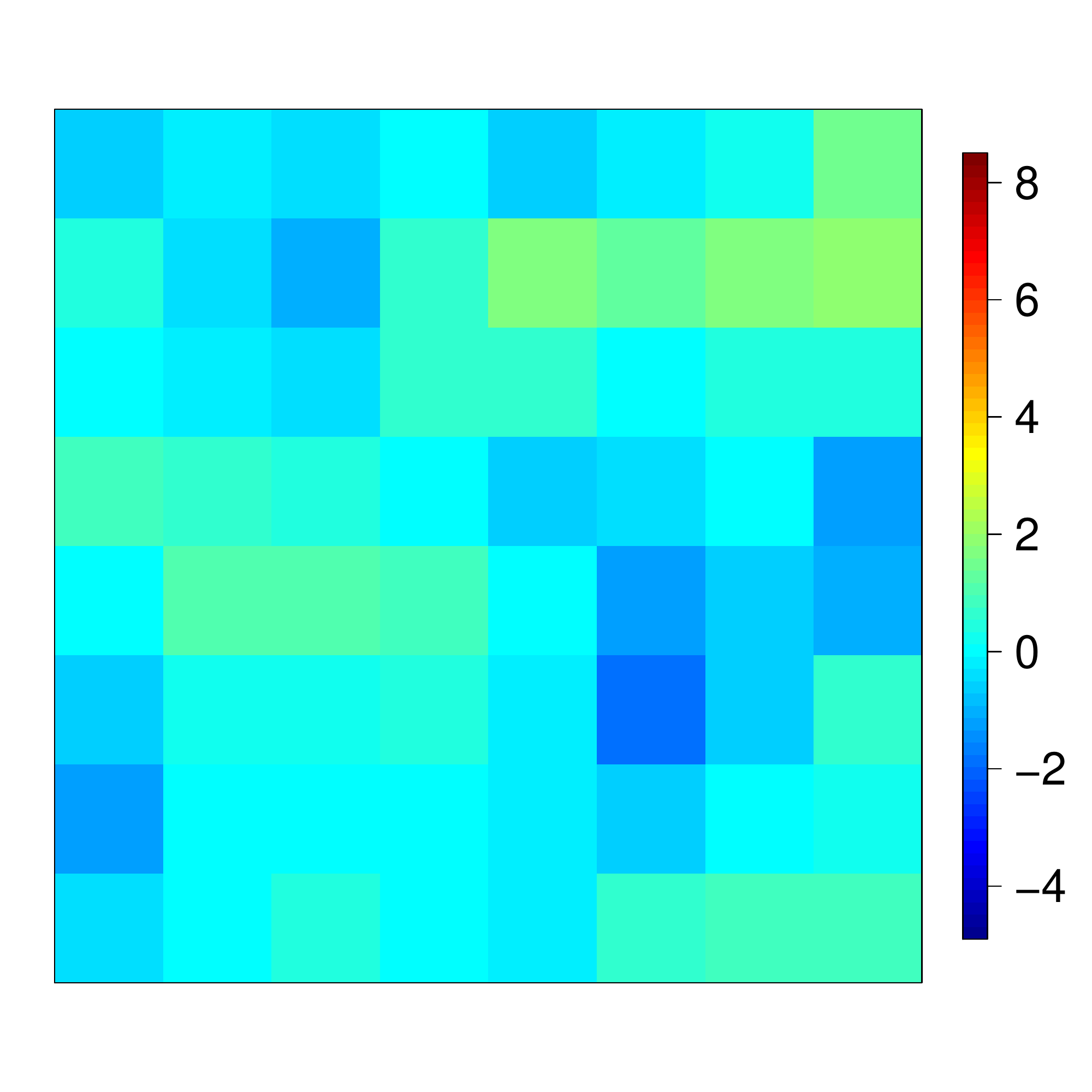} &
\!\!\!\!\!\!\includegraphics[scale=0.155,trim={1cm 2.5cm 2.5cm 2cm},clip]{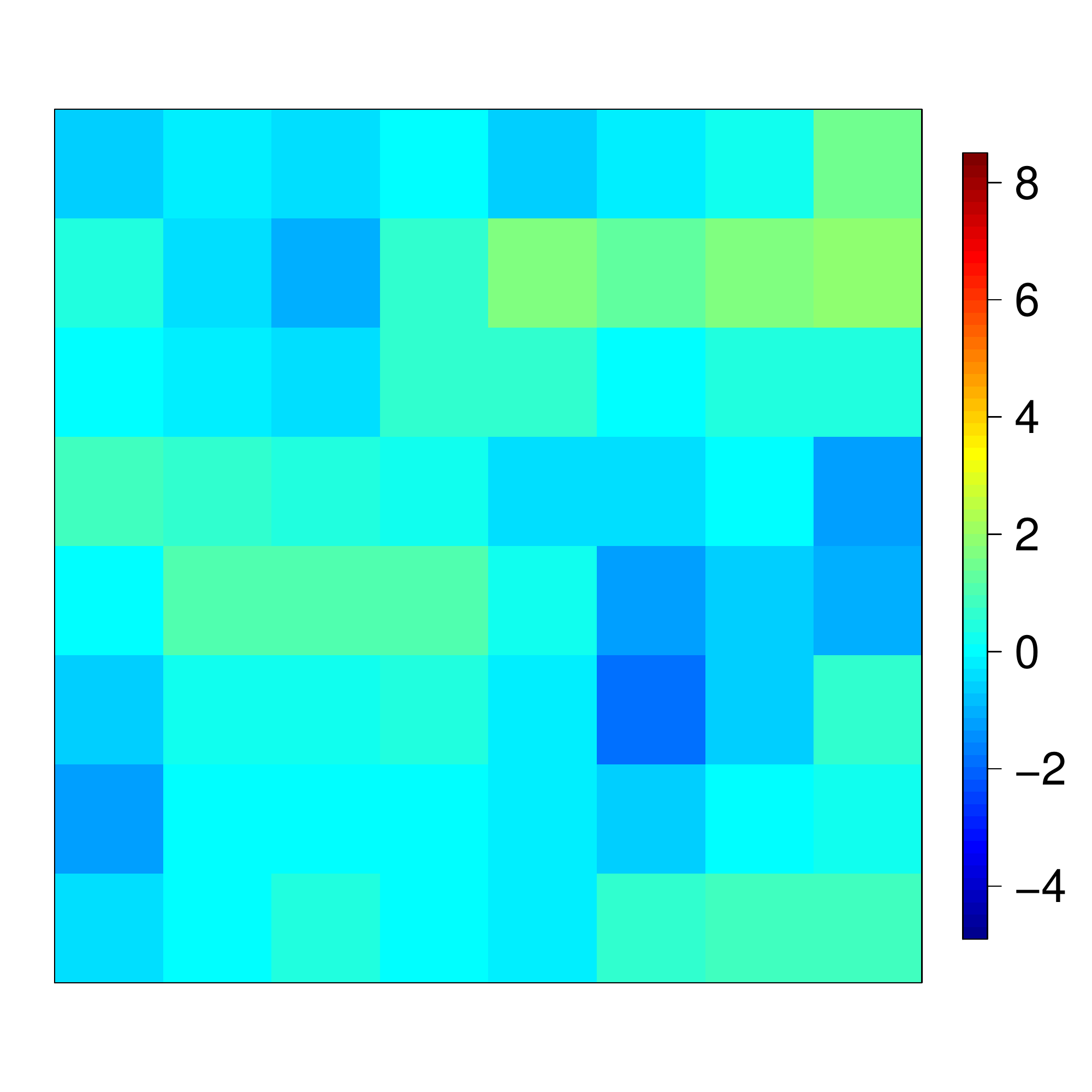} &
\!\!\!\!\!\!\includegraphics[scale=0.155,trim={1cm 2.5cm 2.5cm 2cm},clip]{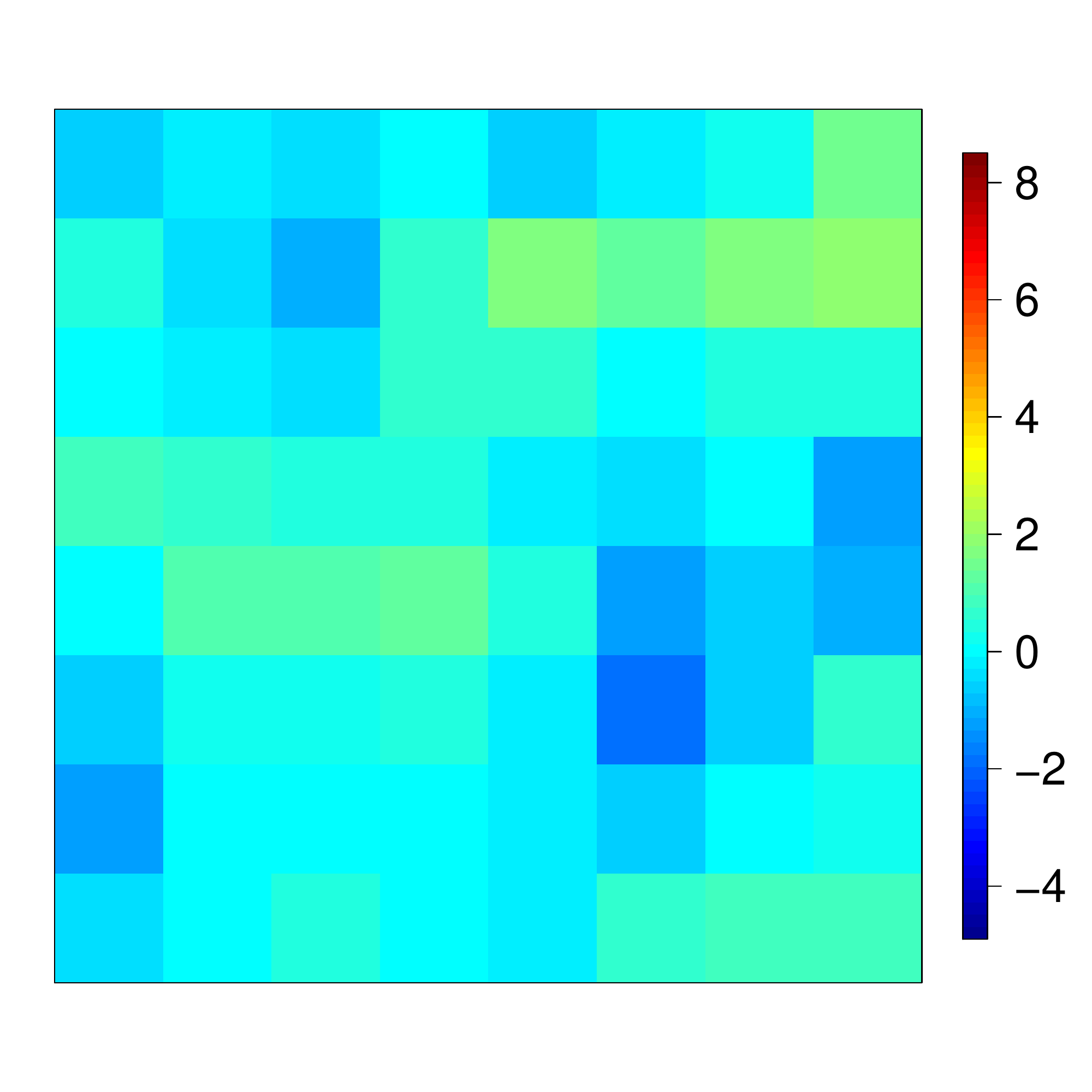} \\
\rotatebox{90}{$\quad \quad r=8$}~~\includegraphics[scale=0.155,trim={1cm 2.5cm 2.5cm 2cm},clip]{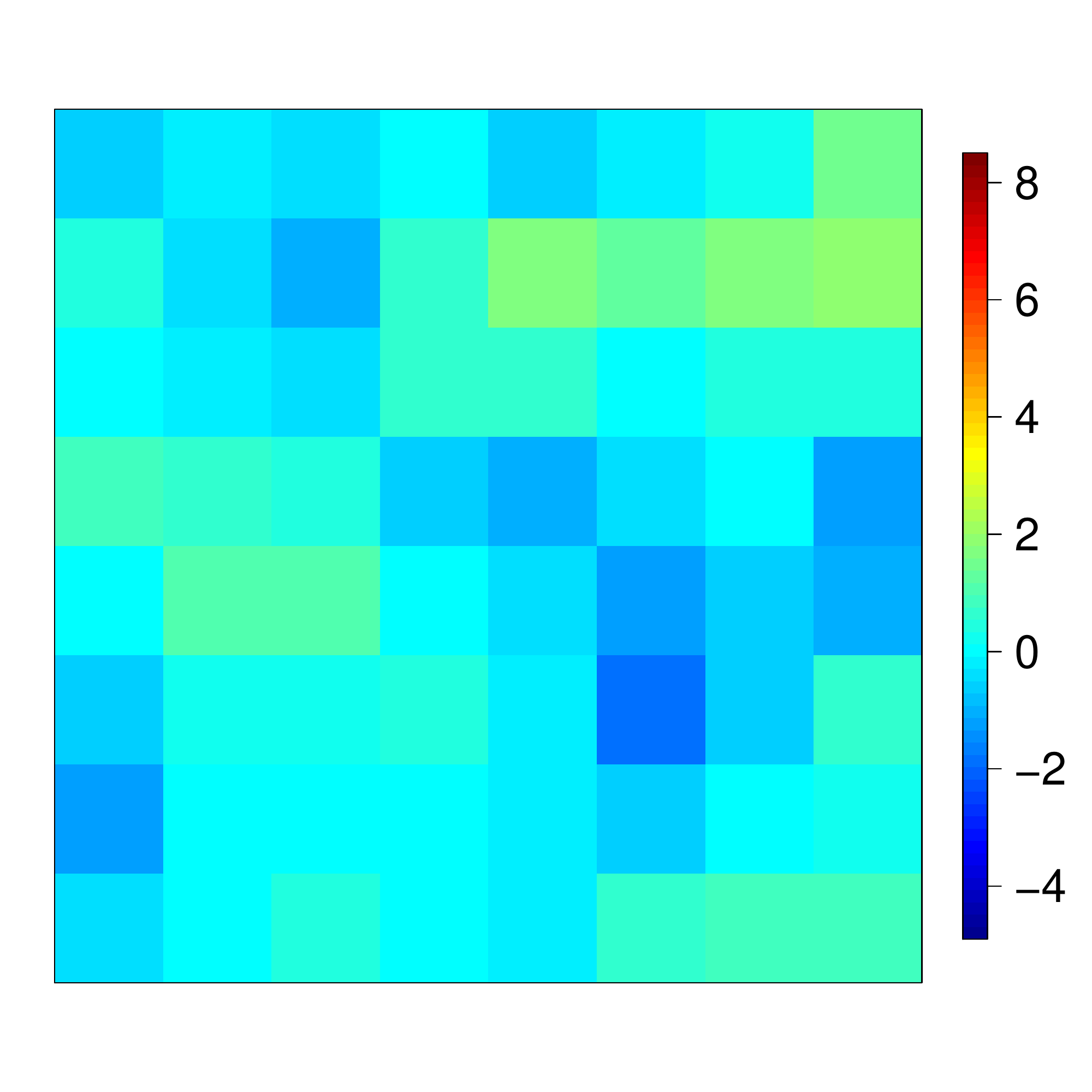} &
\!\!\!\!\!\!\includegraphics[scale=0.155,trim={1cm 2.5cm 2.5cm 2cm},clip]{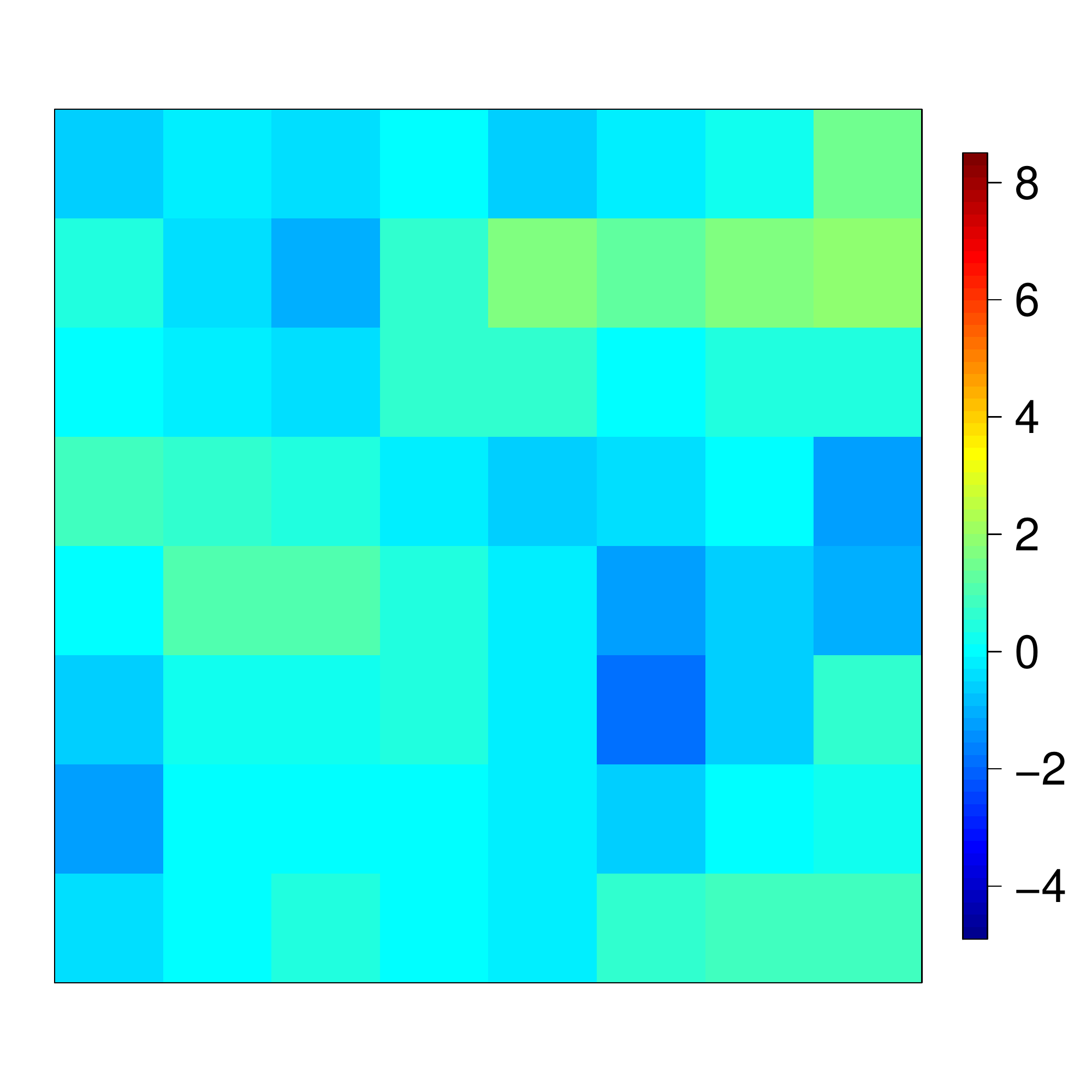} &
\!\!\!\!\!\!\includegraphics[scale=0.155,trim={1cm 2.5cm 2.5cm 2cm},clip]{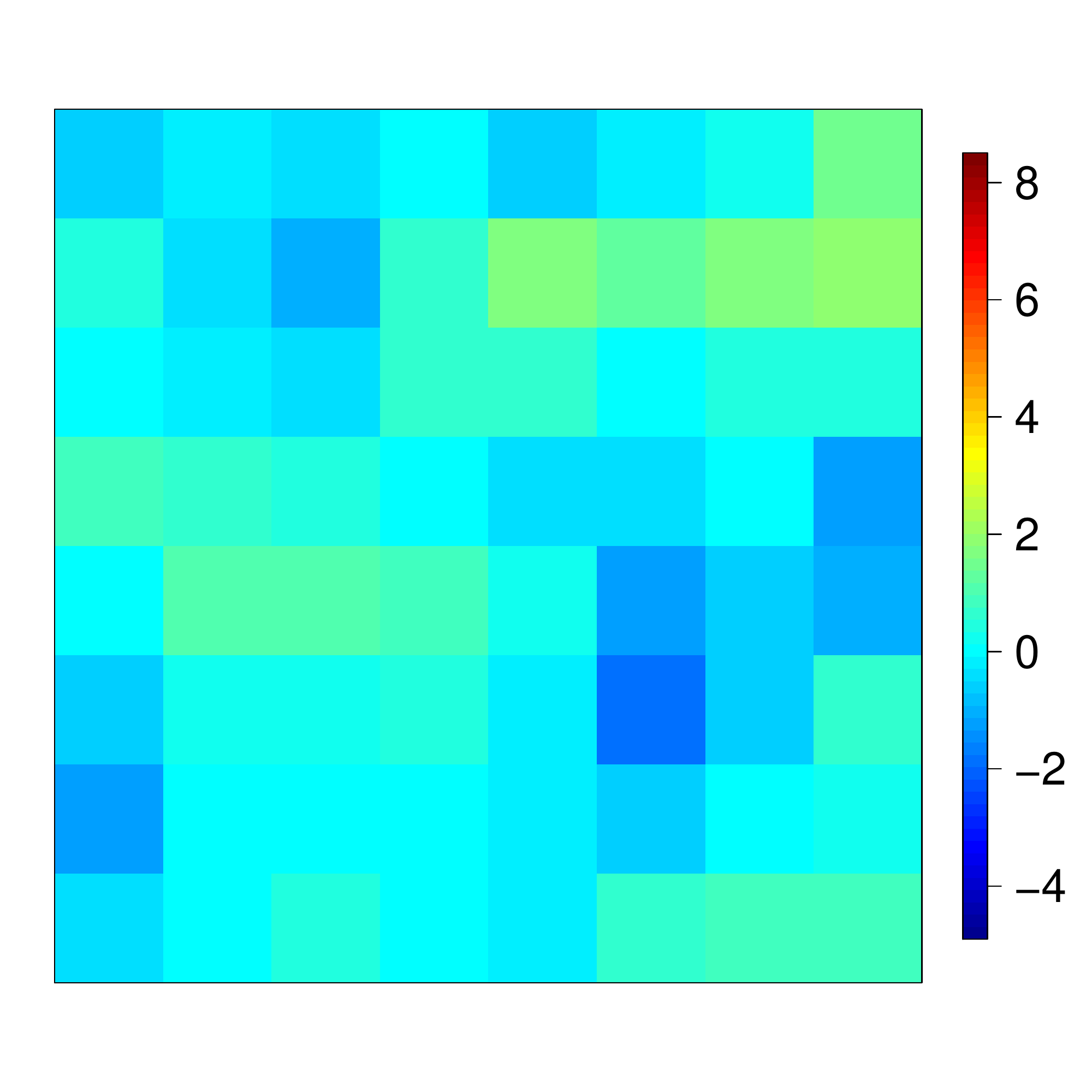} &
\!\!\!\!\!\!\includegraphics[scale=0.155,trim={1cm 2.5cm 2.5cm 2cm},clip]{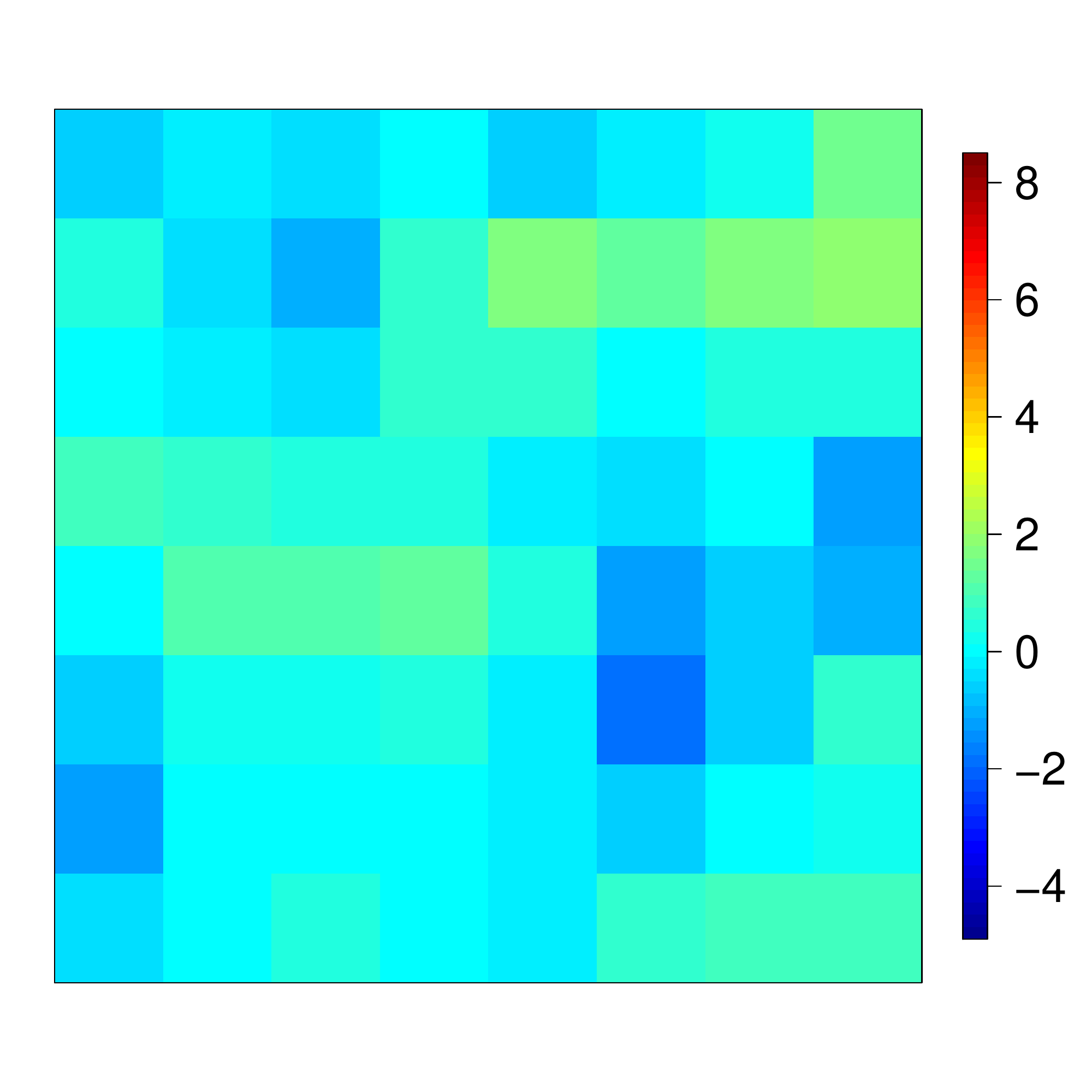} &
\!\!\!\!\!\!\includegraphics[scale=0.155,trim={1cm 2.5cm 2.5cm 2cm},clip]{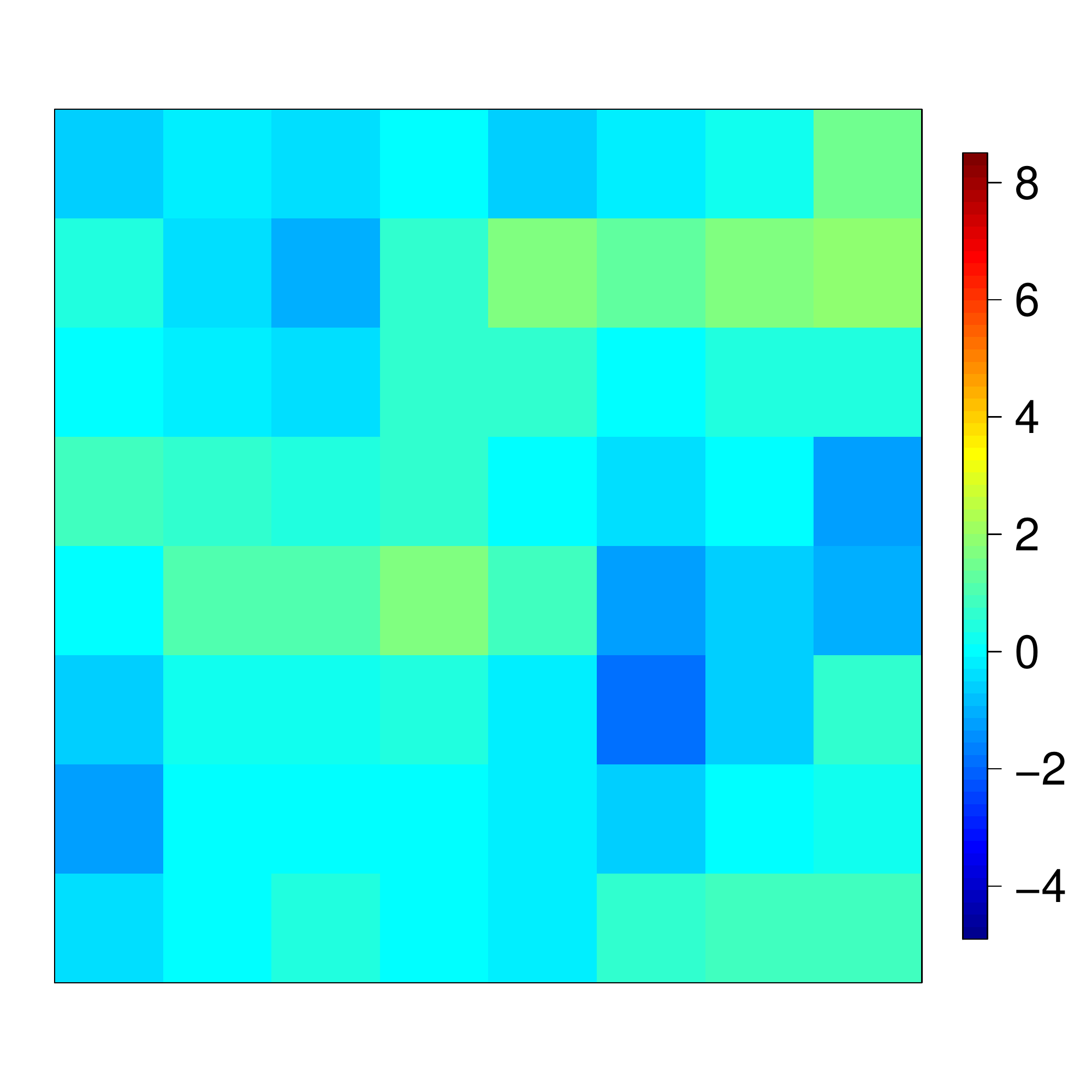} &
\!\!\!\!\!\!\includegraphics[scale=0.155,trim={1cm 2.5cm 2.5cm 2cm},clip]{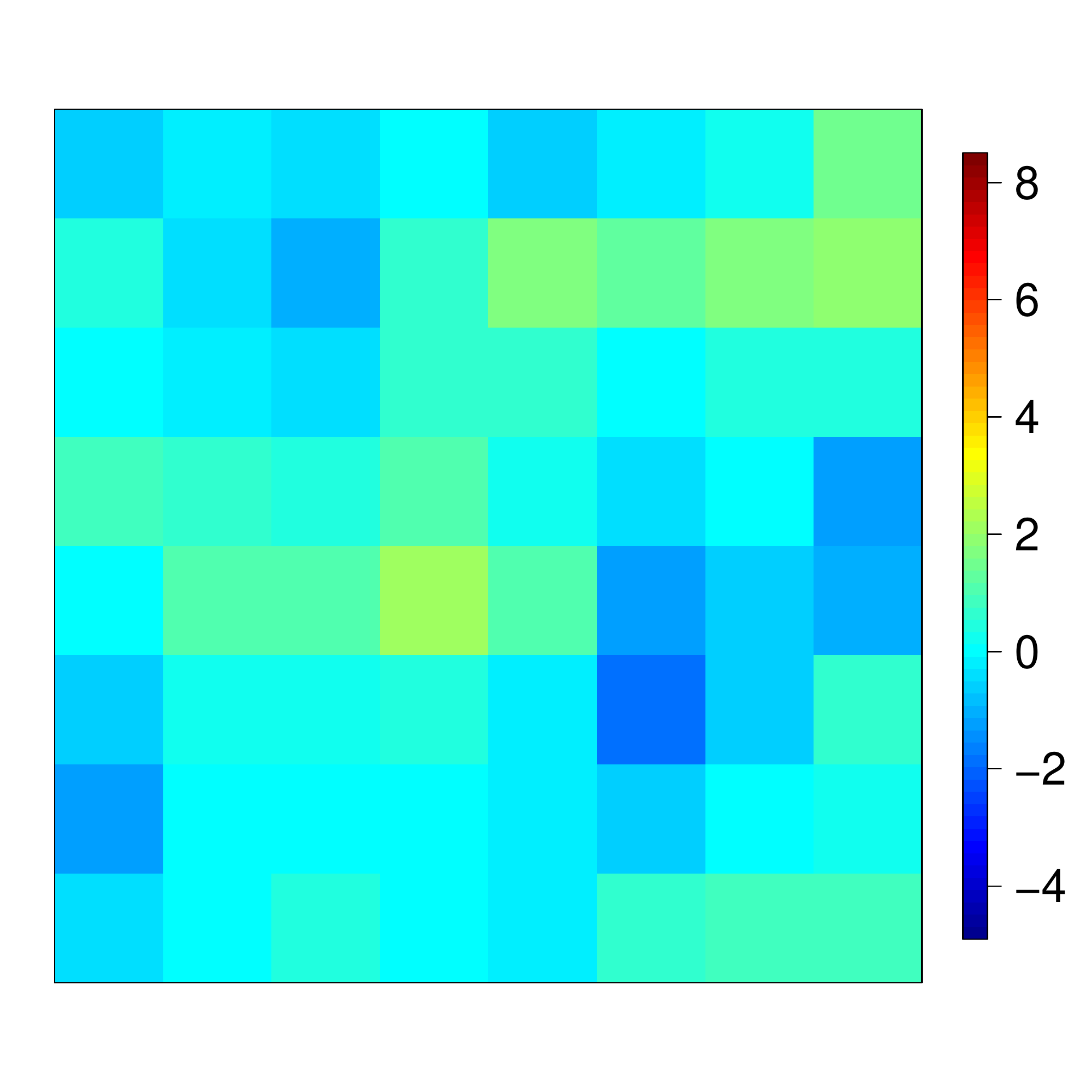} \\
\rotatebox{90}{$\quad \quad r=10$}~~\includegraphics[scale=0.155,trim={1cm 2.5cm 2.5cm 2cm},clip]{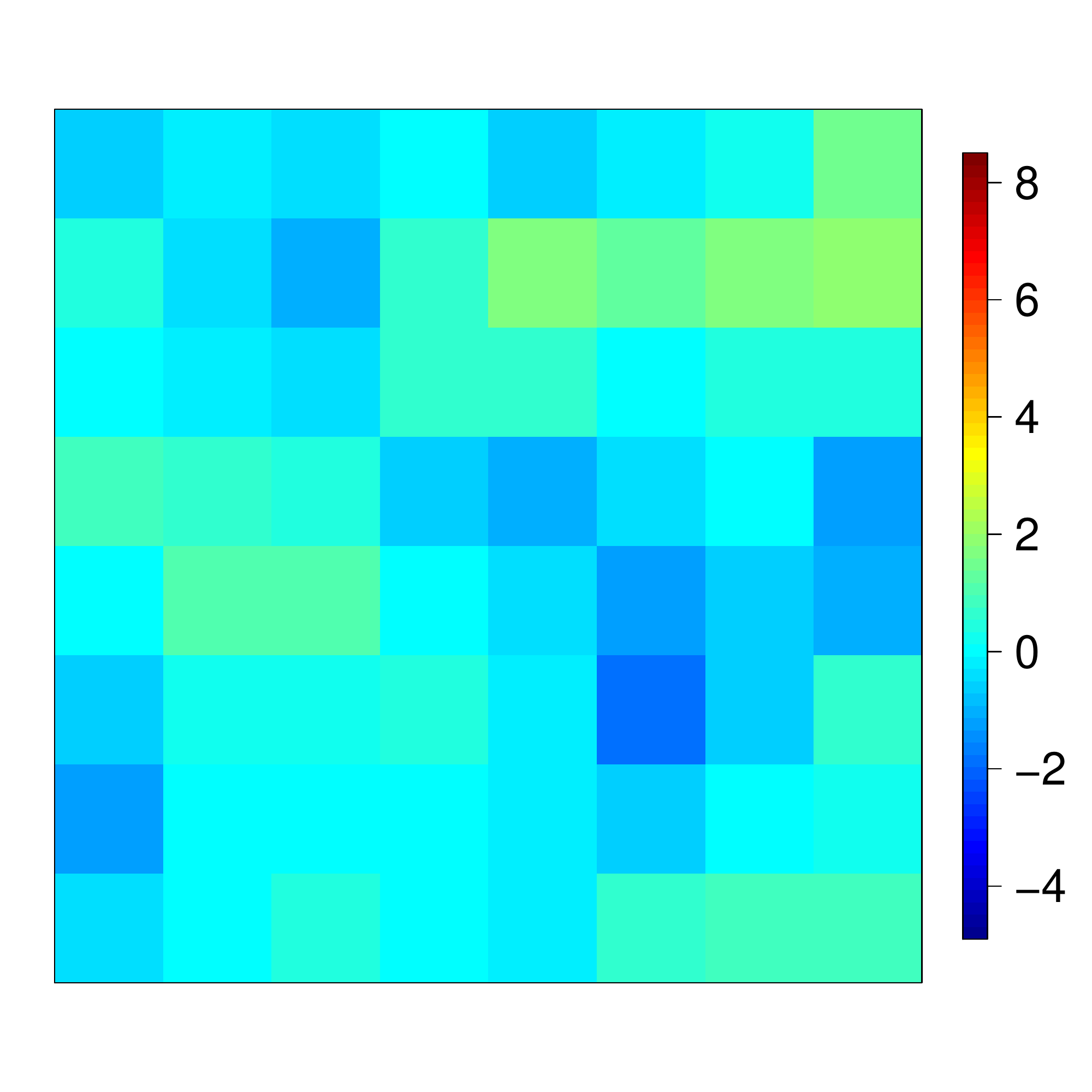} &
\!\!\!\!\!\!\includegraphics[scale=0.155,trim={1cm 2.5cm 2.5cm 2cm},clip]{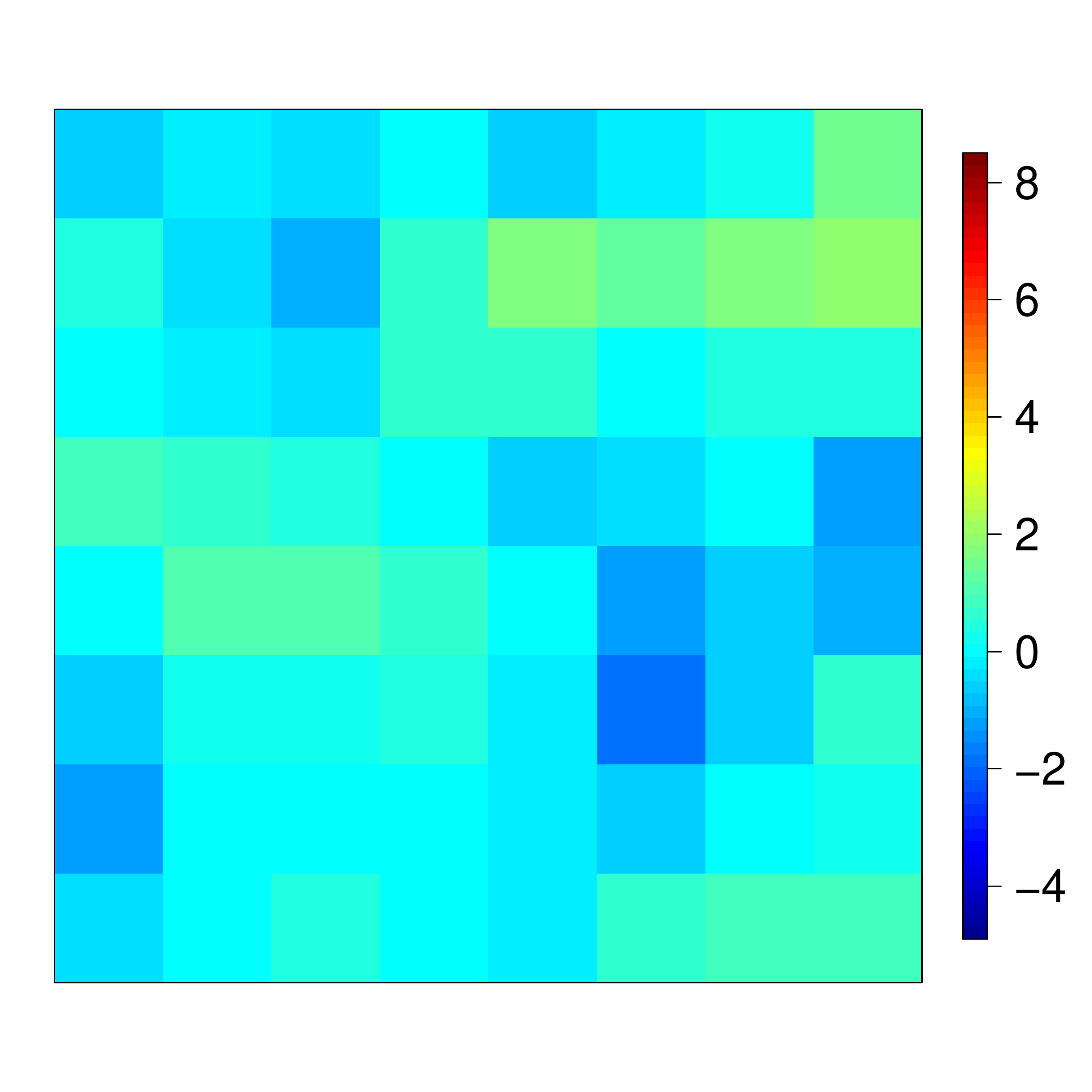} &
\!\!\!\!\!\!\includegraphics[scale=0.155,trim={1cm 2.5cm 2.5cm 2cm},clip]{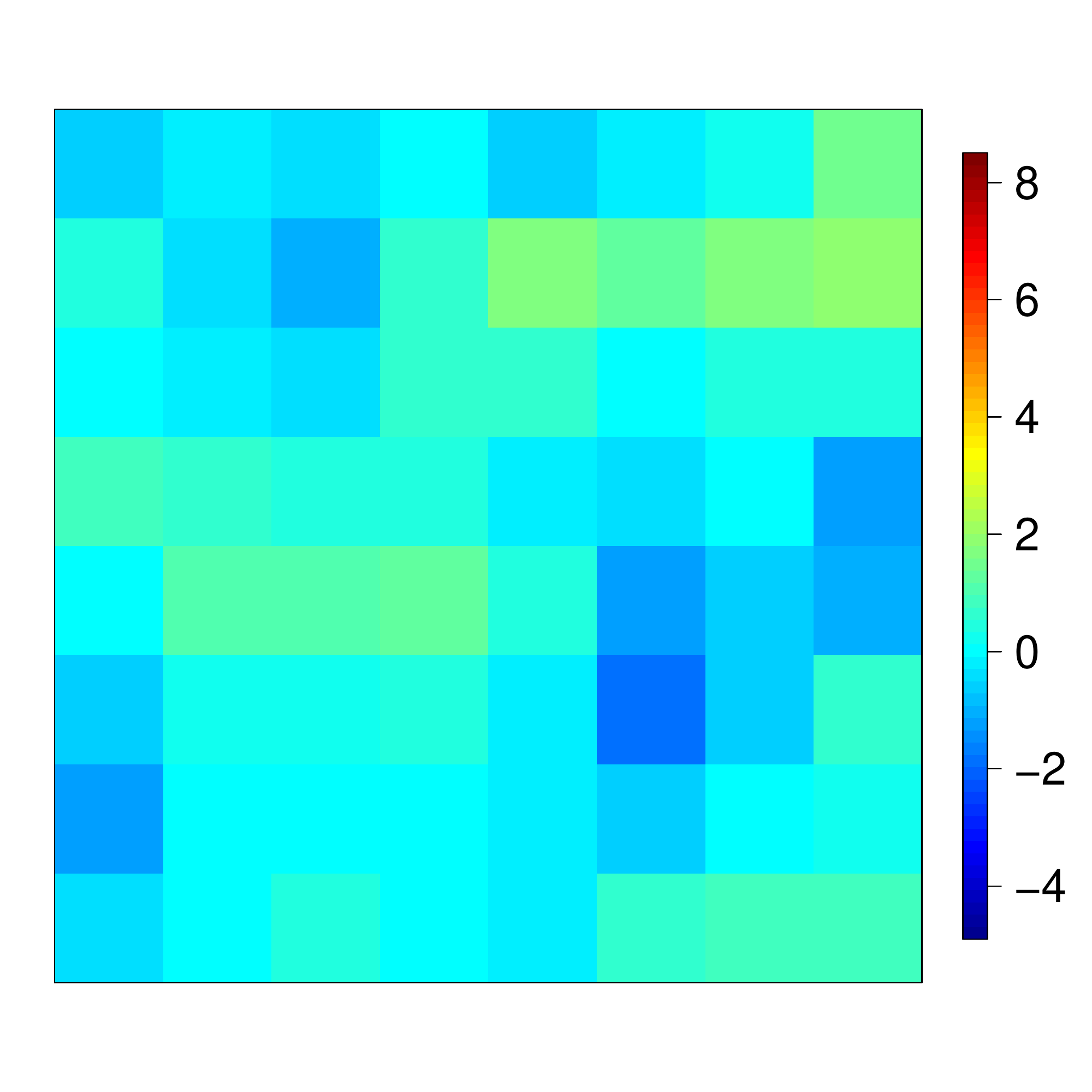} &
\!\!\!\!\!\!\includegraphics[scale=0.155,trim={1cm 2.5cm 2.5cm 2cm},clip]{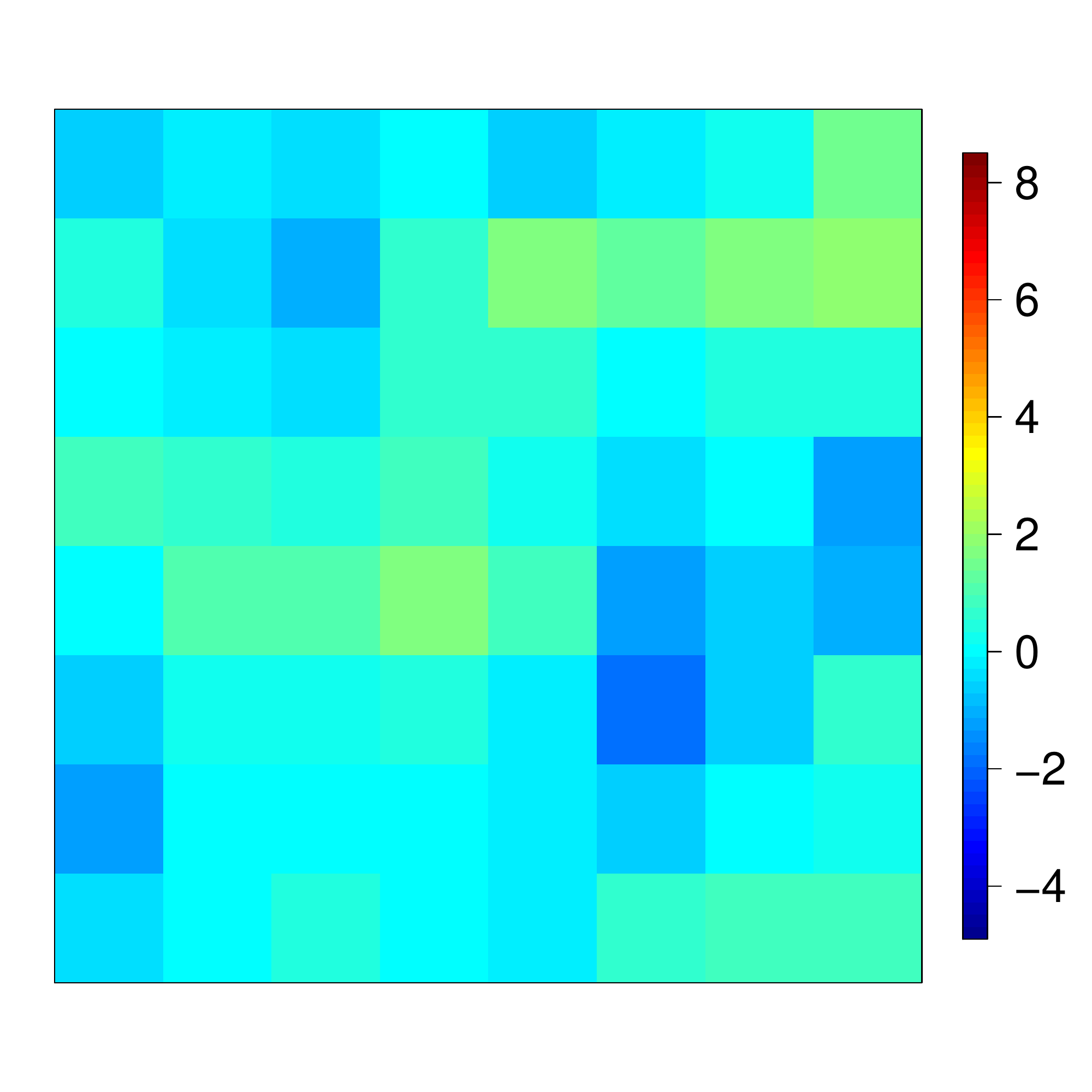} &
\!\!\!\!\!\!\includegraphics[scale=0.155,trim={1cm 2.5cm 2.5cm 2cm},clip]{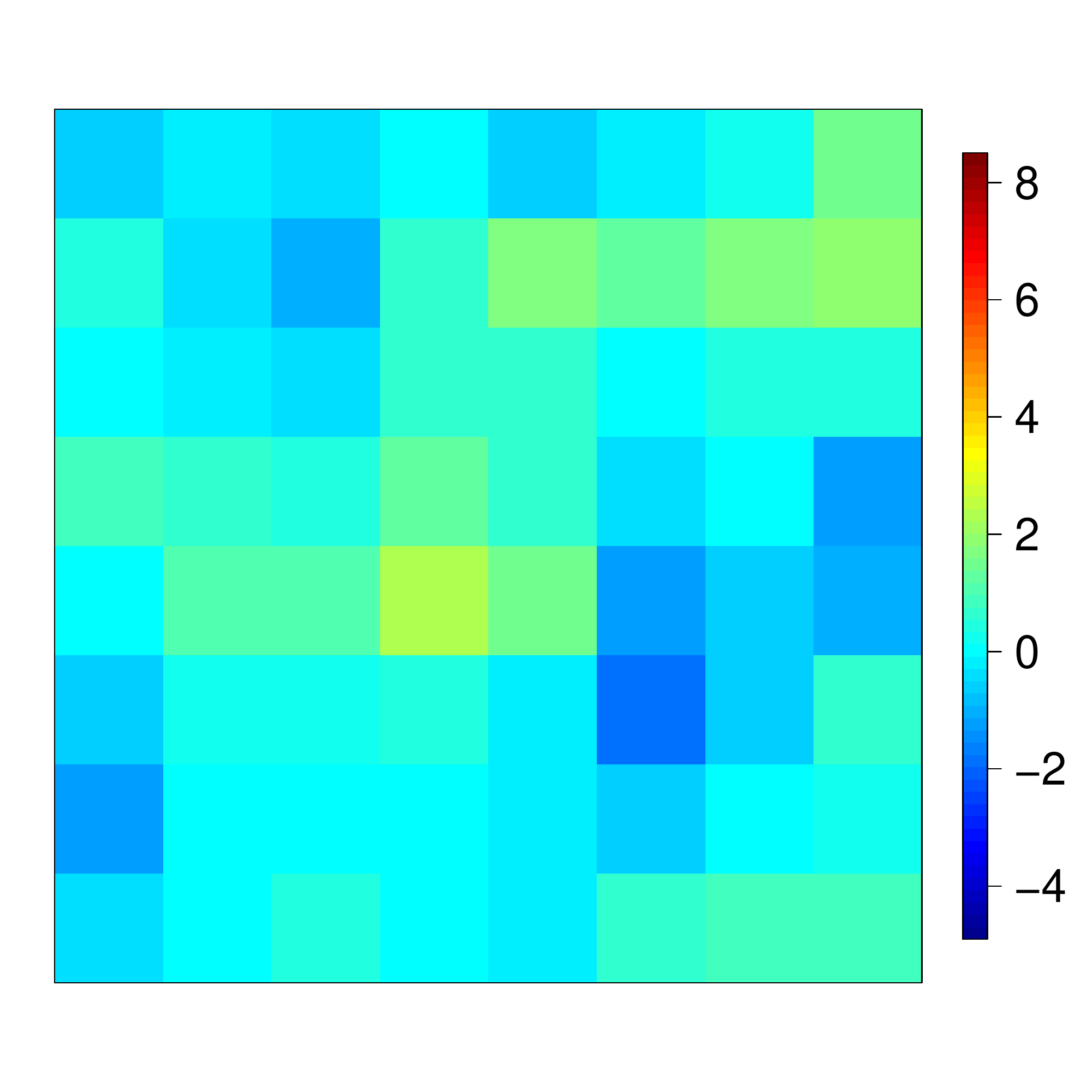} &
\!\!\!\!\!\!\includegraphics[scale=0.155,trim={1cm 2.5cm 2.5cm 2cm},clip]{./signal-r10-h5-grid8} 
\end{tabular}
\caption{Images of $\tilde{\bm{Z}}_{8 \times 8}$ obtained by aggregating $\bm{Z}$ in Figure~\ref{fig:signalonimageraw-all}
into $8\times 8$ blocks resulting in $8\times 8$ grid cells.}
\label{fig:signalonimageaggregated}
\end{figure}

\begin{figure}[!p]\centering
\begin{tabular}{ccc}
~~~~~$\phi=0$ & ~~$\phi=5$ & ~~$\phi=10$
\smallskip\\
\rotatebox{90}{$\quad \quad \quad r=4$}~~
\includegraphics[scale=0.22,trim={0.18cm 0.3cm 1cm 1cm},clip]{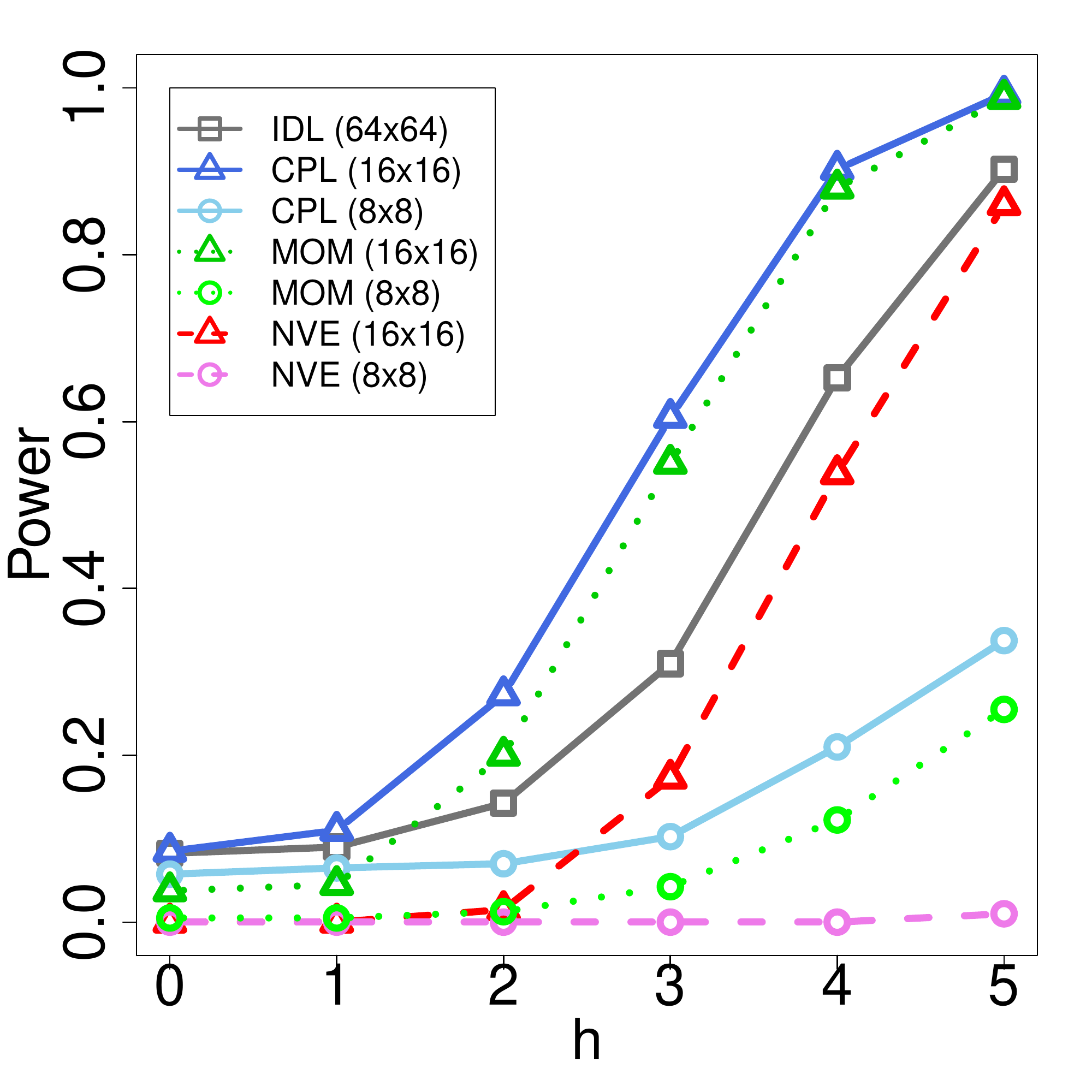} &
\includegraphics[scale=0.22,trim={0.18cm 0.3cm 1cm 1cm},clip]{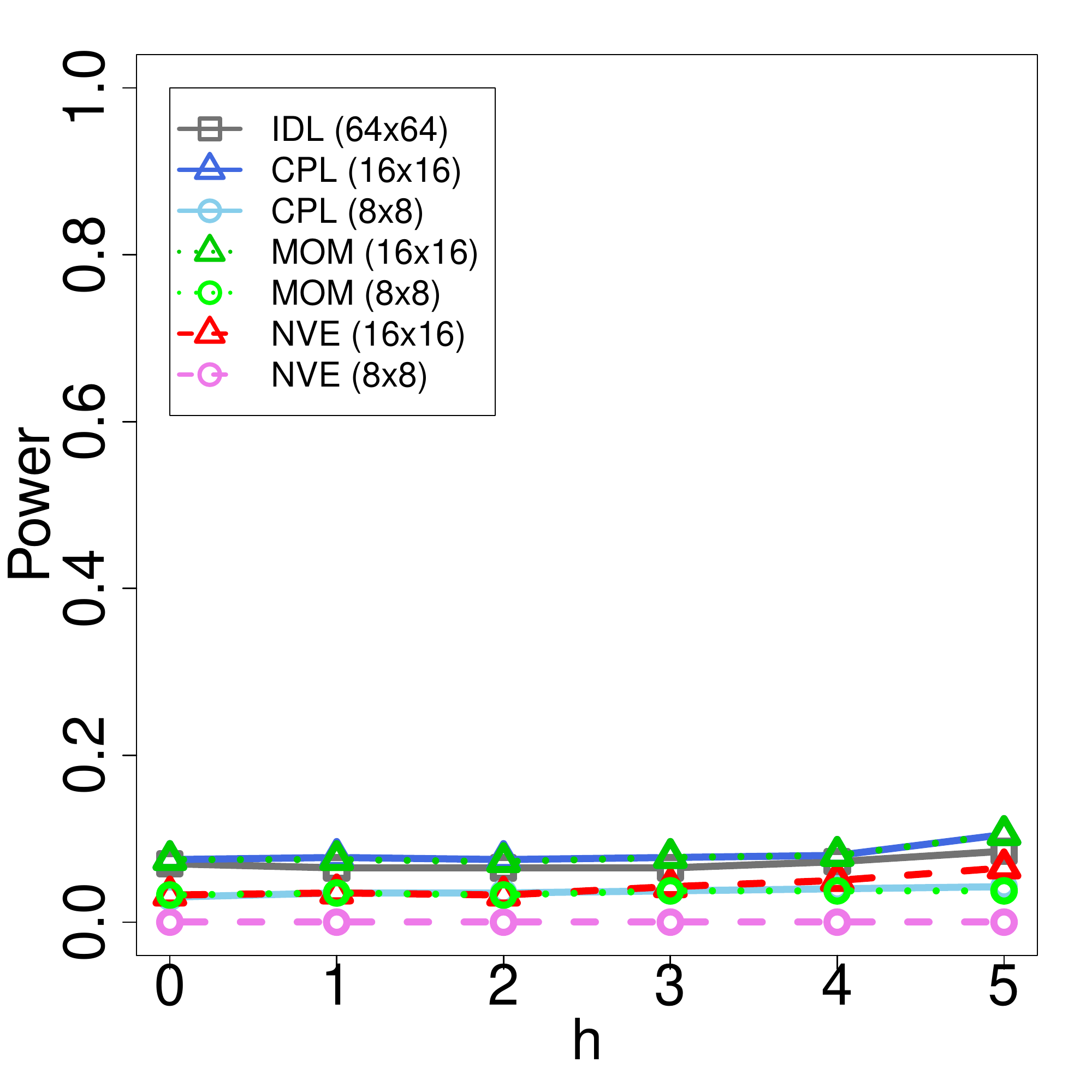} &
\includegraphics[scale=0.22,trim={0.18cm 0.3cm 1cm 1cm},clip]{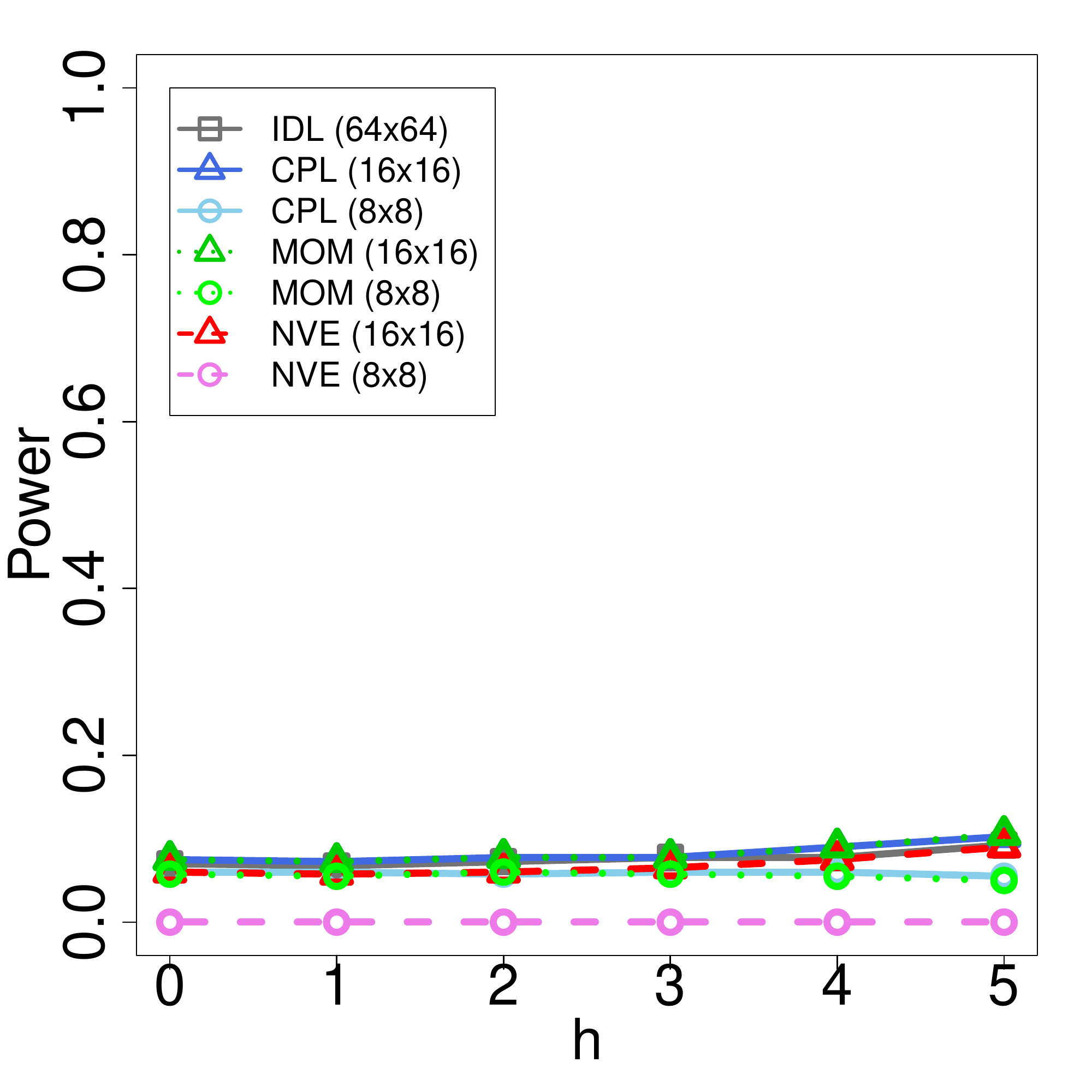} \\
\rotatebox{90}{$\quad \quad \quad r=6$}~~
\includegraphics[scale=0.22,trim={0.18cm 0.3cm 1cm 1cm},clip]{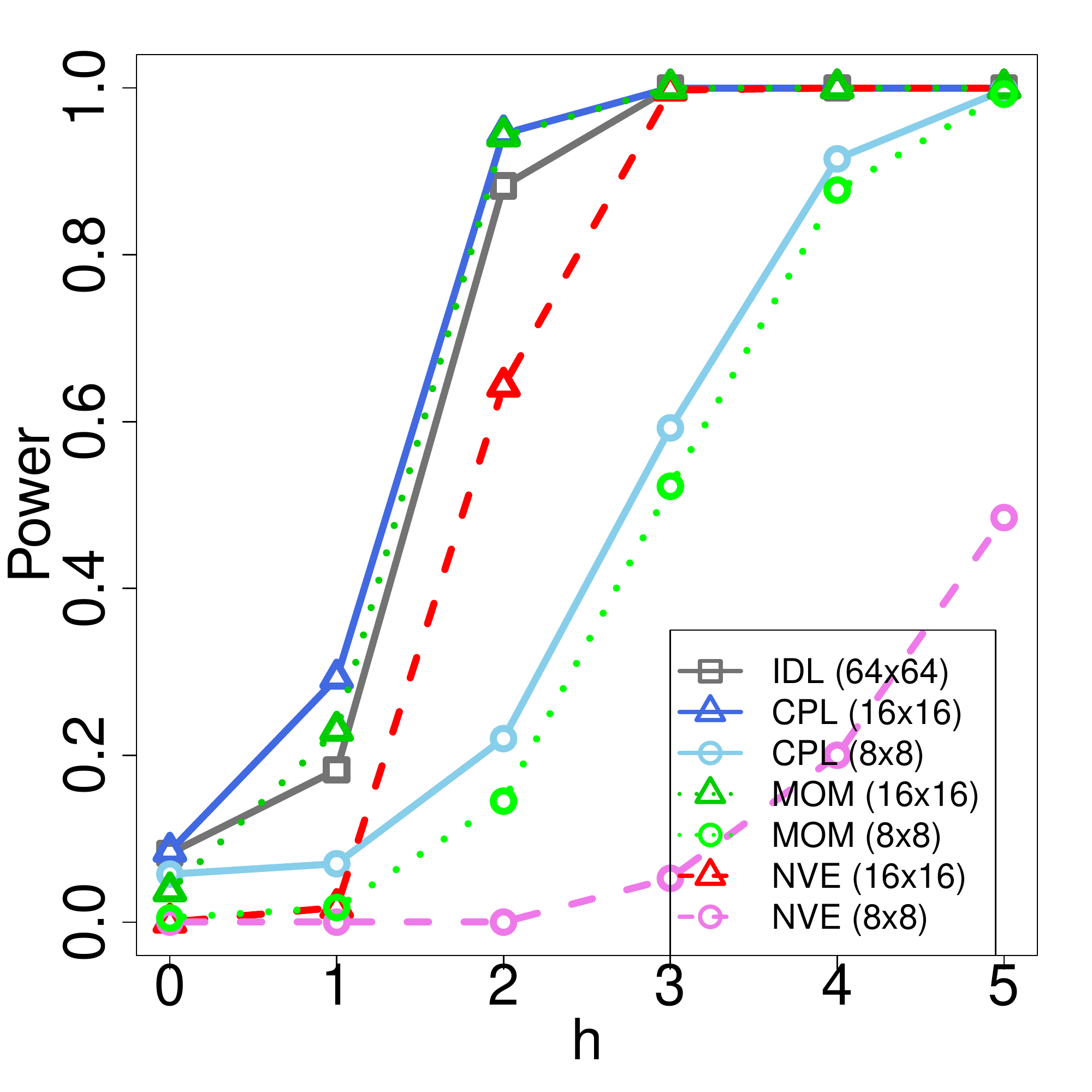} &
\includegraphics[scale=0.22,trim={0.18cm 0.3cm 1cm 1cm},clip]{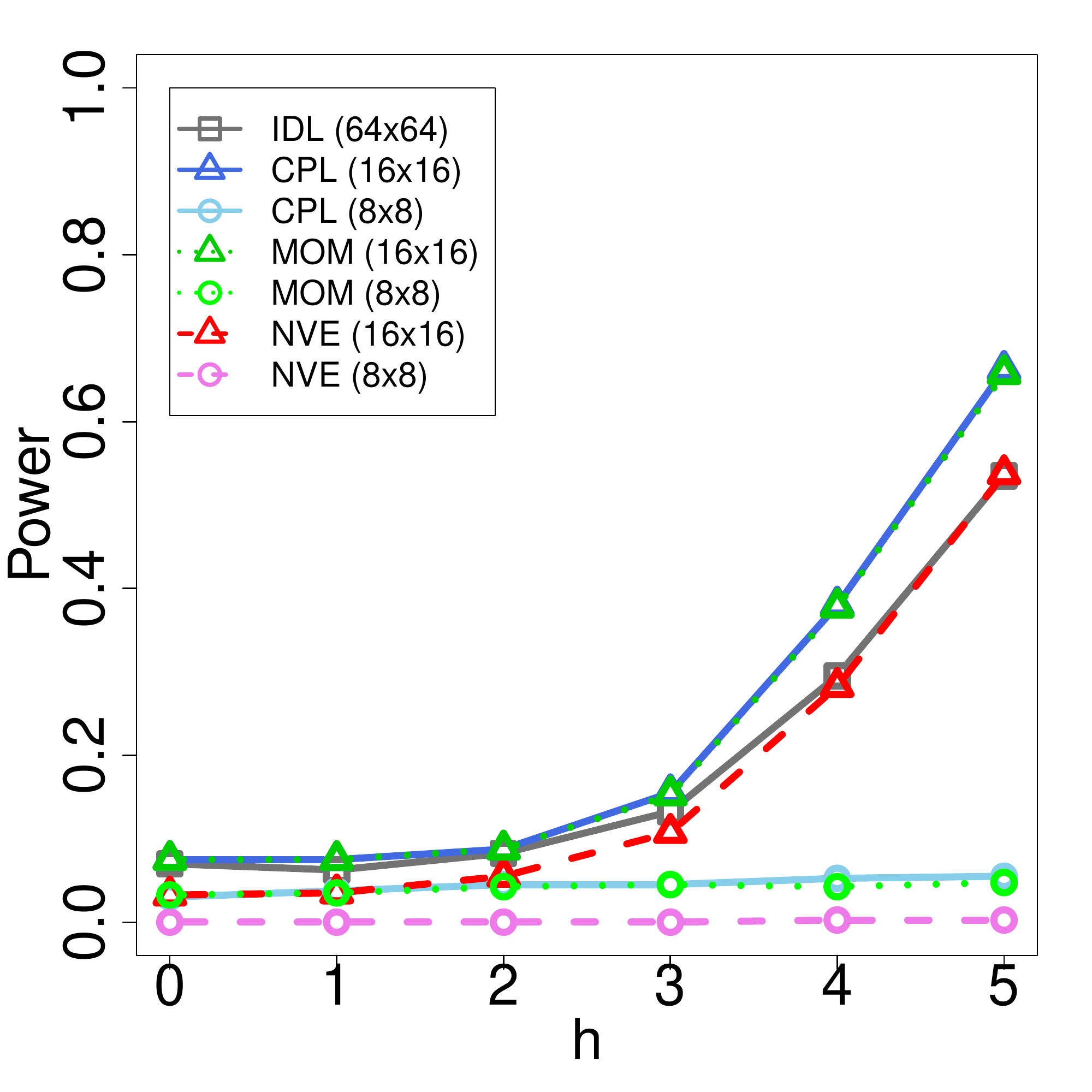} &
\includegraphics[scale=0.22,trim={0.18cm 0.3cm 1cm 1cm},clip]{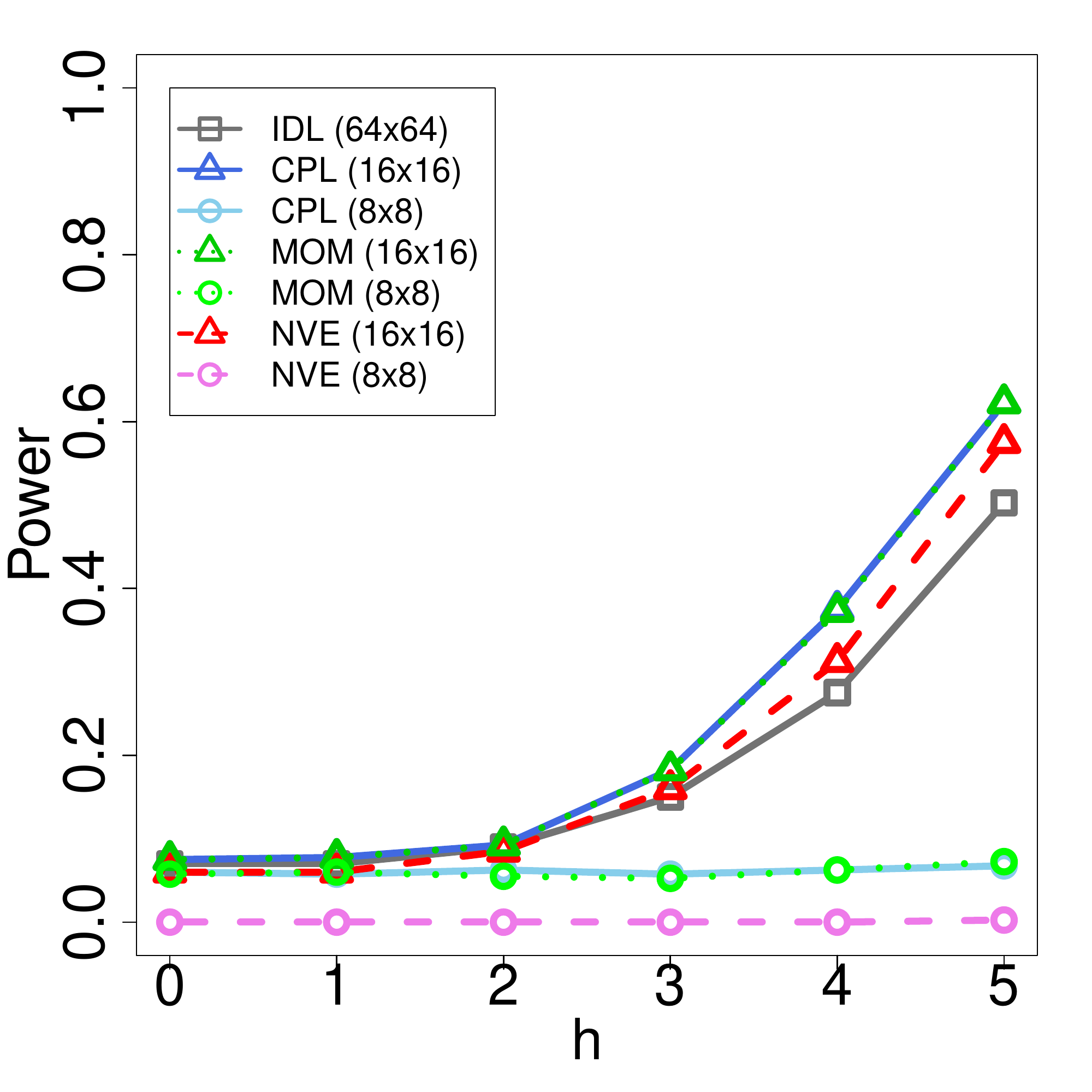} \\
\rotatebox{90}{$\quad \quad \quad r=8$}~~
\includegraphics[scale=0.22,trim={0.18cm 0.3cm 1cm 1cm},clip]{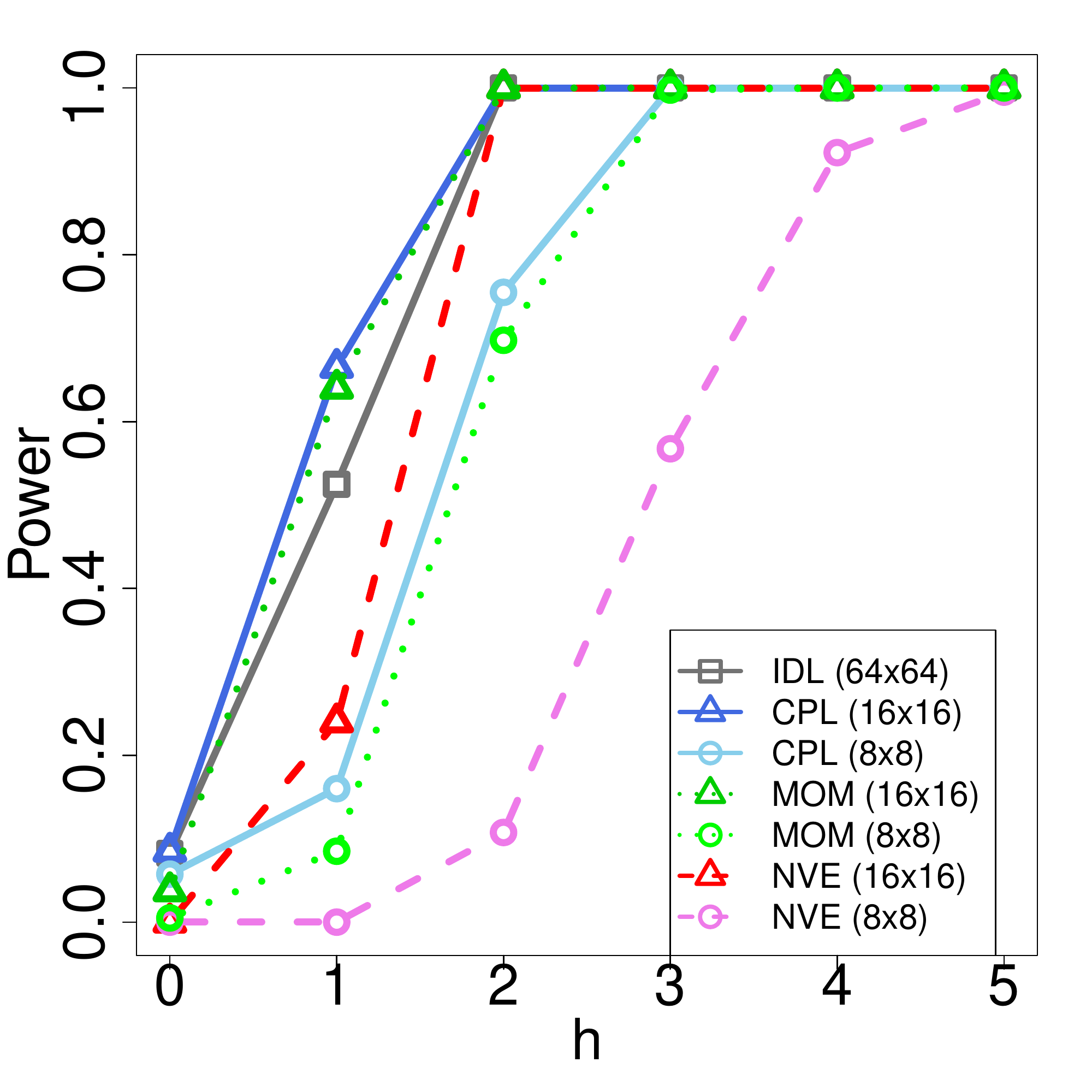} &
\includegraphics[scale=0.22,trim={0.18cm 0.3cm 1cm 1cm},clip]{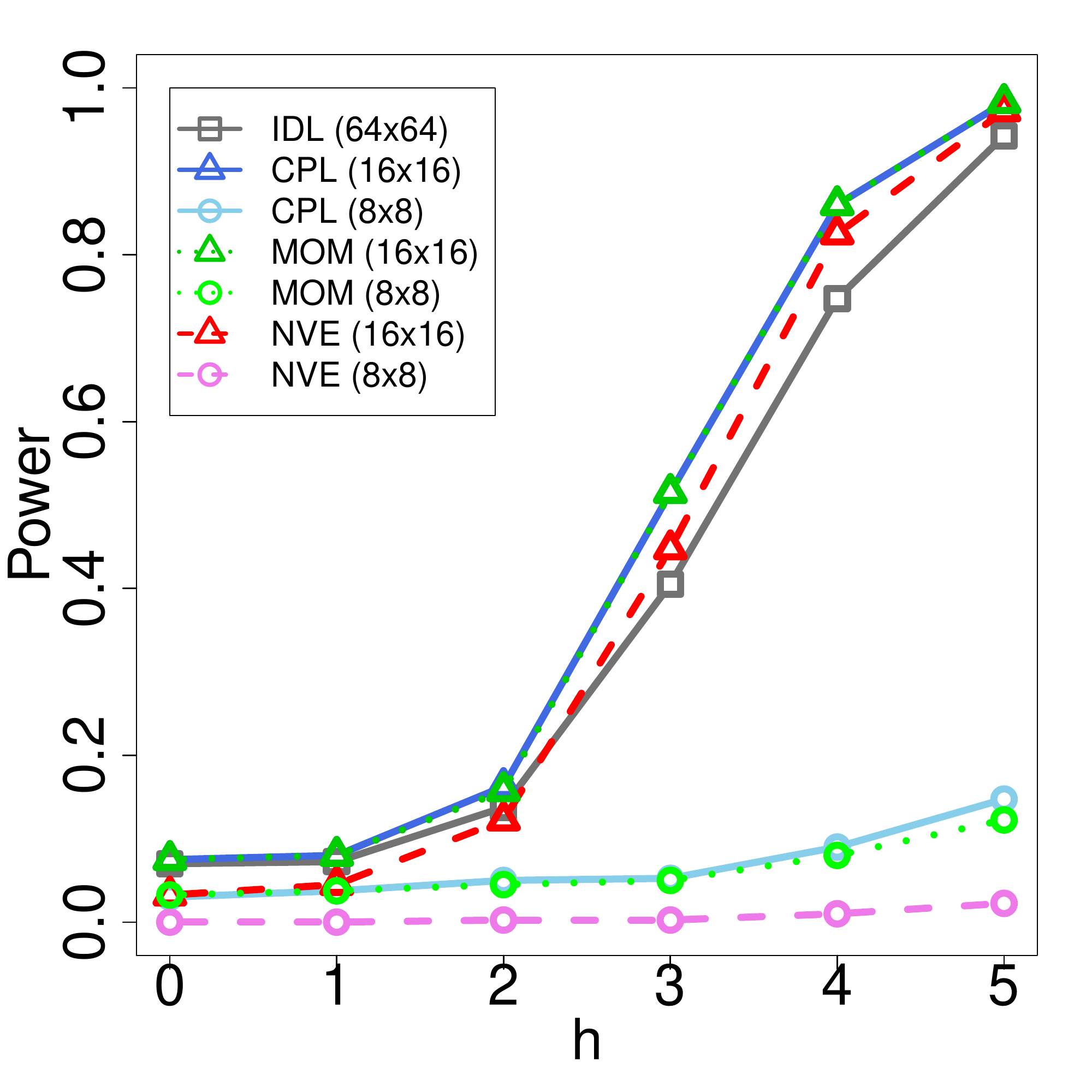} &
\includegraphics[scale=0.22,trim={0.18cm 0.3cm 1cm 1cm},clip]{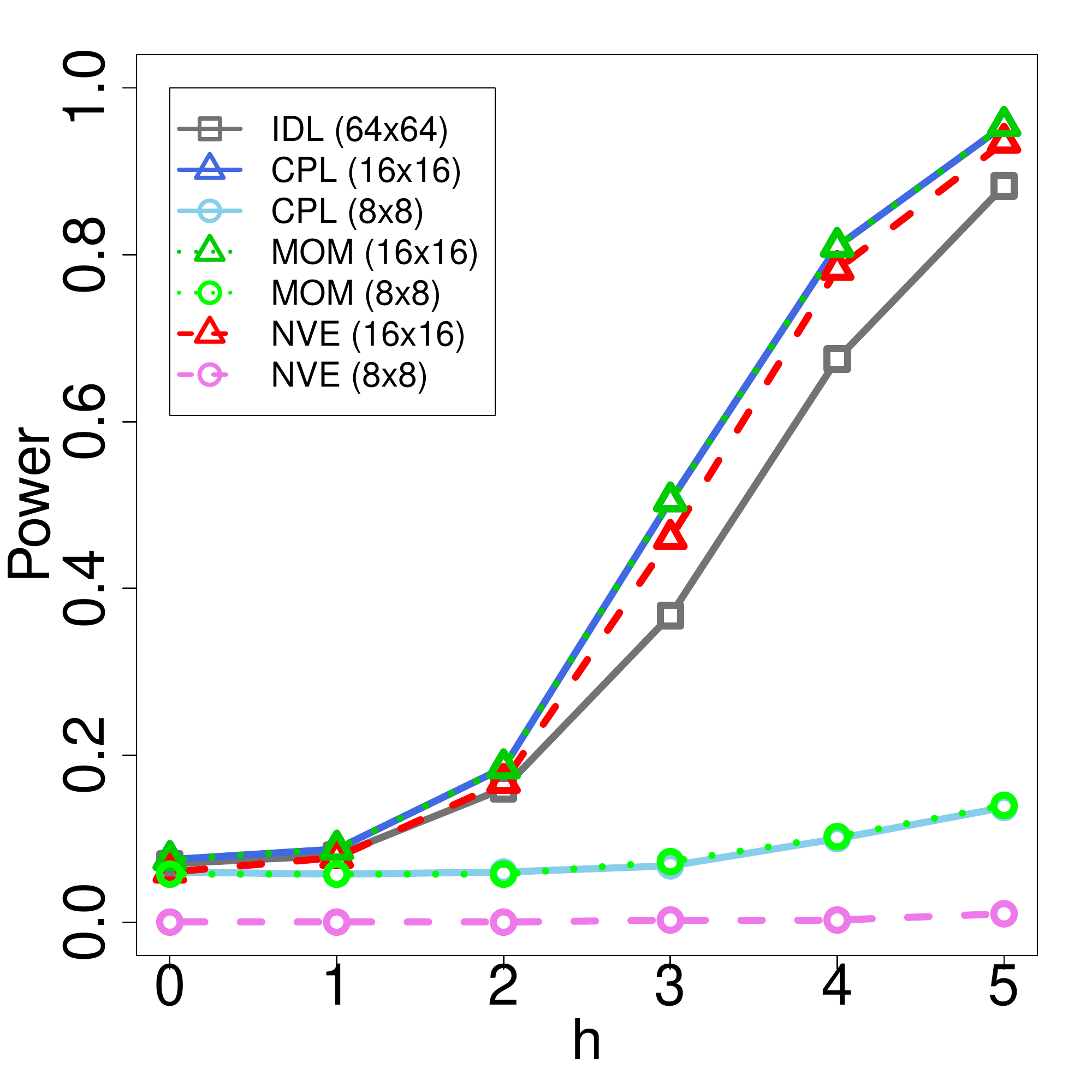} \\
\rotatebox{90}{$\quad \quad \quad r=10$}~~
\includegraphics[scale=0.22,trim={0.18cm 0.3cm 1cm 1cm},clip]{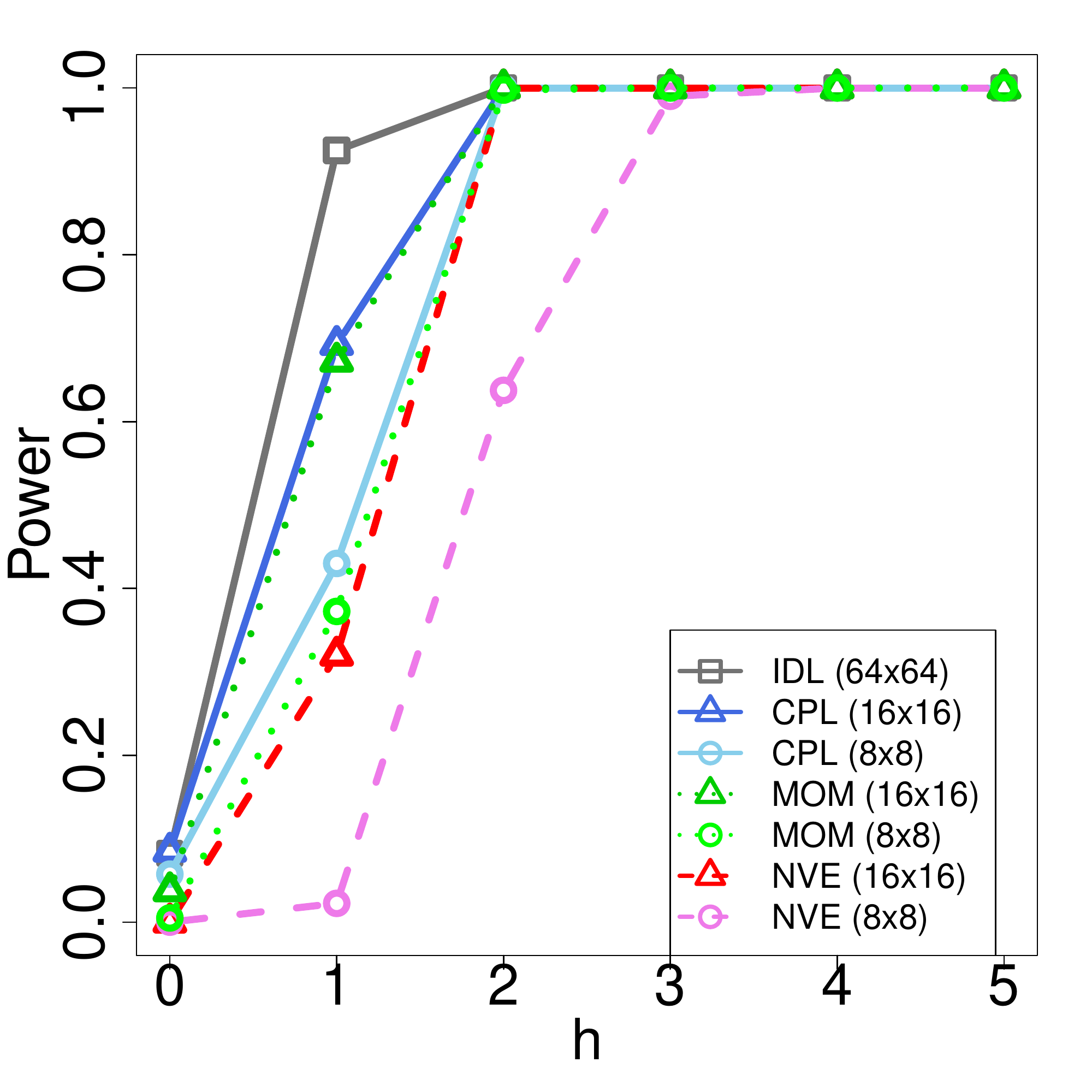} &
\includegraphics[scale=0.22,trim={0.18cm 0.3cm 1cm 1cm},clip]{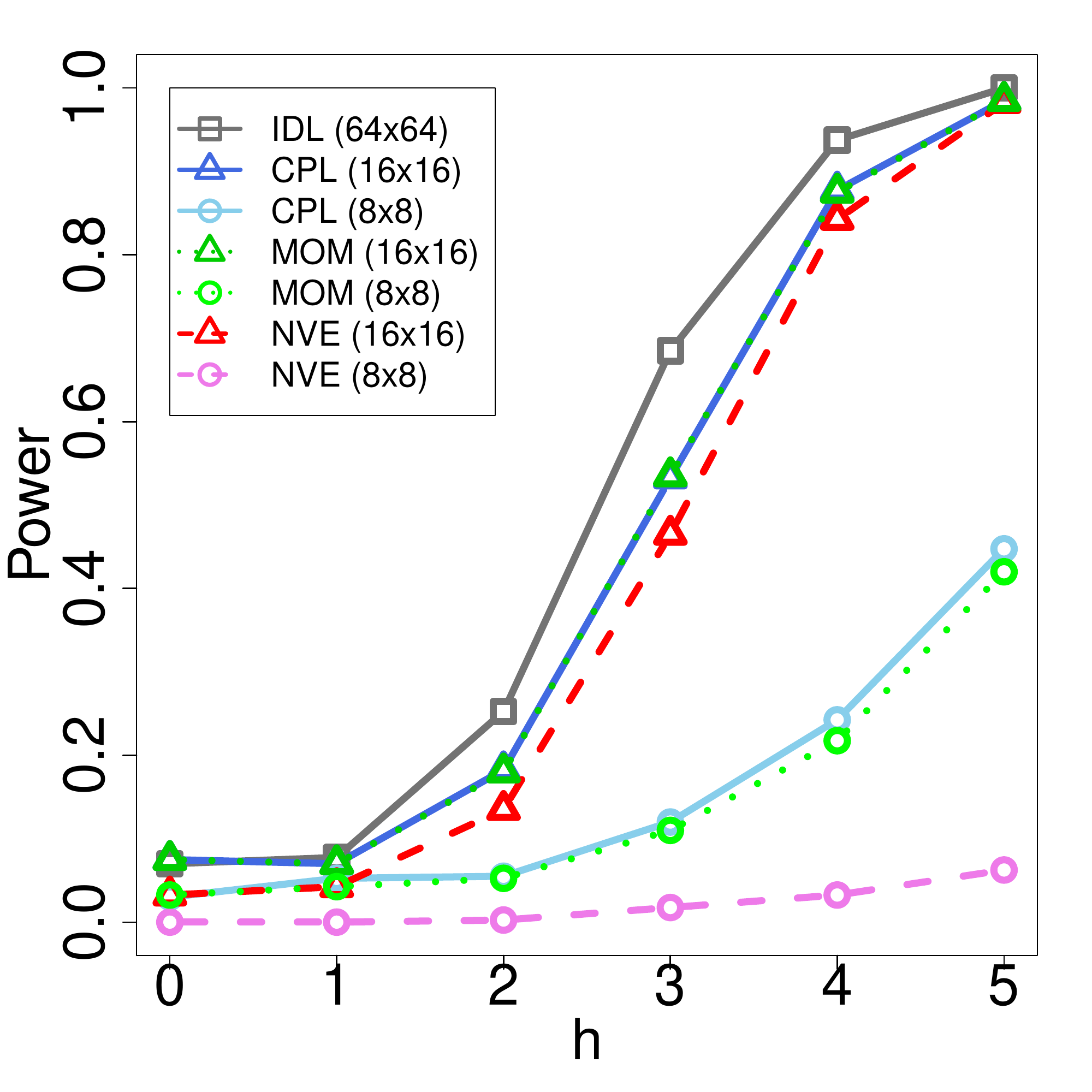} &
\includegraphics[scale=0.22,trim={0.18cm 0.3cm 1cm 1cm},clip]{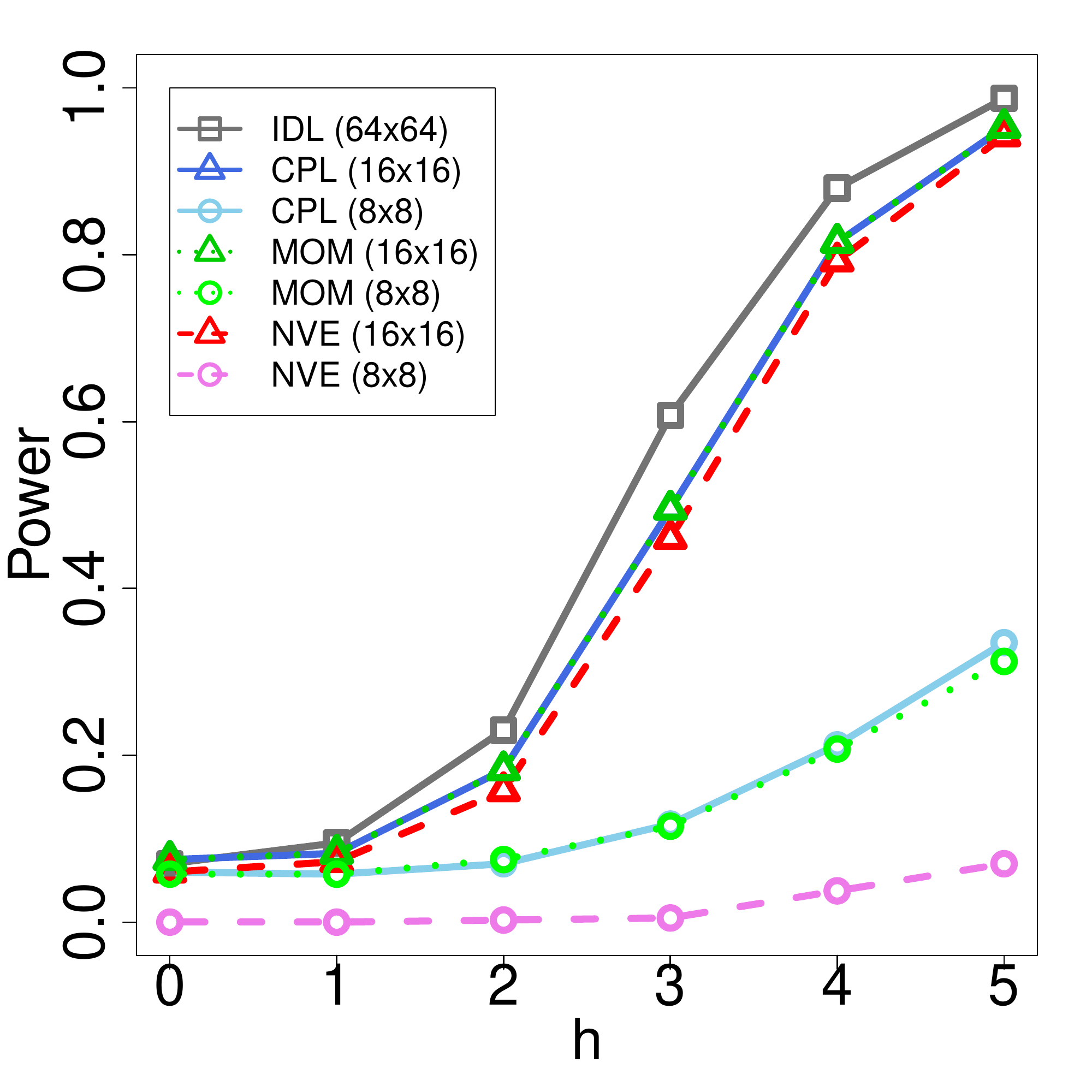} \\
\end{tabular}
\caption{Empirical power curves as a function of the signal's magnitude $h$, for various procedures for testing $H_0$ in Experiment 1
in Section \ref{sec:simulation}.
Down the rows, the curves correspond to different signal extents $r$,
while across the columns, the curves correspond to different spatial-dependence values $\phi$.}
\label{fig:power for experiment 1-2}
\end{figure}

\begin{figure}[!ptbh]\centering
\begin{tabular}{ccc}
~~~~~$\phi=0$ & ~~$\phi=5$ & ~~$\phi=10$
\smallskip\\
\rotatebox{90}{$\quad \quad \quad r=4$}~~
\includegraphics[scale=0.22,trim={0.18cm 0.3cm 1cm 1cm},clip]{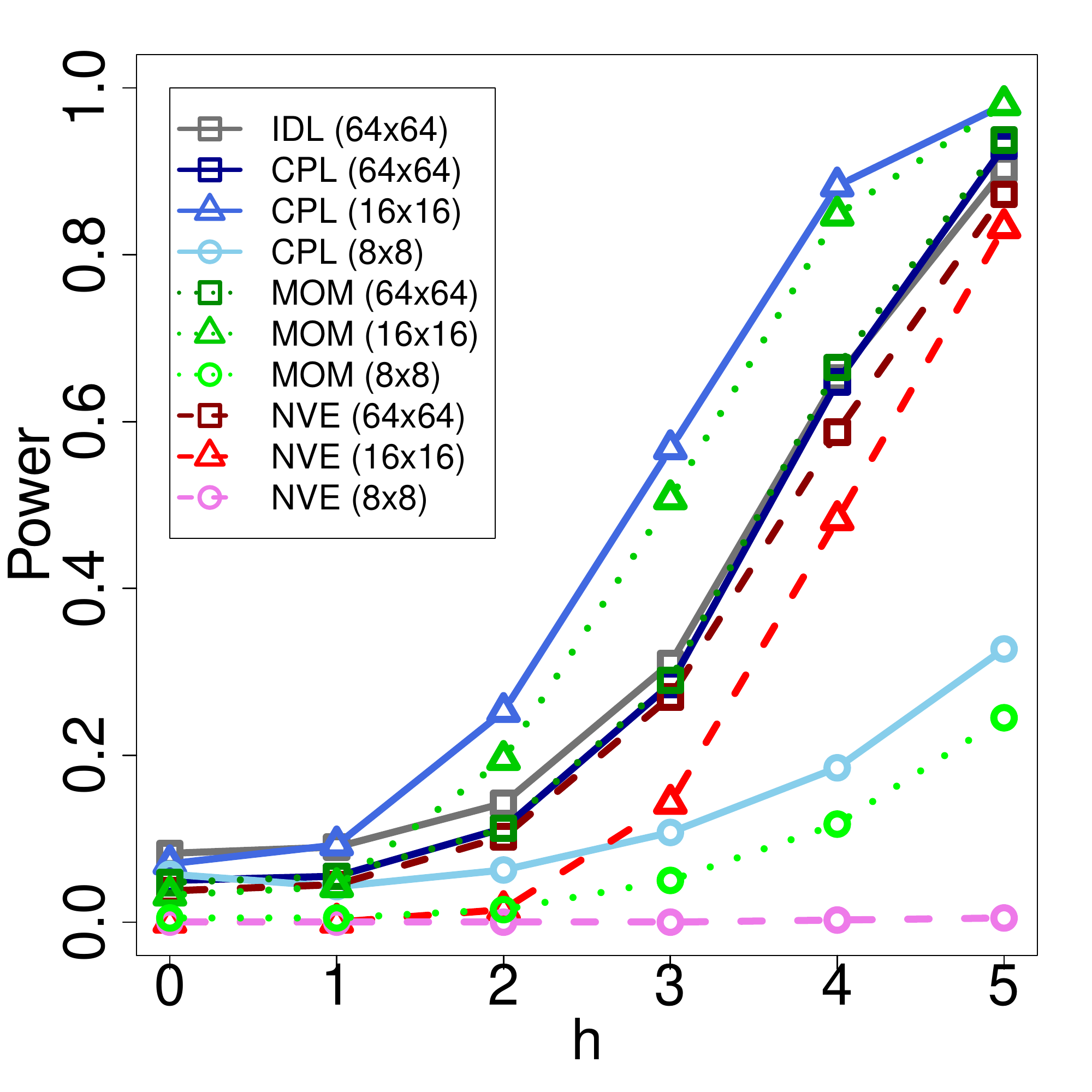} &
\includegraphics[scale=0.22,trim={0.18cm 0.3cm 1cm 1cm},clip]{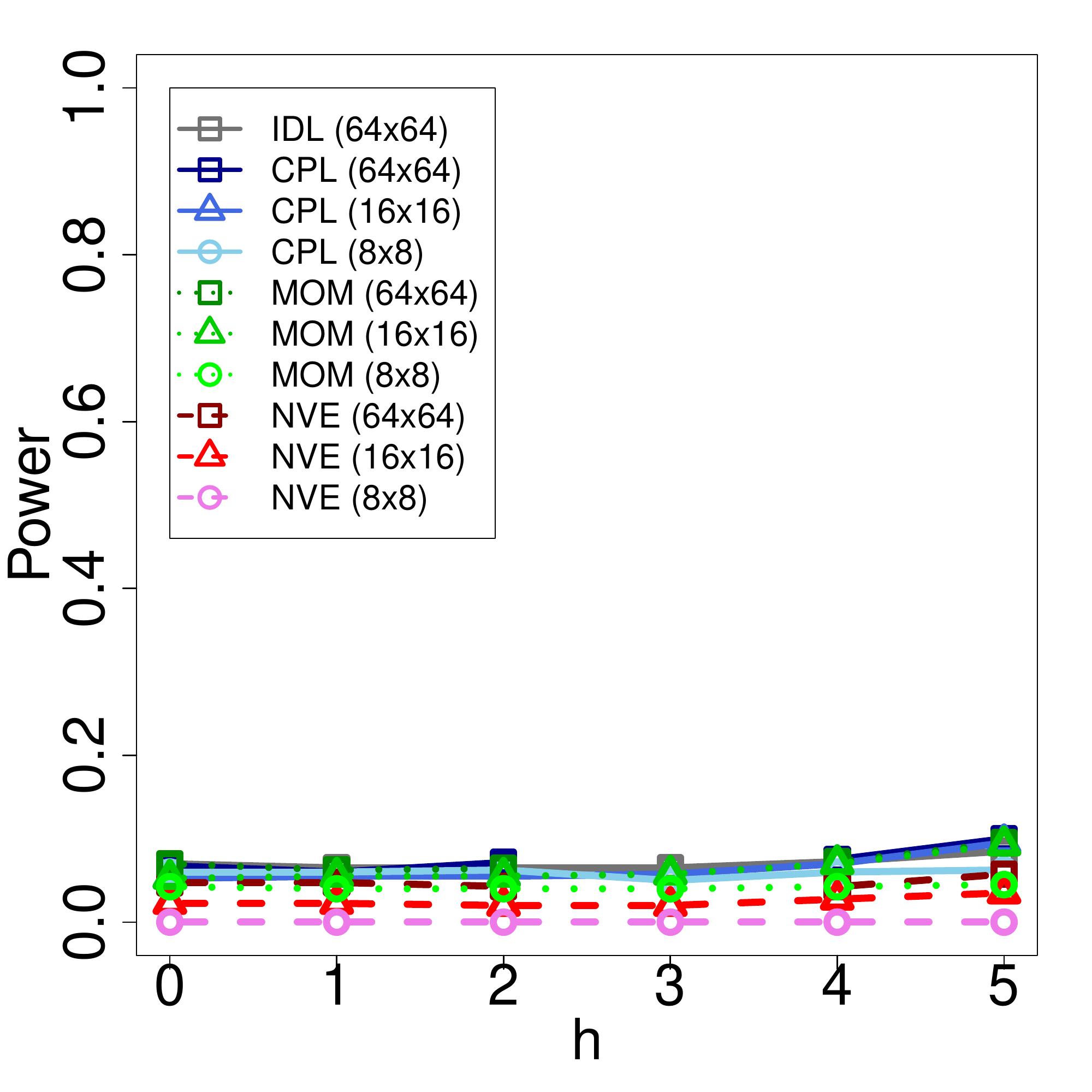} &
\includegraphics[scale=0.22,trim={0.18cm 0.3cm 1cm 1cm},clip]{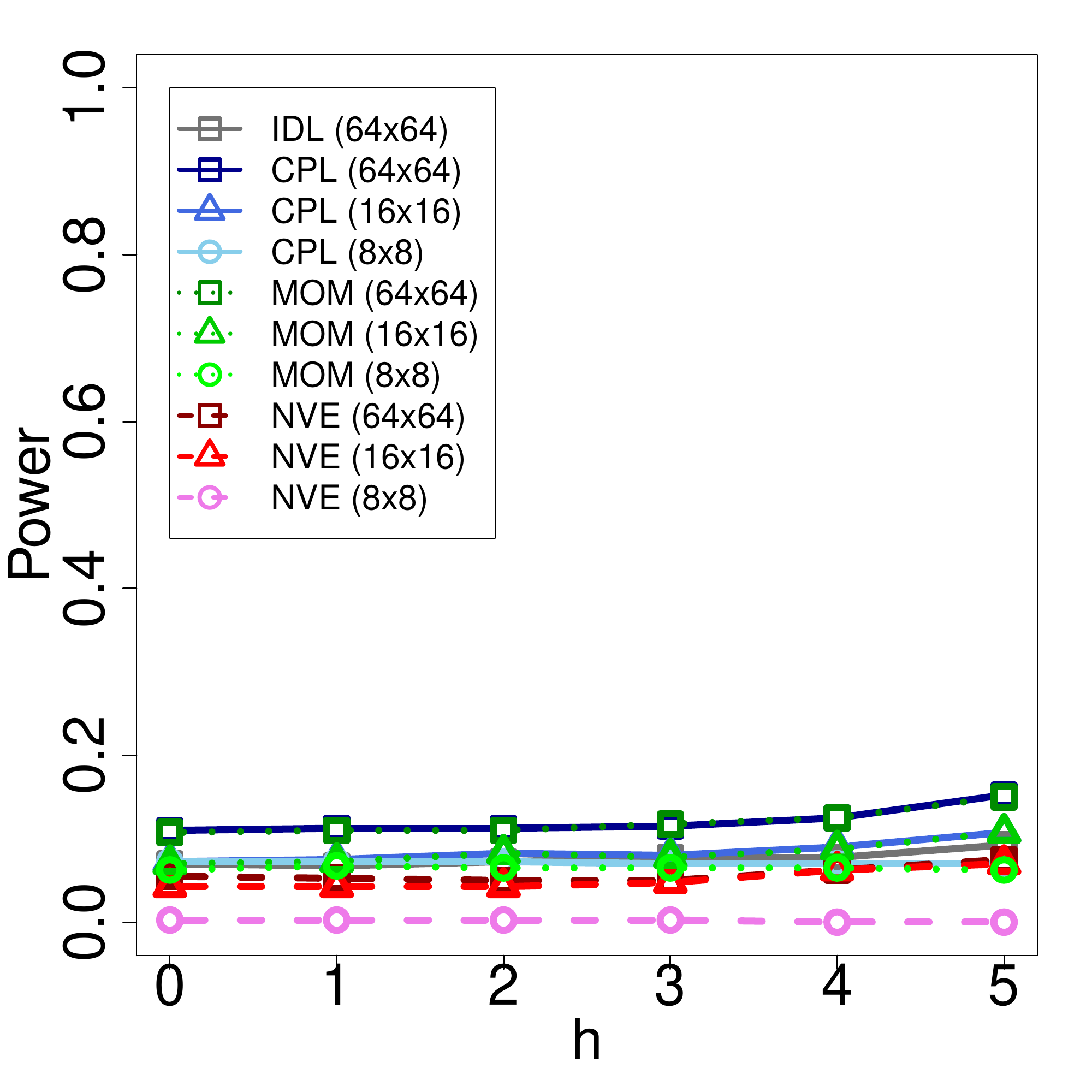} \\
\rotatebox{90}{$\quad \quad \quad r=6$}~~
\includegraphics[scale=0.22,trim={0.18cm 0.3cm 1cm 1cm},clip]{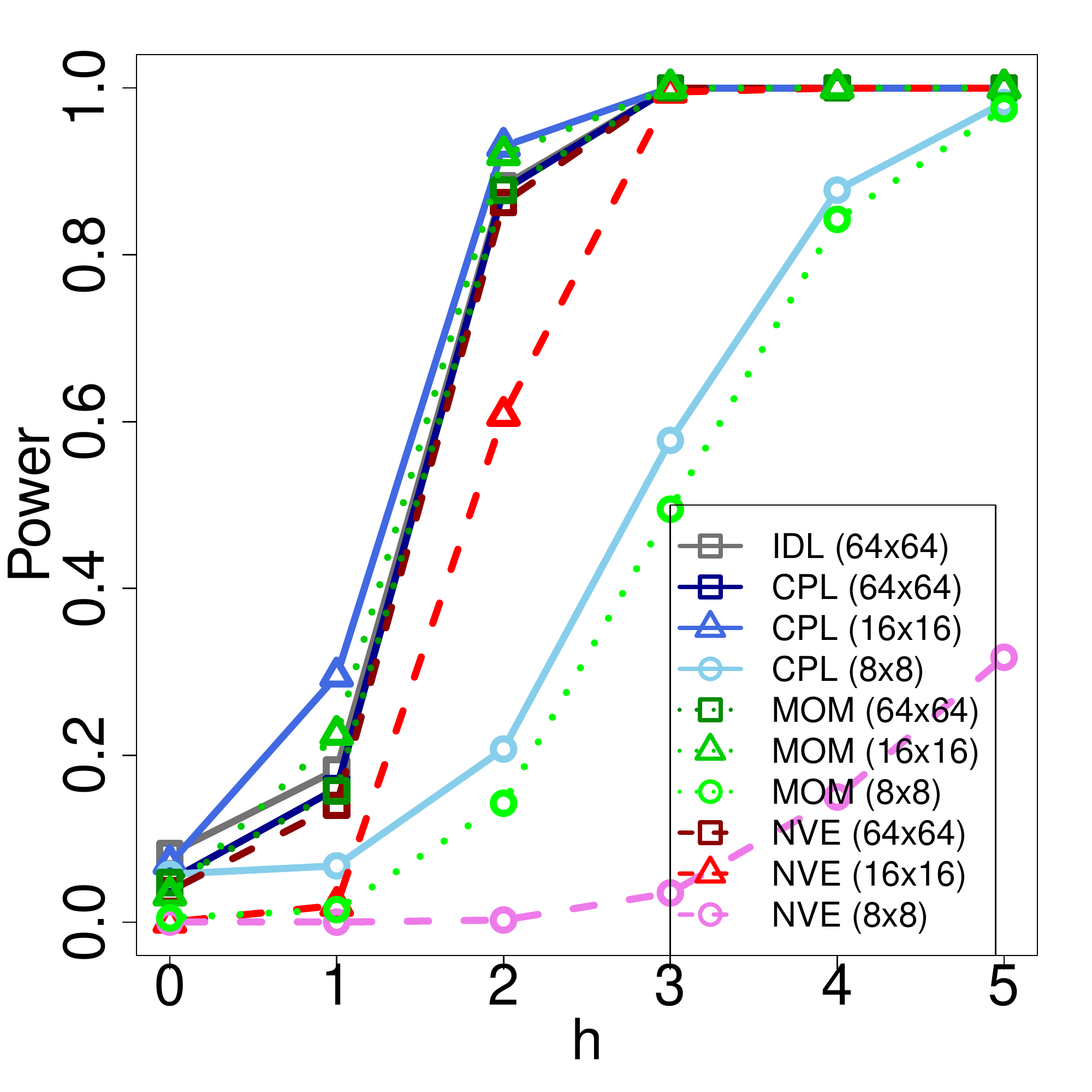} &
\includegraphics[scale=0.22,trim={0.18cm 0.3cm 1cm 1cm},clip]{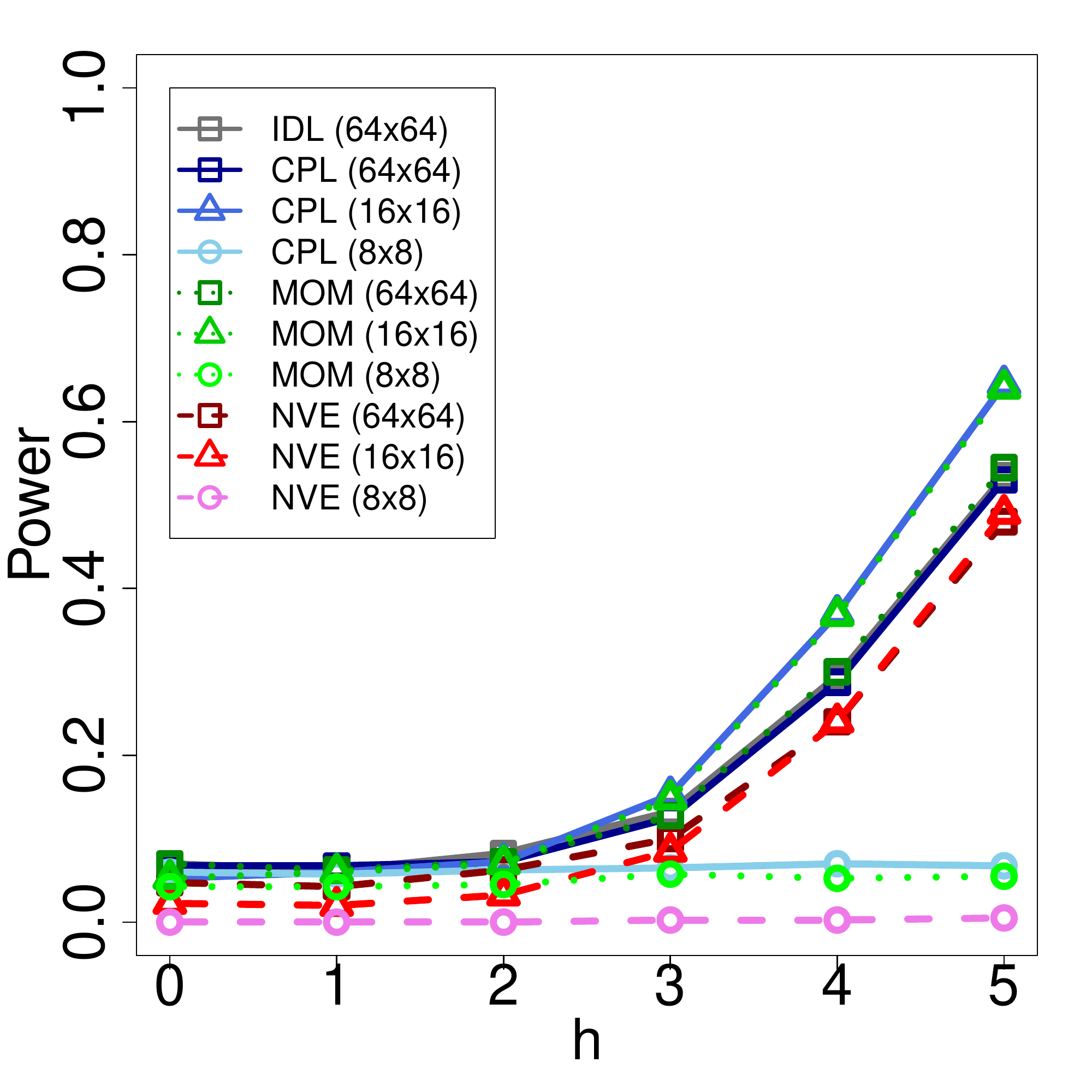} &
\includegraphics[scale=0.22,trim={0.18cm 0.3cm 1cm 1cm},clip]{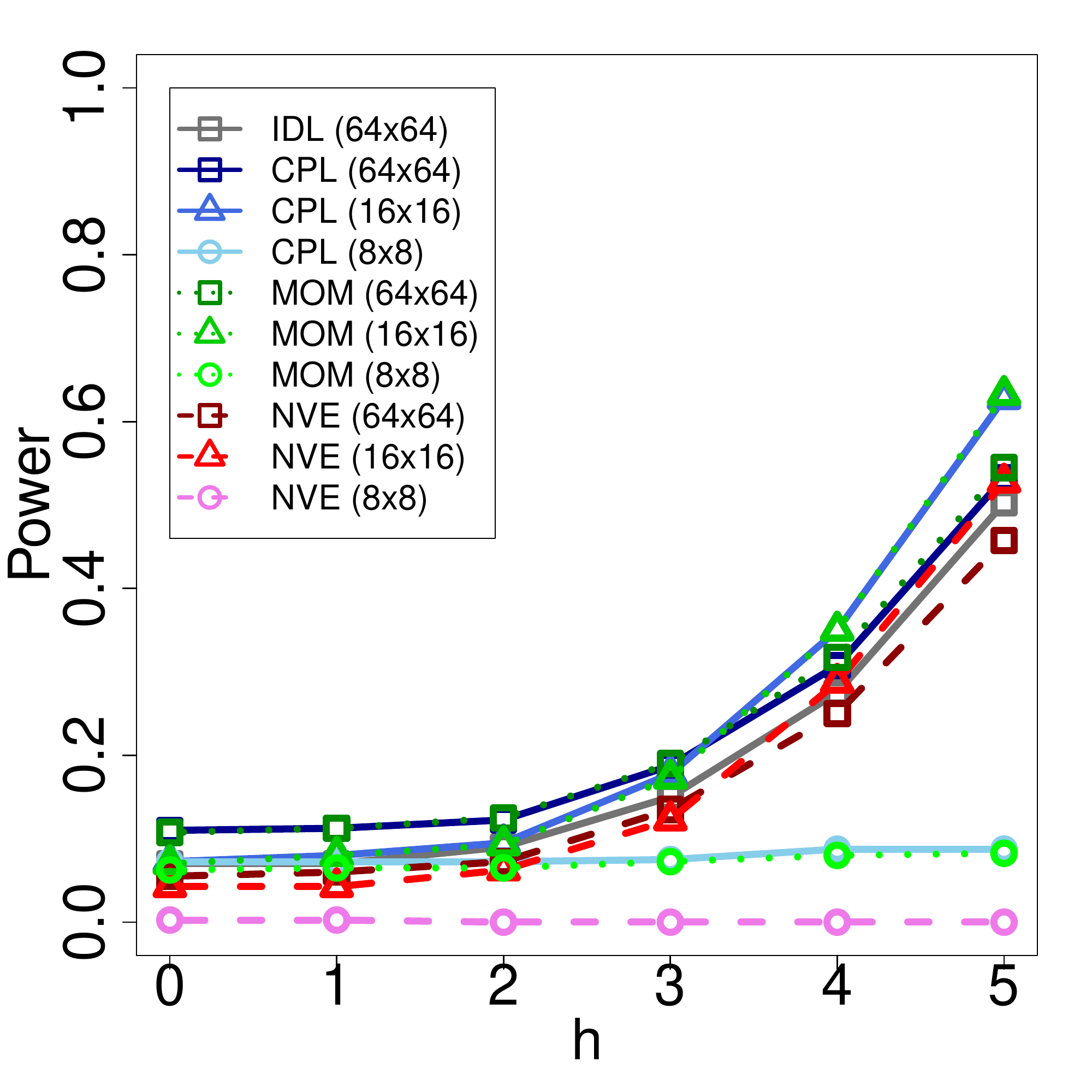} \\
\rotatebox{90}{$\quad \quad \quad r=8$}~~
\includegraphics[scale=0.22,trim={0.18cm 0.3cm 1cm 1cm},clip]{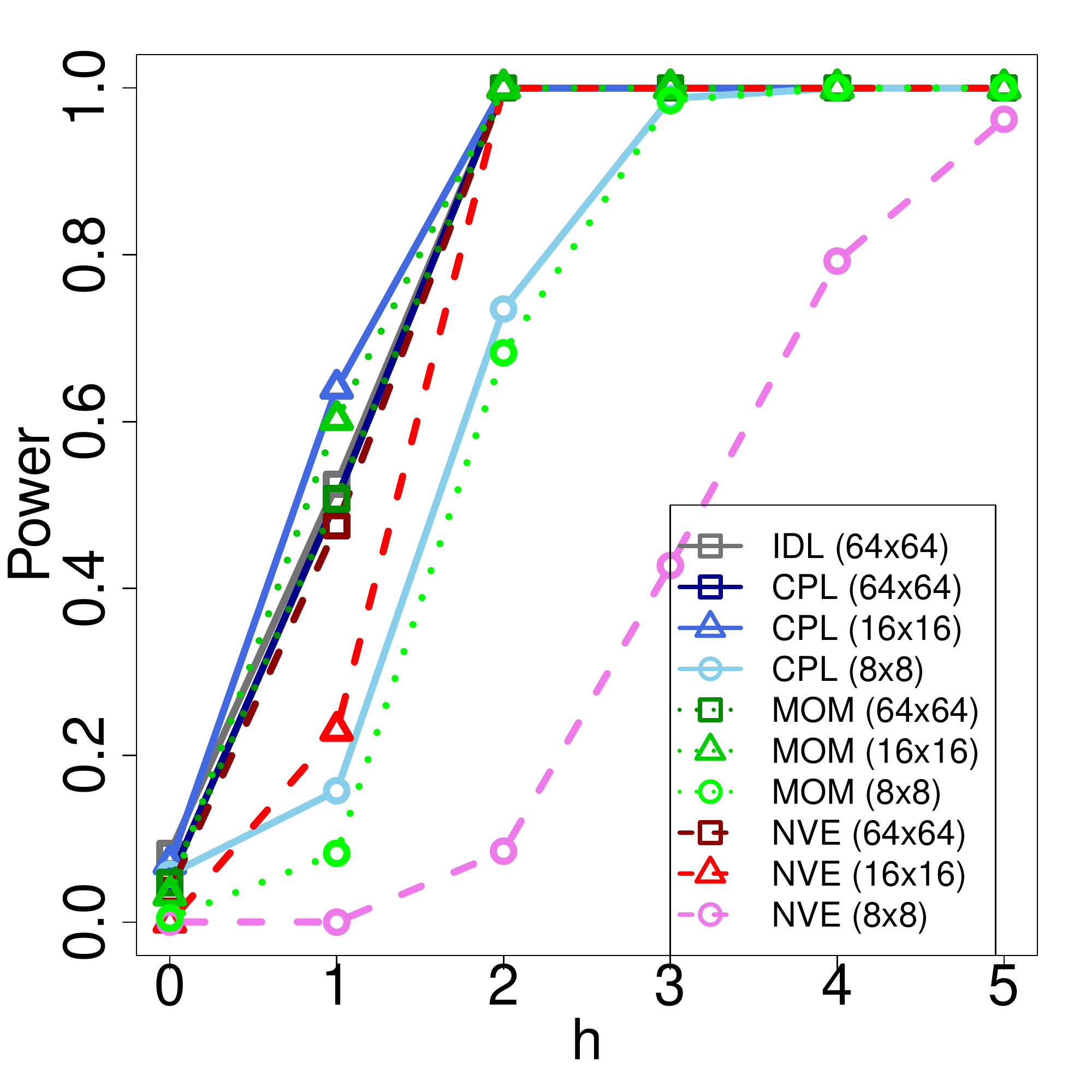} &
\includegraphics[scale=0.22,trim={0.18cm 0.3cm 1cm 1cm},clip]{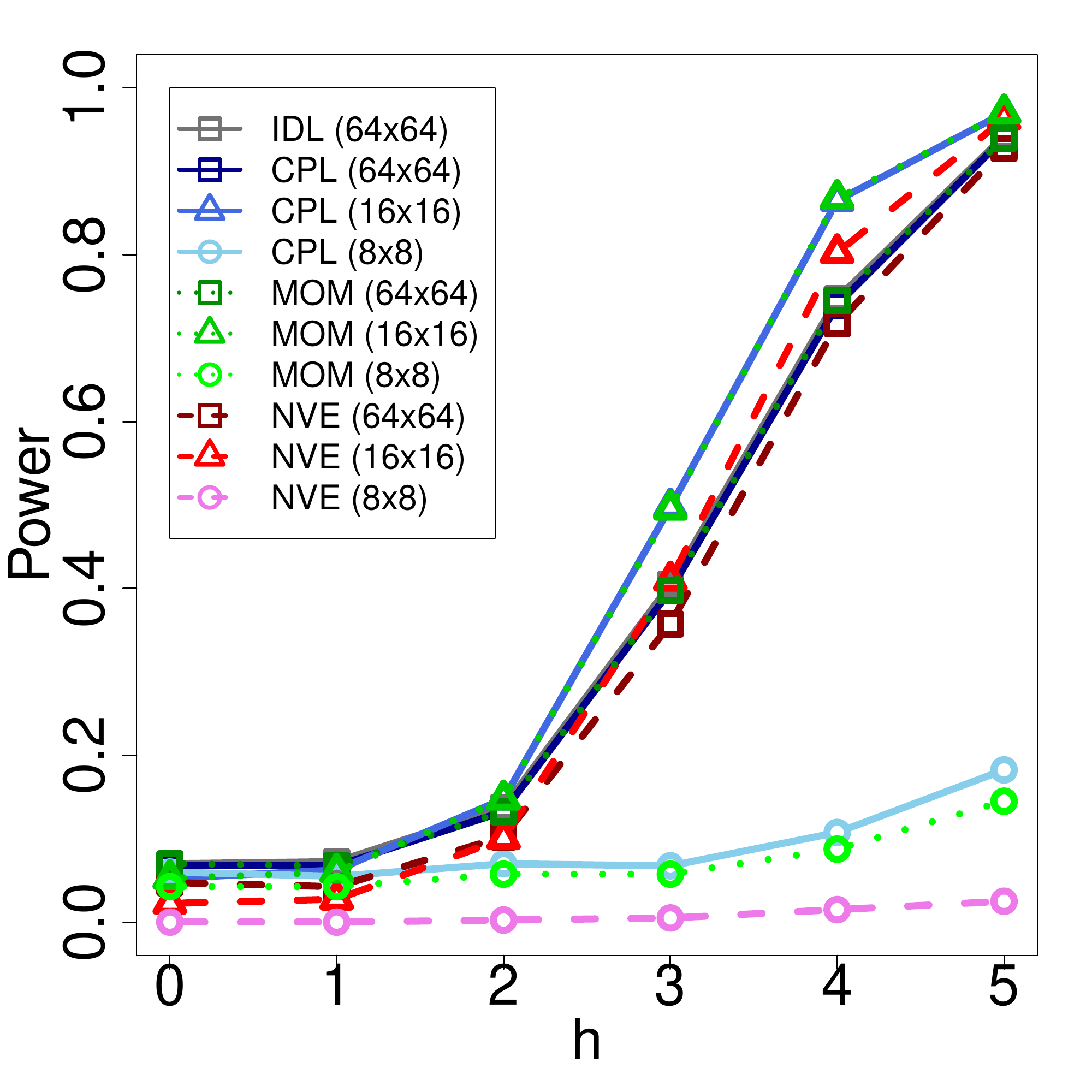} &
\includegraphics[scale=0.22,trim={0.18cm 0.3cm 1cm 1cm},clip]{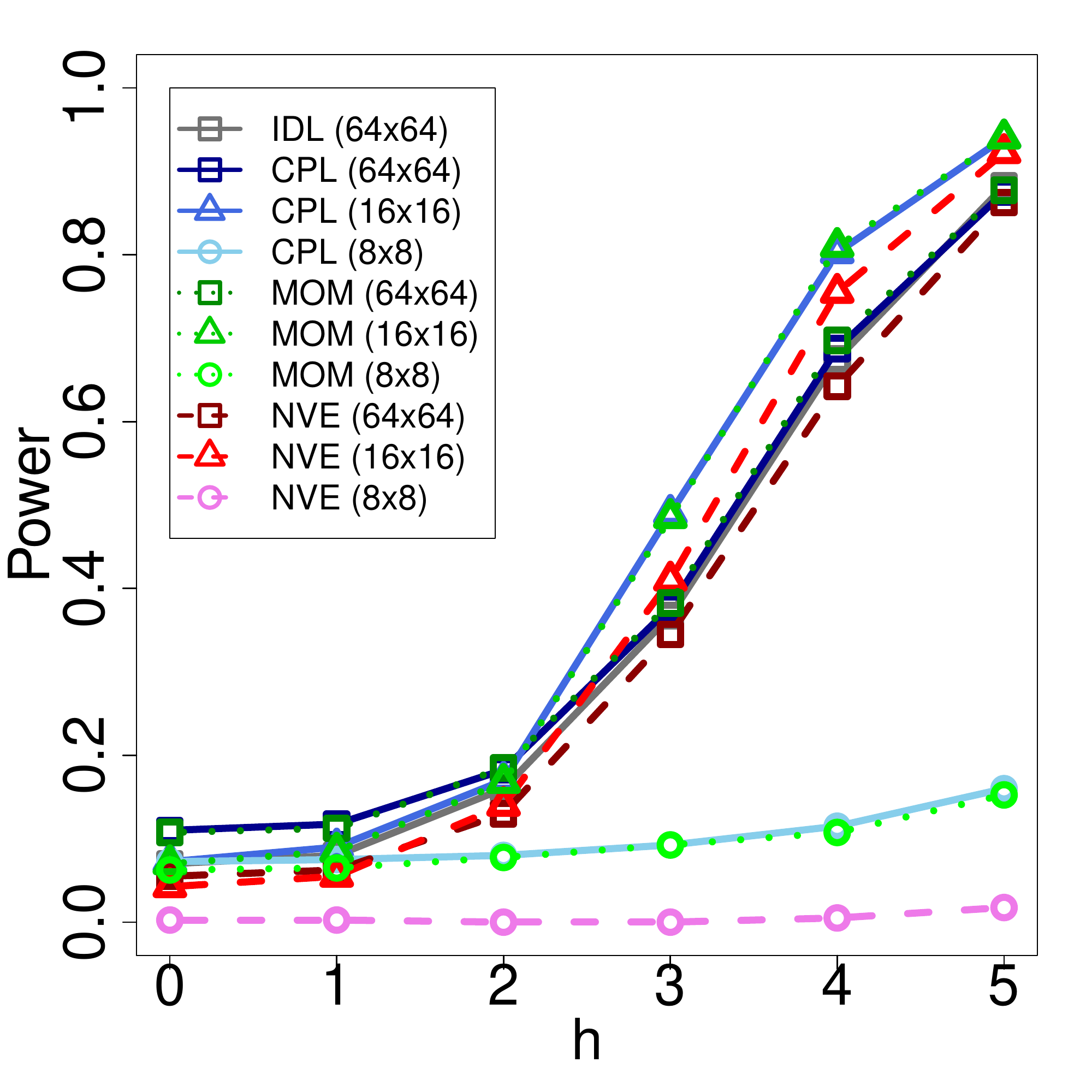} \\
\rotatebox{90}{$\quad \quad \quad r=10$}~~
\includegraphics[scale=0.22,trim={0.18cm 0.3cm 1cm 1cm},clip]{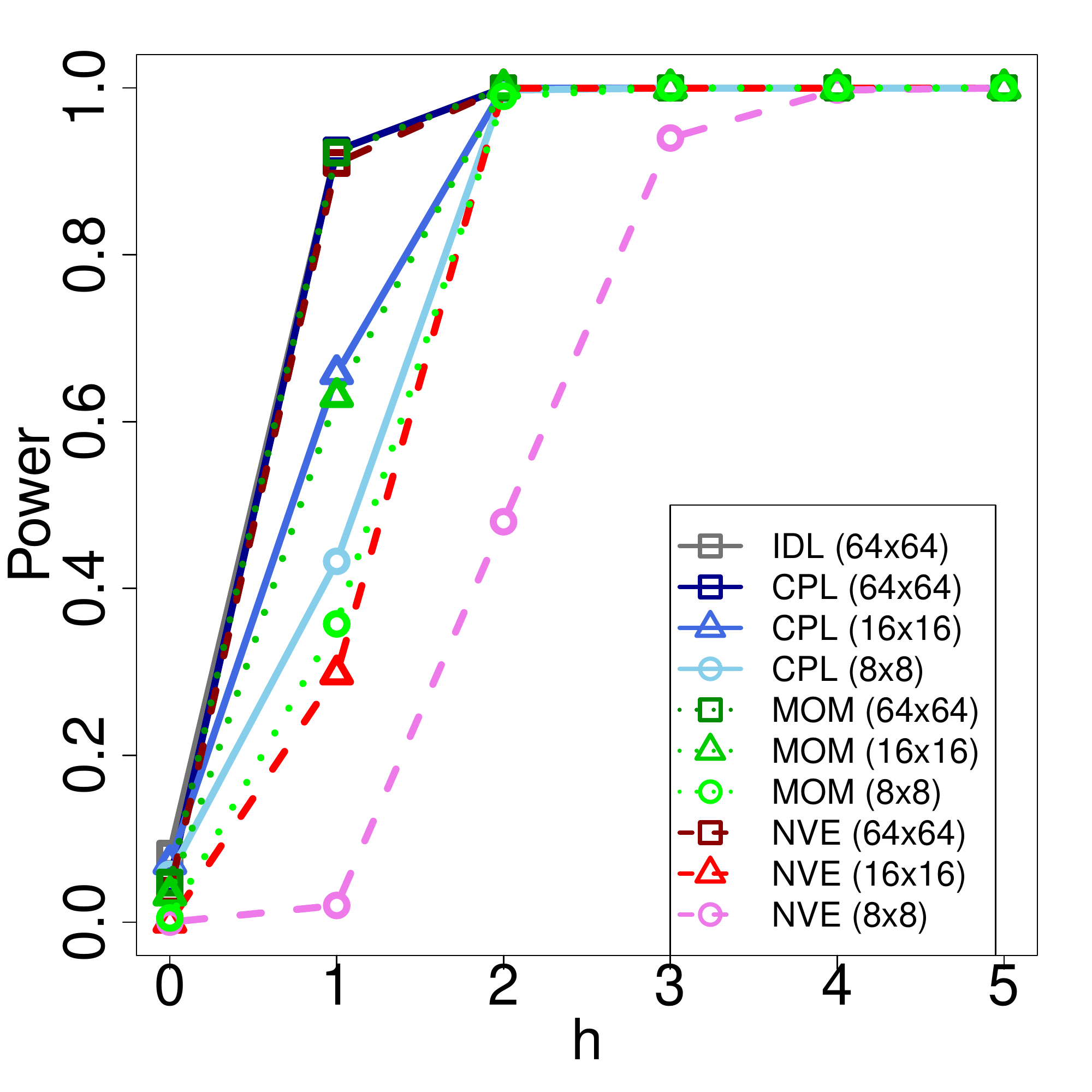} &
\includegraphics[scale=0.22,trim={0.18cm 0.3cm 1cm 1cm},clip]{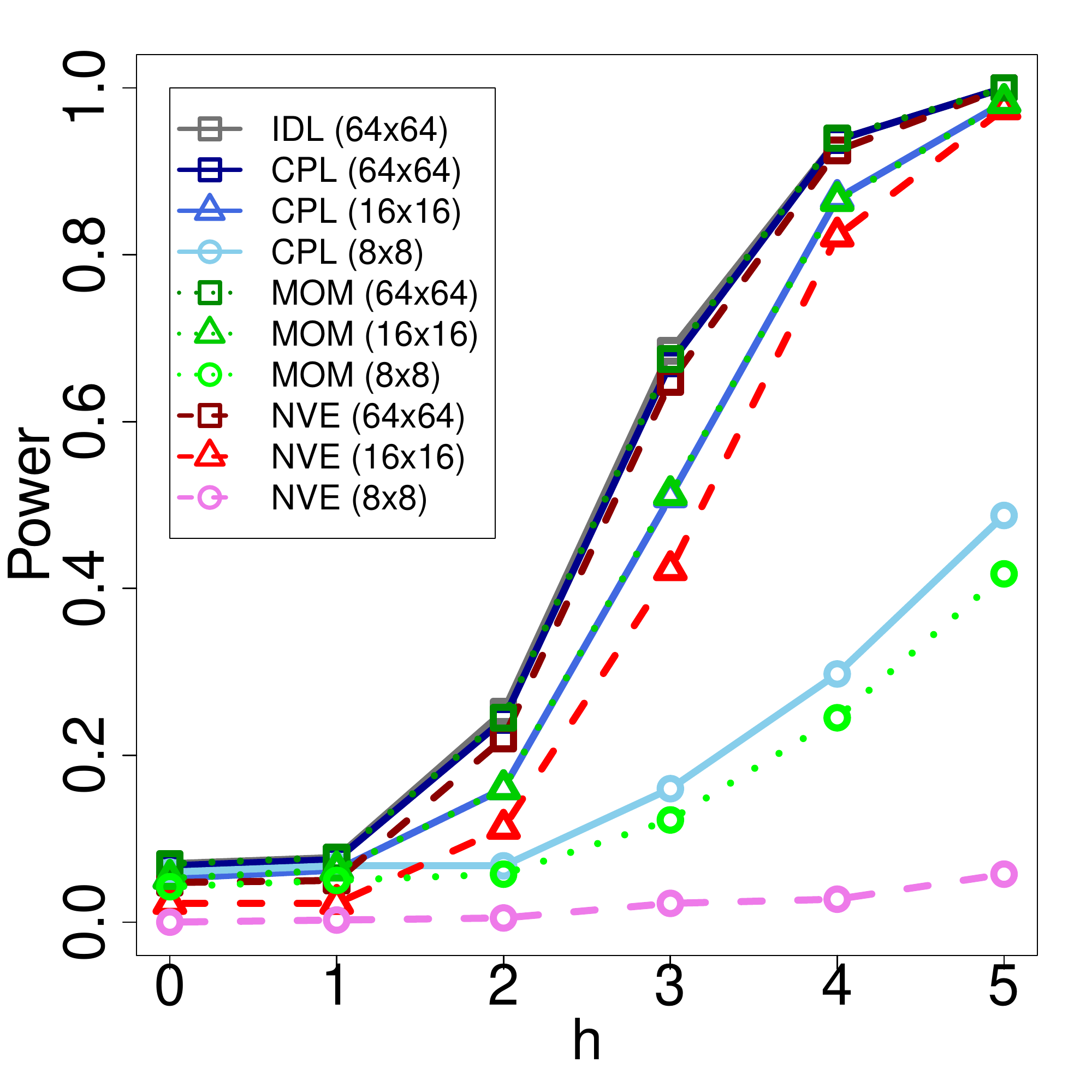} &
\includegraphics[scale=0.22,trim={0.18cm 0.3cm 1cm 1cm},clip]{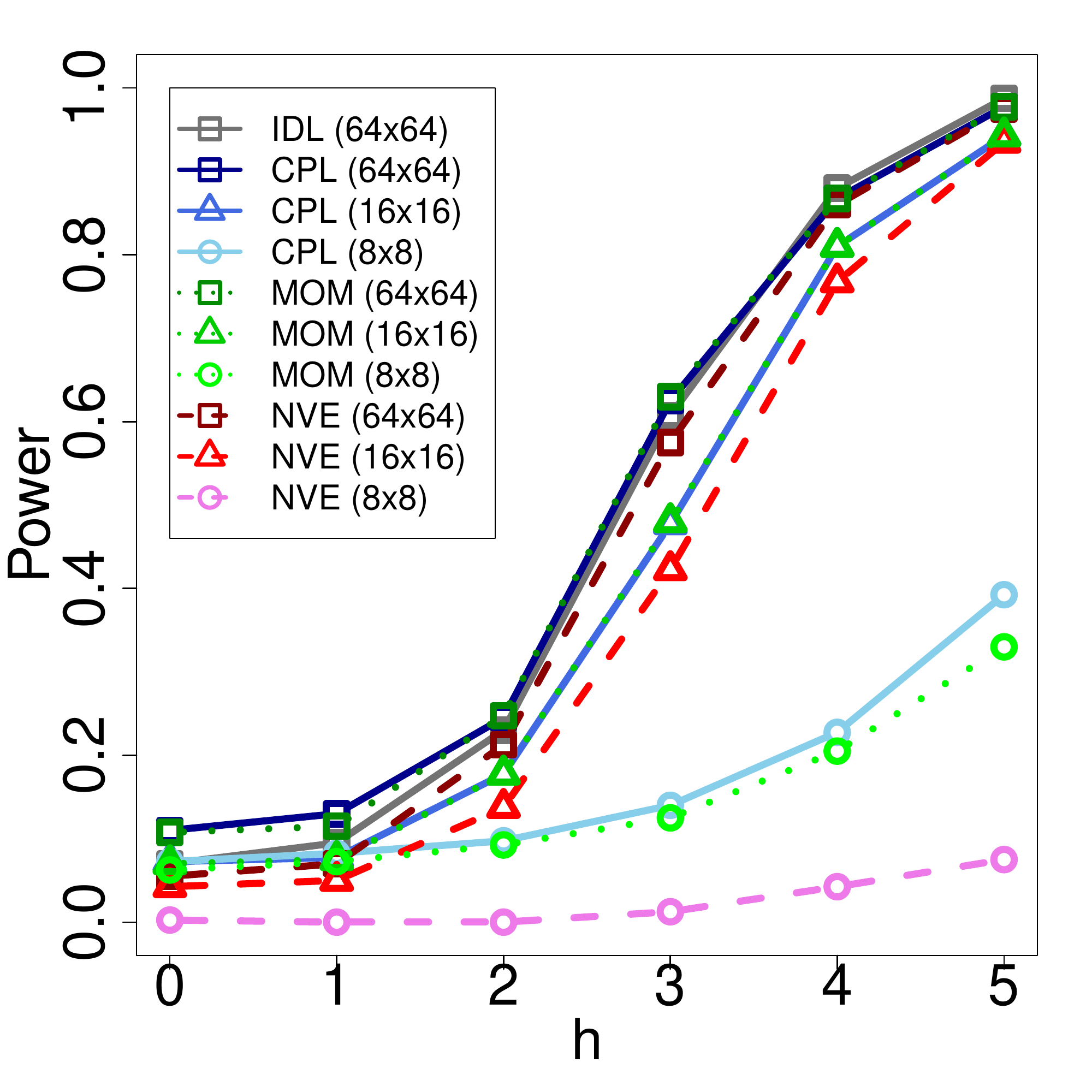} \\
\end{tabular}
\caption{Empirical power curves as a function of the signal's magnitude $h$, for various procedures for testing $H_0$ in Experiment 2
in Section \ref{sec:simulation}.
Down the rows, the curves correspond to different signal extents $r$,
while across the columns, the curves correspond to different spatial-dependence values $\phi$.}
\label{fig:power for experiment 2-2}
\end{figure}

\begin{figure}[!ptbh]\centering
\begin{tabular}{ccc}
~~~~~$\phi=0$ & ~~$\phi=5$ & ~~$\phi=10$
\smallskip\\
\rotatebox{90}{$\quad \quad \quad r=4$}~~
\includegraphics[scale=0.22,trim={0.18cm 0.3cm 1cm 1cm},clip]{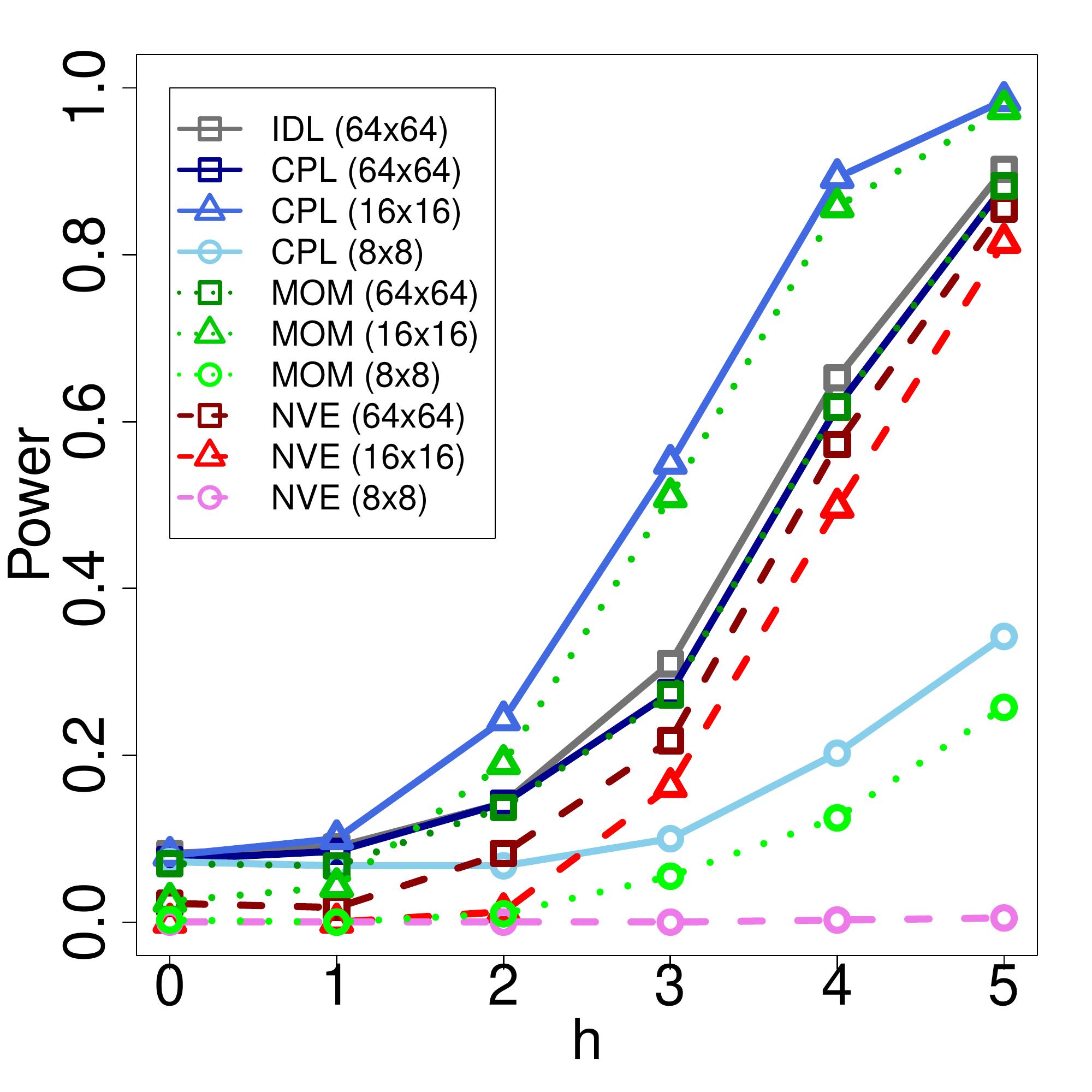} &
\includegraphics[scale=0.22,trim={0.18cm 0.3cm 1cm 1cm},clip]{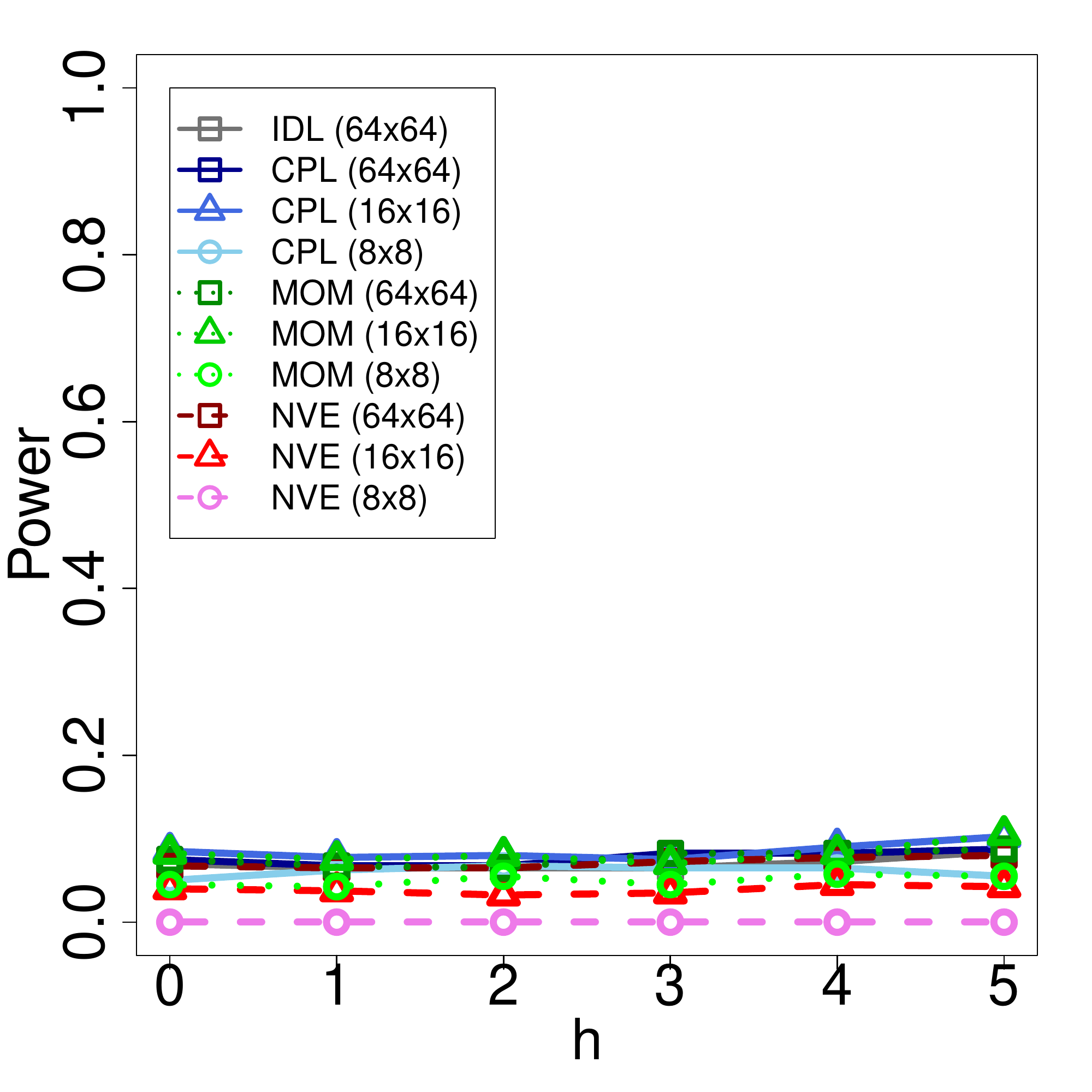} &
\includegraphics[scale=0.22,trim={0.18cm 0.3cm 1cm 1cm},clip]{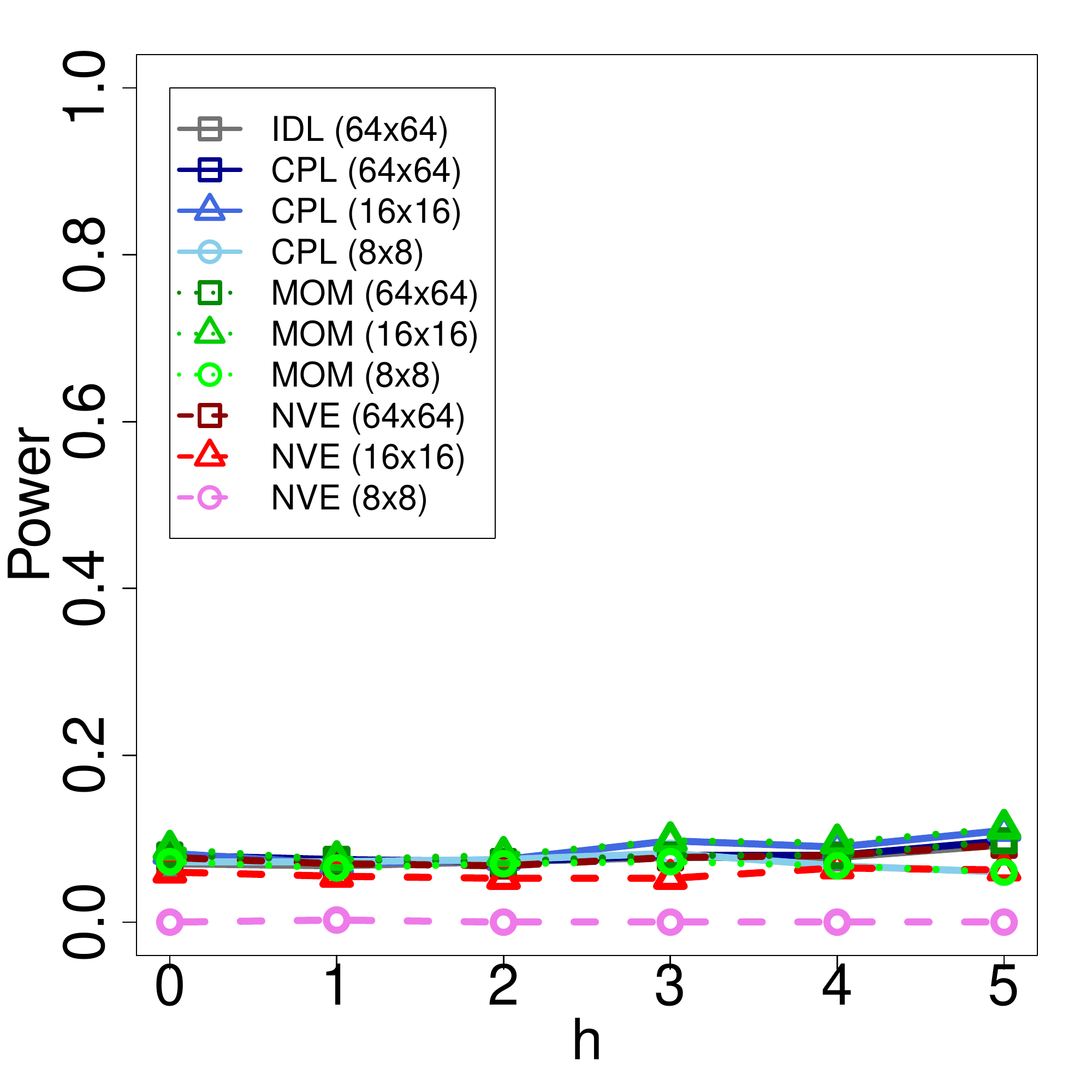} \\
\rotatebox{90}{$\quad \quad \quad r=6$}~~
\includegraphics[scale=0.22,trim={0.18cm 0.3cm 1cm 1cm},clip]{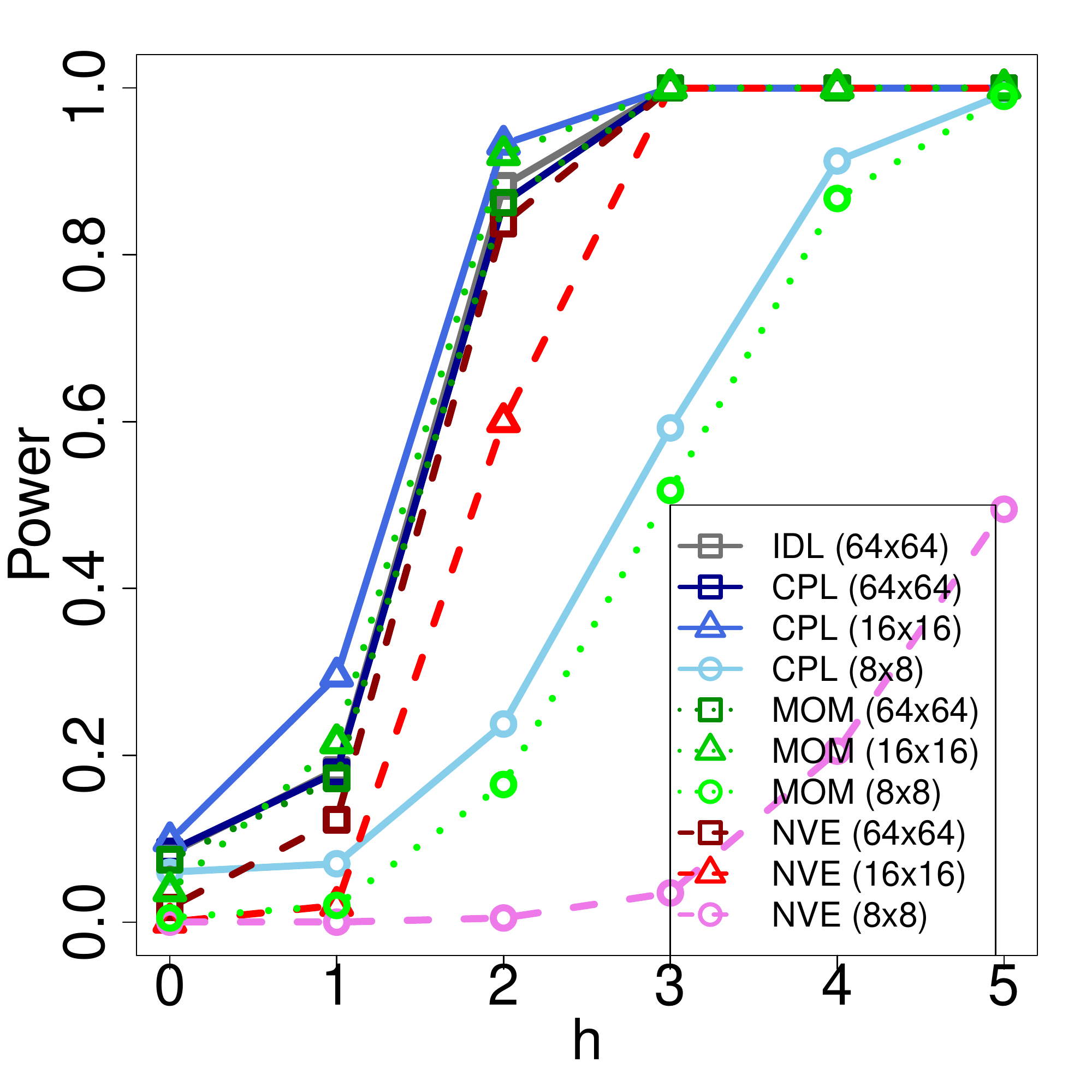} &
\includegraphics[scale=0.22,trim={0.18cm 0.3cm 1cm 1cm},clip]{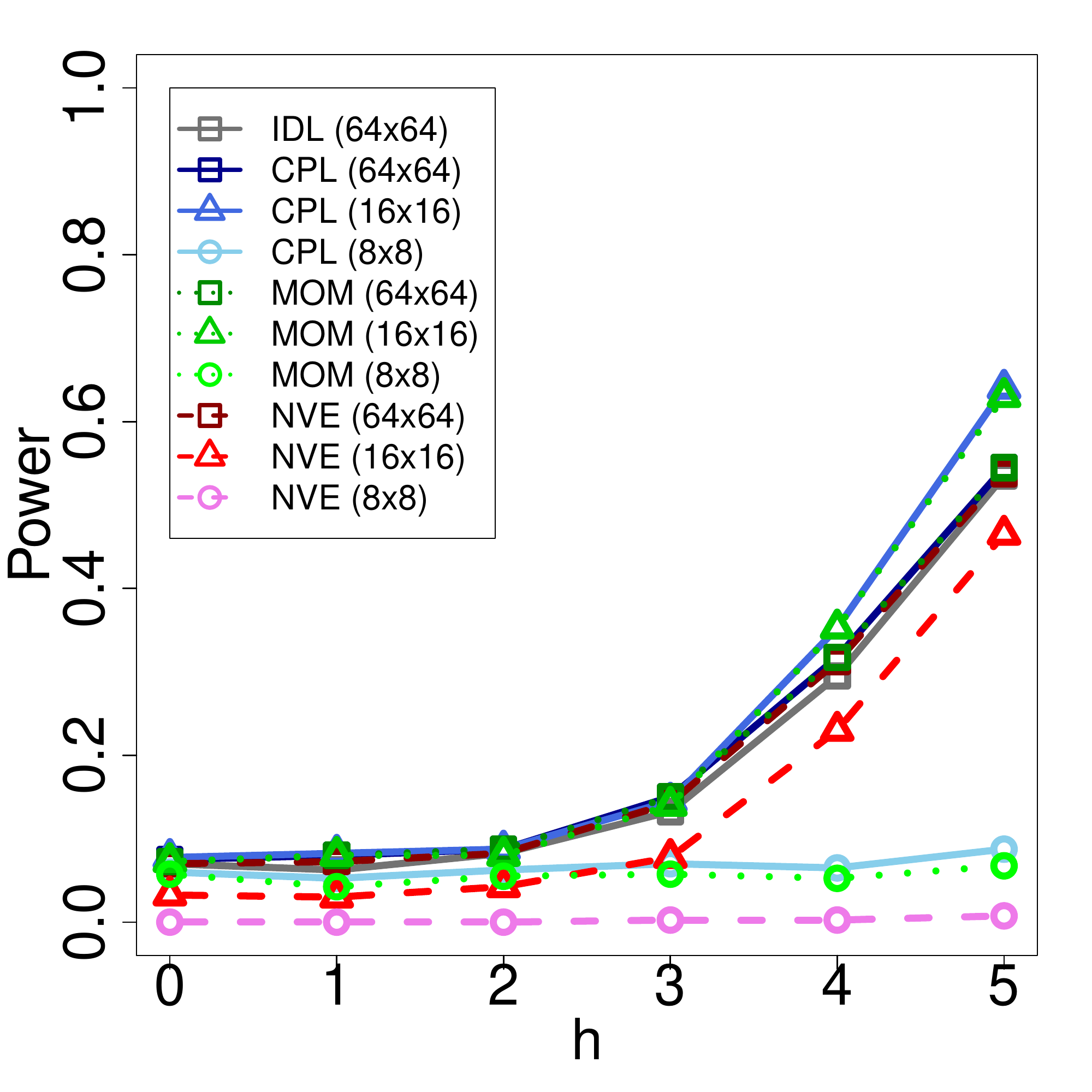} &
\includegraphics[scale=0.22,trim={0.18cm 0.3cm 1cm 1cm},clip]{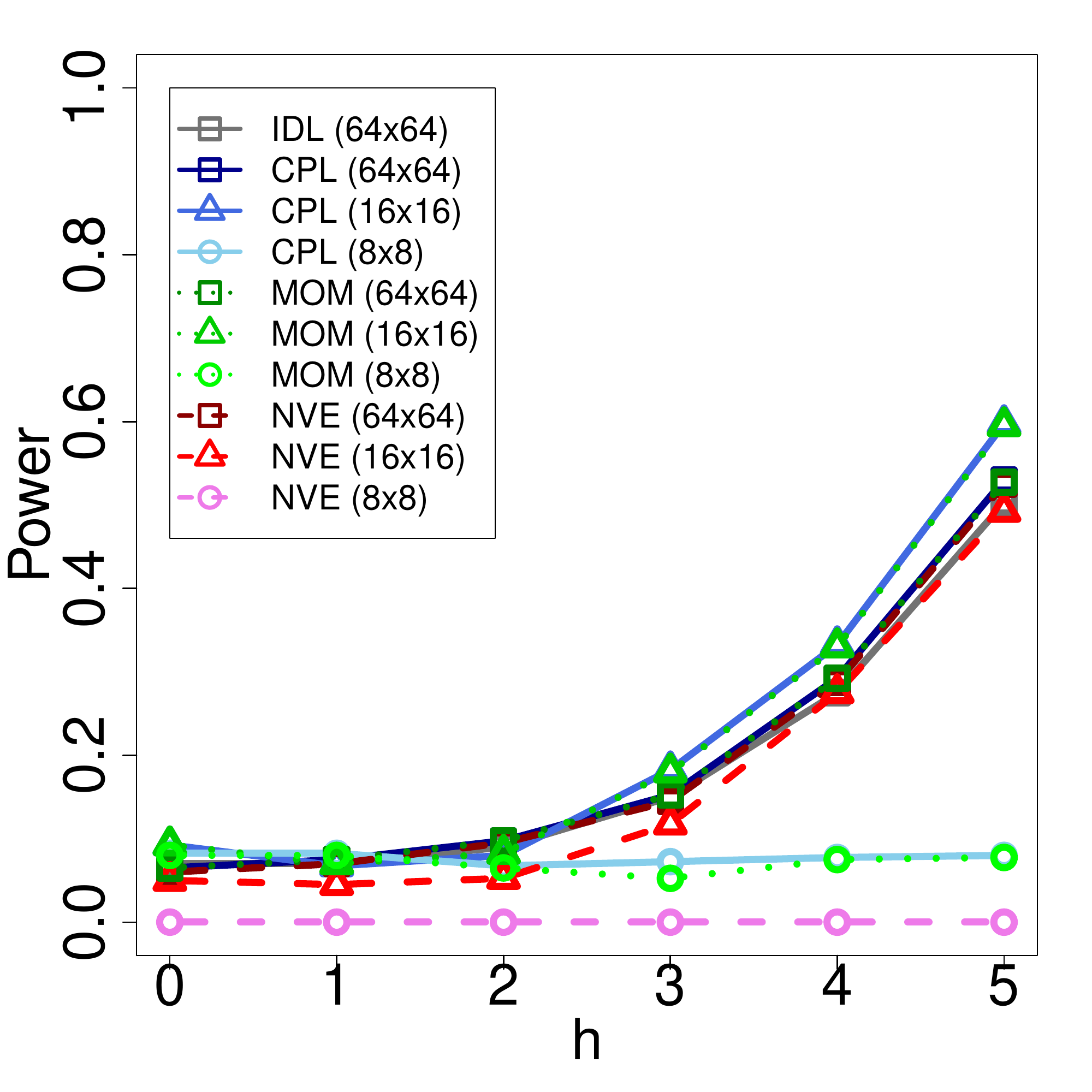} \\
\rotatebox{90}{$\quad \quad \quad r=8$}~~
\includegraphics[scale=0.22,trim={0.18cm 0.3cm 1cm 1cm},clip]{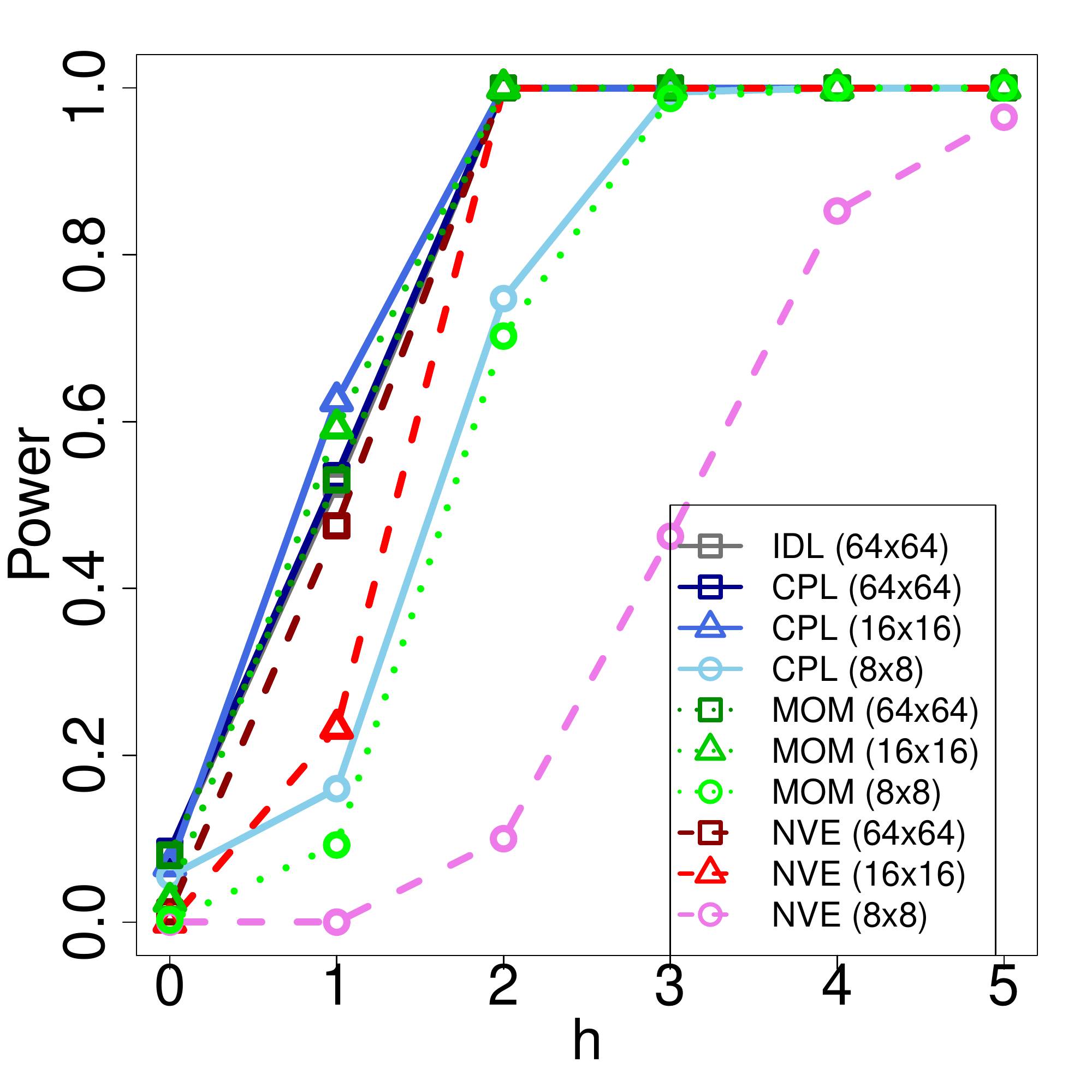} &
\includegraphics[scale=0.22,trim={0.18cm 0.3cm 1cm 1cm},clip]{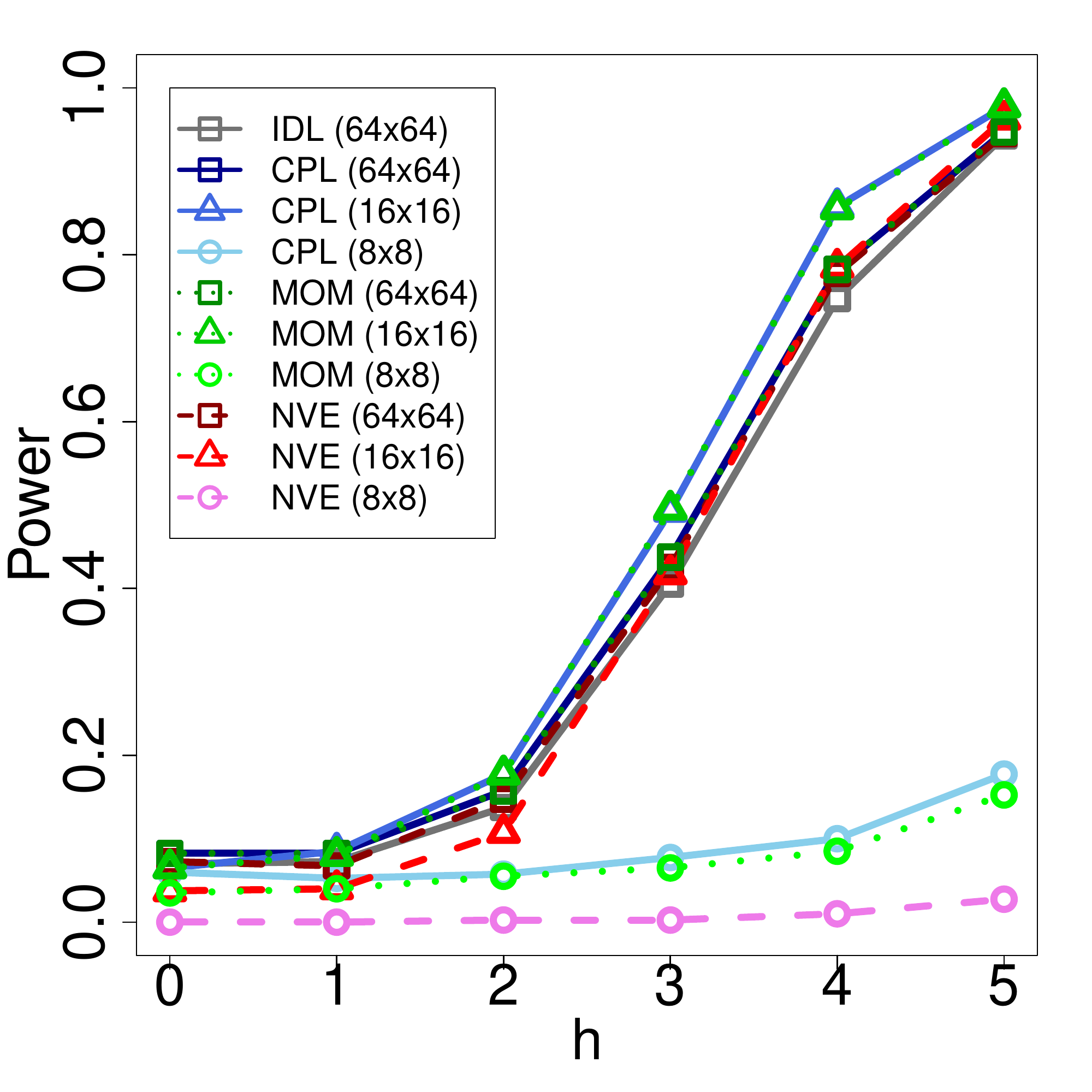} &
\includegraphics[scale=0.22,trim={0.18cm 0.3cm 1cm 1cm},clip]{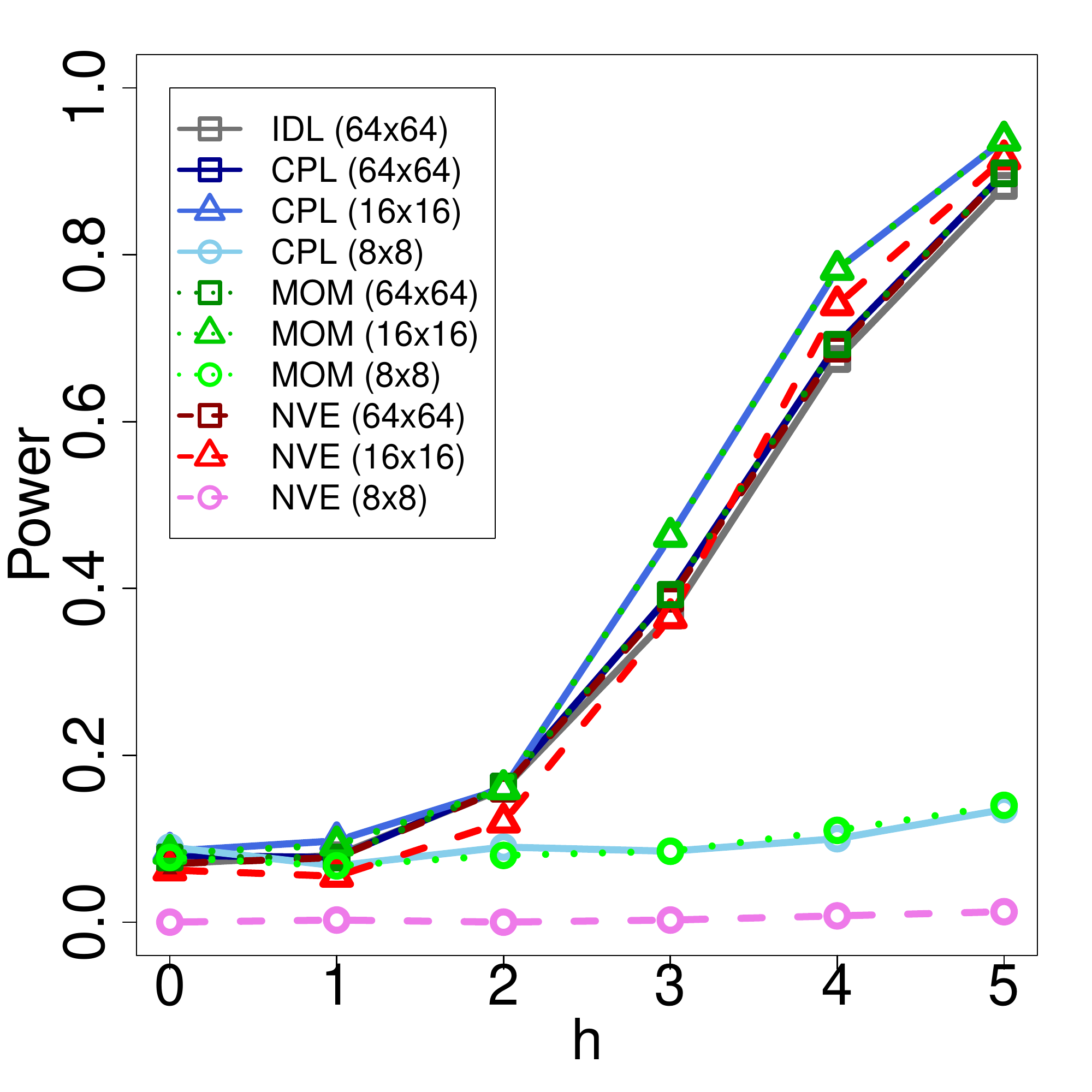} \\
\rotatebox{90}{$\quad \quad \quad r=10$}~~
\includegraphics[scale=0.22,trim={0.18cm 0.3cm 1cm 1cm},clip]{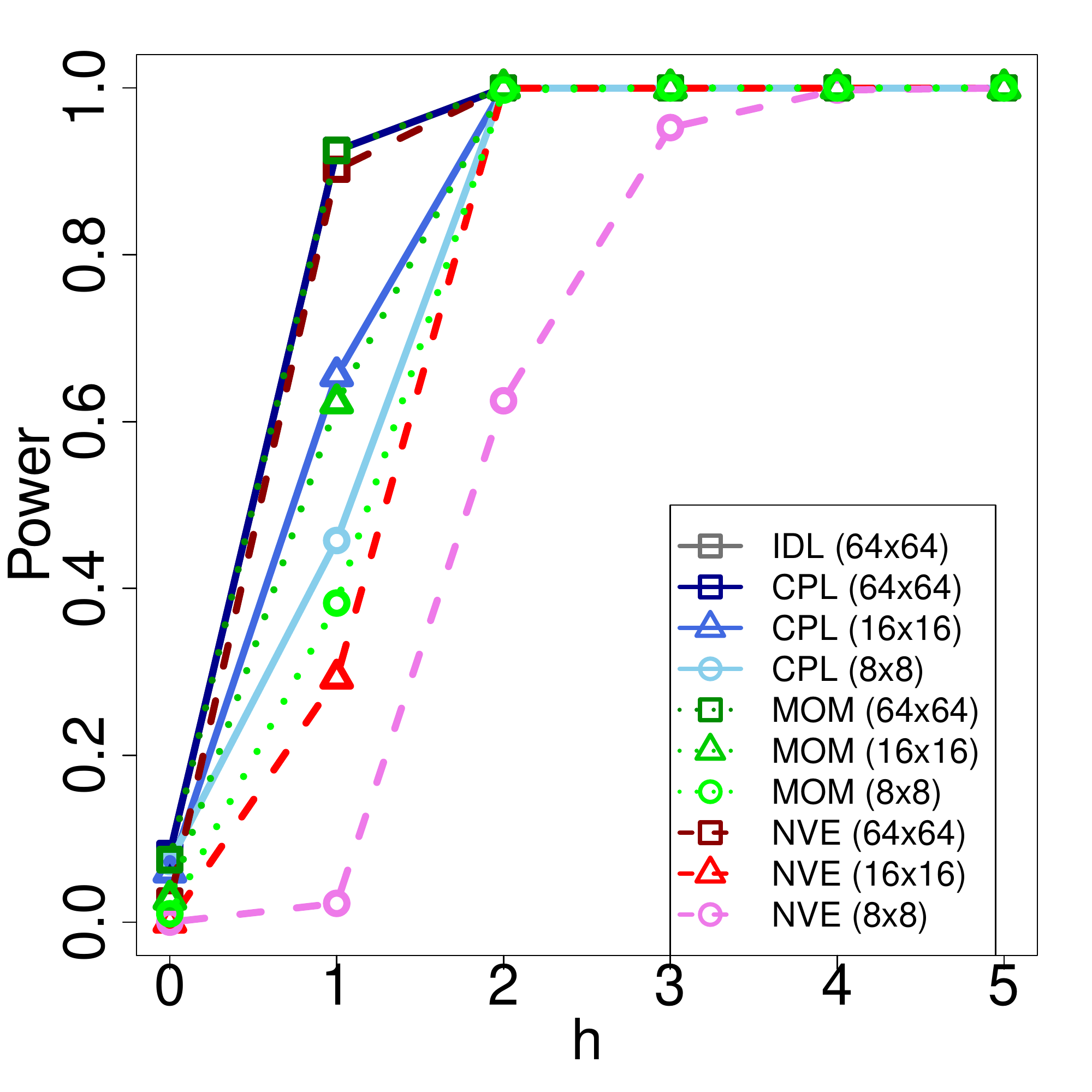} &
\includegraphics[scale=0.22,trim={0.18cm 0.3cm 1cm 1cm},clip]{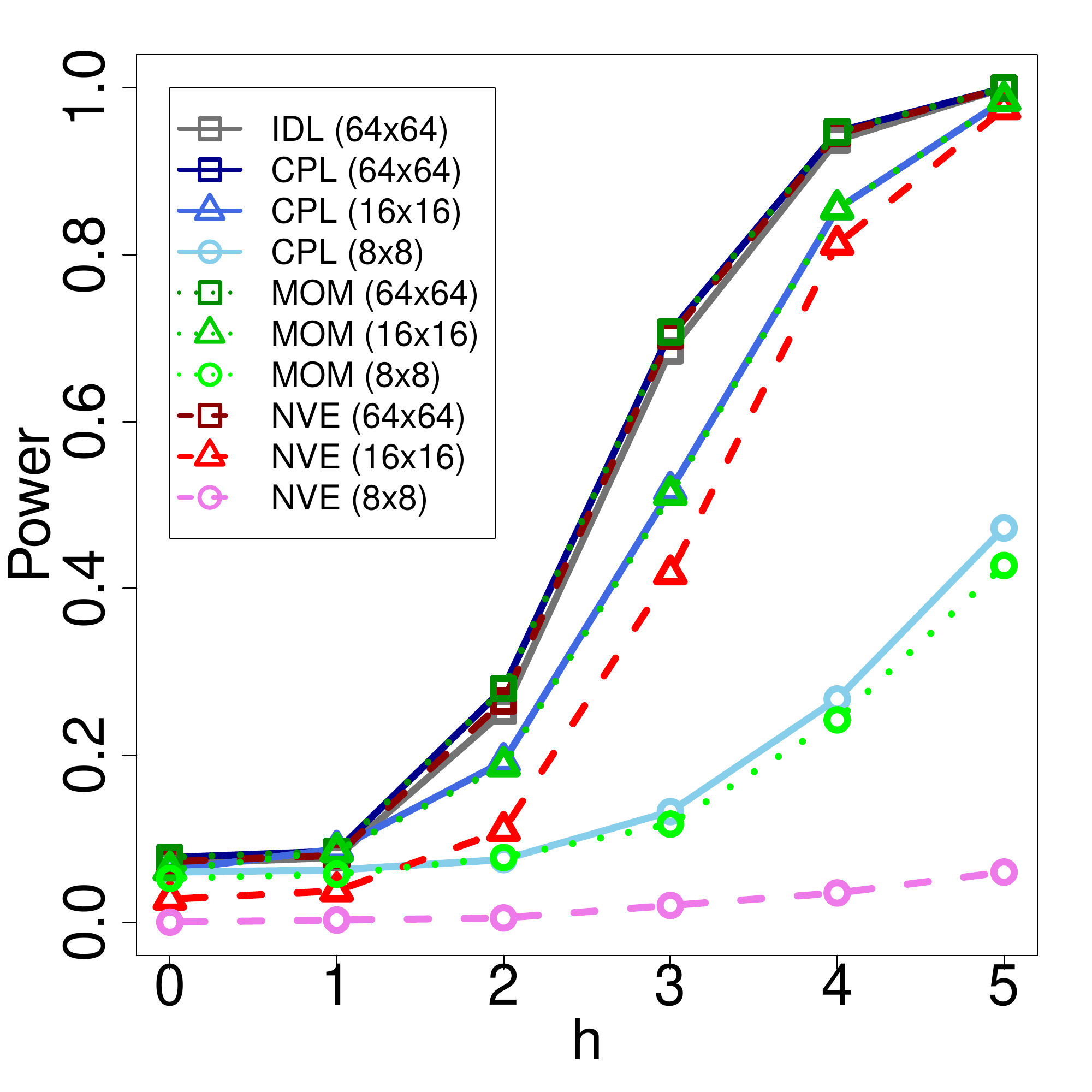} &
\includegraphics[scale=0.22,trim={0.18cm 0.3cm 1cm 1cm},clip]{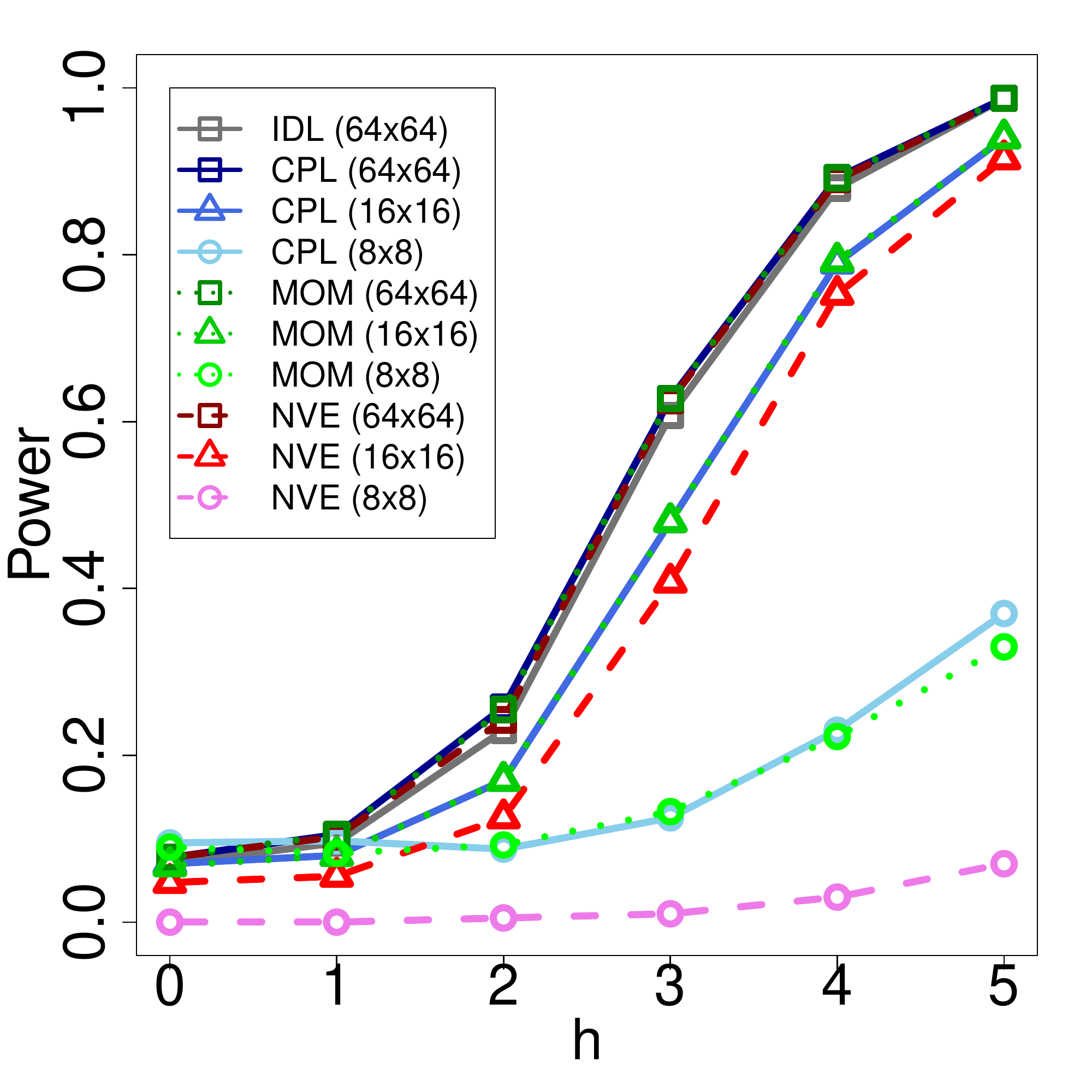} \\
\end{tabular}
\caption{Empirical power curves as a function of the signal's magnitude $h$, for various procedures for testing $H_0$ in Experiment 3
in Section \ref{sec:simulation}.
Down the rows, the curves correspond to different signal extents $r$,
while across the columns, the curves correspond to different spatial-dependence values $\phi$.}
\label{fig:power for experiment 3-2}
\end{figure}

\begin{figure}[!tbh]\centering
\begin{tabular}{ccc}
~~~~~~~IDL & ~~~CPL ($16\times 16$)  & ~~CPL ($8\times 8$) \\
\rotatebox{90}{$\quad \quad \quad  \quad  \quad \phi=0$}~
\includegraphics[scale=0.24,trim={0.2cm 0.2cm 0.2cm 0.25cm},clip]{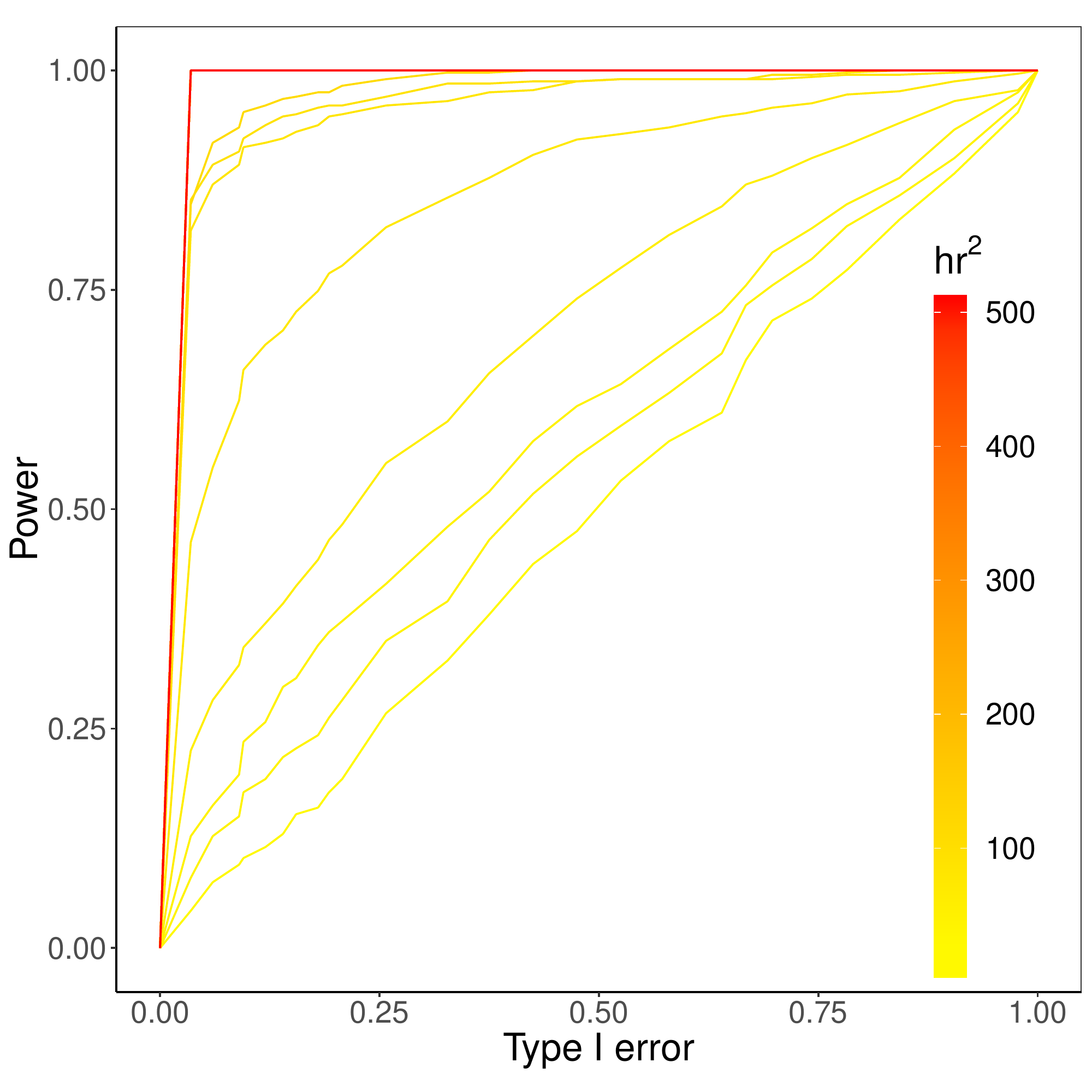} &
\includegraphics[scale=0.24,trim={0.2cm 0.2cm 0.2cm 0.25cm},clip]{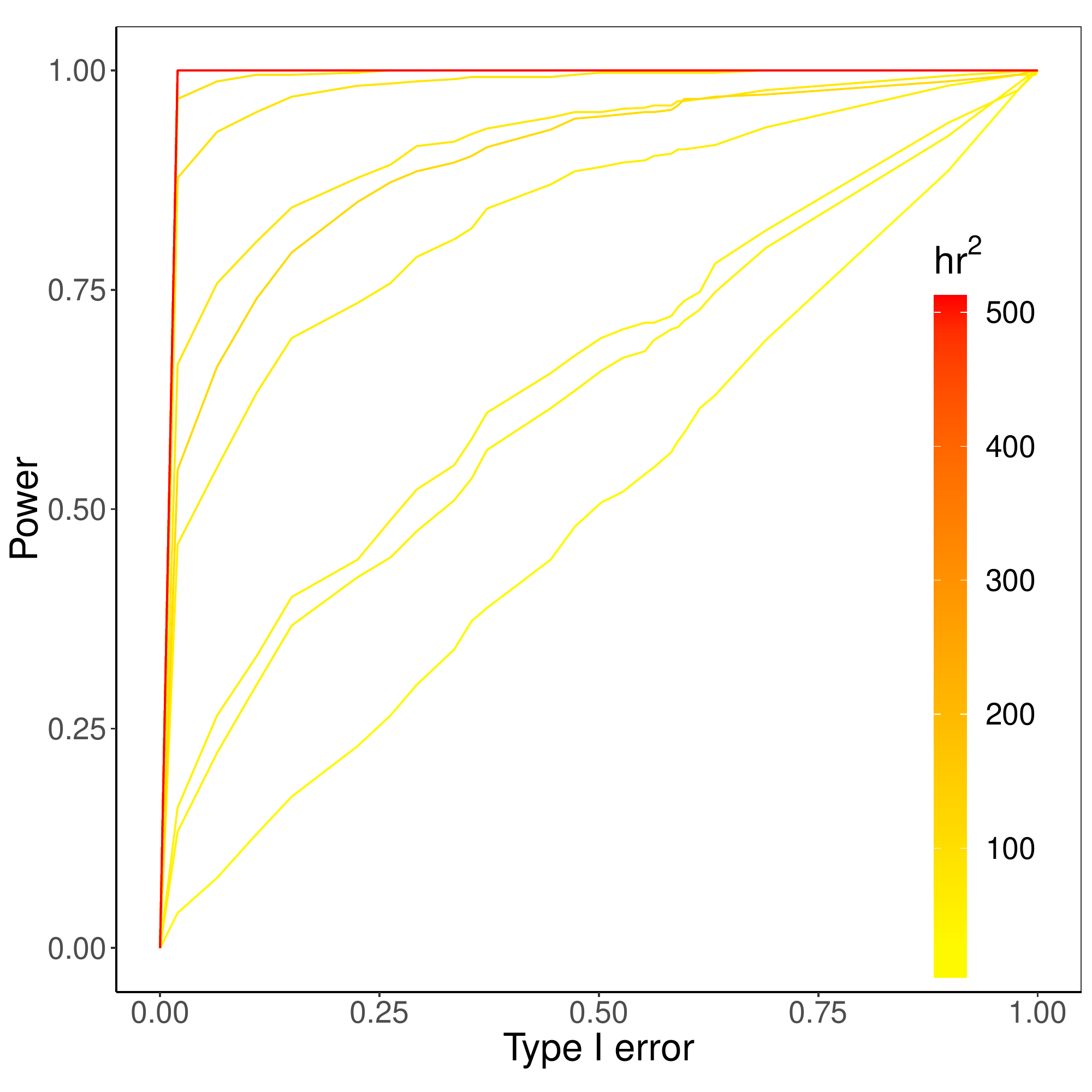} &
\includegraphics[scale=0.24,trim={0.2cm 0.2cm 0.2cm 0.25cm},clip]{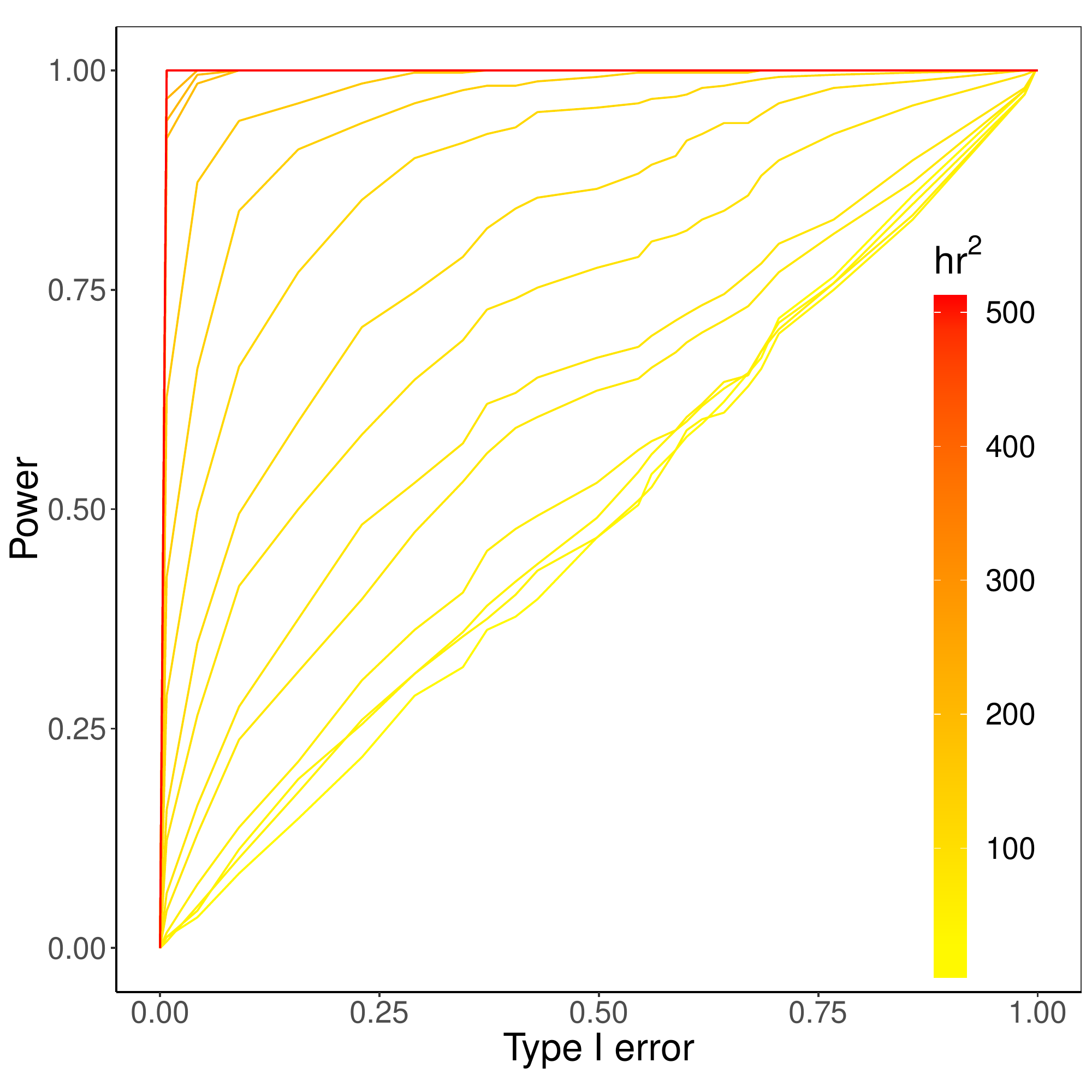} \\
\rotatebox{90}{$\quad \quad \quad  \quad  \quad \phi=5$}~
\includegraphics[scale=0.24,trim={0.2cm 0.2cm 0.2cm 0.25cm},clip]{./Exp1-ROC-grid64-phi5} &
\includegraphics[scale=0.24,trim={0.2cm 0.2cm 0.2cm 0.25cm},clip]{./Exp1-ROC-grid16-phi5} &
\includegraphics[scale=0.24,trim={0.2cm 0.2cm 0.2cm 0.25cm},clip]{./Exp1-ROC-grid8-phi5} \\
\rotatebox{90}{$\quad \quad \quad  \quad  \quad \phi=10$}~
\includegraphics[scale=0.24,trim={0.2cm 0.2cm 0.2cm 0.25cm},clip]{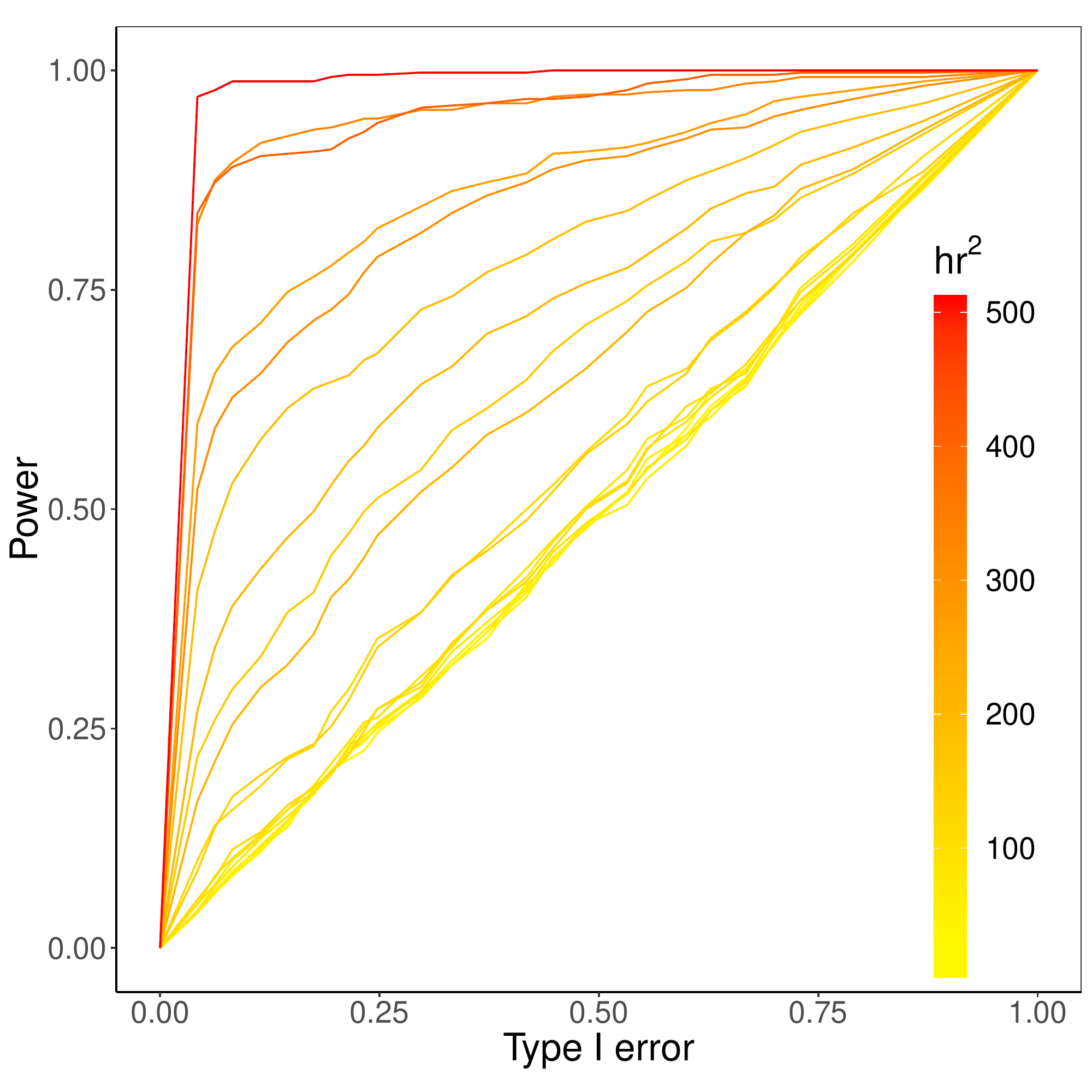} &
\includegraphics[scale=0.24,trim={0.2cm 0.2cm 0.2cm 0.25cm},clip]{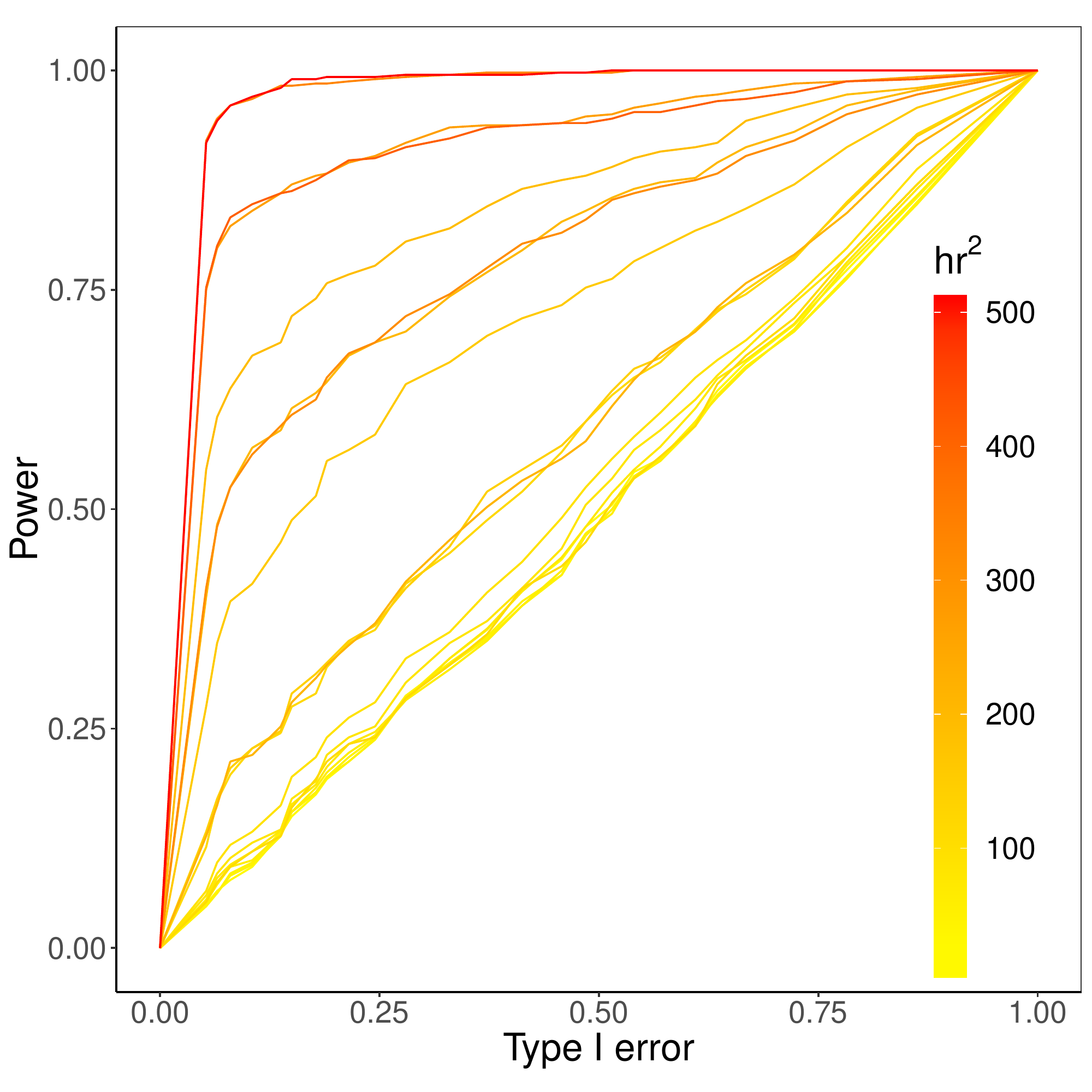} &
\includegraphics[scale=0.24,trim={0.2cm 0.2cm 0.2cm 0.25cm},clip]{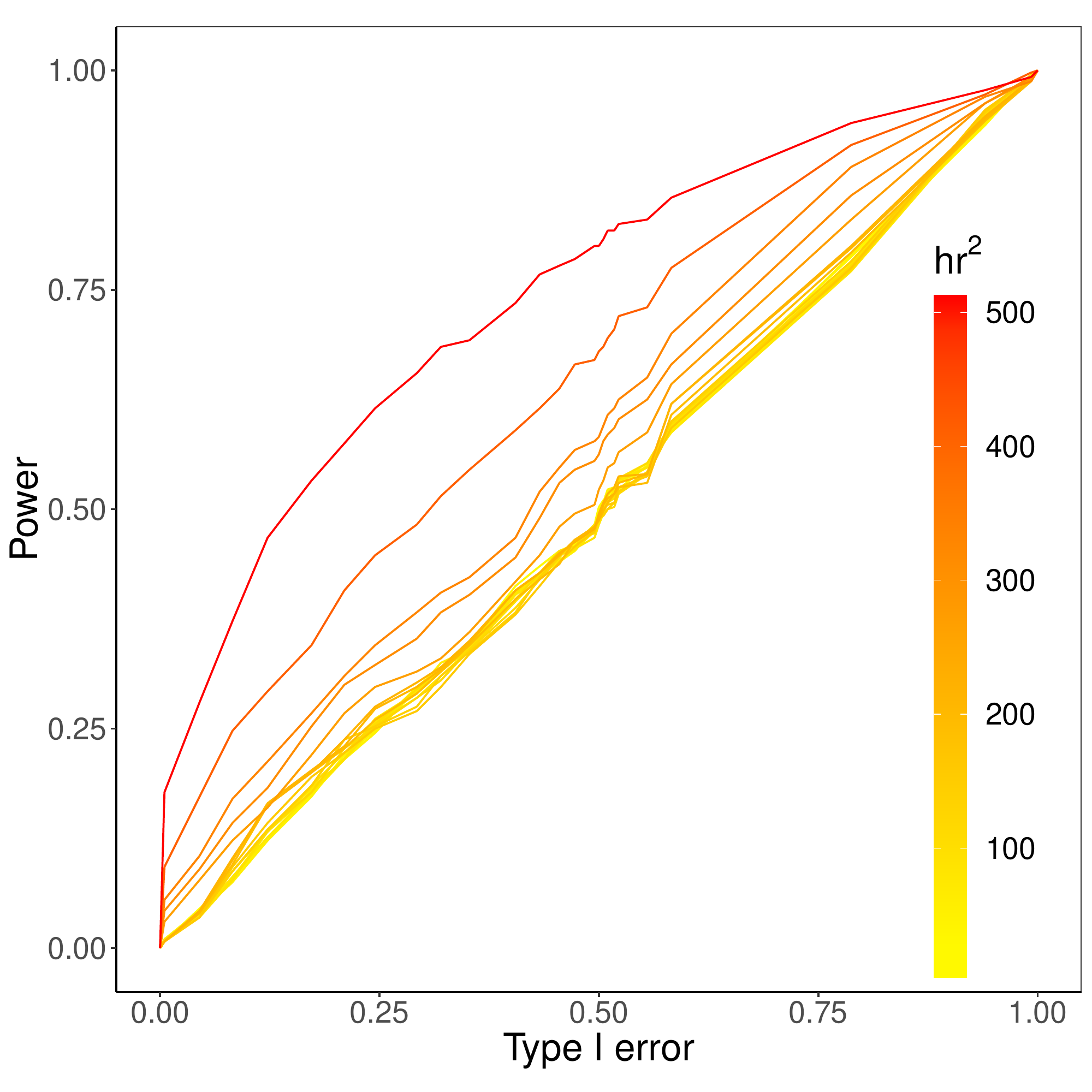} 
\end{tabular}
\caption{Empirical ROC curves for IDL, CPL ($16\times 16$), and CPL ($8\times 8$) in Experiment 1
in Section \ref{sec:simulation}.
colored according to the volume $h r^2$ of the signal obtained from 24 combinations of $r\in\{4,6,8,10\}$ and $h\in\{0,1,2,3,4,5\}$ in each plot.
Down the rows, the curves correspond to different spatial-dependence values $\phi$.
Each curve was obtained by varying $\alpha$ on the right-hand side of \eqref{eq:test}.}
\label{fig:ROC for experiment 1-2}
\end{figure}

\begin{figure}[!tbh]\centering
\begin{tabular}{ccc}
~~~~~~~CPL ($64\times 64$) & ~~~CPL ($16\times 16$)  & ~~CPL ($8\times 8$) \\
\rotatebox{90}{$\quad \quad \quad  \quad  \quad \phi=0$}~
\includegraphics[scale=0.24,trim={0.2cm 0.2cm 0.2cm 0.25cm},clip]{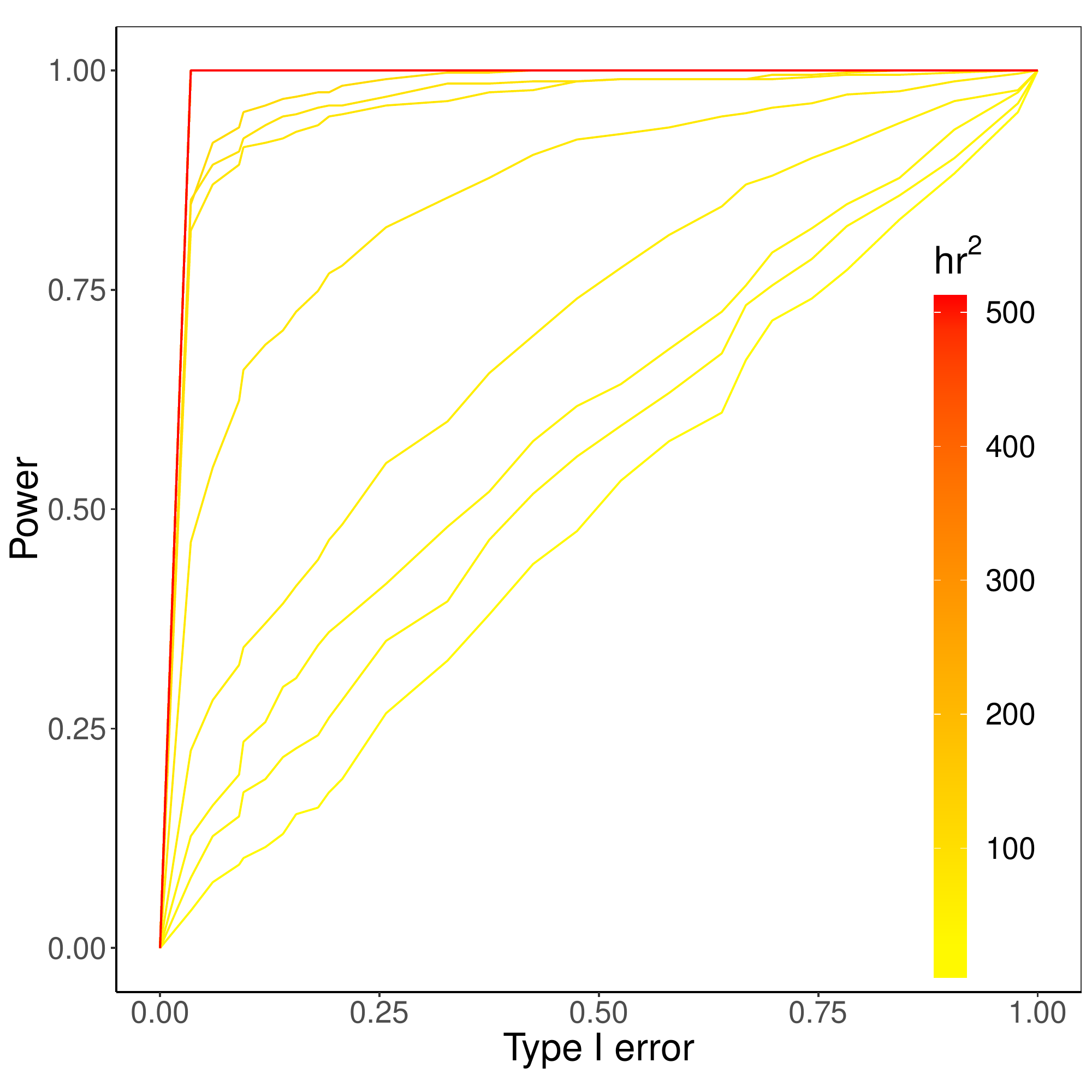} &
\includegraphics[scale=0.24,trim={0.2cm 0.2cm 0.2cm 0.25cm},clip]{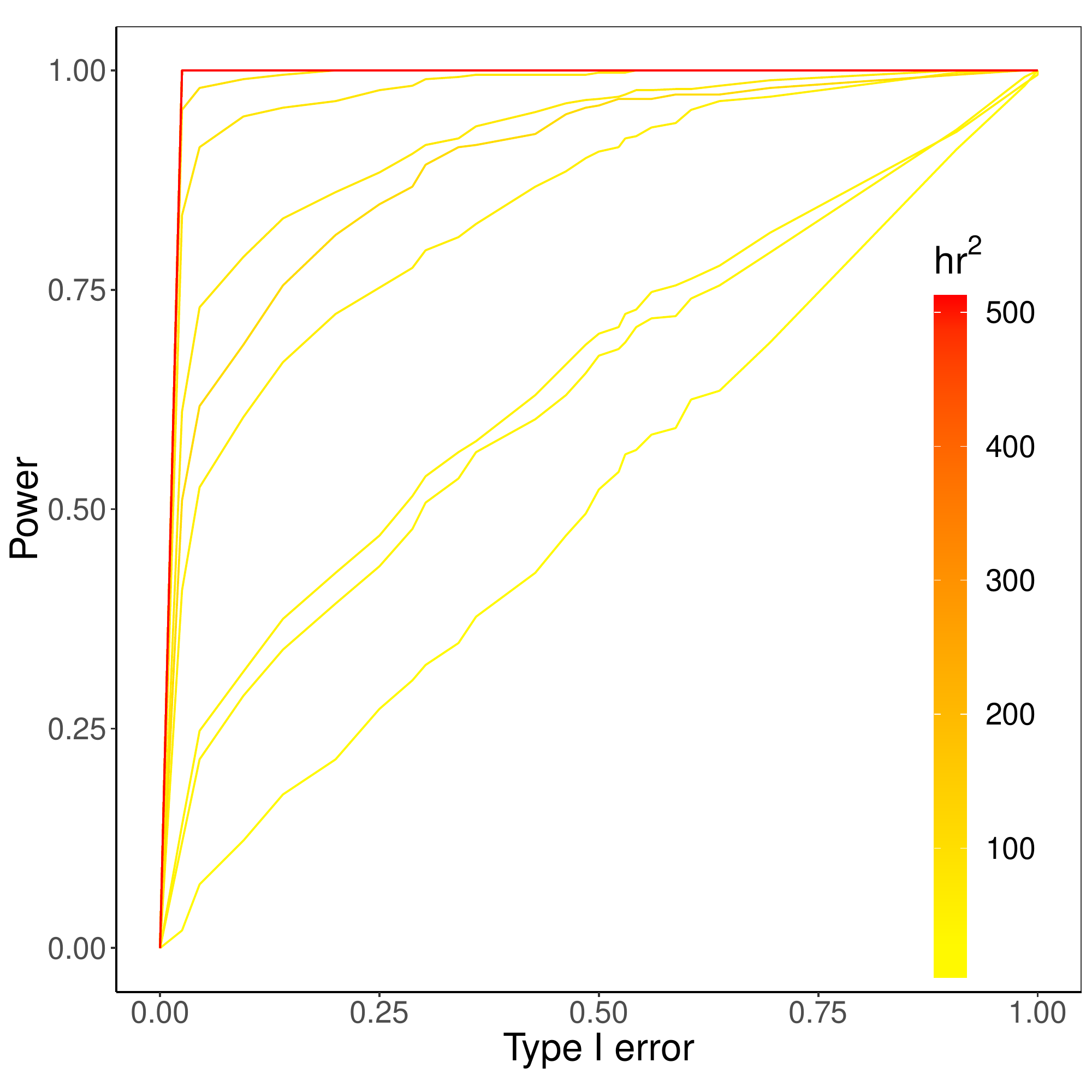} &
\includegraphics[scale=0.24,trim={0.2cm 0.2cm 0.2cm 0.25cm},clip]{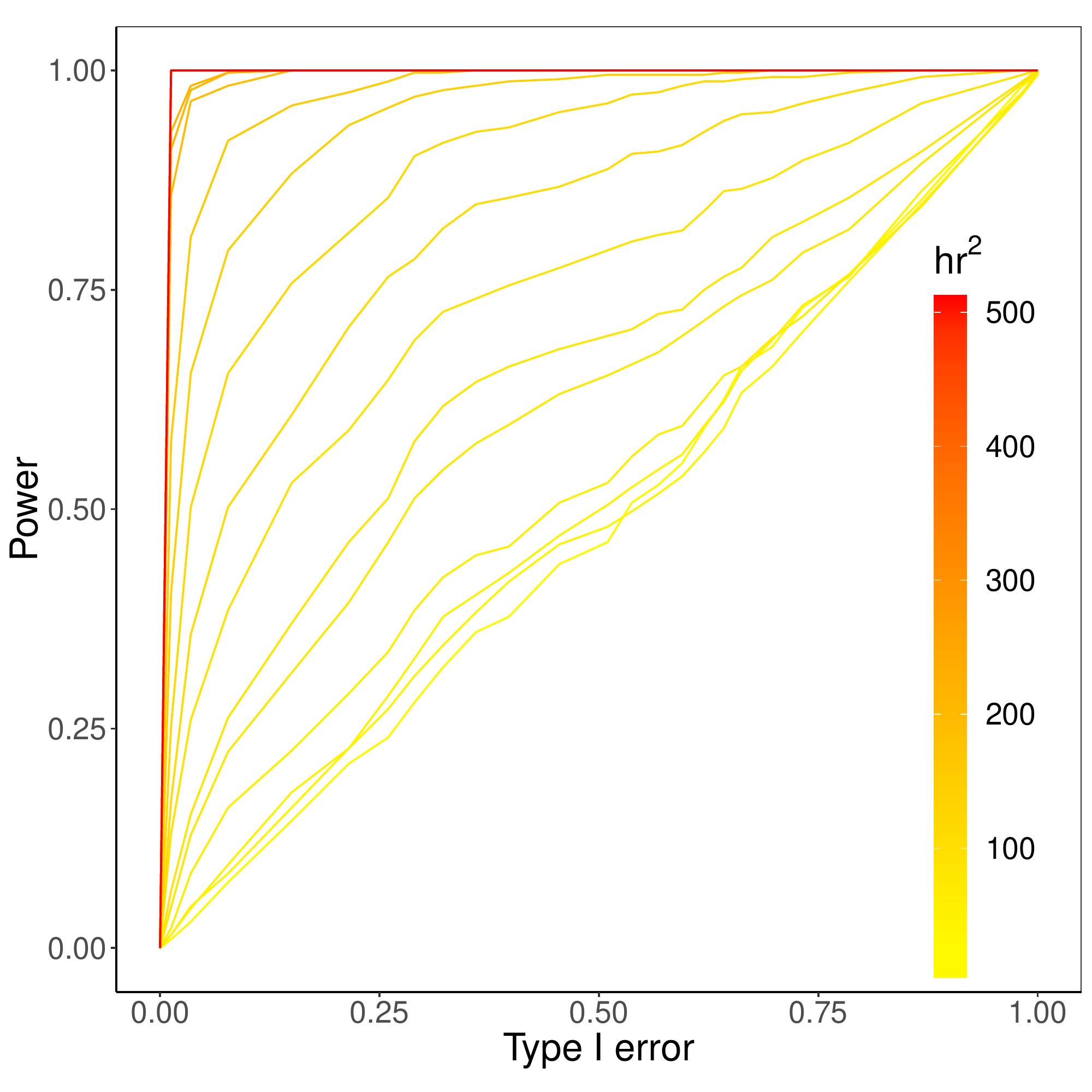} \\
\rotatebox{90}{$\quad \quad \quad  \quad  \quad \phi=5$}~
\includegraphics[scale=0.24,trim={0.2cm 0.2cm 0.2cm 0.25cm},clip]{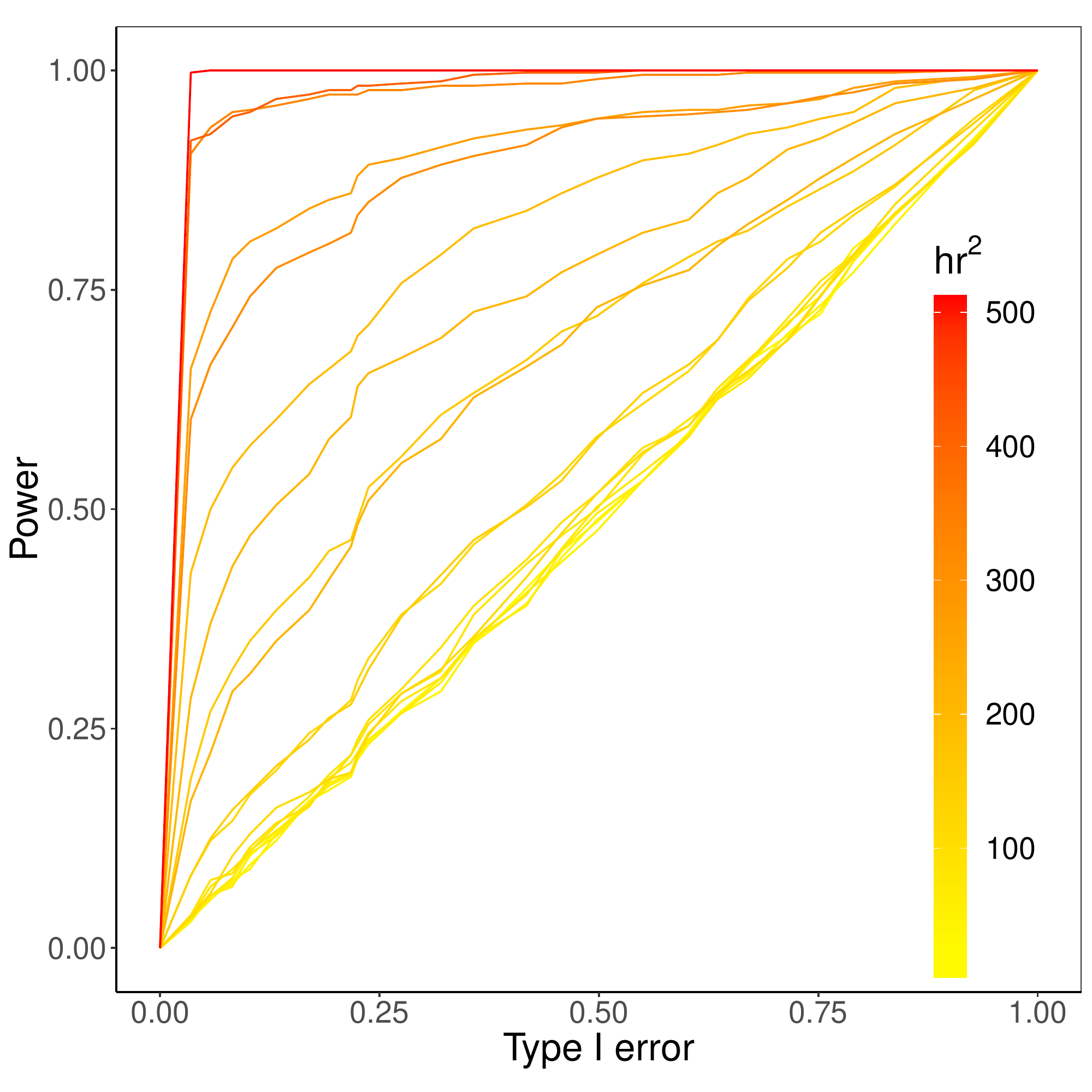} &
\includegraphics[scale=0.24,trim={0.2cm 0.2cm 0.2cm 0.25cm},clip]{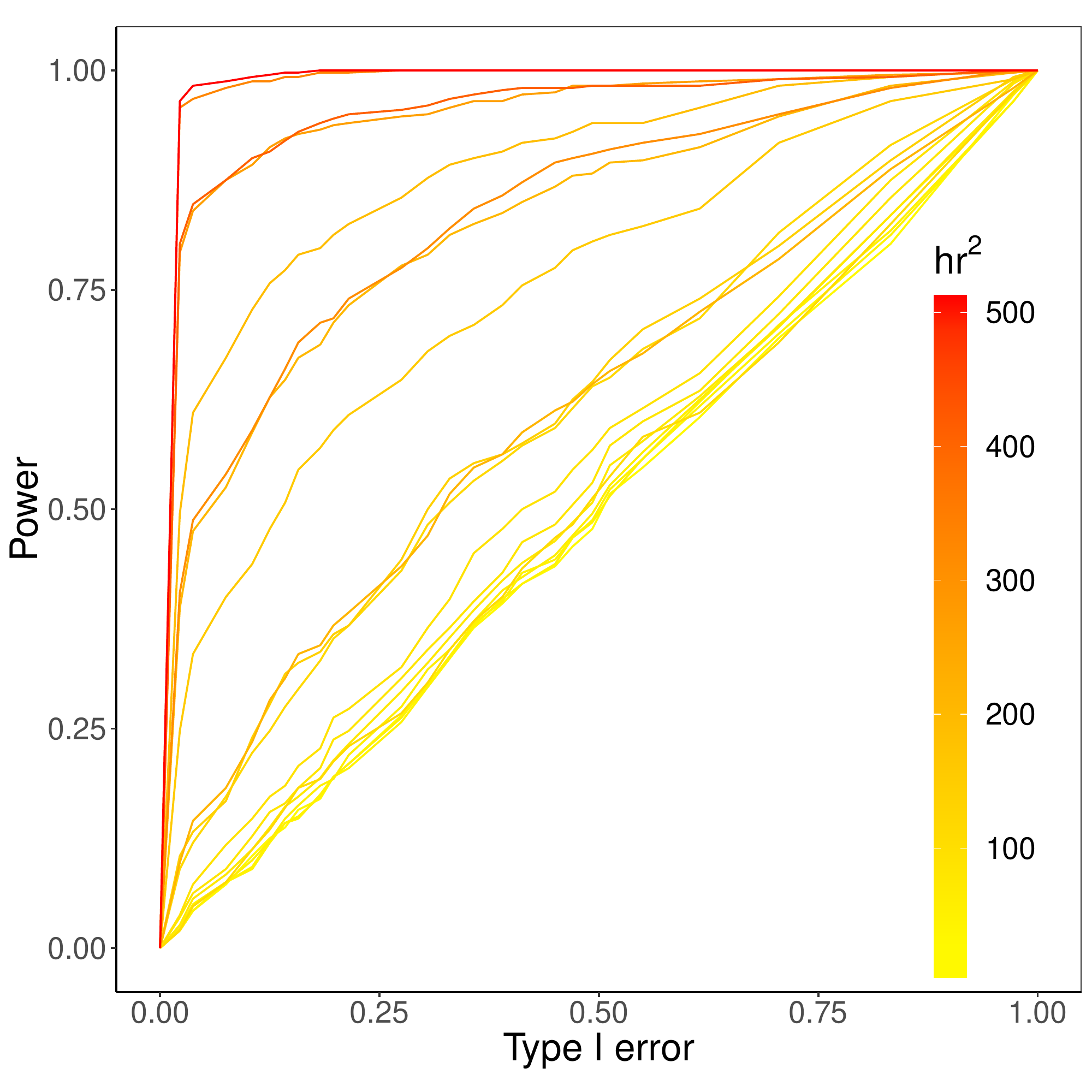} &
\includegraphics[scale=0.24,trim={0.2cm 0.2cm 0.2cm 0.25cm},clip]{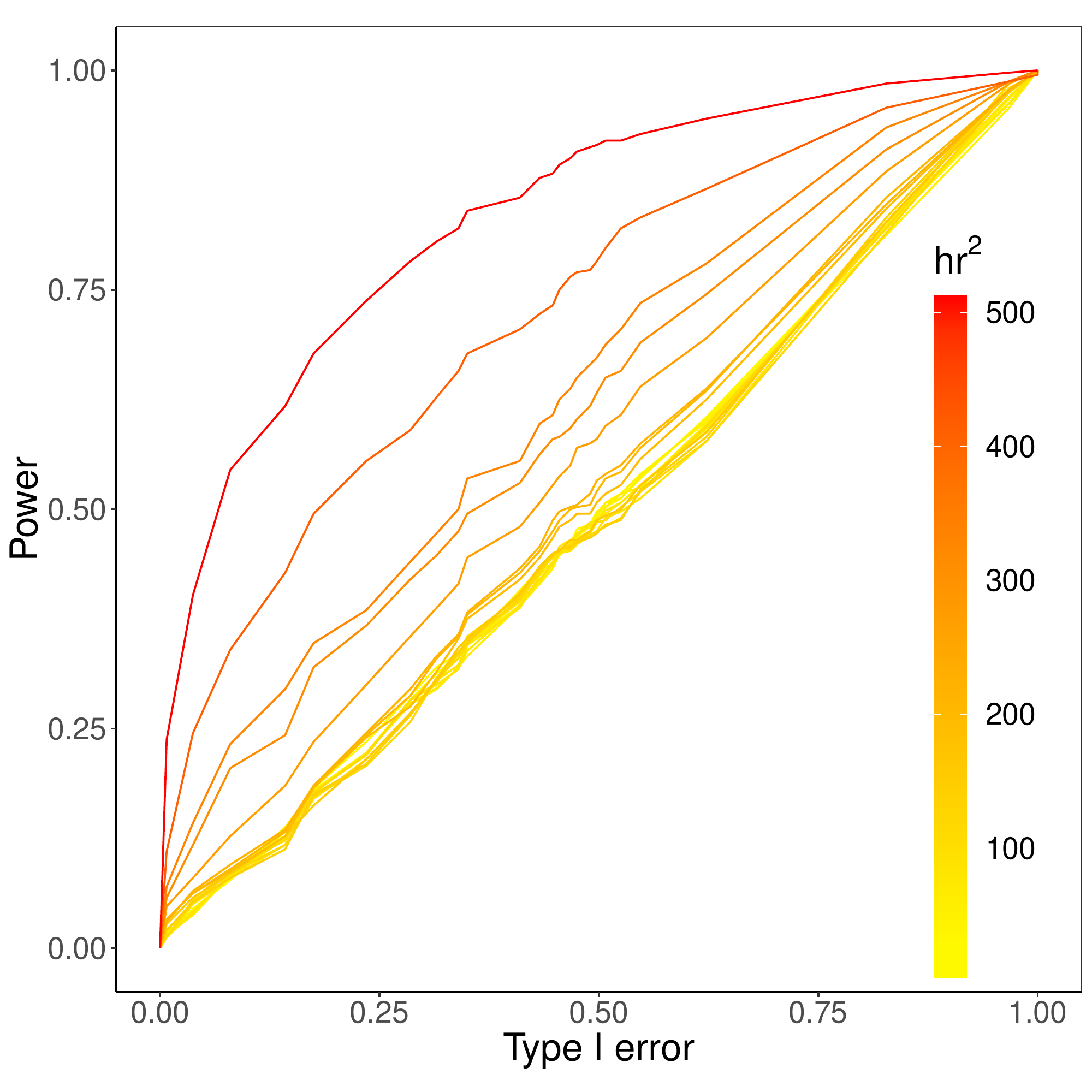} \\
\rotatebox{90}{$\quad \quad \quad  \quad  \quad \phi=10$}~
\includegraphics[scale=0.24,trim={0.2cm 0.2cm 0.2cm 0.25cm},clip]{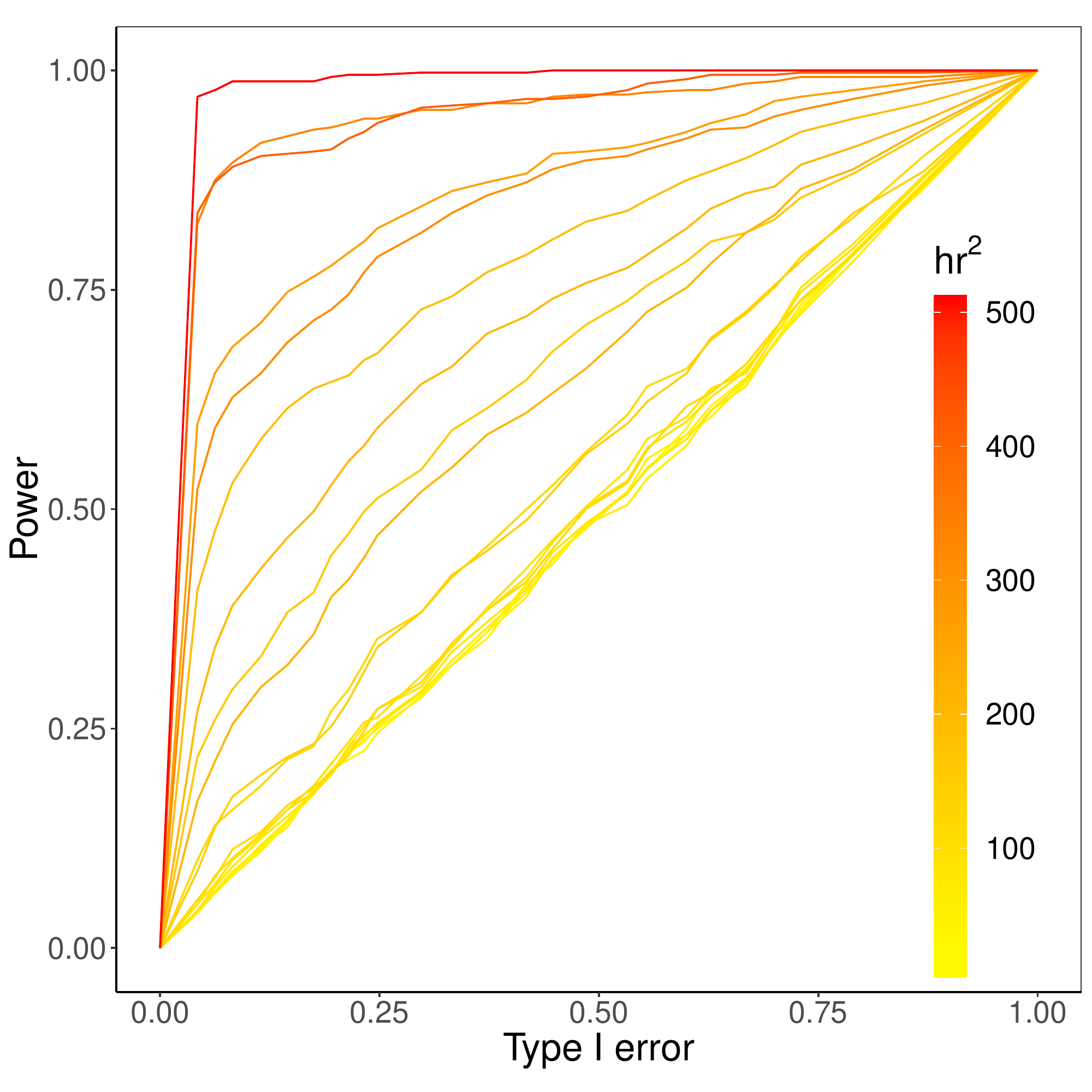} &
\includegraphics[scale=0.24,trim={0.2cm 0.2cm 0.2cm 0.25cm},clip]{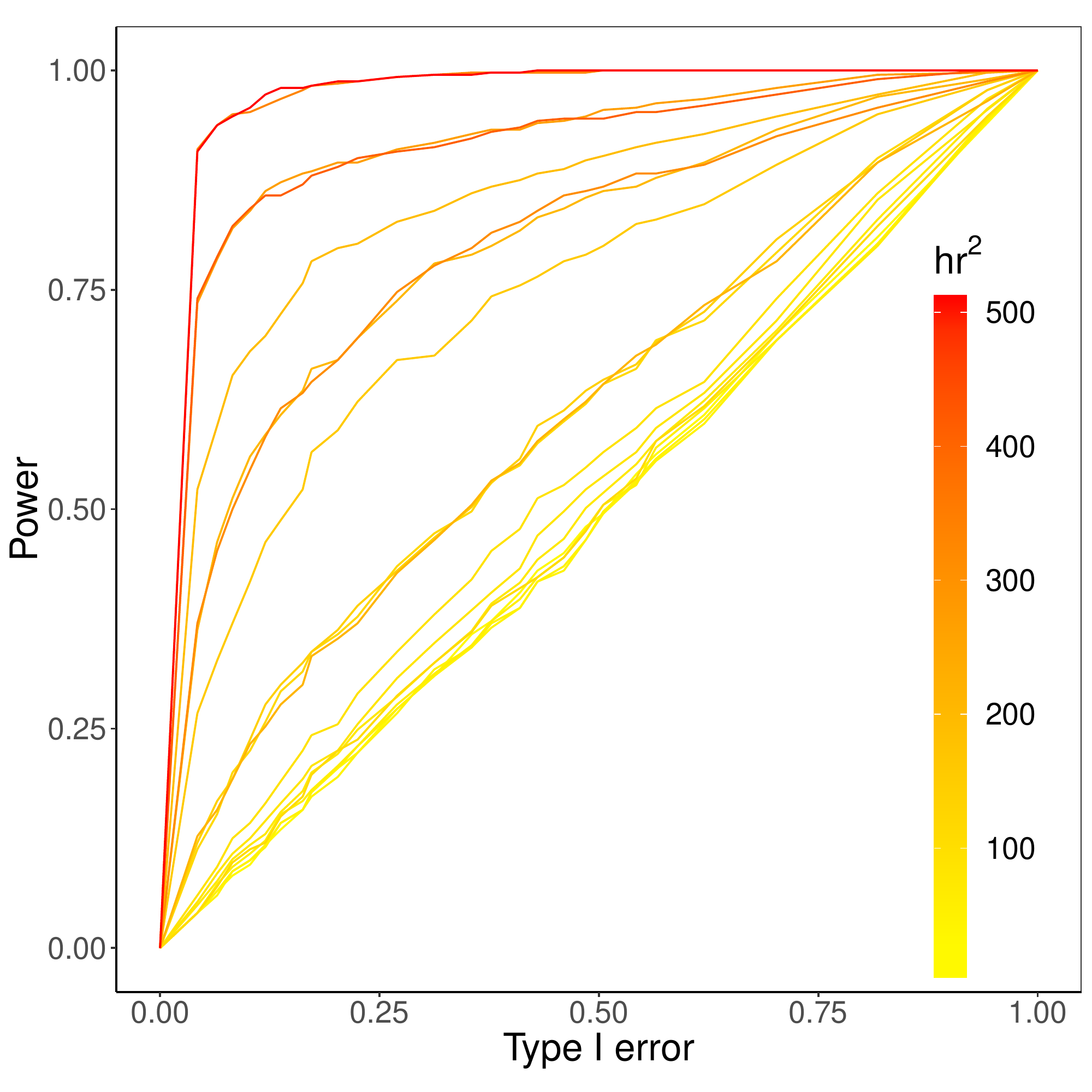} &
\includegraphics[scale=0.24,trim={0.2cm 0.2cm 0.2cm 0.25cm},clip]{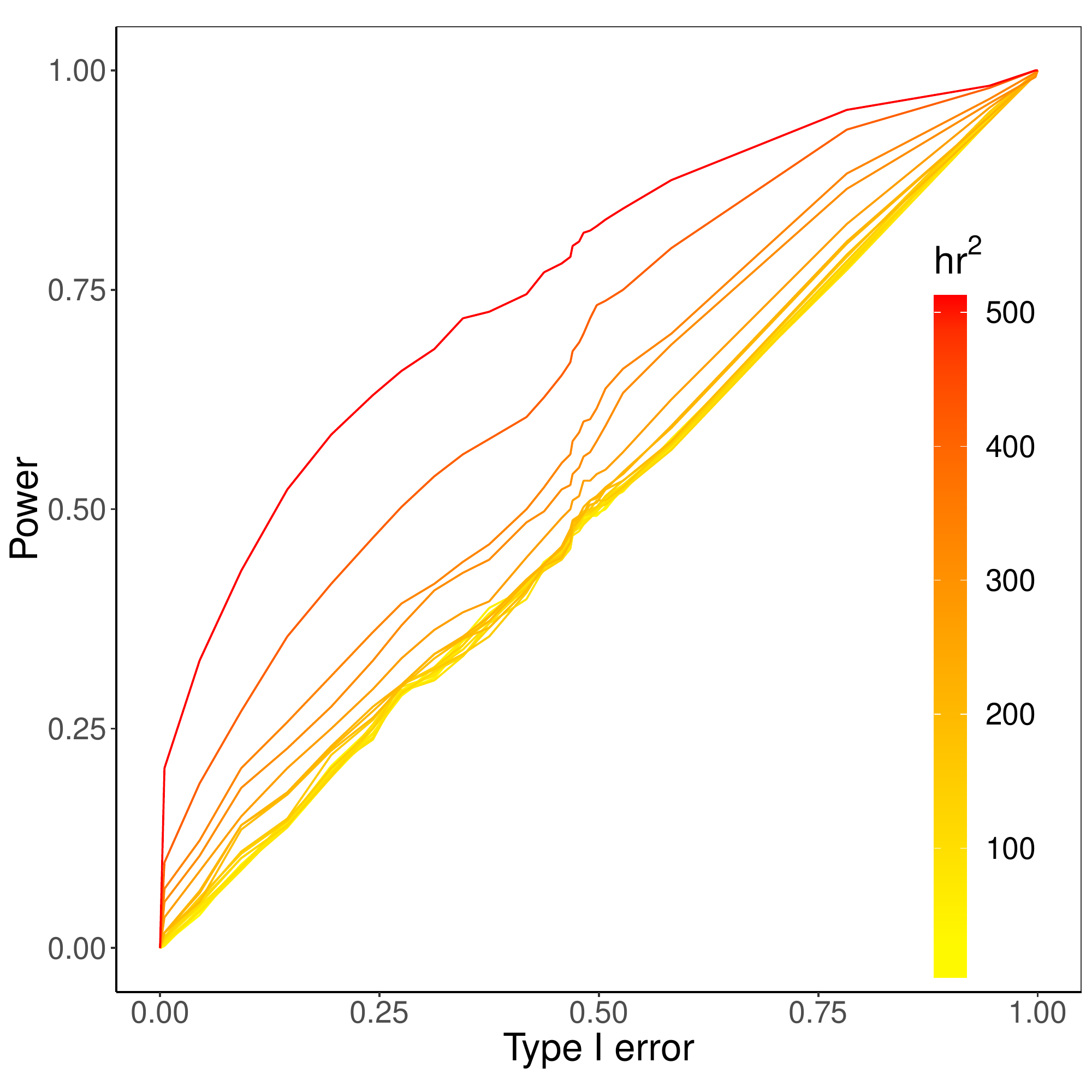} 
\end{tabular}
\caption{Empirical ROC curves for CPL ($64\times 64$), CPL ($16\times 16$), and CPL ($8\times 8$) in Experiment 2
in Section \ref{sec:simulation}.
colored according to the volume $h r^2$ of the signal obtained from 24 combinations of $r\in\{4,6,8,10\}$ and $h\in\{0,1,2,3,4,5\}$ in each plot.
Down the rows, the curves correspond to different spatial-dependence values $\phi$.
Each curve was obtained by varying $\alpha$ on the right-hand side of \eqref{eq:test}.}
\label{fig:ROC for experiment 2-2}
\end{figure}

\begin{figure}[!tbh]\centering
\begin{tabular}{ccc}
~~~~~~~CPL ($64\times 64$) & ~~~CPL ($16\times 16$)  & ~~CPL ($8\times 8$) \\
\rotatebox{90}{$\quad \quad \quad  \quad  \quad \phi=0$}~
\includegraphics[scale=0.24,trim={0.2cm 0.2cm 0.2cm 0.25cm},clip]{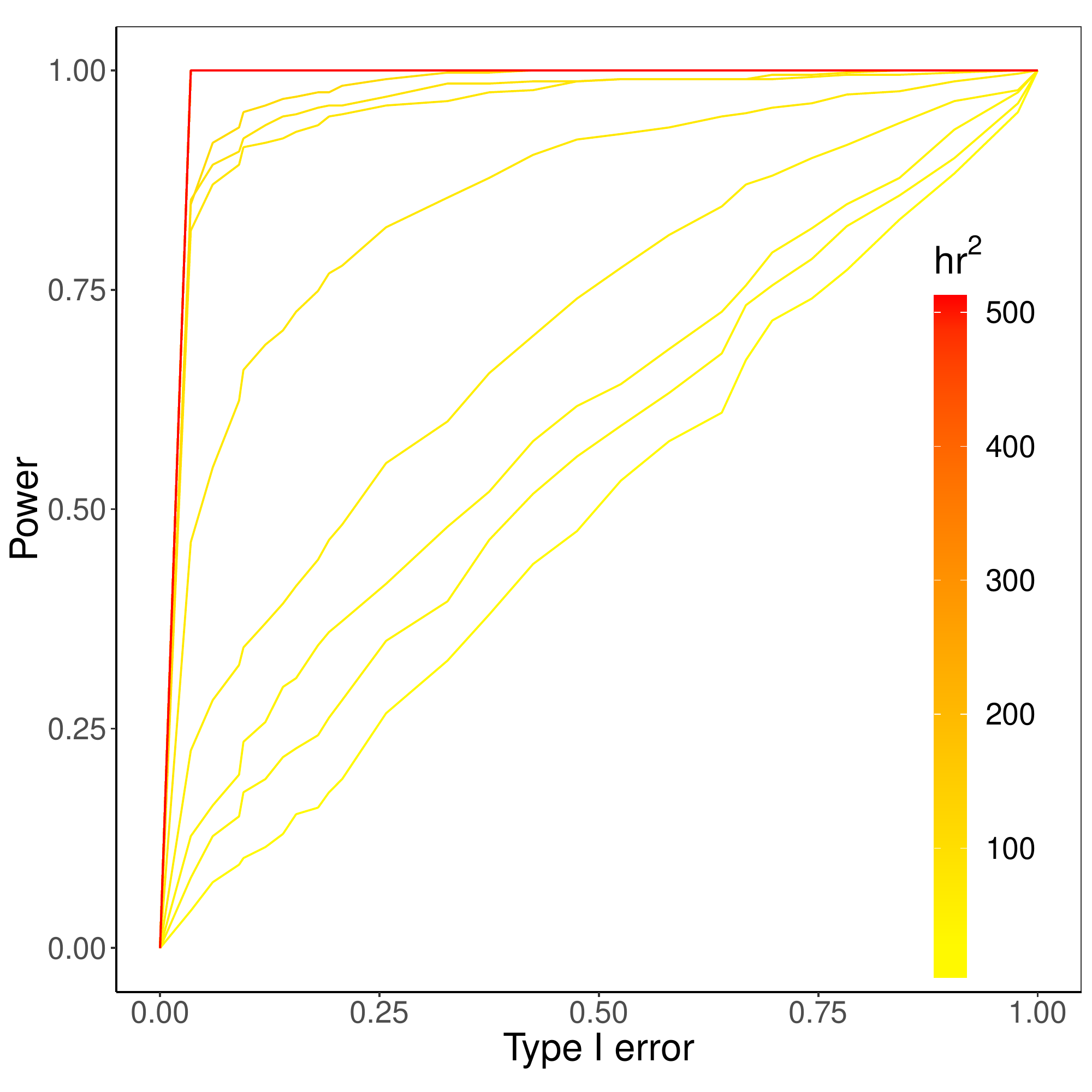} &
\includegraphics[scale=0.24,trim={0.2cm 0.2cm 0.2cm 0.25cm},clip]{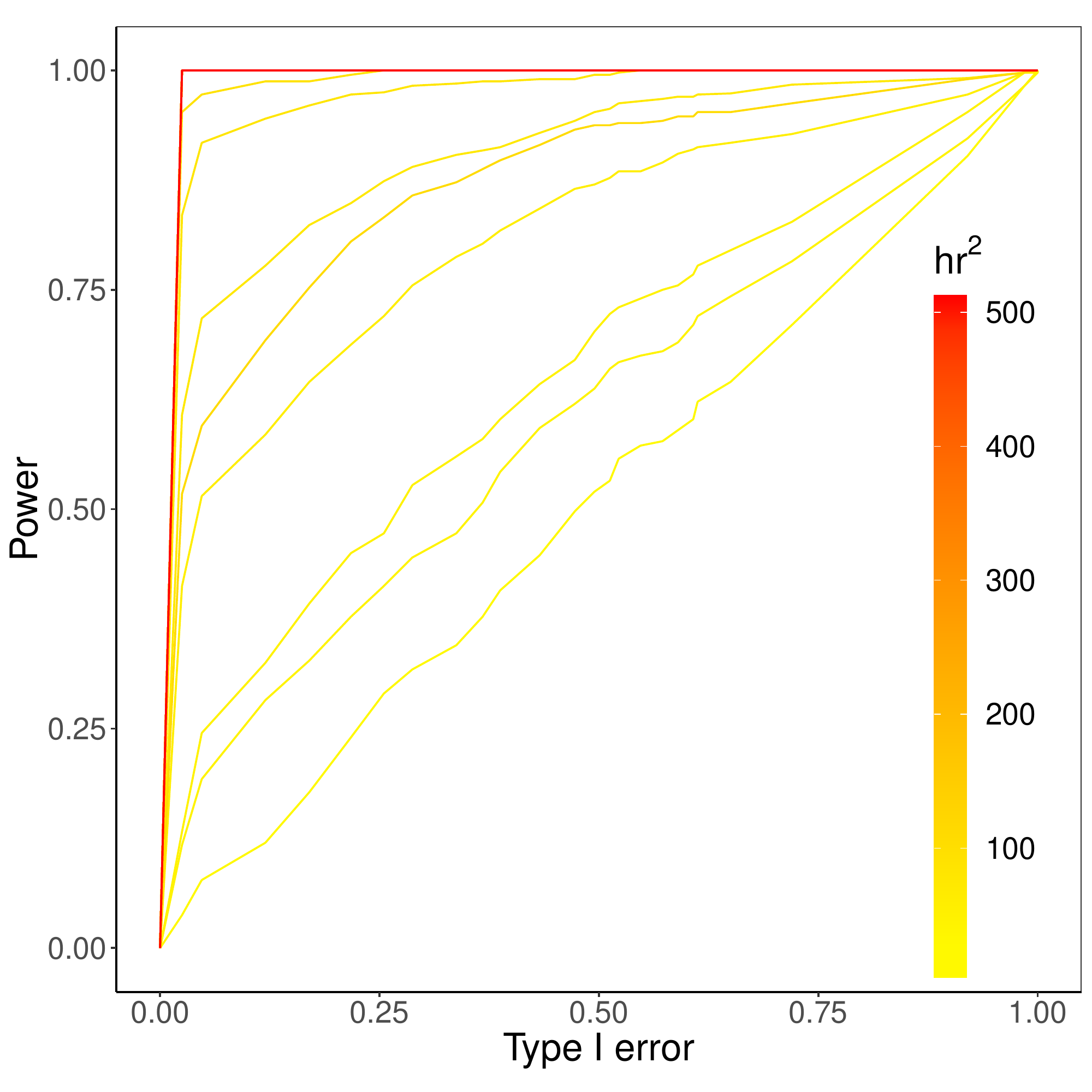} &
\includegraphics[scale=0.24,trim={0.2cm 0.2cm 0.2cm 0.25cm},clip]{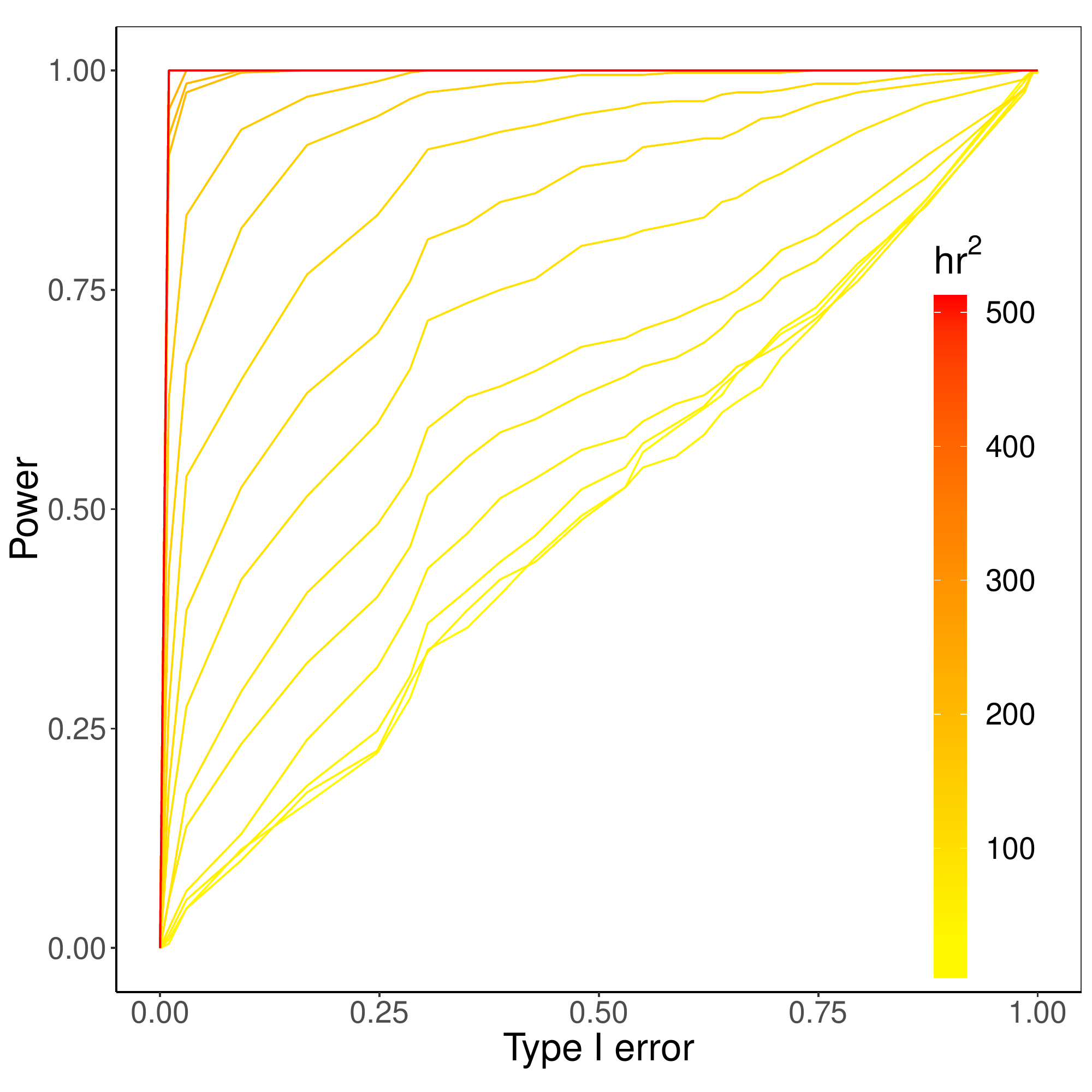} \\
\rotatebox{90}{$\quad \quad \quad  \quad  \quad \phi=5$}~
\includegraphics[scale=0.24,trim={0.2cm 0.2cm 0.2cm 0.25cm},clip]{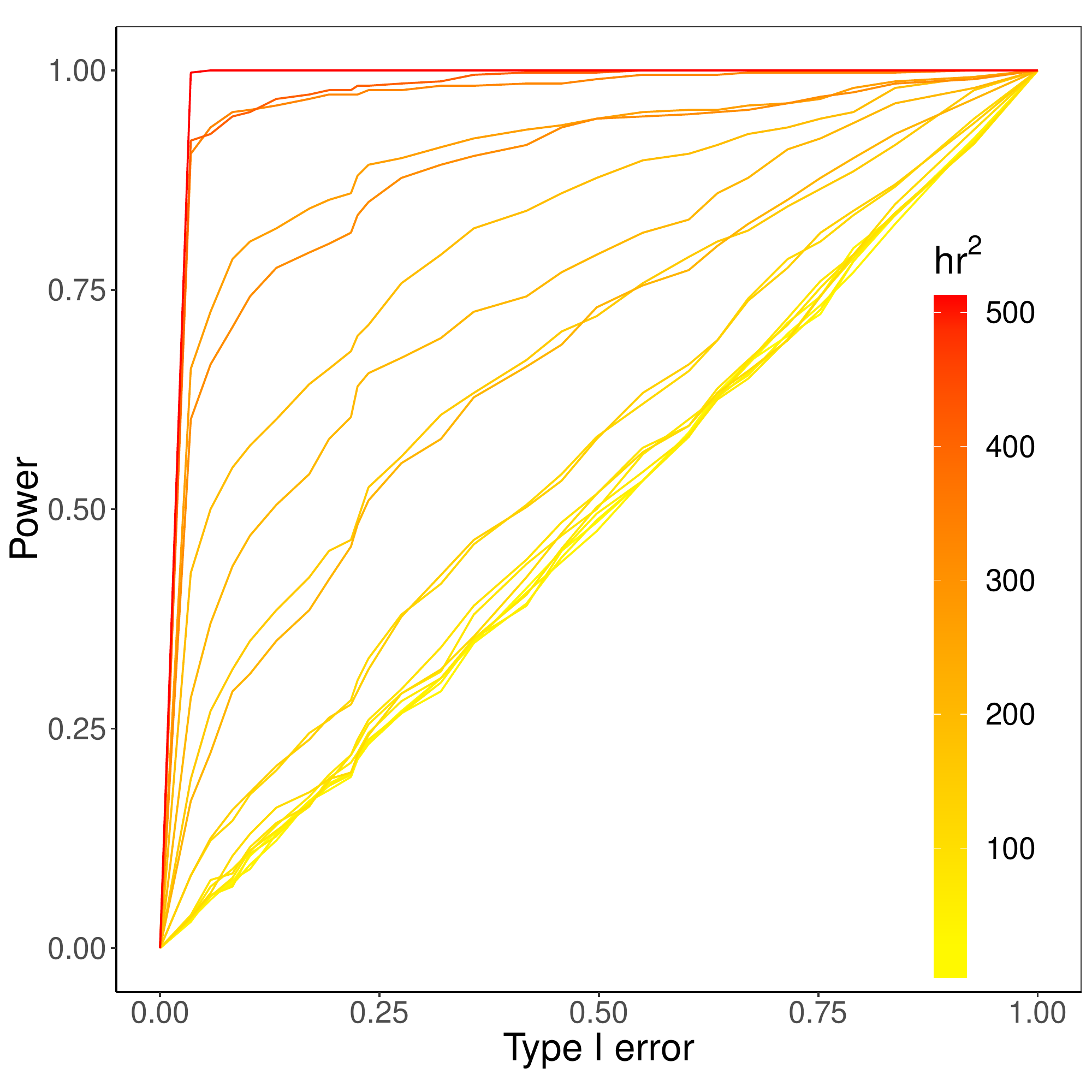} &
\includegraphics[scale=0.24,trim={0.2cm 0.2cm 0.2cm 0.25cm},clip]{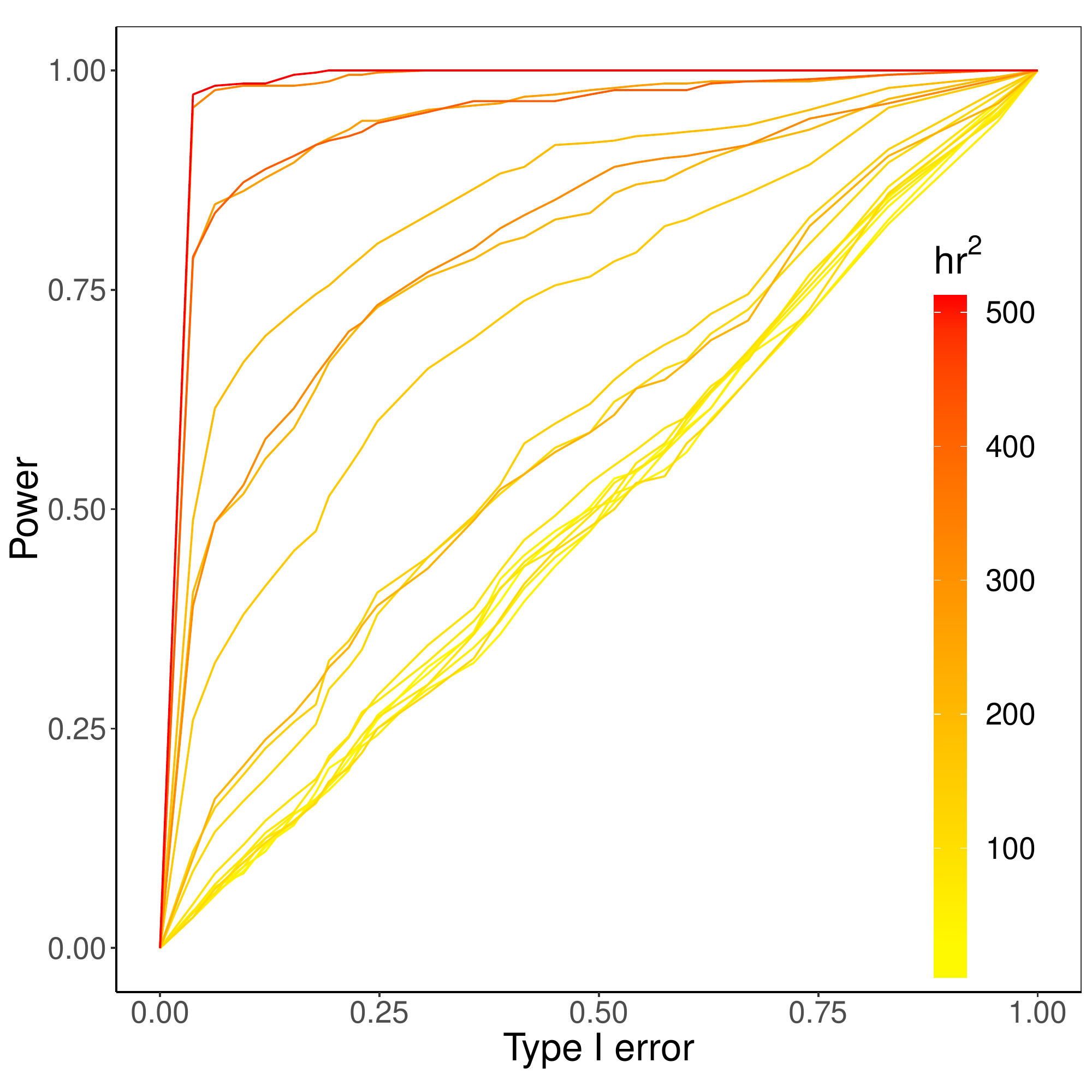} &
\includegraphics[scale=0.24,trim={0.2cm 0.2cm 0.2cm 0.25cm},clip]{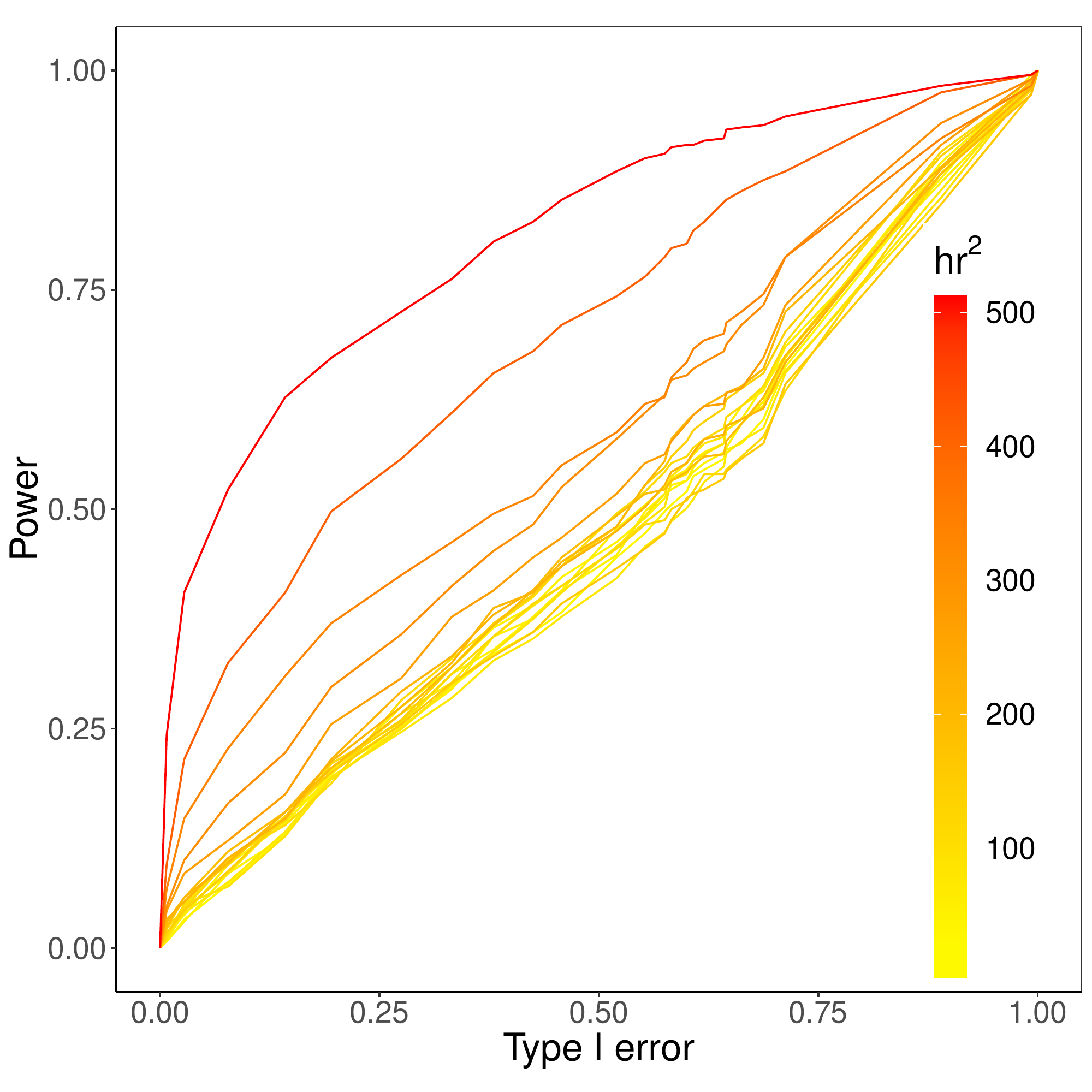} \\
\rotatebox{90}{$\quad \quad \quad  \quad  \quad \phi=10$}~
\includegraphics[scale=0.24,trim={0.2cm 0.2cm 0.2cm 0.25cm},clip]{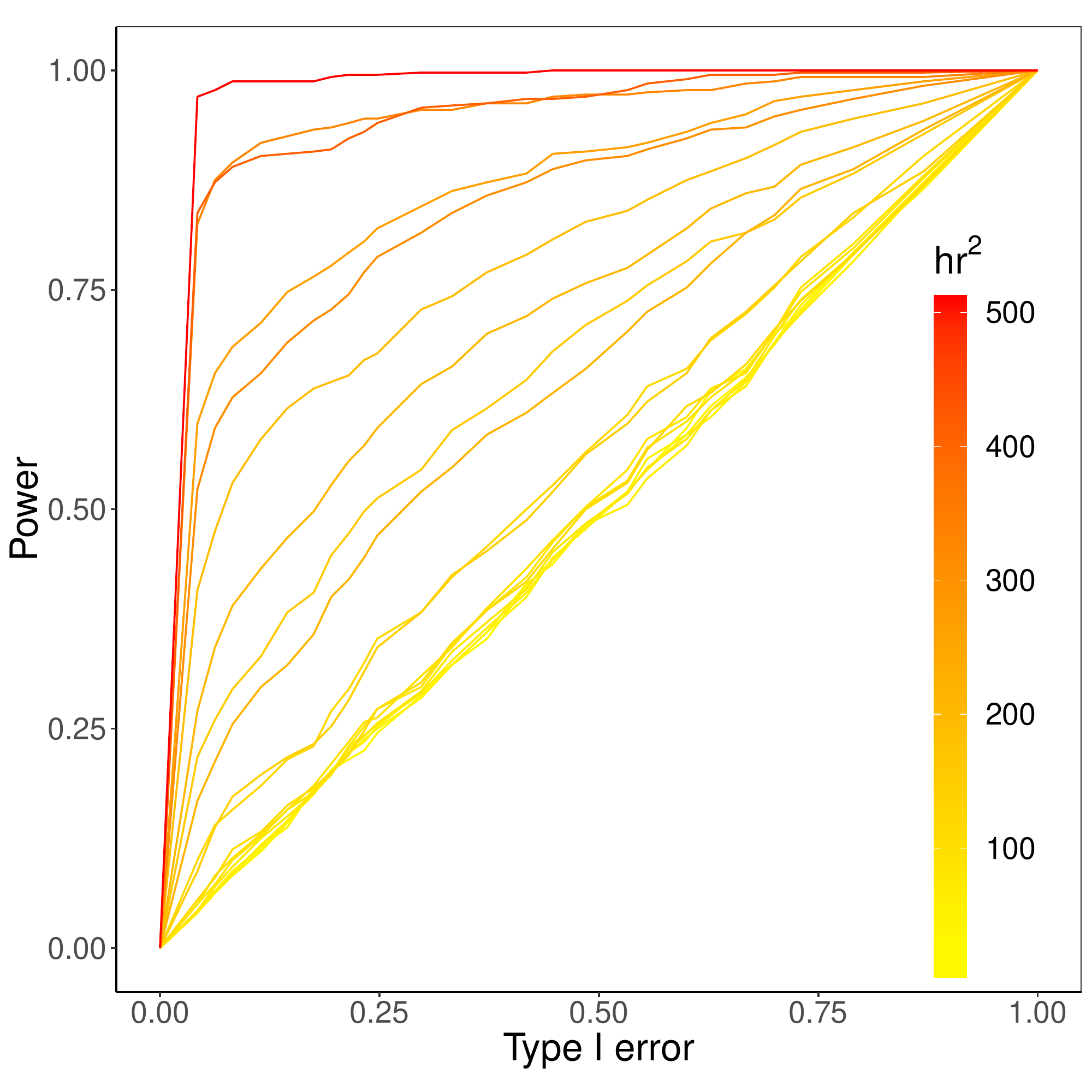} &
\includegraphics[scale=0.24,trim={0.2cm 0.2cm 0.2cm 0.25cm},clip]{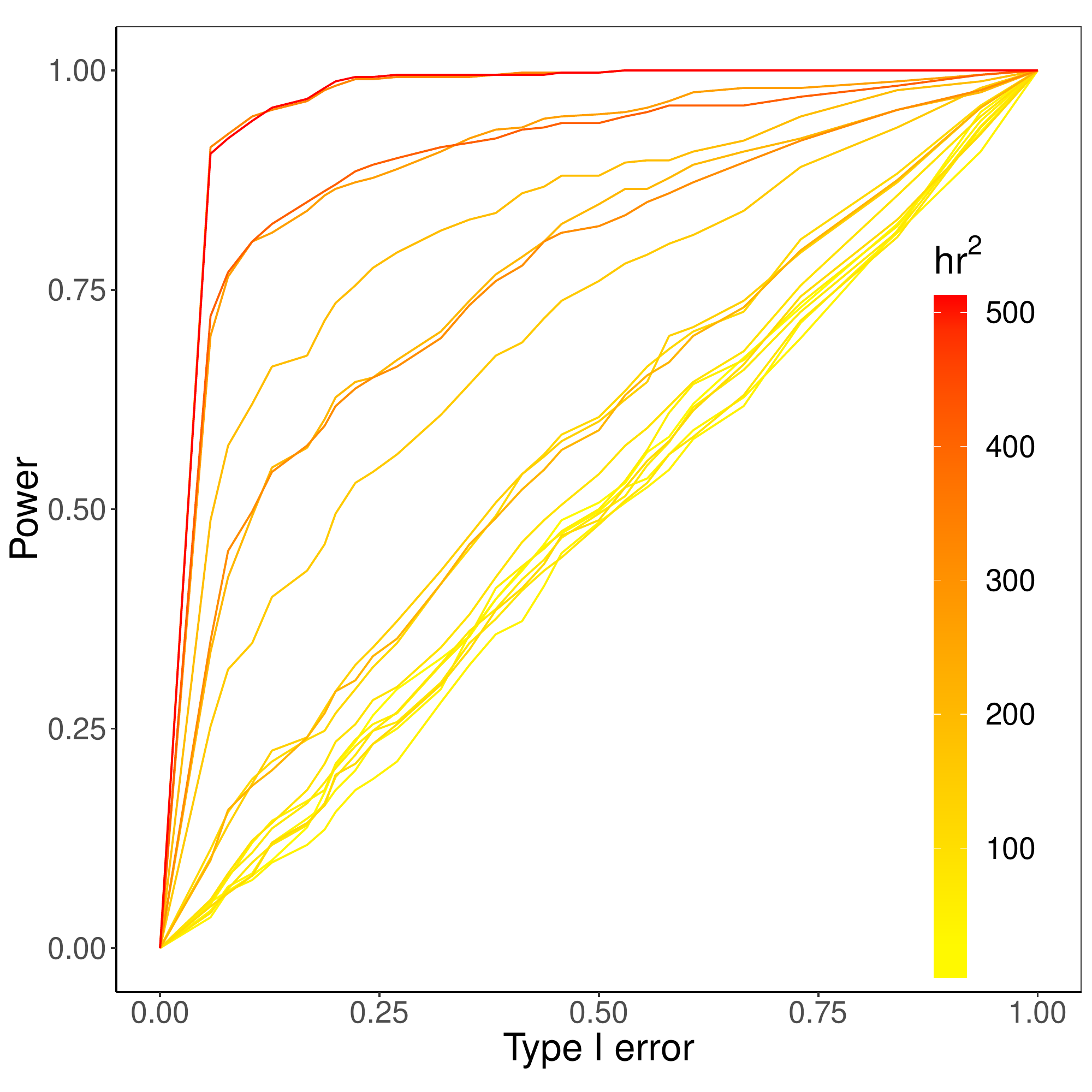} &
\includegraphics[scale=0.24,trim={0.2cm 0.2cm 0.2cm 0.25cm},clip]{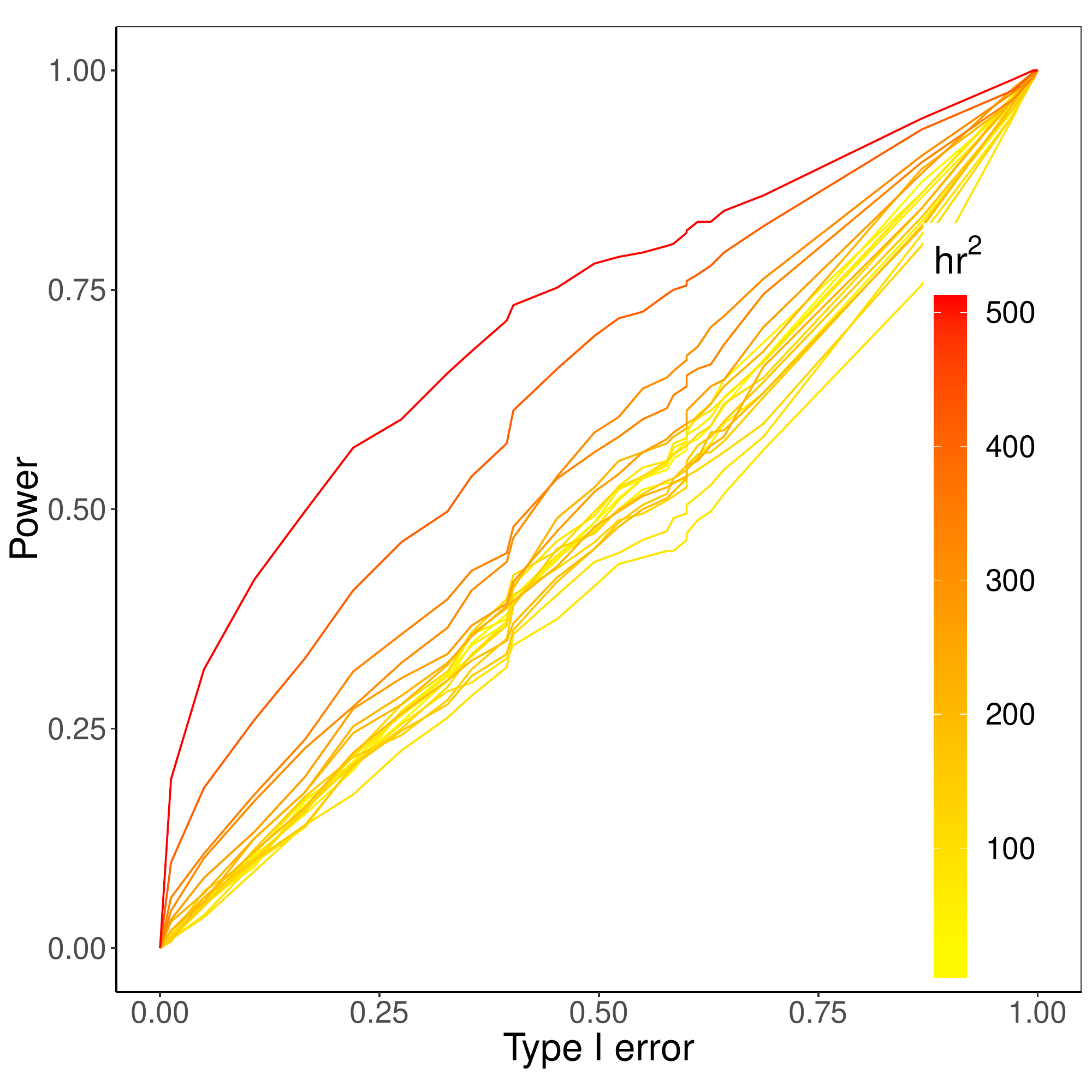} 
\end{tabular}
\caption{Empirical ROC curves for CPL ($64\times 64$), CPL ($16\times 16$), and CPL ($8\times 8$) in Experiment 3
in Section \ref{sec:simulation}.
colored according to the volume $h r^2$ of the signal obtained from 24 combinations of $r\in\{4,6,8,10\}$ and $h\in\{0,1,2,3,4,5\}$ in each plot.
Down the rows, the curves correspond to different spatial-dependence values $\phi$.
Each curve was obtained by varying $\alpha$ on the right-hand side of \eqref{eq:test}.}
\label{fig:ROC for experiment 3-2}
\end{figure}

\end{document}